\documentclass[a4paper, 12pt]{report}

\usepackage[textwidth = 430 pt, textheight = 630 pt]{geometry}
\usepackage{epigraph}

\usepackage{amsfonts}
\usepackage{latexsym}
\usepackage{slashed}
\usepackage{upgreek}
\usepackage{mathrsfs}
\usepackage{bbm}
\usepackage{relsize}
\usepackage{comment}

\usepackage{dsfont}
\usepackage{xfrac} 
\usepackage{amsmath}
\usepackage{amssymb}
\usepackage{xcolor} 
\usepackage{epigraph}
\usepackage[applemac]{inputenc}
\usepackage{graphicx}
\usepackage{comment}
\usepackage{setspace}
\usepackage[toc,page]{appendix}

\definecolor{MyDarkBlue}{rgb}{0.15,0.15,0.45}
\usepackage[linktocpage=true]{hyperref}  
\hypersetup{
colorlinks=true,
citecolor=MyDarkBlue,
linkcolor=MyDarkBlue,
urlcolor=MyDarkBlue,
pdfauthor={Sebastian Lautz },
pdftitle={ }, 
pdfsubject={hep-th}
} 

\usepackage{cite}

\numberwithin{equation}{section}

\usepackage[english]{babel}
\usepackage[Bjornstrup]{fncychap}
\usepackage{fancyhdr}
\usepackage[small]{caption}
\usepackage{pifont} 
\usepackage{multirow}
\usepackage{booktabs}

\usepackage{tikz}
\usetikzlibrary{automata,
                fit,
                positioning,
                quotes,
                shapes.multipart
                }

\usepackage{amsthm}
\theoremstyle{named}
\newtheorem*{namedtheorem}{Theorem}

\usepackage{bm}
\usepackage{mathtools}

\def\bbe{{\bf{e}}}
\def\bbl{{\bf{\ell}}}
\font\mybb=msbm10 at 11pt

\def\bb#1{\hbox{\mybb#1}}

\def\bZ {\bb{Z}}
\def\bR {\bb{R}}

\def\bC {\bb{C}}

\def\gX{\Gamma\mkern-4.0mu X}
\def\gom{\Gamma\mkern-4.0mu \omega}
\def\gY{\Gamma\mkern-2.0mu Y}
\def\gF{\Gamma\mkern-4.0mu F}
\def\gQ{\Gamma\mkern-4.0mu Q}
\def\gH{\Gamma\mkern-4.0mu H}

\def\sX{\slashed {X}}
\def\sgX{\slashed {\gX}}
\def\sY{\slashed {Y}}
\def\sgY{\slashed {\gY}}

\def\sF{\slashed {F}}
\def\sgF{\slashed {\gF}}
\def\sQ{\slashed {Q}}
\def\sgQ{\slashed {\gQ}}

\def\sH{\slashed{H}}
\def\sgH{\slashed{\gH}}

\newcommand{\bea}{\begin{eqnarray}}
\newcommand{\eea}{\end{eqnarray}}

\newcommand{\tI}{\text{\tiny $I$}}

\newcommand{\tA}{\text{\tiny $A$}}

\newcommand*{\defeq}{\mathrel{\vcenter{\baselineskip0.5ex \lineskiplimit0pt
                     \hbox{\scriptsize.}\hbox{\scriptsize.}}}%
                     =}
                  
\begin{document}

\begin{titlepage}

\begin{center}

\vspace*{1cm}

\includegraphics[scale=.9]{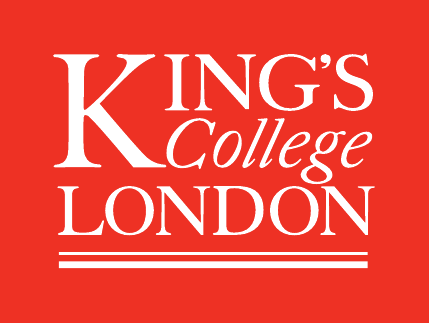}
\vspace{2cm}
 
\Huge
\textbf{Geometrical Aspects of AdS/CFT}

\vspace{1.5cm}

\Large
\textbf{Wolfgang Sebastian Lautz}


\vspace{3cm}

\normalsize
A thesis submitted in partial fulfilment \\ \vspace{0.5cm} of the requirements for the degree of \\ \vspace{0.5cm} Doctor of Philosophy

\vspace{1cm}
in the \\
\textit{Department of Mathematics} \\
\textit{King's College London} \\
\vspace{1cm}

\large
September 2019

\end{center}
\end{titlepage}

\pagenumbering{arabic}

\abstract

\setcounter{page}{2}

\noindent In this thesis, we investigate all warped AdS$_4$ and AdS$_3$ backgrounds with the most general allowed fluxes that preserve more than 16 supersymmetries in 10- and 11-dimensional supergravities. Assuming either that the internal manifold is compact without boundary or that the isometry algebra of the background decomposes into that of AdS and that of the transverse space, we find that there are no AdS$_4$ backgrounds in IIB supergravity. Similarly, we find a unique such background with 24 supersymmetries in IIA supergravity, locally isometric to $AdS_4\times \mathbb{CP}^3$. In 11-dimensional supergravity all more than half BPS AdS backgrounds are shown to be locally isometric to the maximally supersymmetric $AdS_4\times S^7$  solution. Furthermore, we prove a non-existence theorem for AdS$_3$ solutions preserving more than 16 supersymmetries.  
Finally, we demonstrate that warped Minkowski space backgrounds of the form $\mathbb{R}^{n-1,1}\times_w M^{D-n}$ ($n\geq 3,D=10,11$) in 11-dimensional and type II supergravities preserving strictly more than 16 supersymmetries and with fields, which may not be smooth everywhere, are locally isometric to the Minkowski vacuum $\mathbb{R}^{D-1,1}$. In particular, all such flux compactification vacua of these theories have the same local geometry as the maximally supersymmetric vacuum $\mathbb{R}^{n-1,1}\times T^{d-n}$.

\newpage

\chapter*{Acknowledgements}

\epigraph{\textit{``It was the best of times, it was the worst of times, it was the age of wisdom, it was the age of foolishness, it was the season of Light, it was the season of Darkness, it was the spring of hope, it was the winter of despair, we had everything before us, we had nothing before us, we were all going direct to Heaven, we were all going direct the other way''}}{Charles Dickens, A Tale of Two Cities}

First and foremost, I would like to thank my supervisor George Papadopoulos for his continued guidance and support throughout these four years of my PhD. His patience, understanding, and profound knowledge helped me grow both as a person and a researcher. This thesis would not have been possible without him!

On this note, I would like to thank my collaborator Alexander Haupt for helpful discussions and insights. I would also like to extend many thanks to the academic and administrative staff in the Department of Mathematics for being extraordinarily welcoming and helpful every step of the way.

Further, I want to thank my fellow PhD students. The above Dickens quote is an apt summary of our shared plight, and it has been a privilege to procrastinate with them. A special mention goes to Manya Sahni for being the most kind, warm-hearted and genuine sister in arms, as well as my younger academic brother Jake Phillips for giving me invaluable feedback on early drafts of the manuscript.

Now is the time to thank all the people who had to endure seemingly endless monologues about string theory and supergravity: my friends and (chosen) family (at least those I have not already mentioned). I wish to thank by name Louise Anderson for being a true friend and colleague, Lucy Dove for eating my pizza despite her severe lactose intolerance, Magdalena Gebhart for being my younger somewhat high-strung sister, Colin Gorrie for being a true mensch and fellow ``p.j.'', Sebastian Hoof for jointly surviving Kennington Lane, Frances and Stephen Lipman for all their help and support, Tome Perrin for counteracting four years of sedentary lifestyle, and Maria Stillfried for being my younger somewhat calmer sister. I would like to thank Markus Dollmann for being the best friend anyone could wish for, he is a veritable (self-declared) bantersaurus rex I can always count on, and for that I am eternally grateful.  

I could not have completed this work without the unfaltering emotional support and love of Natasha Lipman. I thank her for being my muse, ``voluntary'' proofreader, despite her inexplicable dislike of Fraktur, and for always calming me down when it all seemed overwhelming. 

Finally, I would like to dedicate this thesis to my mother and late father, I owe them everything. My Maman encouraged me to follow my own interests from a young age, and I am grateful for knowing I can always count on her unwavering support and love. My Papschi is the person who first fostered my passion for physics and mathematics. I am sure he would have been delighted to read this thesis, and probably not understand a word of it. 

This work was supported by an NMS Faculty Studentship in the Department of Mathematics.

\tableofcontents
\newpage

\listoffigures
\begingroup
\let\clearpage\relax
\listoftables
\endgroup


\chapter*{Introduction}
\phantomsection
\addcontentsline{toc}{chapter}{Introduction}
\fancyhead[L]{\slshape Introduction}

The two major breakthroughs that unequivocally shaped most of theoretical physics in the past century were general relativity and quantum field theory. Quantum field theory is the essential framework for describing our Universe at the very smallest of scales in the form of the Standard Model. The Standard Model, in particular, has celebrated unparalleled success in withstanding ever more precise experimental tests time after time. General relativity, on the other hand, is critical to our current understanding of the large-scale behaviour of the Universe. It accurately predicts corrections to Newtonian gravity, accounting for the perihelion precession of Mercury and the deflection of light by the Sun, as well as the existence of black holes and gravitational waves, of which the latter have eluded measurement a little over a hundred years after being first conjectured.  

Unification and striving to describe Nature by ever fewer fundamental principles has historically been a driving force in the development of physics. Some of the most prominent examples include Maxwell's theory of electromagnetism, Einstein's concept of spacetime, and the electroweak theory by Glashow, Salam and Weinberg. Therefore, it is of little surprise that ever since their discovery, physicists have sought to reconcile gravity with the remaining three fundamental forces captured by the Standard Model. However, what is the need for such a quantum theory of gravity beyond aesthetic aspiration? Consider a black hole for instance, this has a spacetime singularity at its centre where the curvature ``blows up'', it diverges. Now, this poses a problem, since a black hole can be formed from a perfectly smooth initial matter distribution, i.e.\ there are dynamical processes within general relativity that take us out of the regime of validity of the classical theory. As long as we do not fall into the black hole and get to feel these infinite tidal forces ourselves, we need not worry about this too much from a computational and practical perspective, since the singularity is hidden behind a horizon. Unfortunately, this becomes much more serious when both gravitational and quantum mechanical effects are important, such as in the early Universe. In four dimensions the coupling constant of gravity, i.e.\ Newton's constant, has mass dimension minus two, and therefore by a simple power counting argument, Einstein gravity is non-renomarlisable! 

To date, one of the most promising candidates to resolve this conundrum and to provide a consistent quantum theory of gravity is string theory. String theory was originally developed in the late 1960s to describe the strong interaction, but was soon succeeded by quantum chromodynamics in this purpose. However, not long after its inception it became apparent that string theory contains a massless spin two state, the gauge boson of gravity or the 	graviton. In fact, as we will see, the two basic topologies of a string - open and closed - correspond to gauge and gravitational degrees of freedom. Since open strings can close up, gravity and gauge theory are inextricably connected in string theory, thereby potentially providing a unified theory of all known fundamental forces.


The discovery of the AdS/CFT correspondence in the late 1990s \cite{maldacenab,witten,Gubser:1998bc}, which states that certain string theories on $(d+1)$-dimensional Anti-de Sitter (AdS$_{d+1}$) backgrounds are dual to $d$-dimensional conformal field theories (CFT), marks string theory finding a way back to its historical roots in the following sense. With this being a strong-weak coupling duality, string theory offers a framework for understanding strongly coupled quantum field theories which would otherwise be inaccessible to conventional perturbative methods. Initially, AdS backgrounds were studied in the context of supergravity compactifications or Freund-Rubin solutions \cite{fr,duff} in the 1980s. Over the past twenty years, with the inception of the AdS/CFT correspondence, the problem of classifying AdS backgrounds has sparked renewed interest, cf.\ \cite{Gran:2018ijr} for a recent review and references therein. Indeed, this together with the fact that the best understood examples of the correspondence are highly supersymmetric, e.g. \cite{maldacenab,abjm}, constitutes the main motivation for the work described in this thesis.

This thesis is arranged as follows: In chapter \ref{introAdSSUGRA}, we aim to give a lightning review of strings, supergravity and AdS/CFT to put our main results into context. Furthermore, we introduce some of the key developments that enabled the analysis in this thesis, e.g. the integration of the Killing spinor equations along AdS, the homogeneity theorem, as well as the classification of Killing superalgebras. In chapter \ref{ads4}, we present our findings on the classification of more than half maximally supersymmetric warped AdS$_4$ backgrounds in 11-dimensional, (massive) type IIA and type IIB supergravity which were published in \cite{ads4Ngr16}. In chapter \ref{ads3}, we present our proof of a non-existence theorem for more than half BPS warped AdS$_3$ backgrounds in 10- and 11-dimensional supergravities, based on work published in \cite{ahslgp2}. In chapter \ref{minkngr16}, we demonstrate that all more than half maximally supersymmetric warped Minkowski backgrounds in 10 and 11 dimensions are necessarily locally isometric to the maximally supersymmetric flat vacuum. This work was published in \cite{Lautz:2018qcb}. Finally, in chapter \ref{concluions}, we provide a conclusion to this thesis.

\fancyhead[L]{\slshape\nouppercase{\leftmark}}
\chapter{AdS/CFT and Supergravity Backgrounds}\label{introAdSSUGRA}
In this chapter, we would like to put the main results of this work in chapters \ref{ads4}, \ref{ads3} and \ref{minkngr16} into a wider context, and along the way introduce some of the tools and techniques which facilitated our analysis. We proceed by giving a short introduction to string theory, and consequently supergravity as its low energy limit, thereby turning the historical development of the subject on its head. By virtue of space(time) limitations our outline has to be selective, omitting many details and important contributions, however we aim to give historical references for some of the milestones in the field. A comprehensive and much more complete review of the fundamentals of strings, branes and supergravity may be found in \cite{Green:2012oqa,Green:2012pqa,Polchinski:1998rq,Polchinski:1998rr,Becker:2007zj,Freedman:2012zz,Blumenhagen:2013fgp}, which heavily influenced the following section.    

\section{From Superstrings to Supergravity}

One of the two most successful developments of 20th century physics, quantum field theory (QFT), is fundamentally preoccupied with studying excitations of fields that may be interpreted as point-like particles. String theory, as its very name suggests, takes this one step further and declares one-dimensional extended objects called strings to be the fundamental entities of matter. Incidentally, and somewhat enticingly, this approach also yields a consistent theory of quantum gravity, thereby marrying QFT with the second great achievement of 20th century physics, General Relativity (GR).

However, from a historical point of view this was an accident. String theory was originally conceived in the late 1960s by Veneziano, who constructed an amplitude satisfying the then suspected ``duality'' symmetry under the exchange of $s$- and $t$-channel diagrams in hadronic scattering processes \cite{Veneziano:1968yb}. Unfortunately, this turned out not to be a particularly fruitful approach for the study of the strong interaction, and with the discovery of quantum chromodynamics soon became obsolete\footnote{Arguably, with the conception of the AdS/CFT correspondence, string theory has come full circle by being used as a tool to understand strongly coupled gauge theories.}. Nonetheless, Veneziano's amplitude was not forgotten, and would be reinterpreted as the scattering of relativistic bosonic strings by Nambu \cite{Nambu:1997wf}, Nielsen \cite{nielsenalmost} and Susskind \cite{Susskind:1970xm}. Subsequently, excitement grew as people realised the spectrum of the theory contained a massless spin 2 state corresponding to the graviton, the gauge boson of gravity \cite{Yoneya:1974jg,Scherk:1974mc}. Nevertheless, some obvious problems remained, namely the theory did not include fermioninc matter, and the vacuum featured a tachyonic state, signalling instability. Enter supersymmetry, a symmetry relating bosonic degrees of freedom to fermionic ones, which solves both of these problems at once. This is where we shall begin.

\subsection{String Theory}\label{stringtheory}
A string sweeps out a $(1+1)$-dimensional worldsheet $\Sigma$ in spacetime, which we assume to be $D$-dimensional Minkowski space with metric $\eta_{MN}$ for now, generalisation to curved backgrounds is achieved straightforwardly via minimal coupling to $g_{MN}$. The embedding of the string into the target space is given by the functions $X^M(\tau,\sigma)$, where $\tau$ corresponds to proper time and $\sigma$ is the spatial coordinate of the string, and is usually fixed to take values in $[0,2\pi]$ for closed and $[0,\pi]$ for open strings. Just like the action of a point particle is given by the length of its worldline, one takes the action of the string to be proportional to its worldsheet, thus manifestly ensuring diffeomorphism invariance
\begin{align}
    S_{NG} = -\frac{1}{2\pi\alpha'} \int\limits_{\Sigma} d^2\xi \sqrt{-\det\left(\partial_{\alpha} X^M \partial_{\beta} X^N \eta_{MN}\right)}~,
\end{align}
with $(\xi^0,\xi^1) = (\tau,\sigma)$ and the constant $\alpha'$ refers to the inherent length scale of the problem, the string length $l_s^2 = \alpha'$. This is the so-called Nambu-Goto action \cite{nambu:1970ngaction,Goto:1971ce,Hara:1971ur}. However, the square root makes it rather hard to quantise, so one usually considers an alternative form, the Polyakov action\footnote{The so-called Polyakov action was actually discovered independently by Brink, DiVecchia and Howe, and by Deser and Zumino. Polyakov later used it in the context of the path integral quantisation of the bosonic string.} \cite{Deser:1976rb,Brink:1976sc,Polyakov:1981rd}, where one introduces a worldsheet metric $h_{\alpha\beta}$ as an auxiliary field
\begin{align}
    S_{P} = -\frac{1}{4\pi\alpha'} \int\limits_{\Sigma} d^2\xi \sqrt{-h} \, h^{\alpha\beta} \partial_{\alpha} X^M \partial_{\beta} X^N \eta_{MN}~.
\end{align}
The Polyakov action can be viewed as a collection of $D$ worldsheet scalars coupled to two dimensional gravity. Since gravity in two dimensions is trivial, and by our construction $h_{\alpha\beta}$ is a Lagrange multiplier, it is not surprising that the associated energy momentum tensor, i.e. the equations of motion for $h_{\alpha\beta}$, vanish identically. This gives rise to the so called Virasoro constraints. In fact, one may integrate out $h_{\alpha\beta}$ this way to recover the Nambu-Goto action at the classical level. Turning to the symmetries of the Polyakov action, we observe that it is invariant under $D$ dimensional Poincar\'e transformations, diffeomorphisms of the worldsheet, and Weyl transformations $h_{\alpha\beta} \rightarrow e^{2\omega(\xi)} h_{\alpha\beta}$, i.e. local rescalings. In that sense, the bosonic string may be viewed as a gauge theory of the two dimensional conformal group. We may use diffeomorphism invariance to fix the gauge to the so-called conformal gauge
\begin{align}
    h_{\alpha\beta} = e^{2\omega(\xi)} \, \eta_{\alpha\beta}~.
\end{align}
With the gauge fixed in this manner and introducing a collection of $D$ worldsheet Majorana-Weyl spinors $\Psi^M= (\psi^M_-,\psi^M_+)^T$, the supersymmetrised Polyakov action \cite{Brink:1976sc,Ramond:1971gb,Neveu:1971rx} reads
\begin{align}\label{susyPoly}
    S&=-\frac{1}{4\pi\alpha'} \int d^2\xi \, \eta^{\alpha\beta} \left( \partial_{\alpha} X^M \partial_\beta X^N + \Bar{\Psi}^M \rho_{\alpha} \partial_\beta \Psi^N \right)\eta_{MN} \notag\\
    &= \frac{1}{2\pi \alpha'} \int d^2\xi \, \left(2\partial_+ X^M \partial_- X_M +  i\psi^M_+\partial_- \psi_{+,M} + i\psi^M_-\partial_+ \psi_{-,M}\right)~,
\end{align}
where $\xi^\pm = \tau \pm \sigma$, $\partial_\pm = \frac{1}{2}(\partial_\tau \pm \partial_\sigma)$, and we have chosen the explicit representation
\begin{align}
    \rho^0 = \begin{pmatrix*}[r] 0 & -1 \\ 1 & 0 \end{pmatrix*} = -i\sigma_2~,\quad \rho^1 = \begin{pmatrix*}[r] 0 & 1 \\ 1 & 0 \end{pmatrix*} = \sigma_1
\end{align}
of the two dimensional Clifford algebra. The signs appearing in the subscript of the components of $\Psi^M$ refer to the eigenvalue with respect to the chirality matrix $\rho_* = \rho^0\rho^1$. Note that the supersymmetry variations with parameter $\epsilon$ are given by
\begin{align}
	\delta_\epsilon X^M= \bar{\epsilon} \Psi^M~, \quad \delta_\epsilon \Psi^M = \rho^\alpha \partial_\alpha X^M \epsilon ~.
\end{align}
Varying this action, we find for the equations of motion
\begin{align}
	\partial_+\partial_- X^M = 0~, \quad \partial_+ \psi_-^M = 0 ~, \quad \partial_-\psi^M_+ = 0~,
\end{align}
i.e. the bosonic degrees of freedom satisfy a free wave equation and the fermions the free Dirac equation in two dimensions. As always in a free QFT, we could now diagonalise the Laplacian and promote the modes to creation/annihilation operators, giving rise to a collection of bosonic and fermionic harmonic oscillators. However, before acting prematurely, we need to think carefully about possible boundary conditions for the string, which we have glossed over in deriving the field equations so far. In fact, when varying the action by $\delta X^M$ and $\delta\Psi^M$, one obtains the following boundary terms from integration by parts
\begin{align}\label{bcstring}
	\delta S &= -\frac{1}{2\pi\alpha'} \int d\tau \, \left(\partial_\sigma X^M \delta X_M \right) \Big\vert_{0}^{\pi,2\pi} + \frac{i}{4\pi\alpha'} \int d\tau \, (\psi_+^M \, \delta\psi_{+,M} \, - \notag\\
	&- \psi_-^M \, \delta\psi_{-,M} )\Big\vert_{0}^{\pi,2\pi}~,
\end{align}
where the limits $\pi$ and $2\pi$ refer to open and closed strings, respectively. Focusing on the bosonic boundary term for now, we note that for a closed string, we have to pick periodic boundary conditions $X^M(\tau,\sigma)=X^M(\tau,\sigma+2\pi)$ anyway and so this term vanishes without further ado. In the open string case on the other hand, we have the choice between von Neumann ($\partial_\sigma X^M|_{0 \text{ and/or }\pi} = 0$) or Dirichlet ($\delta X^M|_{0\text{ and/or }\pi}=0$) boundary conditions independently for each endpoint and direction. The former imply that the ends move freely and no momentum flows off the string, whereas Dirichlet boundary conditions imply that the corresponding end of the string is fixed to a hypersurface that can exchange momentum with the open string. These dynamical and, as it turns out, non-perturbative objects are called D-branes. 

As for the fermionic term in \eqref{bcstring}, in the open string sector the terms at $\sigma=0$ and $\sigma=\pi$ have to vanish separately, which relates the components of the Majorana-Weyl fermion as
\begin{align}
	\psi_+(\tau,\sigma)|_{0 \text{ or } \pi} = \pm \psi_-(\tau,\sigma)|_{0 \text{ or } \pi}~.
\end{align} 
By convention the overall sign is chosen such that $\psi_+(\tau,0)=\psi_-(\tau,0)$, which leaves us with the following possbilities
\begin{align}
	\psi_+(\tau,\pi)=\psi_-(\tau,\pi)~,
\end{align}
the Ramond (R) sector with integer modes\footnote{This is only true for NN or DD boundary conditons, otherwise there is a sign flip at one of the boundaries and the modes are half integer} or
\begin{align}
	\psi_+(\tau,\pi)=-\psi_-(\tau,\pi)~,
\end{align}
the Neveu-Schwarz (NS) sector with half-integer modes\footnote{Again this only applies for NN or DD conditions, for ND or DN the modes are integer.}.

One then proceeds to promote both the bosonic and fermionic modes to operators and imposes (anti-)commutation relations. As usual, the Hilbert space is built up by acting with creation operators on a vacuum, which is defined as the state being annihilated by all annihilation operators. Furthermore, one needs to impose physical state conditions similar to the Gupta-Bleuler quantisation procedure in quantum electrodynamics in the ``weak'' sense, i.e. in expectation values. More concretely, these are the (super-)Virasoro constraints, arising classically as the modes of the worldsheet energy momentum tensor and supercurrent. The critical dimension in superstring theory $D=10$ and a possible normal ordering constant $a$ as well as the aforementioned physical state conditions can be obtained in several different ways, be it the absence of negative norm states in covariant quantisation \cite{Brower:1973iz,Goddard:1972iy}, or enforcing target space Lorentz invariance in light-cone quantisation \cite{Goddard:1973qh}. A particularly elegant approach is the so-called BRST quantisation, in which physical states are identified with cohomology classes of a nilpotent operator $Q_{BRST}$, related to the gauge symmetry of the underlying theory (in our case superconformal invariance on the worldsheet). It can be shown that $Q_{BRST}$ is nilpotent if and only if the superconformal anomaly (central term in super-Virasoro algebra) is cancelled, which depends on the target space dimension \cite{Kato:1982im,Hwang:1982mc,freemanBRST}.

 Let us turn our attention to the spectrum in the open string case. In the NS-sector, one finds that the ground state is once again tachyonic as for the original bosonic string, and the first excited state is a massless ten dimensional Lorentz vector i.e. it transforms under the little group $SO(8)$ in the fundamental representation $\mathbf{8_V}$. In the R-sector on the other hand, the vacuum is massless but highly degenerate in return, since the oscillators corresponding to the zero modes take one ground state into another. They form a representation of the $(9+1)$-dimensional Clifford algebra, and so the ground state is a spacetime spinor with spin $1/2$. Under $SO(8)$ or its cover $Spin(8)$ this 16-dimensional Majorana-Weyl spinor can be reorganised in the two  8-dimensional irreducible spinor representations of opposite chirality $\mathbf{8}$ and $\mathbf{8'}$. Now, in order to get rid of the tachyon and to obtain a spacetime supersymmetric theory, we define the $G$-parity operator as
 \begin{align}
 	G &:= (-1)^{F+1} \quad \text{(NS)}~, \notag\\
 	G &:= \Gamma_{11}(-1)^{F} \quad \text{(R)}~,
 \end{align} 
where $\Gamma_{11}$ is the highest rank Clifford algebra element in 10 dimensions and $F$ is the worldsheet fermion number operator. One can perform a consistent truncation of the spectrum, known as GSO projection after Gliozzi, Scherk and Olive \cite{Gliozzi:1976qd}, by only keeping states with positive $G$-parity, thereby discarding the tachyon in the NS-sector and the $\mathbf{8'}$ in the R-sector. Thus we are left with a $D=10,~\mathcal{N}=1$ gauge multiplet at the massless level.

In the closed string sector we have four different choices of boundary conditions, since we can impose these separately on left and right moving modes. Therefore, we have to consider NS-NS, R-R, NS-R and R-NS sectors, with the former two giving rise to spacetime bosons, whereas the mixed sectors are spacetime fermions. Again, one can perform a GSO projection, which gives rise to only four consistent closed superstring theories - quite a remarkable cull of the a priori thousands of possibilities. Only two of these are physically distinct, the so called type IIA and type IIB theories which are given by
\begin{center}
\begin{tabular}{ccccc}
    Type IIA: & (NS$_+$, NS$_+$) & (R$_+$, R$_-$) & (R$_+$, NS$_+$) & (NS$_+$, R$_-$)  \\
    Type IIB: & (NS$_+$, NS$_+$) & (R$_+$, R$_+$) & (R$_+$, NS$_+$) & (NS$_+$, R$_+$),
\end{tabular}
\end{center}
where the subscripts refer to $G$-parity. We find the bosonic field content by taking tensor products of two of the three eight dimensional representations of $SO(8)$ corresponding to the above sectors, and decomposing them into irreducible representations of the little group again (cf. table \ref{table:closedSTspectrum}). A similar procedure in the mixed R-NS and NS-R sectors yields dilatinos and gravitinos, i.e.\ the superpartners of dilaton and graviton. 

\begin{table}
    \renewcommand{\arraystretch}{1.3}
    \centering
    \caption[Decomposition of massless closed string states into irreducible representations of $SO(8)$]{Decomposition of massless closed string states into irreducible representations of $SO(8)$. Here, $\phi$ denotes the dilaton, $C_{(0)}$ the IIB axion, $B_{(2)}$ the Kalb-Ramond field, $g_{MN}$ the spacetime metric. Other R-R-potentials are collectively denoted by $C_{(p)}$, where $p$ is the degree of the form. The subscript ``$+$'' indicates self-duality.}
    \begin{tabular}{ll}
        \toprule
        (NS$_+$, NS$_+$) & $\mathbf{8_V} \otimes \mathbf{8_V} = \underbrace{\mathbf{1}}_{\phi} \oplus \underbrace{\mathbf{28}}_{B_{(2)}} \oplus \underbrace{\mathbf{35}}_{g_{MN}} $\\
        (R$_+$, R$_-$) & $\mathbf{8} \otimes \mathbf{8'} = \underbrace{\mathbf{8_V}}_{C_{(1)}} \oplus \underbrace{\mathbf{56}}_{C_{(3)}}$   \\
        (R$_+$, R$_+$) &  $\mathbf{8} \otimes \mathbf{8} = \underbrace{\mathbf{1}}_{C_{(0)}} \oplus \underbrace{\mathbf{28}}_{C_{(2)}} \oplus \underbrace{\mathbf{35_+}}_{C_{(4)}}$    \\
        \bottomrule
    \end{tabular}

    \label{table:closedSTspectrum}
\end{table}

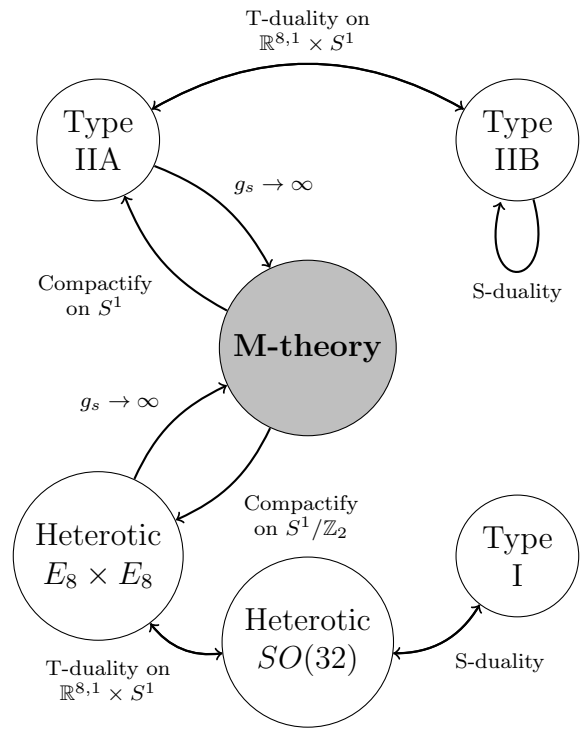
\begin{figure}
	\begin{center}
	\begin{tikzpicture}[
node distance = 3.9cm,
    MP/.style = {circle split, draw, inner sep=1pt,
                 align=center, font=\scriptsize},
state/.append style = {fill=#1, align=center},
state/.default = white,
every edge quotes/.append style={font=\scriptsize, align=center, auto}
    ]
  \tikzstyle{every state}=[fill=white,shape=circle,text=black]

  \node (M)     [state=gray!50]                 {\textbf{M-theory}};
  \node (IIA)     [state,above left of=M]        {Type\\IIA};
  \node (IIB)     [state,above right of=M]       {Type\\IIB};
  \node (hete8)     [state, below left of=M]      {Heterotic\\ $E_8\times E_8$};
  \node(hetso32)     [state,below of=M]    {Heterotic\\ $SO(32)$};
  \node (I)         [state, below right of=M]    {Type \\ I};

  \path[thick, ->]  
        (IIA)    edge [bend left,"T-duality on \\ $\mathbb{R}^{8,1}\times S^1$"]  (IIB)
        (IIB)   edge [bend right]                (IIA)
        (IIB) edge [loop below,"S-duality"]  (IIB)
        (hete8)    edge  [bend right]                  (hetso32)
        (hetso32)   edge [bend left,"T-duality on \\ $\mathbb{R}^{8,1}\times S^1$"]  (hete8)
        (IIA)  edge[bend left=20, "$g_s\rightarrow\infty$"] (M)
        (M) edge[bend left=20, "Compactify \\ on $S^1$"] (IIA)
        (hete8)  edge[bend left=20, "$g_s\rightarrow\infty$"] (M)
        (M) edge[bend left=20, "Compactify \\ on $S^1/\mathbb{Z}_2$"] (hete8)
        (hetso32) edge  [bend right] (I)
        (I) edge   [bend left,"S-duality"] (hetso32)
        
        ;
  
	\end{tikzpicture}
	\caption{Web of string dualities}
	\label{figure:strduals}
	\end{center}
\end{figure}

There are three more consistent superstring theories in 10 dimensions, type I which is made up of both open and closed unoriented strings with gauge group $SO(32)$ and the two heterotic string theories with gauge groups $SO(32)$ and $E_8 \times E_8$. The name heterotic refers to the fact that one combines the left-moving  degrees of freedom of a 10-dimensional superstring theory with the right-movers of a 26-dimensional bosonic string theory. These five different string theories are not independent though. Instead they are related to one another via a web of dualities, cf. figure \ref{figure:strduals}, and as it was realised in the mid 1990's, are all different limits of a unique 11-dimensional theory of branes containing no strings, namely M-theory \cite{Hull:1994ys,Townsend:1995kk,Witten:1995ex}. In figure \ref{figure:strduals} we mention two types of dualities, T- and S-duality. T-duality relates a string theory (e.g. IIA) compactified on a circle of radius $R$ with another theory (e.g. IIB) compactified on a circle of radius $\frac{\alpha'}{R}$, with momentum and winding modes interchanged. S-duality on the other hand is a strong/weak coupling duality. The string coupling $g_s$ governing string perturbation theory is dynamically determined by the theory itself as the vacuum expectation value of the dilaton $g_s = e^{\langle\phi\rangle}$. S-duality then states that a theory, e.g. type I, with coupling $g_s$ is equivalent to a theory with inverse coupling $1/g_{s}$, e.g. heterotic $SO(32)$.

\subsection{Supergravity as low energy effective theory} \label{EFT}

String theory ultimately aspires to provide a unified description of all fundamental interactions, including gravity. In the previous section we found promising hints from the spectrum in that direction, as it contained a massless spin two particle, a prime candidate for the graviton. However, the correct particle spectrum alone is meaningless without knowing how these couple to one another. Besides, at low energies\footnote{By low energy we mean that $\alpha'k^2 \ll 1$, where $k$ denotes the momentum of an external on-shell state.} the massive tower of string modes freezes out and only the massless modes survive. So, in the appropriate limit, we should recover the Einstein Hilbert action as well as gauge theory. How do we go about finding this low energy effective theory?

One possibility is to write down the free Lagrangian for the massless fields that are present in the spectrum, and add in interactions that reproduce the amplitudes computed in string perturbation theory. Further restrictions arise from spacetime symmetries, such as general coordinate invariance and supersymmetry, which can be inferred from the structure of the worldsheet theory, e.g.\ from conserved currents.
Alternatively, requiring worldsheet superconformal invariance\footnote{That is an anomaly free gauge symmetry on the worldsheet.}, the $\beta$-functions of the two-dimensional (non-linear) sigma model coupled to target space background fields, the massless modes, have to vanish. This yields the equations of motion and one may then write down an action that reproduces these. 

After GSO projection both type IIA and IIB contain an equal number of on-shell bosonic and fermionic degrees of freedom (128+128) at the massless level. As advertised this is a sign of spacetime supersymmetry which is made manifest in the Green Schwarz formalism \cite{Green:1983wt}, an alternative approach to the Ramond-Neveu-Schwarz formalism outlined in the previous section to obtain superstring theories. We mentioned that the mixed R-NS and NS-R sectors in both type IIA and IIB contain two gravitinos $\psi^M_I$ ($I=1,2$), i.e.\ massless spin $3/2$ fields. In the case of IIA these are of opposite chirality, whereas IIB is chiral.  Now, just like the presence of a massless vector boson implies a gauge symmetry in a consistent QFT, the occurrence of a gravitino, the superpartner of the graviton, implies that spacetime supersymmetry has to be local. Therefore, the low energy effective theories will be \emph{supergravities}. Furthermore, the presence of two independent gravitinos is linked to the existence of two independent $\mathcal{N}=2$ supersymmetry algebras\footnote{Hence the name type II.} in 10 dimensions, differing by the chirality of their two Majorana-Weyl supercharges. In fact, it turns out this is also the maximum number of supersymmetry one can have in 10 dimensions, if we demand not to exceed spin two in the supermultiplets \cite{Nahm:1977tg}. The effective action to lowest order in $\alpha'$ of the bosonic fields in type IIA \cite{Campbell:1984zc,Giani:1984wc,Huq:1983im} and IIB\footnote{In the case of IIB, this is really more of a pseudo-action, since one still has to impose self-duality of the 5-form $\tilde{F}_{(5)} = *\tilde{F}_{(5)}$ at the level of the equations of motion.} \cite{Schwarz:1983wa,Schwarz:1983qr,Howe:1983sra} (in the string frame) is of the form
\begin{align}
	S_{\text{IIA}/\text{IIB}} = S_{NS} + S_{R} + S_{CS}~,
\end{align}    
where $S_{NS}$ is the common sector of both theories given by
\begin{align}
	S_{NS} = \frac{1}{2\tilde{\kappa}_{10}^2} \int d^{10}x \, \sqrt{-g} \, e^{-2\phi} \left(R + 4 \partial_M\phi \, \partial^M\phi - \frac{1}{12} H_{MNL} \, H^{MNL} \right)~,
\end{align}
and $H_{(3)}= dB_{(2)}$ is the field strength of the Kalb-Ramond field. $\tilde{\kappa}_{10}^2$ denotes the 10-dimensional gravitational constant
\begin{align}
	2\tilde{\kappa}_{10}^2 = (2\pi)^7 \alpha'^4~.
\end{align} 
Moreover, $S_R$ contains the kinetic terms of the R-R-gauge fields and $S_{CS}$ are Chern-Simons terms, so of a topological nature. In the case of type IIA one finds
\begin{align}
	S_R &= -\frac{1}{4\tilde{\kappa}_{10}^2} \int \left(F_{(2)} \wedge *F_{(2)} + \tilde{F}_{(4)} \wedge *\tilde{F}_{(4)} \right) \notag\\
	 S_{CS} &= -\frac{1}{4\tilde{\kappa}_{10}^2} \int B_{(2)} \wedge F_{(4)} \wedge F_{(4)}~,
\end{align}
where 
\begin{align}
	F_{(2)} = dC_{(1)}~, \quad F_{(4)} = dC_{(3)}~, \quad \tilde{F}_{(4)} = F_{(4)} - C_{(1)} \wedge H_{(3)}~.
\end{align}
For IIB this reads
\begin{align}
	S_R &= -\frac{1}{4\tilde{\kappa}_{10}^2} \int \left(F_{(1)} \wedge *F_{(1)} + \tilde{F}_{(3)} \wedge *\tilde{F}_{(3)} + \tilde{F}_{(5)} \wedge *\tilde{F}_{(5)} \right) \notag\\
	 S_{CS} &= -\frac{1}{4\tilde{\kappa}_{10}^2} \int C_{(4)} \wedge H_{(3)} \wedge F_{(3)}~,
\end{align}
with
\begin{align}
	F_{(1)}&= dC_{(0)}~, \quad F_{(3)}= dC_{(2)}~, \quad F_{(5)}= dC_{(4)}~, \quad \tilde{F}_{(3)} = F_{(3)} - C_{(0)} H_{(3)}~, \notag \\
	\tilde{F}_{(5)} &= F_{(5)} - \frac{1}{2} C_{(2)} \wedge H_{(3)} + \frac{1}{2} B_{(2)} \wedge F_{(3)}~.
\end{align}
The action of type IIA supergravity may also be derived via dimensional reduction of 11-dimensional supergravity. Supergravity in 11 dimensions is unique in the sense that this is the maximal spacetime dimension in which the massless multiplets only contain particles of spin less or equal than two. The bosonic part of the action of 11-dimensional supergravity is given by \cite{Cremmer:1978km}
\begin{align}
	S_{11} = \frac{1}{2\kappa_{11}^2}\left[ \int d^{11}x \, \sqrt{-g} \,  \left(R - \frac{1}{48} F_{(4)}^2 \right) - \frac{1}{6} \int A_{(3)} \wedge F_{(4)} \wedge F_{(4)}\right]~,
\end{align}
where $A_{(3)}$ is a 3-form gauge potential with field strength $F_{(4)}= dA_{(3)}$ and $\kappa_{11}$ denotes the gravitational coupling in 11 dimensions. Just like the UV completions of type IIA and type IIB supergravity are type IIA and type IIB string theory, respectively, the strong coupling limit of 11-dimensional supergravity was conjectured to be M-theory \cite{Townsend:1995kk,Witten:1995ex,Schwarz:1995jq}.

\subsection{The two faces of branes}\label{2facesbranes}
So far we have been preoccupied with the quantisation of open and closed strings, as well as the low energy effective theory of closed strings. In section \ref{stringtheory}, we have briefly alluded to the fact that D-branes are not merely hyperplanes on which open strings can end, but are dynamical objects in their own right \cite{Dai:1989ua,Leigh:1989jq,Polchinski:1995mt}. In fact, they gravitate by coupling to closed strings in the NS-NS sector, that is they have a mass and hence deform spacetime. On the other hand, they are naturally charged under R-R form potentials, and one finds a gauge theory living on the worldvolume.

The derivation of the low energy effective action of open strings proceeds along similar lines as the one of closed strings in section \ref{EFT}, except that now boundary conditions specified by D-branes have to be taken into account. The endpoints of a string are charged, and so couple to a gauge field $A_a$ living on the D-brane. Again one has to impose constraints, namely the worldsheet energy momentum tensor has to vanish. One obtains restrictions on the equations of motion, and thus the form of the gauge field. These may be reformulated as the field equations of the D-brane action. To this end, let $\xi^a$ be coordinates on the worldvolume of a Dp-brane. Analogously to how the Nambu-Goto action encoded the area swept out by a string, the bosonic part of the Dirac-Born-Infeld (DBI) action of a $(p+1)$-dimensional brane is given by
\begin{align}
	S_{DBI} = -T_p \int d^{p+1}\xi \, e^{-\phi} \sqrt{-\det\left(g_{ab} + B_{ab} + 2\pi \alpha' F_{ab} \right)}~,
\end{align}
where $g_{ab}$ and $B_{ab}$ denote the pullback of the target space metric and Kalb-Ramond field
\begin{align}
    g_{ab} = \frac{\partial X^M}{\partial\xi^a} \frac{\partial X^N}{\partial\xi^b} g_{MN}~, \quad B_{ab} = \frac{\partial X^M}{\partial\xi^a} \frac{\partial X^N}{\partial\xi^b} B_{MN}~,
\end{align}
and the coupling constant $T_p = (2\pi)^{-p} \alpha'^{-\frac{p+1}{2}}$ denotes the brane tension. The field strength $F_{ab}$ is that of the $U(1)$ gauge field $A_a$ propagating along a single Dp-brane. If we embed such a Dp-brane into flat space with vanishing Kalb-Ramond field and constant dilaton, i.e. $g_s = e^\phi$, and expand the square root to leading order, we obtain the kinetic term of Maxwell theory or put differently, (Abelian) Yang-Mills (YM) theory
\begin{align}\label{dbiym}
	S_{DBI} = -(2\pi \alpha')^2 \frac{T_p}{4g_s} \int d^{p+1} \xi \, F_{ab}F^{ab}~,
\end{align}
which matches open string scattering computations. From \eqref{dbiym} we may read off the YM coupling
\begin{align}
	g_{YM}^2 = \frac{g_s}{T_p (2\pi \alpha')^2} = (2\pi)^{p-2} g_s \alpha'^{\frac{p-3}{2}}~.
\end{align}
To lowest order, an R-R-form field $C_{(p+1)}$ (pulled back to the brane) naturally couples to a Dp-brane $\Sigma_{p+1}$ as 
\begin{align}
	S_p = \mu_p \int\limits_{\Sigma_{p+1}} C_{(p+1)}~,
\end{align}
with charge $\mu_p = \frac{T_p}{g_s}$. Note that this is invariant under Abelian gauge transformations of the form $\delta C_{(p+1)}= d\Lambda_{(p)}$ for some $p$-form $\Lambda_{(p)}$. Finally, the full (bosonic) action of a Dp-brane also contains a Chern-Simons term 
\begin{align}\label{dpbraneaction}
	S_{Dp} = S_{DBI} + S_{CS}~,
\end{align}
which is of the form
\begin{align}
	S_{CS} = \mu_p \int\limits_{\Sigma_{p+1}} \sum_{k} C_{(k+1)} \wedge e^{B+2\pi \alpha' F}~. 
\end{align}
It is understood that all forms are pulled back appropriately and the exponential power series has to be expanded in terms of the wedge product. In order to obtain non-Abelian gauge theories, we would have to look  at several coincident branes, for instance a stack of $N$ coincident D-branes gives rise to $U(N)$ YM theory.

Now, let us turn our attention to the second viewpoint of branes as gravitational solitons, that is higher dimensional analogues of black holes \cite{Horowitz:1991cd}. These (extremal) black branes solve the equations of motion of a particular supergravity theory and are so-called half BPS solutions, i.e.\ they preserve half of the original supersymmetry of the (flat) background (cf. section \ref{sugrabackks}). In 10 dimensions, the presence of a  Dp-brane breaks 10-dimensional Poincar\'e invariance $ISO(9,1)$ down to $ISO(p,1) \times SO(9-p)$. It can be shown that
\begin{align}
    ds^2 &= H_p(r)^{-\frac{1}{2}} \, \eta_{\mu\nu} \, dx^\mu dx^\nu + H_p(r)^{\frac{1}{2}}\, \delta_{ij}\, dy^i dy^j~, \notag\\
    C_{(p+1)} &= \left(\frac{1}{H_p(r)}-1 \right) \, dx^0\wedge dx^1 \wedge ... \wedge dx^p~,\notag\\
    e^\phi &= g_s H_p(r)^{\frac{3-p}{4}}~, \quad B_{(2)} = 0
\end{align}
solves the equations of motion, see \cite{Stelle:1998xg} and references therein. Here, $x^\mu$ with $\mu=0,1,...,p$ denote coordinates along the brane and $y^i$ with $i=p+1,...,9$ are the transverse coordinates. The radial coordinate $r$ is defined as usual as $r^2 = \delta_{ij}y^iy^j$ and the function $H_p(r)$ has to be harmonic
\begin{align}
    \nabla^2_y \, H_p(r) = 0~,
\end{align}
and thus may be written as
\begin{align}
    H_p(r) = 1 + \left(\frac{\ell}{r}\right)^{7-p}~.
\end{align}
Note that this spacetime is asymptotically flat, i.e. as $r\rightarrow \infty$, we recover Minkowski space. The characteristic length scale $\ell$ may be determined from the R-R-charge of the Dp-brane, which can be computed by integrating the R-R-flux through an $(8-p)$- dimensional sphere at infinity surrounding the brane. One can show by a Dirac string-type argument that such a charge has to be quantised \cite{Nepomechie:1984wu,Teitelboim:1985ya,Teitelboim:1985yc} and basically counts the number $N$ of branes in units of $\mu_p$. Thus one finds
\begin{align}
    \ell^{7-p} = (4\pi)^\frac{5-p}{2} \, \Gamma\left(\frac{7-p}{2}\right) \, g_s N \alpha'^{\frac{7-p}{2}}~.
\end{align}
In type IIA or IIB string theory, Dp-branes with $p$ even or odd, respectively, are stable since the R-R-fields $C_{(p+1)}$ present in the their spectra can couple to the branes. From the supergravity persepective, Dp-branes are extremal in the sense that they saturate the BPS bound, since their mass $M = vol(\mathbb{R}^{p,1})\cdot N \cdot \mu_p$ is poportional to their charge $Q=N\cdot \mu_p$.

In 11-dimensional supergravity, we only have one 3-form field $A_{(3)}$ at our disposal that could couple to branes. Indeed, one finds that there are two different supergravity solitons, the so-called M2-branes which couple electrically to $A_{(3)}$, and M5-branes which couple magnetically to the 3-form. A stack of $N$ coincident M2-branes is given by \cite{Duff:1990xz}
\begin{align}
    ds^2 &= H(r)^{-\frac{2}{3}} \, \eta_{\mu\nu} \, dx^\mu dx^\nu + H(r)^{\frac{1}{3}}\, \delta_{ij}\, dy^i dy^j~, \notag\\
    A_{(3)} &= \frac{1}{H(r)} dx^0\wedge dx^1 \wedge dx^2~,
\end{align}
where the harmonic function $H(r)$ is given by
\begin{align}
    H(r) = 1+ \left(\frac{\ell}{r}\right)^6~, \quad \ell^6 = 32\pi^2 N \, l_p^6~.
\end{align}
Similarly, the fields sourced by a stack of their magnetic duals, namely M5-branes \cite{Gueven:1992hh}, are
\begin{align}
    ds^2 &= H(r)^{-\frac{1}{3}} \, \eta_{\mu\nu} \, dx^\mu dx^\nu + H(r)^{\frac{2}{3}}\, \delta_{ij}\, dy^i dy^j~, \notag\\
    A_{(6)} &= \frac{1}{H(r)} dx^0\wedge dx^1 \wedge ... \wedge dx^5~,
\end{align}
with
\begin{align}
    H(r) = 1+ \left(\frac{\ell}{r}\right)^3~, \quad \ell^3 = \pi N \, l_p^3~,
\end{align}
and $A_{(6)}$ is the magnetic dual of $A_{(3)}$, that is $dA_{(6)} = *dA_{(3)}$.

In section \ref{d3branelimits}, we will get back to stacks of (D3-)branes and in particular investigate the near-horizon limit $r\ll \ell$ in more detail, which will ultimately lead us to the AdS/CFT correspondence.

\section{AdS/CFT}\label{adscftsection}

Most of this thesis is dedicated to studying AdS supergravity backgrounds which are of paramount importance for the AdS/CFT correspondence. Thus, we devote this section to collect some of the necessary tools on the supergravity side to hopefully make sense of most words in the previous sentence. Further, we will give a crude and by no means exhaustive introduction to the original correspondence. In preparing this outline, we found the textbooks \cite{Ammon:2015wua,Nastase:2015wjb} as well as the review \cite{maldacena} most helpful, from which we have freely drawn inspiration.

\subsection{AdS geometry} \label{adsgeom}

Anti-de Sitter spacetime (AdS) is the maximally symmetric spacetime with negative cosmological constant, i.e.\ constant negative scalar curvature. $n$-dimensional AdS$_n$ can be constructed as a hypersurface in $\mathbb{R}^{n-1,2}$ as follows
\begin{align}\label{adsembedding}
	-T_1^2 - T_2^2 + X_1^2 + X_2^2 + ... + X_{n-1}^2 = -\ell^2~,
\end{align}
where $\ell$ is the radius of AdS. Observe that this construction is the non-Euclidean analogue of how one would embed an n-dimensional hyperboloid in $\mathbb{R}^{n+1}$.  As for the isometries of AdS$_n$, we note that \eqref{adsembedding} is manifestly invariant under $SO(n-1,2)$, in particular, this justifies our earlier claim that AdS$_n$ is maximally symmetric, since $\dim SO(n-1,2) = \frac{1}{2}n(n+1)$. Furthermore, given the isometry group of AdS$_n$, it may alternatively be identified with the coset space $SO(n-1,2)/SO(n-1,1)$. In order to see this, consider $(\ell,0,0,...,0)$ as a base point without loss of generality. This satisfies \eqref{adsembedding}, and we may reach any other point in AdS$_n$ by acting with an element of $SO(n-1,2)$. Finally, the point $(\ell,0,0,...,0)$ is invariant under the left action of $SO(n-1,1)$, i.e.\ $SO(n-1,1)$ is the stabiliser of a point in AdS$_n$.  
Let us consider some explicit coordinate systems of AdS$_n$. For instance, one may parametrise the hyperboloid \eqref{adsembedding} by
\begin{align}\label{globalcoords}
	T_1 &= \ell \cosh \rho \, \cos \tau~, \notag\\
	T_2 &= \ell \cosh \rho \, \sin \tau~, \notag\\
	X_i &= \ell \, y_i \sinh \rho~,\quad i= 1,...,n-1
\end{align} 
where $\tau \in [0,2\pi)$, $\rho\in \mathbb{R}_+$ and $\sum_i y_i^2 =1$. This coordinate system is usually referred to as global coordinates, since it covers the entire hypersurface \eqref{adsembedding}. Pulling back the flat metric on $\mathbb{R}^{n-1,2}$ to AdS$_n$, one finds
\begin{align}\label{globaladsmetric}
	ds^2 = \ell^2 \left( -\cosh^2\!\rho \, d\tau^2 + d\rho^2 + \sinh^2\!\rho \,d\Omega_{n-2}^2 \right)~,
\end{align}
where $d\Omega_{n-2}^2$ denotes the metric on the sphere $S^{n-2}$. Note that this features a global timelike Killing vector field $\partial_\tau$, however the associated time coordinate is perdiodic, so we can end up with closed timelike curves. Therefore, to avoid a grandfather paradox-like situation, we should really unwrap the time coordinate and allow $\tau\in \mathbb{R}$, thereby going to the universal cover of AdS$_n$.

Another widely used coordinate system, and probably the most useful in the context of AdS/CFT, are Poincar\'e patch coordinates. Introducing $x^\mu = (t,\bm{x}) \in \mathbb{R}^{n-2,1}$ and $r \in \mathbb{R}_+$, the parametrisation is given by
\begin{align}
	T_1 &= \frac{\ell^2}{2r} \left( 1+ \frac{r^2}{\ell^4} \left(\ell^2-t^2+\bm{x}^2\right) \right) \notag\\
	T_2 &= \frac{r t}{\ell} \notag \\
	X_i &= \frac{r x_i}{\ell}~, \quad i=1,...,n-2 \notag\\
	X_{n-1} &= \frac{\ell^2}{2r} \left( 1+ \frac{r^2}{\ell^4} \left(-\ell^2-t^2+\bm{x}^2\right) \right)~.
\end{align}
The metric of AdS$_n$ then reads
\begin{align}\label{poincpatchmetric}
	ds^2 = \frac{r^2}{\ell^2} \eta_{\mu\nu} dx^\mu dx^\nu +\frac{\ell^2}{r^2} dr^2~,
\end{align}
and so we observe that AdS$_n$ is foliated by leaves of $(n-1)$-dimensional Minkowski space, hence the name Poincar\'e patch. In fact, this is our first example of a warped product metric, which we will come back to later in this section. First, let us note that the conformally equivalent metric $d\tilde{s}^2 = \frac{\ell^2}{r^2} ds^2$ exhibits a boundary at $r\rightarrow\infty$ and it is in this sense that Minkowski space is the conformal boundary of AdS. On the other hand, the slice at $r=0$ is a Killing horizon, since the time-like Killing vector $\partial_t$ has vanishing norm on that plane. As convenient as these coordinates are, we want to stress it is not a coincidence that their very name contains the word ``patch''; indeed they only cover half the hyperboloid \eqref{adsembedding}. The plane $r=0$ is just a coordinate singularity though, and the metric may be extended through the Killing horizon, e.g. using the global coordinates \eqref{globalcoords}.

A first hint that a theory on AdS might be holographic comes from the causal structure of AdS. To this end, let us perform the following gedankenexperiment. Imagine we were a stationary observer in the bulk of AdS at $\rho = \rho_0$ in global coordinates \eqref{globalcoords}, and we  sent out a light signal radially towards the boundary where it is reflected. Setting $\tau(\rho_0)=0$, an outward radial null geodesic is parametrised by
\begin{align}
	\tau(\rho) = 2 \left(\arctan(e^\rho) - \arctan(e^{\rho_0})\right) ~,
\end{align}
thus we find that the signal reaches us in finite proper time as measured on our clock, namely
\begin{align}
	\ell \cosh\rho_0 \left(2\pi - 4 \arctan(e^{\rho_0})\right)~.
\end{align} 
Therefore, the conformal boundary is in causal contact with the bulk.

\subsection{Supergravity Backgrounds and Killing spinors}\label{sugrabackks}

A supergravity background is a solution to the classical equations of motion for the metric, fluxes, and scalar fields of a particular theory. One may then treat any fluctuations around this background quantum mechanically. The fermionic fields are all set to zero, since they have to vanish classically\footnote{Consider Minkowski space $\mathbb{R}^{n-1,1}$ for example, fermions are spinors and as such will transform non-trivially under $SO(n-1,1)$. A non-vanishing VEV would break the Lorentz invariance of the vacuum.}. We will mostly be interested in supersymmetric backgrounds, that is solutions preserving some residual (rigid) supersymmetry of the original (local) supergravity, and thus we require that the SUSY variations leave the fields invariant. Since SUSY transforms bosons into fermions, and we have set the latter to zero, this will not give us any new conditions. However, the SUSY variations with parameter $\epsilon$ of the fermions, in particular that of the gravitino $\psi_M$, which is present in every SUGRA theory, and those of any other remaining fermions $\lambda$ yield
\begin{align}\label{generalkses}
	\delta_{\epsilon}\psi_M|_{\psi_M,\lambda = 0} = \mathcal{D}_M \epsilon = 0~, \quad \delta_{\epsilon}\lambda|_{\psi_M,\lambda=0} = \mathcal{A} \epsilon = 0~,
\end{align}
where $\mathcal{D}_M$ denotes the supercovariant derivative
\begin{align}
	\mathcal{D}_M = \partial_M + \frac{1}{4} \, \Omega_{M,AB} \Gamma^{AB} + \mathcal{B}_M~,	
\end{align}
which contains the spin connection $\Omega_{M,AB}$. Here, $M,N$ etc. denote spacetime indices, $A,B$ are frame indices, and $\mathcal{B}_M$ just collectively denotes fluxes in the theory contracted into Clifford algebra elements. Any SUSY parameters $\epsilon$ which solve these first order partial differential equations and possibly algebraic conditions are called Killing spinors and hence the set of equations \eqref{generalkses} are referred to as Killing spinor equations (KSEs). The number $N$ of linearly independent solutions to \eqref{generalkses} is referred to as the number of preserved supersymmetries of the background.

For the most part, we will focus on warped products of AdS or Minkowski space, and some so called internal space. Backgrounds of this type were originally considered in the context of Kaluza Klein and flux compactifications cf. \cite{duff,Grana:2005jc} for a comprehensive review. More recently, there has been renewed interest in such solutions as gravity duals of supersymmetric field theories, which we aim to give a flavour of in section \ref{d3branelimits}. Requiring that these backgrounds be invariant under the isometries of AdS$_n$ or $\mathbb{R}^{n-1,1}$ respectively, the most general metric one can write down is of the form
\begin{align}
	ds^2 = A^2 ds^2(AdS_n) + ds^2(M^{D-n})~, \quad ds^2 = A^2 ds^2(\mathbb{R}^{n-1,1}) + ds^2(M^{D-n})~.
\end{align}
Here $D=10,11$ denotes the dimension of the supergravity theory under consideration and $A$ is the so-called warp factor, which may depend on the coordinates of the transverse space $M^{D-n}$.

It turns out that the metric of warped AdS backgrounds can be written in the form of a black hole near horizon geometry \cite{adshor}, so that techniques from studying black hole horizons \cite{Gutowski:2009wm,Gutowski:2010jk,Gutowski:2013kma,Grover:2013ima,Gran:2013kfa,Gran:2013wca,Gran:2014fsa,Gran:2014yqa} can be carried over to the analysis of AdS solutions \cite{mads,iibads,iiaads}. Near an extremal black hole horizon, one can adapt coordinates such that the metric becomes \cite{Moncrief:1983xua,Friedrich:1998wq}
\begin{align}\label{nearhorizonmetric}
	ds^2 = 2 du(dr + rh - r^2 \frac{\Delta}{2}du) + ds^2(\mathcal{S})~, 
\end{align} 
where the one-form $h$ and function $\Delta$ depend on the coordinates of the horizon section $\mathcal{S}$. In horizon calculations one usually assumes that $\mathcal{S}$ is compact without boundary, but we are not making this assumption here. If we recall the Poincar\'e patch metric on AdS$_n$ \eqref{poincpatchmetric} and let $r = \ell e^{\frac{z}{\ell}}$, we obtain
\begin{align}
	ds^2 = A^2 e^{\frac{2z}{\ell}} \left(-dt^2 + \sum_{i=1}^{n-2} (dx^i)^2 \right) + A^2 dz^2 + ds^2(M^{D-n})
\end{align}
for the metric on the warped product space. For $n>2$ we introduce a kind of rescaled light cone coordinates
\begin{align}
	u=\frac{1}{\sqrt{2}} (t+x^1)~, \quad r = A^2 e^{\frac{2z}{\ell}}(-t+x^1)~,
\end{align}
so that we end up with
\begin{align}
	ds^2 &= 2 du\left(dr - \frac{2r}{\ell} dz -2r d\log A \right) + A^2 dz^2 +\notag\\
	 &+A^2 e^{\frac{2z}{\ell}} \delta_{ab} dx^a dx^b + ds^2(M^{D-n})~,
\end{align}
where $a,b = 1,...,n-3$ refers to the remaining spatial coordinates of the Minkowski leaves. Note that this is indeed of the form \eqref{nearhorizonmetric} with
\begin{align}
	h &= - \frac{2}{\ell} dz - 2 d\log A~,~~ \Delta = 0~,\notag\\ 
	ds^2(\mathcal{S}) &= A^2 dz^2 +A^2 e^{\frac{2z}{\ell}} \delta_{ab} dx^a dx^b + ds^2(M^{D-n})~.
\end{align}
Similarly, one can cast the metric of warped product spaces containing an AdS$_2$-factor in the form of a near horizon metric \cite{adshor}, however we are only concerned with $n>2$ in this thesis, and so we shall omit this here.
 
 It is natural to introduce a pseudo-orthornormal frame as
\begin{align}
	\bbe^+ = du~,~~ \bbe^- = dr - \frac{2r}{\ell} dz -2r \, d\log A~,~~ \bbe^z = A \, dz~,~~ \bbe^a = A \, e^{\frac{z}{\ell}} dx^a~,
\end{align}
so that 
\begin{align}
	ds^2 = 2 \bbe^+ \bbe^- + (\bbe^z)^2 + \delta_{ab} \bbe^a \bbe^b + \delta_{ij} \bbe^i \bbe^j~,	
\end{align}
where $ds^2(M^{D-n}) = \delta_{ij} \bbe^i \bbe^j$. Since we required that all fields in our background be invariant under the isometries of AdS$_n$, any $p$-form fluxes $F^{(p)}$ will generically have to be of the form
\begin{align}
	F^{(p)} &= X^{(p-n)} \wedge d\text{vol}(AdS_n) + Y^{(p)}~,~~ p\leq n \notag\\
	F^{(p)} &= Y^{(p)}~,~~ p>n~,
\end{align}
where $X^{(p-n)}$ and $Y^{(p)}$ are forms on the internal manifold $M^{D-n}$ of degree $p-n$ and $p$, respectively. Taking this setup as a starting point, the authors in \cite{mads,iibads,iiaads} explicitly solved the Killing spinor equations on the AdS subspace in 11-dimensional, type IIB and (massive) IIA supergravity to find
\begin{align}
	\epsilon &= \sigma_+ + \sigma_- - \frac{1}{\ell} \, e^{\frac{z}{\ell}} x^a \Gamma_{az} \sigma_- - \frac{1}{\ell A} \, u \Gamma_{+z} \sigma_- + e^{-\frac{z}{\ell}} \tau_+ - \notag\\
	&-\frac{1}{\ell A} r e^{-\frac{z}{\ell}} \Gamma_{-z} \tau_+ -\frac{1}{\ell} x^a \Gamma_{az} \tau_+ + e^{\frac{z}{\ell}} \tau_-~,
\end{align}
where the spinors $\sigma_\pm$,$\tau_\pm$ only depend on the transverse space, and satisfy the light cone projection conditions $\Gamma_\pm \sigma_\pm = \Gamma_\pm \tau_\pm =0$. The remaining independent KSEs are schematically a parallel transport equation along the internal manifold (with supercovariant connection $\nabla^{(\pm)}$)
\begin{align}
	\nabla^{(\pm)}_i \sigma_\pm = 0~, \quad \nabla^{(\pm)}_i \tau_\pm = 0~,
\end{align}
restrictions of algebraic KSEs to the transverse space
\begin{align}
	\mathcal{A}^{(\pm)} \sigma_\pm = 0~, \quad \mathcal{A}^{(\pm)} \tau_\pm = 0~,
\end{align}
and an additional algebraic KSE
\begin{align}
	\Xi^{(\pm)} \sigma_\pm = 0~, \quad \left(\Xi^{(\pm)} \pm \frac{1}{\ell} \right) \tau_\pm = 0~,
\end{align}
which arises as an integrability condition from integrating the KSEs along the AdS-factor. Details of the supercovariant connection $\nabla^{(\pm)}$, as well as the Clifford algebra operators $\mathcal{A}^{(\pm)}, \Xi^{(\pm)}$ are theory dependent, and will be introduced as and when we need them in chapters \ref{ads4} and \ref{ads3}. The field equations are for the most part more or less naive reductions on the transverse space with the notable exception of the Einstein equation along AdS, which leads to an equation of motion for the warp factor 
\begin{align}
	\nabla^2 \log A = -n (d\log A)^2 - \frac{n-1}{\ell^2 A^2} + (fluxes)^2.
\end{align}
Depending on the dimension of the AdS subspace, $\sigma_\pm$ and $\tau_\pm$ come in different multiplicities, i.e given some linearly independent solutions of type $\sigma_+$, one can generate $\sigma_-$ and possibly $\tau_\pm$-type spinors via Clifford algebra operators. The number of supersymmetries a background preserves is then given by the number of multiplets times the number of linearly independent spinors in each multiplet.

Lastly, we would like to point out the importance of assuming the transverse space to be compact without boundary as discussed in \cite{desads}. Going back to the hyperboloid \eqref{adsembedding}, we can introduce a slightly different set of global coordinates with $\hat{\rho} \in \mathbb{R}_+$ as
\begin{align}
	T_1 &= \hat{\rho} \cos \tau \notag\\
	T_2 &= \hat{\rho} \sin \tau \notag\\
	X_i &= x_i~,~~ i = 1,...,n-2~, 
\end{align}
so that the metric becomes
\begin{align}
	ds^2(AdS_n) = -\hat{\rho}^2 d\tau^2 -d\hat{\rho}^2 + \sum_{i=1}^{n-2} (dx_i)^2~.
\end{align}
Next, we set
\begin{align}
	\hat{\rho} &= \ell \hat{r} \cosh y~, \notag\\
	x_{n-2} &= \ell \sinh y~, \notag\\
	x_i &= \ell w_i \cosh y~,~~ i= 1,...,n-3~,
\end{align}
where $\hat{r} \in \mathbb{R}_+$, $y\in \mathbb{R}$ and $-\hat{r}^2 + \sum_i (w_i)^2 = -1$, which yields 
\begin{align}
	ds^2(AdS_n) &= \ell^2 dy^2 + \ell^2 \cosh^2\!y \left(-\hat{r}^2 d\tau^2 - d\hat{r}^2 + \sum_{i=1}^{n-3} (dw_i)^2  \right)\notag \\
	&= \ell^2 dy^2 + \ell^2 \cosh^2\!y ~ ds^2(AdS_{n-1})
\end{align}
for the metric. Thus AdS$_n$ may be foliated by AdS$_{n-1}$-leaves, the result being a warped product with the real line and warp factor $\ell^2 \cosh^2\!y$. Repeating this procedure multiple times, the solution AdS$_7 \times S^4$ in 11-dimensional supergravity could then be reinterpreted as an AdS$_4$ background with some non-compact internal space, for instance. Of course, all solutions obtained in this fashion are locally isometric to the parent solution, and therefore do not represent a genuinely new background. Hence, in order to exclude such ``fake'' AdS vacua and to make it unambiguously clear what constitutes distinct backgrounds, we usually assume the internal manifold $M^{D-n}$ to be compact without boundary. As we will see in section \ref{ksaintro}, compactness also guarantees that certain Killing spinor bilinears vanish and thus the bosonic part of the Killing superalgebra decomposes into a direct sum of the isometry algebra of AdS$_n$ and that of the the transverse space $M^{D-n}$.

\subsection{Stacks of D3-branes - a tale of two limits}\label{d3branelimits}

Finally, we have collected all necessary ingredients to formulate and at least ``justify'' the AdS/CFT correspondence, and so without further ado, let us see what this is all about. In its strongest conjectured form, the AdS/CFT correspondence relates a superstring theory on an AdS background to a (super-) conformal field theory living on the conformal boundary of AdS. The original and most prominent example of the correspondence is that type IIB superstring theory on $AdS_5\times S^5$ with radius $\ell$ and $N$ units of five-form flux is dual to $\mathcal{N}=4$ Super Yang-Mills theory on $\mathbb{R}^{3,1}$ and gauge group $SU(N)$ \cite{maldacenab,Gubser:1998bc,witten}. The coupling constants of these theories are identified as
\begin{align}\label{mappingcouplings}
	g_{YM}^2 = 2\pi g_s~, \quad 2g_{YM}^2 N = \frac{\ell^4}{\alpha'^2}~.
\end{align}
Note that this is a strong-weak coupling duality; for instance in the limit $g_s\rightarrow 0$ and $\alpha'/\ell^2 \rightarrow 0$, we recover supergravity on the AdS side, whereas on the CFT side this corresponds to the limit of large 't Hooft coupling $\lambda = g_{YM}^2 N \rightarrow \infty$, i.e. a strongly coupled gauge theory. 

In order to motivate this correspondence, let us consider a stack of $N$ coincident D3-branes extending in the $x^0$, $x^1$, $x^2$ and $x^3$ direction in type IIB. As we mentioned in section \ref{2facesbranes}, we may take two points of view. The first point of view is that D-branes are the endpoints of open strings and we know how to treat these perturbatively provided that $g_s\ll 1$. Furthermore, we saw that at low energies $E\ll \sqrt{\alpha'}$, i.e. when we can neglect the massive tower of string excitations, the effective theory of the massless open string modes is a supersymmetric Yang-Mills theory with gauge group $U(N)$ and effective coupling $g_sN$ living on the worldvolume. Therefore, this open string perspective is really valid for $g_sN\ll1$.

The effective action of this configuration is of the form 
\begin{align}\label{d3eft}
	S = S_{bulk} + S_{brane} + S_{int}~,
\end{align}
where $S_{bulk}$ and $S_{brane}$ describe closed and open string modes, respectively. Further, $S_{int}$ encompasses the interaction between these two sectors. The bulk action $S_{bulk}$ is that of 10-dimensional supergravity plus higher order derivative terms. Expanding the metric around the flat Minkowski background, that is $g=\eta + \kappa h$, schematically to leading order one has
\begin{align}
	S_{bulk} = -\frac{1}{2} \int d^{10}x \, (\partial h)^2 + \mathcal{O}(\kappa)~,
\end{align}
with the gravitational coupling constant $\kappa$ given by $2\kappa^2 = (2\pi)^7\alpha'^4 g_s^2$. Note that introducing a stack of coincident branes breaks half the supersymmetry and so the massless open string excitations can be collected in a $(3+1)$-dimensional $\mathcal{N}=4$ gauge multiplet consisting of a gauge field $A_\mu$, six real scalars $\varphi^i$ ($i=1,...,6$) and fermionic superpartners. Coordinates transverse to the brane are identified with the worldvolume scalars as 
\begin{align}\label{branevevs}
	x^{i+3} = 2\pi \alpha' \varphi^i~, \quad i=1,...,6~.
\end{align}
Now, the remaining contributions to the effective action \eqref{d3eft} follow from plugging \eqref{branevevs} into the DBI-action plus a Wess-Zumino term as given in \eqref{dpbraneaction} and expanding $e^{-\phi}$ and $g=\eta + \kappa h$ so that only terms to leading order in $\alpha'$ remain. One finds the following expressions
\begin{align}\label{open_and_int}
	S_{brane} &= \frac{1}{2\pi g_s} \int d^4x ~ \mathrm{tr} \!\left( \frac{1}{2} F_{\mu\nu}F^{\mu\nu} + D_{\mu} \varphi^{i}D^\mu \varphi^{i} + \frac{1}{2} \sum_{i,j} [\varphi^i,\varphi^j]^2\right) + \mathcal{O}(\alpha') \notag\\
	S_{int} &= \mathcal{O}(\alpha')~.
\end{align}
Thus, we may identify the leading term of $S_{brane}$ in \eqref{open_and_int} as the bosonic part of the action of $\mathcal{N}=4$ Super Yang-Mills with coupling
\begin{align}
	g_{YM}^2 = 2\pi g_s~,
\end{align}
as in \eqref{mappingcouplings}. Due to the fact that the interaction between the open and closed sector is $\mathcal{O}(\alpha')$, we are now tempted to take the limit $\alpha'\rightarrow 0$, in which we obtain $\mathcal{N}=4$ SYM on the worldvolume and free IIB supergravity on $\mathbb{R}^{9,1}$. However, we have to be a bit more careful than that due to eqn. \eqref{branevevs}. In fact, the vacuum expectation values (VEVs) of the scalars parametrise the transverse position of the brane, and as such should be kept fixed when taking the decoupling limit. Therefore we should really be taking the so-called Maldacena limit
\begin{align}\label{maldacenalimit}
	\alpha'\rightarrow 0~, \quad U = \frac{r}{\alpha'} = const.~,
\end{align}
where $U$ is kept fixed and $r$ is some distance which will play a role in the closed string perspective of our brane setup.

The second point of view is that Dp-branes can be identified with extremal p-brane solutions in supergravity, i.e. the low energy effective theory of closed strings. We found that these D-branes viewed as gravitational solitons curve the spacetime around them. In order for the (classical) supergravity approximation to hold, that is so we do not have to invoke $\alpha'$-corrections, we should consider the regime in which the characteristic length scale $\ell$ is large and so curvature is weak. More precisely, we require $\ell/\sqrt{\alpha'} \gg 1$. Since for our stack of D3-branes $g_sN \sim \ell^4/\alpha'^2$, this closed string perspective may be employed when $g_sN\gg 1$.

In section \ref{2facesbranes}, we found the solution for general Dp-branes, and in our particular case, that is a stack of $N$ coincident D3-branes, this reads
\begin{align}\label{d3branemetric}
	ds^2 &= H(r)^{-\frac{1}{2}} \, \eta_{\mu\nu} dx^\mu dx^\nu + H(r)^{\frac{1}{2}} \left(dr^2 + r^2 d\Omega_5^2 \right)\notag\\
	F_{(5)} &= -(1+*) \, d\text{vol}(\mathbb{R}^{3,1}) \wedge dH^{-1}\notag\\
	e^{2\phi(r)} &= g_s^2~,
\end{align} 
where
\begin{align}
	H(r) = 1+\frac{\ell^4}{r^4}~, \quad \ell^4 = 4\pi g_s N \alpha'^2~,
\end{align}
and we have introduced spherical polar coordinates transverse to the brane. Note that this manifestly preserves $ISO(3,1)\times SO(6)$ and one can verify that half of the original 32 real Poincar\'e supercharges are broken again. The solution \eqref{d3branemetric} interpolates between two different regions. For $r\gg \ell$, $H(r)\sim 1$ and we recover the $(9+1)$-dimensional Minkowski vacuum. If on the other hand $r\ll \ell$, the harmonic function $H$ is approximately $H(r) \sim \ell^4/r^4$ and the metric becomes that of $AdS_5\times S^5$ (cf. eqn. \eqref{poincpatchmetric}):
\begin{align}
	ds^2 &= \frac{r^2}{\ell^2} \, \eta_{\mu\nu} \, dx^\mu dx^\nu + \frac{\ell^2}{r^2} \left( dr^2 + r^2 \, d\Omega_5^2 \right)\notag\\
	&= \underbrace{\frac{r^2}{\ell^2} \, \eta_{\mu\nu} \, dx^\mu dx^\nu + \frac{\ell^2}{r^2} dr^2}_{AdS_5} + \underbrace{\ell^2 d\Omega_5^2}_{S^5}~.
\end{align}
Thus, there are really two different kinds of closed string modes in this background; those propagating in flat space far away from the branes and those propagating in the vicinity of the AdS-throat. Upon taking the Maldacena limit, these two types of closed strings decouple from one another. In order to see this, we consider a closed string excitation with energy $E_r \sim \alpha'^{-\frac{1}{2}}$ at some fixed radial position $r$. Even if $\sqrt{\alpha'}E_r\gg 1$ in the near-horizon region, we cannot just integrate these modes out at low energies, since the energy $E_\infty$ measured by an observer at infinity is red-shifted as
\begin{align}
	\sqrt{\alpha'}E_\infty = \sqrt{-g_{00}} \cdot \sqrt{\alpha'}E_r \sim \frac{r}{\ell} \sqrt{\alpha'} E_r \rightarrow 0~,
\end{align}
for $r\ll \ell$ but fixed $\sqrt{\alpha'}E_r$. Hence an observer at infinity sees both supergravity (i.e. massless string modes) modes propagating in $\mathbb{R}^{9,1}$, as well as closed string excitations in the throat region $AdS_5\times S^5$. Similarly to the field theory side, these decouple when we take the Maldacena limit, since
\begin{align}
	\frac{\ell^4}{r^4} = 4\pi g_s \, N \, \frac{\alpha'^2}{r^4} = 4\pi g_s \, N\, \frac{\alpha'^4}{r^4} \cdot \frac{1}{\alpha'^2} \rightarrow \infty
\end{align}
as $\alpha'\rightarrow 0$ with $\frac{\alpha'}{r}$ kept fixed. Thus, what we are really doing here, is zooming in on the near-horizon region, hence the alternative name near-horizon limit for \eqref{maldacenalimit}.  

All in all, we discovered two decoupled low energy effective theories in both pictures, namely type IIB supergravity on $AdS_5\times S^5$ and type IIB supergravity on $\mathbb{R}^{9,1}$ in the closed string picture, and (3+1)-dimensional $\mathcal{N}=4$ Super Yang-Mills and type IIB supergravity on $\mathbb{R}^{9,1}$ in the open string picture. These two perspectives should yield equivalent descriptions of the same underlying physics and, since IIB supergravity in flat space is present in both of them, this led Maldacena to conjecture that type IIB superstring theory on $AdS_5\times S^5$ is dual to $\mathcal{N}=4$ Super Yang-Mills\footnote{A little caveat here is what happened to the $U(1)$ in $U(N)$. We originally stated that the gauge group is $SU(N)$ and not $U(N)$. The reason for this is that the $U(1)$ multiplet is free but in IIB on $AdS_5\times S^5$ everything has to couple to gravity and no field is free. From the gravity perspective, the $U(1)$ modes are singletons only living on the boundary of AdS and unable to propagate into the bulk\cite{Gibbons:1993sv,Witten:1998wy,maldacena}.} in four dimensions.

As a first consistency check of the correspondence, let us compare the symmetries on both sides, since these should match up. $\mathcal{N}=4$ SYM is superconformal, i.e. the operators form representations of the conformal group in four dimensions $SO(4,2)$. Furthermore, as the theory is also supersymmetric, there are $4\times4=16$ Poincar\'e supercharges associated with translations and 16 superconformal charges associated with conformal boosts, that is we have 32 supersymmetries in total. Last but not least, the R-symmetry group rotating these supercharges amongst one another is $SU(4)$. On the gravity side, we find both $SO(4,2)$ and $SO(6)\approx SU(4)$ (at least on the Lie algebra level) as the isometry groups of AdS$_5$ and the internal manifold $S^5$. Besides, $AdS_5\times S^5$ is maximally supersymmetric \cite{Schwarz:1983qr,maxsusy}, which is curious considering we started out with a half BPS brane configuration. Such supersymmetry enhancement near the horizon is a very common feature and has been studied in \cite{Gutowski:2013kma,Gran:2013wca,Gran:2014fsa,Gran:2014yqa}.

One may then go on to exploit the representations of these symmetry groups to establish the dictionary between operators on the field theory side and the spectrum of IIB on the gravity side. This allows us for example to compute correlation functions in supergravity with relative ease, whose field theory dual would be computationally inaccessible by other methods. Let us give an example of how this matching on both sides works. An important class of gauge invariant operators are the half BPS chiral primaries of conformal dimension $\Delta$ given by
\begin{align}
	\mathcal{O}_{\Delta}(x) = C^{\Delta}_{i_1...i_\Delta} \, \mathrm{tr}\left(\varphi^{i_1}(x)...\varphi^{i_\Delta}(x)\right)~,
\end{align}
where the elementary scalars $\varphi^i$ transform in the $\mathbf{6}$ of $\mathfrak{so}(6) = \mathfrak{su}(4)$, and the coefficients $C^{\Delta}_{i_1...i_\Delta}$ form the rank $\Delta$ symmetric traceless tensor representation of $\mathfrak{so}(6)$. Their conformal dimension is protected from any quantum corrections by supersymmetry, hence these operators provide a non-trivial check of the correspondence, as in that case we have control over both sides. Now, on the supergravity side, we have a compact space with transitive group $SO(6)$, namely $S^5$, and so we can expand the fields into Kaluza-Klein modes. In our case, these are the spherical harmonics $Y^{I_n}(\Omega_5)$ of $S^5$ constituting irreducible representations of $\mathfrak{so}(6)$. In particular, viewing $S^5$ as embedded in $\mathbb{R}^6$ with coordinates $x^i$, these can be written as
\begin{align}
	Y^{I_n} = C^{I_n}_{i_1...i_n} \, x^{i_1}...\,x^{i_n}~,
\end{align}
where the $C^{I_n}_{i_1...i_n}$ are again symmetric traceless tensors of $\mathfrak{so}(6)$. On a five-sphere of radius $\ell$, the spherical harmonics diagonalise the Laplacian
\begin{align}
	\nabla^2_{S^5} Y^{I_n}(\Omega_5) = -\frac{1}{\ell^2} n(n+4) Y^{I_n}(\Omega_5)~.
\end{align}
Suppressing any spacetime indices, a supergravity field $\Phi$ has a Kaluza-Klein expansion of the form
\begin{align}
	\Phi(x^\mu,r,\Omega_5) = \sum_{n}\sum_{I_n} \Phi_{n}^{I_n}(x^\mu,r)\,Y^{I_n}(\Omega_5)~,
\end{align}
where $(x^\mu,r)$ are the coordinates on AdS$_5$ and $\Omega_5$ denotes those on $S^5$. Plugging this ansatz into the SUGRA equations of motion \cite{Kim:1985ez}, one finds masses $m$ (and couplings) for the AdS$_5$ fields $\Phi_{n}^{I_n}$ which correspond to operators $\mathcal{O}_\Delta$ in $\mathcal{N}=4$ SYM with scaling dimension $\Delta =n$ \cite{witten}. The mass $m$ and conformal dimension $\Delta$ are related by
\begin{align}
	\Delta = 2+ \sqrt{4+m^2\ell^2} ~.
\end{align}

A similar heuristic derivation to the one we have outlined here can be performed for stacks of M2- and M5-branes in 11-dimensional supergravity, whose near-horizon limits are the maximally supersymmetric $AdS_4\times S^7$ and $AdS^7\times S^4$ M-theory backgrounds, resepectively.

\section{The Homogeneity Theorem}\label{homogentheorem}

The common theme of this thesis is to classify supergravity backgrounds in 10 and 11 dimensions preserving more than half maximal supersymmetry. In this section, we are going to address why such a setup is particularly tractable, and will allow us to classify a large class of supergravity backgrounds.
\\
In particular, we will be concerned with the so-called homogeneity theorem, which states the following:
\begin{namedtheorem}[Homogeneity Theorem]
	All 10- and 11-dimsensional supergravity backgrounds preserving $N>16$ supersymmetries are locally Lorentzian homogeneous spaces.
\end{namedtheorem} 
This is an absolutely crucial result for our analysis and was proven in \cite{homogen}, which we will be following in our outline of the proof. Before we proceed though, we should clarify what the main statement entails. Homogeneity in this context means that there is a Lie group acting transitively\footnote{A Lie group $G$ acts transitively on a manifold $M$, if it only possesses a single group orbit. More concretely, $\forall x,y \in M$ there is a group element $g\in G$ that satisfies $gx = y$.} on the spacetime and preserving all the bosonic fields present in the background, e.g. the metric and fluxes. However, the truly relevant notion for our purposes is that of local homogeneity, i.e.\ at the level of the Lie algebra, since all our computations are performed locally. This translates to the Killing vectors spanning the tangent space at every point and leaving the bosonic fields invariant, in the sense that their Lie derivatives with respect to any of these isometries vanish.

\subsection{Outline of the proof}

From a mathematical persepective, the main ingredients of a supergravity background are a Lorentzian spin manifold $(M,g)$ and a spinor bundle $\mathcal{S}\rightarrow M$ with a connection $D$, which depends on the bosonic fields (contracted into Clifford algebra operators) of the particular theory. $\mathcal{S}$ is a vector bundle associated with a representation $S$ of the spin group. The tangent bundle $TM$ is the vector bundle associated to the vector representation $\mathcal{V}$ of the orthogonal group corresponding to the spin group. A Killing spinor is then defined as a section of $\mathcal{S}$ which is parallel with respect to the supercovariant connection $D$. In line with how we introduced Killing spinors in section \ref{sugrabackks}, this corresponds to setting the supersymmetry variaton of the gravitino equal to zero. Depending on the background in question, there might be additional algebraic Killing spinor equations arising from the variation of additional fermionic fields, e.g. the dilatino. We note that the Killing spinor equations are linear, so the Killing spinors form a vector space which we shall denote by $K$. As the KSEs are at most first order in the derivatives, a Killing spinor is specified by its value at a point $p\in M$. Therefore, having chosen a point $p$, the vector space $K$ can be regarded as a subspace of the fibre $\mathcal{S}_p$, which in turn can be identified with the representation $S$. In that sense, the space of Killing spinors $K$ is a vector subspace of $S$.

Probably one of the most important features in analysing supergravity backgrounds is the fact that given a spin invariant, real valued inner product $\langle \cdot, \cdot \rangle_S$ on $S$ and some Killing spinors $\epsilon_1,\epsilon_2 \in K$, bilinears of the form
\begin{align}
	\langle \epsilon_1, \Gamma^M \epsilon_2 \rangle_S ~ \bbe_M
\end{align}
are Killing vectors, where $\bbe_M$ ($M=1,...\dim M -1$) is a basis of $T_pM$. With that in mind, in order to prove the homogeneity theorem, we have to show that these Killing spinor bilinears span the tangent space $T_pM$ at every point, if there are more than $16 = \frac{1}{2} \dim S$ of them, i.e.
\begin{align}
	\omega: S \times S \rightarrow T_pM~, ~~ (\psi_1,\psi_2) \mapsto \omega(\psi_1,\psi_2) =\langle \psi_1, \Gamma^M \psi_2 \rangle_S ~ \bbe_M
\end{align}
restricted to $K$ is onto $T_pM$, if $\dim K > \frac{1}{2} \dim S$. To proceed, we assume that $\omega|_K$ is not onto, then there exists a vector $n \in S$, $n\neq 0$ such that
\begin{align}\label{normK}
	n^M \langle \epsilon_1, \Gamma_M \, \epsilon_2 \rangle_S = 0 \quad \forall\epsilon_1,\epsilon_2 \in K~,
\end{align}
so $n$ is orthogonal to the image of $\omega|_K$. We may now take the point of view that $\slashed{n}$ defines a linear map $\slashed{n}: K \rightarrow K^\bot$, taking Killing spinors and mapping them to $K^\bot$, the orthognal complement of $K$ in $S$ with respect to the inner product $\langle \cdot, \cdot\rangle_S$. Let us first consider the case that $n$ is space-like or time-like, that is $n^2 \neq 0$. Then, as
\begin{align}
	\slashed{n}^2 = n^2 \, \mathds{1} \neq 0~,
\end{align}
$\slashed{n}$ has trivial kernel and as such is an injection. However, this leads to a contradiction with our assumption that $\dim K > \frac{1}{2} \dim S$, and since $\dim K + \dim K^\bot = \dim S$, $\slashed{n}$ cannot be one-to-one. Hence $\omega|_K$ must be onto.

All that remains to consider is the case that $n$ is null, for which the proof depends on the explicit form of the inner product $\langle \cdot, \cdot\rangle_S$ of a particular supergravity theory. The argument in all cases is very similar though, and so we shall illustrate the general idea in the framework of 11-dimensional supergravity for definiteness. In 11 dimensions, $S$ is the 32-dimensional Majorana representation of $Spin(10,1)$ and we can choose $\omega$ to be 
\begin{align}
	\omega(\psi_1,\psi_2) = \langle \psi_1, \Gamma^0 \Gamma^M \psi_2\rangle \, \bbe_M ~,
\end{align}
with $\psi_{1/2} \in S$ and $\bbe_M$ a pseudo-orthonormal frame on $M$ defined by $ds^2 = \eta_{MN} \bbe^M \bbe^N$. The inner product on the space of spinors is defined by
\begin{align}
	\langle \psi_1, \Gamma^0 \psi_2 \rangle = \psi_1^\dagger\Gamma^0\psi_2~.
\end{align}
Taking $X := \omega(\epsilon,\epsilon) = \langle \epsilon, \Gamma^0 \Gamma^M \epsilon \rangle \, \bbe_M$ for $\epsilon \in K$, observe that this has a non-vanishing component along $\bbe_0$
\begin{align}\label{x0notspacelik}
	X^0 = \langle \epsilon, \Gamma^0 \Gamma^0 \epsilon \rangle  = - \parallel \epsilon \parallel^2 < 0 ~.
\end{align}  
Therefore X is either null or time-like, since if X were space-like, one could perform a Lorentz transformation such that $X^0 = 0$. If $n$ in eqn. \eqref{normK} is null, the orthogonal complement of the image of $\omega|_K$ is a totally null subspace of $T_pM$. As the tangent space is Lorentzian, any such subspace is at most one-dimensional and must be spanned by $n$. Without loss of generality, let us choose a basis $\{\bbe_+,\bbe_-,\bbe_i \}$ of $T_pM$ such that $n=\bbe_+$. Thus, the image of $\omega|_K$ is spanned by $\bbe_+$ and $\bbe_i$. Now, as eqn. \eqref{x0notspacelik} implies that for all $\epsilon\in K$, $\omega(\epsilon,\epsilon)$ cannot be space-like, this must lie along $\bbe_+$, i.e.\ $\omega(\epsilon,\epsilon) = \alpha(\epsilon) \, \bbe_+$, for some function $\alpha: K \rightarrow \mathbb{R}$. Note that $\omega$ is a symmetric bilinear map and as such satisfies the polarisation identity which gives
\begin{align}
	\omega(\epsilon_1,\epsilon_2) &= \frac{1}{2} \, \left( \omega(\epsilon_1 + \epsilon_2,\epsilon_1 + \epsilon_2) - \omega(\epsilon_1,\epsilon_1) - \omega(\epsilon_2,\epsilon_2) \right) \notag\\
	&= \frac{1}{2} \, \left(\alpha(\epsilon_1 + \epsilon_2) - \alpha(\epsilon_1) - \alpha(\epsilon_2) \right) \, \bbe_+~.
\end{align}
Hence, we find that the image of $\omega|_K$ is contained in the linear null subspace spanned by $\bbe_+$. This, however, is a contradiction again, since we assumed that the original complement of the image of $\omega|_K$ is one dimensional and spanned by $n= \bbe_+$. Therefore, $\omega|_K$ must be onto and the homogeneity theorem holds.

\section{Killing Superalgebras}\label{ksaintro}

Having established that backgrounds preserving $N>16$ supersymmetries are homogeneous spaces, this naturally raises the question which Lie groups or, since we are always working locally, which Lie algebras act on the spacetime. This leads us to the notion of a Killing superalgebra (KSA) in this section.

The Killing spinors, and their associated Killing vectors, from taking bilinears, form a representation of a residual (global) supersymmetry algebra, superalgebra for short, of the background. A (Lie) superalgebra is a $\mathbb{Z}_2$-graded algebra $\mathfrak{g} = \mathfrak{g}_0 \oplus \mathfrak{g}_1$ consisting of two vector subspaces, the even $\mathfrak{g}_0$, a Lie subalgebra, and the odd subspace $\mathfrak{g}_1$, a representation of $\mathfrak{g}_0$ \cite{Kac:1977qb}. If $X,Y,Z \in \mathfrak{g}$ with grading $x,y,z$, the bracket $[\cdot,\cdot]_\mathfrak{g}$ is defined as
\begin{align}\label{superbracket}
	[X,Y]_\mathfrak{g} := XY- (-1)^{xy} \, YX~,
\end{align}
and it satisfies the super-Jacobi identity
\begin{align}
	[[X,Y]_\mathfrak{g}, Z]_\mathfrak{g} + (-1)^{z(x+y)} \, [[Z,X]_\mathfrak{g}, Y]_\mathfrak{g} + (-1)^{x(y+z)} \, [[Y,Z]_\mathfrak{g}, X]_\mathfrak{g} = 0.
\end{align}
If both generators in \eqref{superbracket} are fermionic (odd), this becomes an anti-commutator $\{\cdot,\cdot\}$; in all other cases, the bracket $[\cdot,\cdot]_\mathfrak{g}$ is just the usual commutator $[\cdot,\cdot]$. Specifically, in the case of a Killing superalgebra \cite{Gauntlett:1998kc,Figueroa-OFarrill:1999klq}, the odd subspace $\mathfrak{g}_1$ is spanned by $Q_{\epsilon_m}$, where every fermionic generator is associated with a linearly independent Killing spinor $\epsilon_m$ $m=1,...,N$ of the background. Analogously, the bosonic generators $V_{K_{mn}}$ span $\mathfrak{g}_0$, where each of them is associated with a Killing spinor bilinear of the form
\begin{align}\label{ksbilin}
	K_{mn} := \langle \epsilon_m, \Gamma^M \epsilon_n \rangle_S ~ \bbe_M = K_{nm}~.
\end{align}
As in section \ref{homogentheorem}, $\langle\cdot,\cdot\rangle_S$ denotes a suitable spin invariant inner product on the space of spinors, such that the last equality in \eqref{ksbilin} holds; usually we will take $\langle (\Gamma_+-\Gamma_-)\cdot,\cdot\rangle$ or the real part thereof. The brackets are computed geometrically 
\begin{align}
	\{Q_{\epsilon_m},Q_{\epsilon_n}\} = V_{K_{mn}}~, ~~ [V_{K_{mn}},Q_{\epsilon_p}] = Q_{\mathcal{L}_{K_{mn}}\epsilon_p}~, ~~ [V_{K_{mn}},V_{K_{pq}}] = V_{[K_{mn},K_{pq}]},
\end{align}
where $[K_{mn},K_{pq}]$ denotes the usual Lie bracket of vector fields and the spinorial Lie derivative with respect to a vector field $V$ is defined as
\begin{align}
	\mathcal{L}_V \epsilon = \nabla_V \epsilon + \frac{1}{8} \, dV_{MN} \, \Gamma^{MN} \, \epsilon~.
\end{align} 
Closure of the KSAs in 11-dimensional and IIB SUGRA has been verified in \cite{Figueroa-OFarrill:2004qhu,Figueroa-OFarrill:2007omz} and it is expected that the KSAs in all supergravity theories satisfy the super-Jacobi identity.

For what follows, we will outline the main results in \cite{superalgebra,ads2}, where the authors classified the Killing superalgebras of all warped AdS$_n$, $n>2$ backgrounds. Let $\mathfrak{g}$ be the KSA of a warped AdS$_n$ background. Calculating bilinears of the form \eqref{ksbilin}, one finds that the bosonic subalgebra of $\mathfrak{g}$ contains $\mathfrak{so}(n-1,2)$ corresponding to the isometries of AdS$_n$, as is to be expected. One may further suspect that the bosonic subalgebra decomposes\footnote{This is actually only true for $n>3$. AdS$_3$ is locally a group manifold and thus the algebra decomposes into left and right superalgebras $\mathfrak{g}_L \oplus \mathfrak{g}_R$ corresponding to the left and right action. We will get back to this point.} into a direct sum of the isometries of AdS$_n$, and those on the transverse space $\mathfrak{t}_0$ as $\mathfrak{g}_0 = \mathfrak{so}(n-1,2) \oplus\mathfrak{t}_0$. However, this does not necessarily have to be the case. For instance, consider what would happen, if we did not require the transverse space be compact. Following the argument at the end of section \ref{sugrabackks}, an AdS$_n \times_w M^{D-n}$ solution may be rewritten as an AdS$_k \times_w \mathbb{R}^{n-k} \times_w M^{D-n}$ background with $k<n$. Since generally one cannot write $\mathfrak{so}(n-1,2) = \mathfrak{so}(k-1,2) \oplus \mathfrak{h}$ for some Lie algebra $\mathfrak{h}$, a decomposition into a direct sum of isometries on AdS$_k$ and those on the transverse space is not possible from the perspective of the lower dimensional AdS$_k$ subspace. We also pointed out in section \ref{sugrabackks} that in order to avoid these ``fake'' vacua, one should really consider closed, i.e.\ compact without boundary, transverse spaces. Indeed, it is this assumption together with a Hopf maximum principle type argument, like the one following eqn. \eqref{hopf1}, that implies
\begin{align}
	\langle \sigma'_+, \Gamma_{iz} \sigma_+ \rangle = 0~, \quad \langle \tau_+, \Gamma_{iz} \sigma_+ \rangle = 0~,\quad \parallel \sigma_+ \parallel = const.~, \quad \parallel \tau_+ \parallel = const. ~,
\end{align} 
for all Killing spinors $\sigma_+$, $\sigma'_+$ and $\tau_+$. Note that for $n>3$, spinors of type $\tau_+$ can be generated from the $\sigma_+$, which renders some of the above conditions redundant. Furthermore, the algebraic KSE implies that $\sigma_+$ and $\tau_+$ are orthogonal (cf.\ eqn. \eqref{iibortho2})
\begin{align}
	\langle \tau_+, \sigma_+\rangle=0~,
\end{align}
and one may construct a Killing spinor bilinear $W$ along the internal manifold, such that
\begin{align}
	\mathcal{L}_W A = 0~, \quad \nabla_i W_j + \nabla_j W_i = 0~,
\end{align} 
where $\nabla$ denotes the Levi-Civita connection on $M^{D-n}$. The latter means that the warp factor $A$ is invariant under the entire bosonic subalgebra $\mathfrak{g}_0$ and $W$ is Killing on the transverse space. One may then proceed to show that these conditions guarantee that the Killing spinor bilinears split into components along AdS$_n$ and the internal manifold $M^{D-n}$, which mutually commute and are independently Killing. Therefore, one finds the aforementioned decomposition
\begin{align}
	\mathfrak{g}_0 = \mathfrak{so}(n-1,2) \oplus\mathfrak{t}_0~,
\end{align}
as a consequence of the transverse space being compact without boundary. Equivalently, one could also simply assume that the KSA of the background decomposes in this fashion to exclude ``fake'' vacua.

Let us first consider AdS$_n$ backgrounds with $n>3$. Since the KSEs have been explicitly solved along the AdS factor, and so the Killing spinors' explicit dependence on the coordinates of the AdS subspace is known, one may directly calculate the brackets
\begin{align}
	\{\mathfrak{g}_1, \mathfrak{g}_1\} = \mathfrak{so}(n-1,2) \oplus \mathfrak{t}_0~, \quad [\mathfrak{so}(n-1,2),\mathfrak{g}_1] \subseteq \mathfrak{g}_1~.
\end{align}
One might expect the commutator $[\mathfrak{t}_0,\mathfrak{g}_1]$ to be the trickiest to compute, as one would need knowledge of the underlying geometry of the transverse space. Quite remarkably, this turns out not to be the case, and the super-Jacobi identity completely fixes both $[\mathfrak{t}_0,\mathfrak{g}_1]$ as well as the last remaining bracket $[\mathfrak{t}_0,\mathfrak{t}_0]$. However, we shall not dwell on how these are explicitly obtained, and refer the curious reader to \cite{superalgebra} for the nitty gritty details. The results are displayed in table \ref{table:AdSngr3KSA}.
\begin{table}
    \renewcommand{\arraystretch}{1.3}
	\centering
	\caption[Killing superalgebras of warped AdS$_n$ ($n>3$) backgrounds in $D=10,11$ supergravity]{Killing superalgebras of warped AdS$_n$ ($n>3$) backgrounds preserving $N$ supersymmetries in $D=10,11$ supergravity. $\mathfrak{f}^*(4)$ denotes a different real form of $\mathfrak{f}(4)$ which appears in the AdS$_3$ case. The real form of the KSA is completely determined by the real form of the bosonic subalgebra $\mathfrak{g}_0 = \mathfrak{so}(n-1,2) \oplus \mathfrak{t}_0$.}
	\begin{tabular}{ccccccccc}
		\toprule
		\multirow{2}{*}{$N$} & \multicolumn{2}{c}{AdS$_4$} & \multicolumn{2}{c}{AdS$_5$} & \multicolumn{2}{c}{AdS$_6$} & \multicolumn{2}{c}{AdS$_7$} \\
		\cmidrule{2-9}
		 & $\mathfrak{g}$ & $\mathfrak{t}_0$ &  $\mathfrak{g}$ & $\mathfrak{t}_0$ &  $\mathfrak{g}$ & $\mathfrak{t}_0$ &  $\mathfrak{g}$ & $\mathfrak{t}_0$ \\
		\midrule	
		4 & $\mathfrak{osp}(1|4)$ & $\{0\}$ & \textcolor{red}{\ding{55}} & \textcolor{red}{\ding{55}} & \textcolor{red}{\ding{55}} & \textcolor{red}{\ding{55}} & \textcolor{red}{\ding{55}} & \textcolor{red}{\ding{55}} \\
		8 & $\mathfrak{osp}(2|4)$ & $\mathfrak{so}(2)$ & $\mathfrak{sl}(1|4)$ & $\mathfrak{u}(1)$ & \textcolor{red}{\ding{55}} & \textcolor{red}{\ding{55}} & \textcolor{red}{\ding{55}} & \textcolor{red}{\ding{55}} \\
		12 & $\mathfrak{osp}(3|4)$ & $\mathfrak{so}(3)$ & \textcolor{red}{\ding{55}} & \textcolor{red}{\ding{55}} & \textcolor{red}{\ding{55}} & \textcolor{red}{\ding{55}} & \textcolor{red}{\ding{55}} & \textcolor{red}{\ding{55}} \\
		16 & $\mathfrak{osp}(4|4)$ & $\mathfrak{so}(4)$ & $\mathfrak{sl}(2|4)$ & $\mathfrak{u}(2)$ & $\mathfrak{f}^*(4)$ & $\mathfrak{so}(3)$ & $\mathfrak{osp}(6,2|2)$ & $\mathfrak{so}(3)$ \\
		20 & $\mathfrak{osp}(5|4)$ & $\mathfrak{so}(5)$ & \textcolor{red}{\ding{55}} & \textcolor{red}{\ding{55}} & \textcolor{red}{\ding{55}} & \textcolor{red}{\ding{55}} & \textcolor{red}{\ding{55}} & \textcolor{red}{\ding{55}} \\
		24 & $\mathfrak{osp}(6|4)$ & $\mathfrak{so}(6)$ & $\mathfrak{sl}(3|4)$ & $\mathfrak{u}(3)$ & \textcolor{red}{\ding{55}} & \textcolor{red}{\ding{55}} & \textcolor{red}{\ding{55}} & \textcolor{red}{\ding{55}} \\
		28 & $\mathfrak{osp}(7|4)$ & $\mathfrak{so}(7)$ & \textcolor{red}{\ding{55}} & \textcolor{red}{\ding{55}} & \textcolor{red}{\ding{55}} & \textcolor{red}{\ding{55}} & \textcolor{red}{\ding{55}} & \textcolor{red}{\ding{55}} \\
		32 & $\mathfrak{osp}(8|4)$ & $\mathfrak{so}(8)$ & $\mathfrak{sl}(4|4)/\mathds{1}_{8\times 8}$ & $\mathfrak{su}(4)$ & \textcolor{red}{\ding{55}} & \textcolor{red}{\ding{55}} & $\mathfrak{osp}(6,2|4)$ & $\mathfrak{so}(5)$ \\
		\bottomrule
	\end{tabular}
	\label{table:AdSngr3KSA}
\end{table}
Most importantly for our analysis in chapters \ref{ads4} and \ref{ads3}, we would like to point out that in order for the KSA to be closed, the isometry algebra $\mathfrak{t}_0$ has to act (almost) effectively on the transverse space $M^{D-n}$. A Lie group acts (almost) effectively on a manifold, if the only group element that leaves a point invariant is the identity or, for an almost effective action, an element in the centre of the Lie group. At the level of the Lie algebra this translates to every generator  being mapped to a non-trivial Killing vector field on the manifold.

Finally, we would like to focus on the somewhat more subtle case of AdS$_3$ backgrounds. First of all, we note that AdS$_3$ is locally a group manifold and so the KSA $\mathfrak{g}$ decomposes into left and right superalgebras $\mathfrak{g}= \mathfrak{g}_L \oplus \mathfrak{g}_R$ with $[\mathfrak{g}_L,\mathfrak{g}_R]_\mathfrak{g}=0$. We choose to associate the left superalgebra with the $\sigma_+$ spinors and the right superalgebra with the $\tau_+$ spinors, resepectively. Since these are independent, and results obtained for $\mathfrak{g}_L$ will also apply to $\mathfrak{g}_R$, we shall focus on the left superalgebras for what follows. Starting with the minimal case of $N=2$ real supercharges, one finds that
\begin{align}
    \{Q_A,Q_B\} = V_{AB}~, \quad [V_{AB}, Q_C] = -\frac{1}{\ell} \, \left( \epsilon_{CA} Q_B + \epsilon_{CB} Q_{A} \right)~,
\end{align}
where $A,B=1,2$. The $V_{AB}$ are the generators of $\mathfrak{g}_0 = \mathfrak{so}(1,2)$ and the $Q_A$ are fermionic generators corresponding to the Killing spinors $\sigma_\pm$. There are no Killing spinor bilinears along the transverse space, i.e.\ the isometry algebra is $\mathfrak{t}_0 = \{0\}$. All in all, we conclude that this KSA is isomorphic to $\mathfrak{osp}(1|2)$. Next, consider the case of $N=2k$ preserved supersymmetries, $k$ being the number of linearly independent $\sigma_+$ type spinors. Similarly to the analysis of higher dimensional AdS backgrounds, we can perform the following calculations explicitly
\begin{align}
    \{Q_{Ar},Q_{Bs}\} = V_{AB} \delta_{rs} + \epsilon_{AB} \tilde{V}_{rs}~, \quad [V_{AB}, Q_{Cr}] = -\frac{1}{\ell} \left(\epsilon_{CA} Q_{Br} + \epsilon_{CB} Q_{Ar} \right)~,
\end{align}
with $r,s=1,...,k$ and $\tilde{V}_{rs}\in \mathfrak{t}_0$. Again the isometries of the internal space acting on the supercharges turns out to be the key bracket to compute. Since the spinorial Lie derivative with respect to the isometries of the transverse space does not affect the dependence of the Killing spinors on the coordinates of AdS$_3$, one finds 
\begin{align}
    [\tilde{V}_{rs},Q_{At}] = -\frac{1}{\ell} \left(\delta_{rt} Q_{As} - \delta_{st} Q_{Ar} \right) + \frac{1}{\ell} \alpha_{rst}{}^u Q_{Au}~,
\end{align}
where the structure constants $\alpha_{rst}{}^u$ still need to be determined. The super-Jacobi identity together with the inner product of the $\sigma_+$ spinors being invariant under the isometries of the transverse space 
\begin{align}
    \langle\mathcal{L}_{\tilde{V}}\sigma_+^r, \sigma_+^s\rangle + \langle\sigma_+^r, \mathcal{L}_{\tilde{V}}\sigma_+^s\rangle = 0~,
\end{align}
imply that $\alpha$ is a 4-form. One can then further show that $\alpha$ is invariant under the representation $D$ of $\mathfrak{t}_0$ acting on the supercharges in $\mathfrak{g}_1$ as
\begin{align}
    D(\tilde{V}_{rs}) Q_{At} := [\tilde{V}_{rs},Q_{At}]~.
\end{align}
The KSA $\mathfrak{g}$ may have central terms $\mathfrak{c}_L := \{\tilde{V}\in\mathfrak{t}_0|D(\tilde{V}) = 0\}$, however the centre is trivial apart from one case where it can be at most 3-dimensional, cf table \ref{table:AdSngr3KSA}. In order to identify possible representations $D$, we take two vectors $u,v\in\mathbb{R}^{\frac{N}{2}}$ and contract them into $\tilde{V}_{rs}$, which yields an element $R(u,v) = u^r v^s\, \tilde{V}_{rs}\in\mathfrak{t}_0$ generating $SO(2)$ rotations in the plane spanned by $u$ and $v$. To see that $R(u,v)$ indeed generates rotations, note that
\begin{align}
    D(R(u,v))(w\cdot Q_A) = [u^r v^s \,\tilde{V}_{rs}, w^t Q_{At}] =  -\frac{1}{\ell} \left((u\cdot w)v^r - (v\cdot w)u^r \right)Q_{Ar}~,
\end{align}
where $w$ lies in the plane spanned by $u,v$ so that $\alpha(u,v,w,\cdot)=0$. Since these $SO(2)$ rotations act transitively on the 2-plane, and one can repeat this argument for arbitrary $u,v\in\mathbb{R}^{\frac{N}{2}}$, it follows that $\mathfrak{t}_0$ acts transitively (and almost effectively so the super-Jacobis are satisfied) on $S^{\frac{N}{2}}\subset \mathbb{R}^{\frac{N}{2}}$. Lie groups acting transitively and effectively on spheres have been classified in \cite{montgomery1943transformation} and have been used in the context of the Berger classification of irreducible simply connected Riemannian manifolds \cite{simons1962transitivity}. The resulting possible KSAs for warped AdS$_3$ backgrounds that the authors of \cite{superalgebra} obtained have been tabulated in table \ref{table:AdS3KSA}.

\begin{table}
	\centering
	\caption[Killing superalgebras of warped AdS$_3$ backgrounds in $D=10,11$ supergravity]{Killing superalgebras of warped AdS$_3$ backgrounds in $D=10,11$ supergravity. $N_L$ denotes the number of left supersymmetries.}
	\begin{tabular}{cccc}
		\toprule
		$N_L$ & $\mathfrak{g}_L/\mathfrak{c}_L$ & $(\mathfrak{t}_L)_0/\mathfrak{c}_L$&$\mathrm{dim}\, \mathfrak{c}_L $ \\
		\midrule
		$2n$& $\mathfrak{osp}(n|2)$ & $\mathfrak{so}(n)$ & 0\\
		$4n,~n>2$ & $\mathfrak{sl}(n|2)$ & $\mathfrak{u}(n)$& 0 \\
		$8n, n>1$ & $\mathfrak{osp}(4|2n)$ & $\mathfrak{sp}(n) \oplus \mathfrak{sp}(1)$ &0\\
		16 & $\mathfrak{f}(4)$ & $\mathfrak{spin}(7)$ &0\\
		14 & $\mathfrak{g}(3)$ & $\mathfrak{g}_2$&0 \\
		8 & $\mathfrak{D}(2,1,\alpha)$ & $\mathfrak{so}(3) \oplus \mathfrak{so}(3)$&0 \\
		8 & $\mathfrak{sl}(2|2)/1_{4\times 4}$ & $\mathfrak{su}(2)$& $\leq 3 $\\
		\bottomrule
	\end{tabular}
	\label{table:AdS3KSA}
\end{table}

The AdS$_2$ Killing superalgebras may be identified with the left superalgebras $\mathfrak{g}_L$ in the AdS$_3$ case \cite{ads2}.

\section{Homogeneous Spaces}\label{homogeneousspaces}

In the following section, we shall collect some useful properties of homogeneous spaces which have facilitated our analysis of the AdS backgrounds in chapters \ref{ads4} and \ref{ads3}. A more detailed review can  be found in e.g. \cite{kobayashi, muellerhossein}.

Consider the left coset space $M=G/H$, where $G$ is a compact connected semisimple Lie group $G$ which acts  effectively from the left on $M=G/H$ and $H$ is a closed Lie subgroup of $G$. Let us denote the Lie algebras of $G$ and $H$ with  $\mathfrak{g}$ and $\mathfrak{h}$, respectively. As there is always an invariant inner product on  $\mathfrak{g}$, it can be used to take the orthogonal complement of  $\mathfrak{h}$ in  $\mathfrak{g}$ and so
\begin{align}
\mathfrak{g}=\mathfrak{h} \oplus \mathfrak{m}~.
\end{align}
Denote   the generators of  $\mathfrak{h}$ by $h_\alpha$, $\alpha=1,2,..., \dim{\mathfrak{h}}$  and a basis in $\mathfrak{m}$ as  $m_A$, $A=1,..., \dim{\mathfrak{g}}-\dim{\mathfrak{h}}$. In this basis,  the brackets of the Lie algebra $\mathfrak{g}$ take the following form
\begin{align}\label{commutation}
[h_\alpha, h_\beta] &= f_{\alpha\beta}{}^\gamma \, h_\gamma~,\quad
[h_\alpha, m_A] = f_{\alpha A}{}^B \, m_B~, \notag\\
[m_A,m_B] &= f_{AB}{}^C \, m_C + f_{AB}{}^\alpha \, h_\alpha~.
\end{align}
If $f_{AB}{}^C=0$, that is $[\mathfrak{m},\mathfrak{m}] \subset \mathfrak{h}$, the space is symmetric.

Let $g: U\subset G/H\rightarrow G$  be a local section of the coset. The decomposition of the Maurer-Cartan form in components along  $\mathfrak{h}$ and  $\mathfrak{m}$ is
\begin{align}
g^{-1} dg = \bbl^A \, m_A + \Omega^\alpha \, h_\alpha~,
\label{MC}
\end{align}
which  defines a local left-invariant frame $\bbl^A$  and a canonical left-invariant connection  $\Omega^\alpha$ on  $ G/H$. The curvature and torsion of the canonical connection are
\bea\label{dei}
&&R^\alpha \equiv d\Omega^\alpha+\frac12 f_{\beta\gamma}{}^\alpha \Omega^\beta\wedge \Omega^\gamma=-\frac12 f_{BC}{}^\alpha \bbl^B\wedge \bbl^C~,
\cr
&&T^A\equiv d\bbl^A+f_{\beta C}{}^A \Omega^\beta\wedge \bbl^C=-\frac12 f_{BC}{}^A \bbl^B\wedge \bbl^C~,
\eea
respectively, where the equalities follow after  taking the exterior derivative of (\ref{MC}) and  using (\ref{commutation}).  If $G/H$ is symmetric, then
the torsion vanishes.

A left-invariant   $p$-form $\omega$ on $ G/H$ can be written as
\begin{align}
\omega = \frac{1}{p!} \, \omega_{A_1 ... A_p} \, \bbl^{A_1} \wedge ... \wedge \bbl^{A_p}~,
\end{align}
where the components   $\omega_{A_1...A_p}$ are constant and satisfy
\begin{align}\label{hinvariance}
f_{\alpha[A_1}{}^B \, \omega_{A_2...A_p]B} =0~.
\end{align}
The latter condition is required for invariance under the right action of $H$ on $G$. All left-invariant forms are  parallel with respect to the canonical connection.

It remains to describe the metrics of $G/H$, which are  left-invariant. These are written as
\bea
ds^2=g_{AB}\, \bbl^A \bbl^B~,
\eea
where the components $g_{AB}$ are constant and satisfy
\begin{align}\label{ginvar}
f_{\alpha A}{}^C \, g_{BC} + f_{\alpha B}{}^C \, g_{AC} = 0~.
\end{align}
For symmetric spaces, the canonical connection coincides with the Levi-Civita connection of the invariant metrics. So all non-vanishing  left-invariant forms are harmonic and
represent non-trivial elements in the de Rham cohomology of  $G/H$.
However, if $G/H$ is strictly homogeneous, this is not the case, since the canonical connection has non-vanishing torsion.

Suppose $G/H$ is homogeneous and equipped with an invariant metric $g$. To describe the results in this thesis, we require the Levi-Civita connection of $g$ and its curvature.
Let $\Phi$ be the Levi-Civita connection in the left-invariant frame. As the difference of two connections is a tensor, we set
\begin{align}
    \Phi{}^A{}_B= \Omega^\alpha f_{\alpha B}{}^A+\bbl^C Q_{C,}{}^A{}_B~.
\end{align}
As $\Phi$ is metric and torsion free, we have
\begin{align}\label{levi-civ}
    \Phi_{AB} + \Phi_{BA} &= 0~,\notag \\
    d\bbl^A + \Phi^A{}_B \, \wedge \, \bbl^B &=0~.
\end{align}
These equations can be solved for $Q$ to find that
\begin{align}
\Phi^A{}_B =  \Omega^\alpha\, f_{\alpha B}{}^A + \frac{1}{2} \left( g^{AD} \, f_{DB}{}^E \, g_{CE} +  g^{AD} \, f_{DC}{}^E \, g_{BE} - f_{BC}{}^A\right) \, \bbl^C~.
\end{align}
In turn, the Riemann  curvature 2-form $R^A{}_B$ is
\begin{align}
R^A{}_B = \frac{1}{2} \left( Q_{C,}{}^A{}_{E} \, Q_{D,}{}^E{}_{B} - Q_{D,}{}^A{}_{E} \, Q_{C,}{}^E{}_{B} - Q_{E,}{}^A{}_{B} \, f_{CD}{}^E - f_{CD}{}^\alpha \, f_{\alpha B}{}^A \right) \, \bbl^C \wedge \bbl^D~.
\end{align}
This  is required for the investigation of the gravitino KSE.  Note that the expression for $\Phi^A{}_B$ is considerably  simplified whenever the coset space is naturally reductive, because in that case the structure constants $f_{ABC}=f_{AB}{}^E \, g_{CE}$ are skew symmetric.

\chapter{\texorpdfstring{AdS$_4$}{AdS4} Backgrounds with \texorpdfstring{$N > 16$}{N greater 16} Supersymmetries in 10 and 11 Dimensions}\label{ads4}

In this chapter, we explore all warped AdS$_4\times_w M^{D-4}$ backgrounds with the most general allowed fluxes that preserve more than 16 supersymmetries  in $D=10$- and $11$-dimensional  supergravities. Assuming that either the internal space $M^{D-4}$
  is compact without boundary or that the isometry algebra of the background decomposes into that of AdS$_4$ and that of $M^{D-4}$, we find that
there are no such backgrounds in IIB supergravity.  Similarly in IIA supergravity,  there is a unique such background with 24 supersymmetries locally isometric to
AdS$_4\times \mathbb{CP}^3$, and  in $D=11$ supergravity  all such backgrounds are locally isometric to the maximally supersymmetric
AdS$_4\times S^7$ solution. This work was conducted in collaboration with Alexander Haupt and George Papadopoulos and was published in \cite{ads4Ngr16}.

\section{Introduction}\label{ads4introduction}

AdS backgrounds in 10 and 11 dimensions that preserve $N$ supersymmetries with $N>16$
have found widespread applications both in supergravity compactifications
and in the AdS/CFT correspondence, for reviews see \cite{duff, maldacena}
and references therein. One of the features of such backgrounds in AdS/CFT \cite{maldacenab}
  is that the CFT R-symmetry group  acts transitively on the internal space of the solution and  this can be used
  to establish the dictionary between some of the operators of the CFT
and spacetime Kaluza-Klein fields \cite{witten}. Therefore the question arises whether it is possible to find all such AdS solutions. Despite the progress that has been made over the years, a complete description of all AdS solutions that
preserve $N>16$ supersymmetries remained an open problem until recently \cite{ads5clas,ads6,ads4Ngr16,ahslgp2,ads2}.

There have been several significant developments which facilitated progress in this direction for a large class
of warped flux AdS solutions. In \cite{mads, iibads, iiaads},  the Killing spinor equations (KSEs) of supergravity theories  have been solved in all generality and the fractions of supersymmetry preserved  by all warped flux AdS backgrounds have been identified.  Furthermore,  global analysis techniques have also been introduced
in the investigation of AdS backgrounds which can be used to  a priori impose properties like the compactness of the internal space and the smoothness of the fields.  Another key development was the proof of the
homogeneity theorem (cf.\ section \ref{homogentheorem}) \cite{homogen} which for the special case of AdS backgrounds states  that all such
backgrounds preserving $N>16$ supersymmetries are Lorentzian homogeneous spaces.

The maximally supersymmetric AdS backgrounds\footnote{The maximally supersymmetric AdS backgrounds are of the type
considered previously in \cite{fr}.} have been classified  in \cite{maxsusy} and it has been found that they are locally isometric
to the  $\text{AdS}_4\times S^7$ \cite{duffpopemax} and  $\text{AdS}_7\times S^4$ \cite{townsend} solutions of  11-dimensional supergravity,  and to the $\text{AdS}_5\times S^5$ solution of  IIB supergravity, see \cite{Schwarz:1983qr} and comment within. There are no  $\text{AdS}_7$ backgrounds
that preserve $16<N<32$ supersymmetries \cite{mads, iibads, iiaads} and no smooth $\text{AdS}_6$ backgrounds that preserve $N>16$ supersymmetries with compact without boundary  internal space \cite{ads6}.  More recently, it has been demonstrated under the same assumptions on the internal space that there are
no smooth  $\text{AdS}_5$ backgrounds  that preserve $16<N<32$ \cite{ads5clas};  see \cite{ferrara, aharony, garcia} for applications to AdS/CFT. Moreover,
it has been shown in \cite{ads2} that there are no smooth AdS$_2$ backgrounds in 10- and 11-dimensional supergravities with compact without boundary internal space that preserve $N>16$
supersymmetries.  Product solutions $\text{AdS}_n\times M^{D-n}$  with $M^{D-n}$ a symmetric space have been classified   in \cite{figueroaa, figueroab, Wulffa, Wulffb}. Furthermore
the geometry of all heterotic AdS$_3$ backgrounds has been investigated  in \cite{adshet} and it has been found that there are no solutions that preserve  $N>8$ supersymmetries.


 The main task in this chapter is to describe all warped  AdS$_4$ backgrounds that admit the most general   fluxes in 10 and 11 dimensions
and preserve more than 16 supersymmetries. It has been shown in  \cite{mads, iibads, iiaads} that such backgrounds preserve $4k$ supersymmetries. Therefore, we shall investigate
the backgrounds preserving 20, 24 and  28 as those with  32 supersymmetries have already been classified in \cite{maxsusy}. In particular, we find that
\begin{itemize}
\item  IIB and massive IIA supergravity do not admit  AdS$_4$ solutions with $N>16$ supersymmetries.
\item  Standard  IIA supergravity admits a unique  solution up to an overall scale preserving 24 supersymmetries,   locally isometric to the AdS$_4\times \mathbb{CP}^3$
background of  \cite{nillsonpope}.
\item  All  AdS$_4$ solutions of 11-dimensional supergravity that preserve  $N>16$  supersymmetries are locally isometric to the maximally
supersymmetric AdS$_4\times S^7$ solution of  \cite{fr, duffpopemax}.
\end{itemize}

These results have been established under certain  assumptions\footnote{Some assumptions are necessary to exclude the possibility that a warped  AdS$_4$
 background is not locally isometric to  an AdS$_n$ background with $n>4$. This has been observed in \cite{strominger} and explored in the context of KSEs in \cite{desads}.}. We begin with a spacetime which is a warped product $AdS_4\times_w M^{D-4}$, for $D=10$ or $11$, and allow for all fluxes which are invariant  under  the isometries of AdS$_4$. Then
we shall assume  that
\begin{enumerate}
\item  either the solutions are smooth and  $M^{D-4}$ is compact without boundary
\item  or that the even part of the Killing  superalgebra of the background
decomposes as a direct sum $\mathfrak{so}(3,2)\oplus \mathfrak{t}_0$, where $\mathfrak{so}(3,2)$ is the Lie algebra of isometries of AdS$_4$ and $\mathfrak{t}_0$ is the Lie algebra of the isometries of $M^{D-4}$.
\end{enumerate}
It has been shown in \cite{superalgebra} that for all AdS backgrounds, the first assumption implies the second. In addition for $N>16$  AdS$_4$ backgrounds\footnote{In what follows,
we use ``$N>16$ AdS backgrounds'' instead of ``AdS backgrounds that preserve $N>16$ supersymmetries'' for short.}, the second assumption implies the first. This is because  $\mathfrak{t}_0$ is the Lie algebra of a compact group and all internal
 spaces are compact without boundary. Smoothness also follows from only considering invariant solutions.

The proof of the main statement of this chapter is firstly based on the results of \cite{mads, iibads, iiaads} that the number of supersymmetries preserved by AdS$_4$ backgrounds are $N=4k$
 and so the solutions under consideration preserve  20, 24, 28 and 32 supersymmetries. Then the homogeneity theorem of \cite{homogen} implies that all such backgrounds
 are Lorentzian homogeneous spaces. Moreover, it has been shown in  \cite{superalgebra} under the assumptions mentioned above that the Killing
superalgebra of warped AdS$_4$ backgrounds preserving $N=4k$ supersymmetries  is isomorphic to $\mathfrak{osp}(N/4 \vert4)$, see also \cite{charles},  and that the even subalgebra $\mathfrak{osp}(N/4 \vert4)_0=\mathfrak{so}(3,2)\oplus \mathfrak{so}(N/4)$ acts  effectively on the spacetime, with $\mathfrak{t}_0=\mathfrak{so}(N/4)$ acting on the internal space.
Thus, together with the homogeneity theorem $\mathfrak{osp}(N/4 \vert4)_0$ acts both {\sl transitively} and {\sl effectively} on the spacetime.
Then we demonstrate in all cases that the warp factor $A$ is constant. As a result all $N>16$  AdS$_4$ backgrounds
are product spaces $AdS_4\times M^{D-4}$.  So the internal space $M^{D-4}$ is a homogeneous space, $M^{D-4}=G/H$, and $\mathfrak{Lie}\, G= \mathfrak{so}(N/4)$.
Therefore, we  have demonstrated the following,
\begin{itemize}
\item The internal spaces of AdS$_4$ backgrounds that preserve $N>16$ supersymmetries are homogeneous spaces admitting a transitive and effective action of a group $G$ with  $\mathfrak{Lie}\, G= \mathfrak{so}(N/4)$.
\end{itemize}
Having established this,
one can use the classification of   \cite{ Castellani:1983yg, klausthesis, niko6dim, niko7dim} to identify all the 6- and 7-dimensional homogeneous spaces that can occur as internal spaces
for $N>16$  AdS$_4$ backgrounds, see also tables\footnote{These tables list the simply connected homogeneous spaces. This suffices for our purpose because we are investigating the  geometry of the backgrounds up to local isometries.   As $\mathfrak{so}(N/4)$ is simple the universal cover of $G/H$ with $\mathfrak{Lie}(G)=\mathfrak{so}(N/4)$ is compact and homogeneous, see e.g.\ \cite{bohmkerr}. So  the internal space can be identified with the universal cover $\tilde G/\tilde H$ of $G/H$ for which $\tilde G$ can be chosen to be simply connected. }
\ref{table:nonlin} and \ref{table:nonlin7}.  Incidentally, this also means that if $N>16$ backgrounds were to exist, the R-symmetry group
of the dual CFT would have to act transitively on the internal space of the solution.


 With regards to the classification  of 6-dimensional homogeneous spaces $G/H$ in table \ref{table:nonlin}, one finds that the a priori possible candidates for internal spaces of AdS$_4$ backgrounds with $N>16$ in 10 dimensions are
\bea
&&\mathrm{Spin}(7)/\mathrm{Spin}(6)~(N=28)~,~~~SU(4)/S(U(1)\times U(3))~(N=24)~,~~~
\cr
&&Sp(2)/U(2)~(N=20)~,~~~Sp(2)/(Sp(1)\times U(1)) ~(N=20)~,
\label{homo166}
\eea
where $N$ denotes the expected number of supersymmetries that can be preserved by the background, and we always take $G$ to be simply connected.  Observe  that there are no maximally supersymmetric AdS$_4$
solutions in 10-dimensional supergravities in agreement with the results of \cite{maxsusy}.
The proof of our result in IIB supergravity is based on a cohomological argument and does not use details of the 6-dimensional homogeneous spaces involved. However in (massive) IIA
supergravity, one has to consider details of the geometry of these coset spaces.
 Solutions with strictly $N=28$ and $N=20$ supersymmetries are ruled out
after a detailed analysis of the KSEs and dilaton field equation. In standard IIA supergravity there is a solution with 24 supersymmetries and an internal space locally isometric to the symmetric space
$SU(4)/S(U(1)\times U(3))=\mathbb{CP}^3$.  This solution   has already been  found in \cite{nillsonpope}.  The homogeneous space $Sp(2)/Sp(1)\times U(1)$, which is diffeomorphic to $\mathbb{CP}^3$, also gives rise to a solution in a special
region  of the moduli space of parameters.  This solution  admits $24$ supersymmetries and  is locally isometric to that with internal space $SU(4)/S(U(1)\times U(3))$.


The classification  of 7-dimensional  homogeneous spaces $G/H$ in table  \ref{table:nonlin7}  reveals that possible candidates for internal spaces of $N>16$ AdS$_4$ backgrounds in 11 dimensions
 are
\bea
&&\mathrm{Spin}(8)/\mathrm{Spin}(7)~(N=32)~,~~\mathrm{Spin}(7)/G_2~(N=28)~,~~SU(4)/SU(3)~(N=24)~,~~~
\cr
&&Sp(2)/Sp(1)_{\text{max}}~(N=20)~, ~~Sp(2)/\Delta(Sp(1))~(N=20)~,~~
\cr
&&Sp(2)/Sp(1)~(N=20)~,
\label{homo167}
\eea
where $Sp(1)_{\mathrm{max}}$  and $\Delta(Sp(1))$ denote the maximal   and diagonal embeddings of $Sp(1)$ in $Sp(2)$, respectively, and $G$ is chosen to be simply connected.
It is known that there is a maximally supersymmetric solution AdS$_4\times S^7$ with internal
space $S^7=\mathrm{Spin}(8)/\mathrm{Spin}(7)$ \cite{fr, duffpopemax}.   After a detailed investigation of the
geometry of the above homogeneous spaces, the solutions of the KSEs and the warp factor field equation, one can also show that the rest of the coset spaces  do not give solutions with strictly 20, 24 and 28 supersymmetries. However, as the homogeneous spaces $Spin(7)/G_2$, $SU(4)/SU(3)$
 and $Sp(2)/Sp(1)$ are diffeomorphic to $S^7$, there is a region in the moduli space of their
  parameters which yields the maximally supersymmetric AdS$_4\times S^7$ solution.

This chapter is organised as follows. In section \ref{ads4iib}, we  show that there are no IIB $N>16$ $AdS_4 \times _w M^6$ solutions. In section \ref{ads4IIA}, we show that up to an overall scale there is a unique solution in IIA supergravity that preserves 24 supersymmetries.  In section \ref{ads411d}, we demonstrate that all $N>16$ AdS$_4$ backgrounds of 11-dimensional supergravity are locally
isometric to the maximally supersymmetric $AdS_4\times S^7$ solution. In section \ref{ads4conclusions} we state our conclusions. Our conventions may be found in appendix \ref{conventions}. Details of the geometry of homogeneous spaces admitting a transitive action of one of the groups with Lie algebra  $\mathfrak{su}(k)$ or $\mathfrak{so}(5)=\mathfrak{sp}(2)$ are outlined in appendices \ref{su(k)}, \ref{bergerspace} and \ref{so5=sp2}.

\section{\texorpdfstring{$N>16$ $AdS_4 \times_w M^6$}{AdS4xM6} solutions in IIB }\label{ads4iib}

To investigate IIB AdS$_4$ backgrounds, we shall use the approach and notation of \cite{iibads} where Bianchi identities,
  field equations and KSEs are first solved along the AdS$_4$ subspace of $AdS_4\times_w M^6$ and then
  the remaining independent conditions along the internal space $M^6$ are identified.   The bosonic fields of IIB supergravity are the metric,
  a complex 1-form field strength $P$, a complex 3-form field strength $G$ and a real  self-dual 5-form $F$. Imposing the symmetry of AdS$_4$ on the fields, one finds  that the metric and form field strengths are given by
\begin{align}
&ds^2 = 2 du (dr+rh) + A^2 (dz^2 +e^{2z/\ell}dx^2) + ds^2(M^6)~, \notag \\
&G = H, \quad P =\xi,  \quad F = A^2 e^{z/\ell} du \wedge (dr + rh) \wedge dz\wedge dx \wedge Y + *_6 Y~,
\end{align}
where the metric has been written as a near-horizon geometry \cite{adshor} with
\begin{align}
h = -\frac{2}{\ell} dz - 2 A^{-1} dA~.
\end{align}
The warp factor $A$ is a function on the internal manifold $M^6$, $H$ is the complex 3-form on $M^6$, $\xi$ is a complex 1-form on $M^6$ and $Y$ is a real 1-form on $M^6$. The $AdS_4$ coordinates are $(u, r, z, x)$ and we introduce the null orthonormal  frame
\begin{align}\label{frameiib}
\bbe^+ = du~, \quad \bbe^- = dr + r h~, \quad \bbe^z = A\, dz~, \quad \bbe^x = A e^{z/\ell}\, dx~, \quad \bbe^i= \bbe^i_I\, dy^I~,
\end{align}
where $ds^2(M^6)=\delta_{ij} \bbe^i \bbe^j$. All gamma matrices are taken with respect to this null orthonormal  frame.

The  Bianchi identities along $M^6$ which are useful
in the  analysis that follows are
\begin{align} \label{bianchi}
d(A^4Y) &= 0, \quad dH = iQ \wedge H - \xi \wedge \overline{H}, \notag\\
\nabla^i Y_i &= -\frac{i}{288} \epsilon^{i_1 i_2 i_3 j_1 j_2 j_3} H_{i_1 i_2 i_3} \overline{H}_{j_1 j_2 j_3}~, \notag \\
dQ &= -i \xi \wedge \bar{\xi}~,
\end{align}
where $Q$ is the pull-back of the canonical connection of the upper-half plane on the spacetime with respect to the dilaton and axion scalars of IIB supergravity. Similarly, the field equations of the warp factor is
\begin{align}
A^{-1}\nabla^2 A = 4Y^2 + \frac{1}{48}  H_{i_1i_2i_3} \overline{ H}^{i_1i_2i_3} - \frac{3}{\ell^2} A^{-2} - 3 A^{-2} (dA)^2~,
\label{einsteinA}
\end{align}
and those of the scalar and 3-form fluxes are
\begin{align}\label{feqn}
\nabla^i \xi_i &= -3 \partial^i \log A \, \xi_i + 2i Q^i \xi_i - \frac{1}{24} H^2~, \notag \\
\nabla^i H_{ijk} &= -3 \partial^i \log A \, H_{ijk} + i Q^i H_{ijk} + \xi^i \overline{H}_{ijk}~.
\end{align}
The full set of Bianchi identities and field equations can be found in \cite{iibads}. Note in particular that \eqref{einsteinA} implies that if $A$ and the other fields are smooth, then $A$ is nowhere vanishing on $M^6$.

\subsection{The Killing spinors}

After solving the KSEs along AdS$_4$, the Killing spinors of the background can be written as
\begin{align}\label{killingspinorsiib}
\epsilon  = \,&\sigma_+ - \ell^{-1} x \Gamma_{xz} \tau_+ + e^{-\frac{z}{\ell}} \tau_+ + \sigma_- + e^{\frac{z}{\ell}} ( \tau_- - \ell^{-1} x \Gamma_{xz} \sigma_-) \notag\\
& - \ell^{-1} u A^{-1} \Gamma_{+z} \sigma_- - \ell^{-1} r A^{-1} e^{-\frac{z}{\ell}} \Gamma_{-z} ~ \tau_+~,
\end{align}
where we have used the light-cone projections
\begin{align}
\Gamma_\pm \sigma_\pm = 0~, \quad \Gamma_\pm \tau_\pm = 0~,
\end{align}
and $\sigma_\pm$ and $\tau_\pm$ are $Spin(9,1)$ Weyl spinors depending only on the coordinates of $M^6$.
The remaining independent KSEs are
\begin{align} \label{kseiib1}
\nabla^{(\pm)}_i \sigma_\pm = 0~, \quad \nabla^{(\pm)}_i \tau_\pm = 0~,
\end{align}
and
\begin{align}\label{ads4algkseiib2}
\left(\frac{1}{24} \sH + \slashed{\xi} C* \right) \sigma_\pm = 0~, \quad \left(\frac{1}{24} \sH + \slashed{\xi} C* \right) \tau_\pm = 0~,
\end{align}
as well as
\begin{align} \label{ads4algkseiib}
\Xi^{(\pm)} \sigma_\pm = 0~, \quad \left(\Xi^{(\pm)} \pm \frac{1}{\ell} \right) \tau_\pm = 0~,
\end{align}
where
\begin{align}
\nabla^{(\pm)}_i &= \nabla_i \pm \frac{1}{2} \partial_i \log A - \frac{i}{2}Q_i \mp \frac{i}{2} \sgY_i \Gamma_{xz} \pm \frac{i}{2} Y_i \Gamma_{xz}  \notag\\
&+\left(-\frac{1}{96} \sgH_i + \frac{3}{32} \sH_i \right) C *~, \\
\Xi^{(\pm)} &= \mp \frac{1}{2\ell} - \frac{1}{2} \Gamma_z \slashed{\partial}A \pm \frac{i}{2} A \Gamma_x \sY + \frac{1}{96} A \Gamma_z \sH C*~,
\end{align}
and $C*$ is the charge conjugation matrix followed by standard complex conjugation. For further explanation of the notation see appendix \ref{conventions}. Equations \eqref{kseiib1} and \eqref{ads4algkseiib2} can be thought of as
   the naive restriction of gravitino and dilatino KSEs of IIB supergravity  on $M^6$, respectively. The equations \eqref{ads4algkseiib} are algebraic and arise as integrability conditions of the integration of the IIB KSEs over the AdS$_4$ subspace of the background. We do not assume that the Killing spinors factorize as Killing spinors on AdS$_4$ and Killing spinors on the internal manifold. It has been observed in \cite{iibads} that if $\sigma_+$ is a Killing spinor,  then
\bea
\tau_+=\Gamma_{zx} \sigma_+~, ~~~\sigma_-=A\Gamma_{-z}\sigma_+~,~~~\tau_-=A \Gamma_{-x}\sigma_+~,
\label{ksrel}
\eea
are also Killing spinors. As a result  AdS$_4$ solutions  preserve $4k$ supersymmetries.


\subsection{The non-existence of \texorpdfstring{$N>16$ $AdS_4$}{N greater 16 AdS4} solutions in IIB}

\subsubsection{Conditions on spinor bilinears}

As it has already been mentioned, the two assumptions we have made in section \ref{ads4introduction} are equivalent for all   IIB, (massive) IIA and 11-dimensional AdS$_4$ backgrounds that preserve $N>16$ supersymmetries.
   Hence in what follows, we shall focus only on the restrictions  on the geometry of the spacetime imposed  by the first assumption which requires that the solutions are smooth
   and the internal space is compact without boundary.

To begin our analysis, note that a consequence of the homogeneity theorem  \cite{homogen} for solutions which preserve $N>16$  supersymmetries is that the IIB
 scalars are constant which in turn implies
\begin{align}
\xi=0~.
\end{align}
As $Q$ is the pull-back of the canonical connection of the upper half plane with respect to the scalars and these are constant,  $Q=0$ as well.

Setting  $\Lambda = \sigma_+ + \tau_+$ and after using the gravitino KSE \eqref{kseiib1}, we find
\begin{align}\label{ads4hopf1}
\nabla_i \parallel \Lambda \parallel^2 = - \parallel \Lambda \parallel^2 A^{-1} \nabla_i A - i Y_i \langle \Lambda, \Gamma_{xz} \Lambda \rangle + \frac{1}{48} \text{Re} \langle \Lambda, \sgH_i C* \Lambda \rangle~.
\end{align}
Next, observe that the algebraic KSE \eqref{ads4algkseiib} implies
\begin{align}
\frac{1}{48}\sH C * \Lambda = \left(A^{-1} \Gamma^j \nabla_j A + i \Gamma^j \Gamma_{xz} Y_j \right) \Lambda + \ell^{-1} A^{-1} \Gamma_z (\sigma_+ - \tau_+)~,
\end{align}
which, when substituted back into \eqref{ads4hopf1}, yields
\begin{align}\label{hopf2}
\nabla_i \parallel \Lambda \parallel^2 = 2 \ell^{-1} A^{-1} \text{Re} \langle \tau_+, \Gamma_{iz} \sigma_+ \rangle~.
\end{align}
However, the gravitino KSE \eqref{kseiib1} also implies that
\begin{align}
\nabla^i \left( A \text{Re} \langle \tau_+, \Gamma_{iz} \sigma_+ \rangle \right) = 0~.
\label{iibortho1}
\end{align}
Thus, in conjunction with \eqref{hopf2}, we obtain
\begin{align}
\nabla^2 \parallel \Lambda \parallel^2 + 2 A^{-1} \nabla^i A \nabla_i \parallel \Lambda \parallel^2 = 0~.
\end{align}
The Hopf maximum principle then implies that $\parallel \Lambda \parallel^2$ is constant, so  \eqref{ads4hopf1} and \eqref{hopf2} give the conditions
\begin{align}\label{hopf1.1}
- \parallel \Lambda \parallel^2 A^{-1} \nabla_i A - i Y_i \langle \Lambda, \Gamma_{xz} \Lambda \rangle + \frac{1}{48} \text{Re} \langle \Lambda, \sgH_i C* \Lambda \rangle = 0~,
\end{align}
and
\begin{align}\label{hopf2.1}
\text{Re} \langle \tau_+, \Gamma_{iz} \sigma_+ \rangle = 0~,
\end{align}
respectively. The above equation can equivalently be written as $\text{Re} \langle \sigma_+, \Gamma_{ix} \sigma_+ \rangle = 0$.

The spinors $\sigma_+$ and $\tau_+$ are linearly independent as can easily be seen from \eqref{ads4algkseiib}.  Moreover as a consequence of (\ref{hopf2.1}),  they are orthogonal
\begin{align}
\text{Re} \langle \tau_+, \sigma_+ \rangle = 0~.
\label{iibortho2}
\end{align}
 To see this take the real part of   $\langle \tau_+, \Xi^{(+)} \sigma_+ \rangle -\langle \sigma_+, (\Xi^{(+)}+ \ell^{-1}) \tau_+ \rangle = 0$.  The conditions (\ref{iibortho1}), (\ref{iibortho2}) as well as $\parallel\Lambda\parallel$ being constant can also be derived from the assumption that the
 isometries of the background decompose into those of AdS$_4$ and those of the internal manifold \cite{superalgebra}.

\subsubsection{The warp factor is constant and the 5-form flux vanishes}

  AdS$_4$ backgrounds preserving $4k$ supersymmetries admit $k$ linearly independent
Killing spinors $\sigma_+$.  For every pair of such spinors $\sigma^1_+$ and $\sigma^2_+$ define the bilinear
\begin{align}\label{KV}
W_i = A \, \text{Re} \langle \sigma^1_+, \Gamma_{iz} \sigma^2_+ \rangle~.
\end{align}
Then the gravitino KSE \eqref{kseiib1} implies that
\begin{align}
\nabla_{(i}W_{j)}=0~.
\end{align}
Therefore W is a Killing vector on $M^6$.

Next consider the algebraic KSE (\ref{ads4algkseiib}) and take the real part of $\langle \sigma^1_+, \Xi^{(+)} \sigma^2_+ \rangle -\langle \sigma^2_+, \Xi^{(+)} \sigma^1_+ \rangle = 0$
to find that
\begin{align}\label{iib44}
	W^i \, \nabla_i A =0~,
\end{align}
where we have used (\ref{hopf2.1}).

Similarly, taking the real part of the difference  $\langle \sigma^1_+, \Gamma_{zx}\Xi^{(+)} \sigma^2_+ \rangle -\langle \sigma^2_+, \Gamma_{zx}\Xi^{(+)} \sigma^1_+ \rangle = 0$
and after using the condition (\ref{iibortho2}), we find
\begin{align}
i_W Y=0~.
\label{iib44a}
\end{align}

The conditions (\ref{iib44}) and (\ref{iib44a}) are valid for  all IIB AdS$_4$ backgrounds.
However if the solution preserves more than 16 supersymmetries,  an argument similar to the one used for the proof of the homogeneity theorem in \cite{homogen} implies that
the Killing vectors $W$ span the tangent spaces of $M^6$ at each point.  As a result, we conclude that
\begin{align}\label{iibvan}
dA=Y=0~.
\end{align}
Therefore the warp factor $A$ is constant and the 5-form flux $F$ vanishes. So, the background is merely a product $AdS_4\times M^6$, and, as was pointed out in section \ref{ads4introduction}, $M^6$ is one of
the homogeneous spaces in \eqref{homo166}.

\subsubsection{Proof of the main statement}
First of all, it has been shown in  \cite{Gran:2009cz} that all IIB AdS backgrounds preserving $N\geq 28$ supersymmetries are locally isometric to the maximally supersymmetric ones.
As there is no maximally supersymmetric AdS$_4$ background in IIB, we conclude that AdS$_4$ solutions which preserve $N\geq 28$ supersymmetries do not exist.

To investigate the $N=20$ and $N=24$ cases, we substitute (\ref{iibvan}) into the Bianchi identities and field equations to find that  $H$ is harmonic and
\begin{align}\label{Hsqu}
H^2 = 0~.
\end{align}
If $H$ were real, this condition  would have implied $H=0$ and, in turn, would have led to a contradiction.  This is  because  the field equation for the warp factor \eqref{einsteinA}  cannot be satisfied.  Thus, we can already exclude the existence of such backgrounds.

Otherwise for solutions to exist,  $M^6$ must be a compact, homogeneous,  six dimensional Riemannian manifold whose de-Rham cohomology  $H^3(M^6)$ has at least two generators and which admits a transitive and effective action of a group with Lie algebra isomorphic to either  $\mathfrak{so}(6)$ or $\mathfrak{so}(5)$ for  $N=24$ and $N=20$, respectively \cite{superalgebra}.
The homogeneous spaces that admit a transitive and effective  action of $\mathfrak{so}(6)$ or $\mathfrak{so}(5)=\mathfrak{sp}(2)$ have already been listed in (\ref{homo166}) and none of them satisfy these
 cohomology criteria. All compact homogeneous 6-manifolds have been classified in \cite{niko6dim} and the complete list of the simply connected ones relevant here is given in table~\ref{table:nonlin}.  Therefore, we conclude that there are no AdS$_4$ backgrounds preserving $N>16$ supersymmetries in IIB supergravity\footnote{Note that the possibility of IIB $AdS_4\times Z\backslash G/H$  backgrounds preserving $N>16$ supersymmetry is  also excluded, where $Z$
 is a discrete subgroup of $G$,  as there are  no   IIB  $AdS_4\times  G/H$ local geometries that  preserve $N>16$ supersymmetries.}.

\begin{table}\renewcommand{\arraystretch}{1.3}
	\caption{6-dimensional compact, simply connected,  homogeneous spaces}
	\centering
	\begin{tabular}{c l}
		\hline
		& $M^6=G/H$  \\  
		\hline
		(1)& $\frac{\mathrm{Spin}(7)}{\mathrm{Spin}(6)}= S^6$, symmetric space\\
        (2) & $\frac{G_2}{SU(3)}$ diffeomorphic to $S^6$\\
		(3)&$\frac{SU(4)}{S(U(1)\times U(3))}=\mathbb{CP}^3$, symmetric space\\
		(4)& $\frac{Sp(2)}{U(2)}$, symmetric space \\
		(5) & $\frac{Sp(2)}{Sp(1)\times U(1)}$ diffeomorphic to $\mathbb{CP}^3$  \\
(6) & $\frac{SU(3)}{T_{max}}$ Wallach space\\
		(7) & $\frac{SU(2)\times SU(2)}{\Delta(SU(2))} \times \frac{SU(2)\times SU(2)}{\Delta(SU(2))} = S^3 \times S^3 $\\
		(8) & $SU(2) \times \frac{SU(2) \times SU(2)}{\Delta(SU(2))}$  diffeomorphic to $S^3\times S^3$\\
        (9) & $SU(2) \times SU(2) $ diffeomorphic to $S^3\times S^3$ \\
(10) & $\frac{SU(2)}{U(1)} \times \frac{SU(2)}{U(1)} \times \frac{SU(2)}{U(1)}= S^2 \times S^2 \times S^2$\\
		(11) & $\frac{SU(2)}{U(1)} \times\frac{\mathrm{Spin}(5)}{\mathrm{Spin}(4)} = S^2 \times S^4$  \\
(12) & $\frac{SU(2)}{U(1)} \times\frac{SU(3)}{S(U(1)\times U(2))} = S^2 \times \mathbb{CP}^2$  \\
 [1ex]
		\hline
	\end{tabular}
	\label{table:nonlin}
\end{table}

\section{\texorpdfstring{$N>16$ $AdS_4 \times_w M^6$}{AdS4xM6} solutions  in (massive) IIA }\label{ads4IIA}

To begin, let us summarise the solution of Bianchi identities, field equations and KSEs for (massive) IIA  $AdS_4 \times_w M^6$ backgrounds as presented in \cite{iiaads}, whose notation we follow.   The bosonic fields
of (massive) IIA supergravity are  the metric,  a 4-form field strength $G$, a 3-form field strength $H$, a 2-form field strength $F$, the dilaton $\Phi$ and  the mass parameter $S$ of massive IIA dressed with the dilaton. Imposing the symmetries of AdS$_4$ on the fields, one finds that
\begin{align}
ds^2&=2 \bbe^+ \bbe^- + (\bbe^z)^2 + (\bbe^x)^2 + ds^2(M^6)~, \quad G = X \bbe^+ \wedge \bbe^- \wedge \bbe^z \wedge \bbe^x + Y~, \notag\\
\quad H &=H~, \quad F=F, \quad \Phi=\Phi~, \quad S=S~,
\end{align}
where $ds^2(M^6)= \delta_{ij} \bbe^i \bbe^j$ and the frame $(\bbe^+, \bbe^-, \bbe^x, \bbe^z, \bbe^i)$ is defined as in \eqref{frameiib}. Note that the fields $H$, $F$, $\Phi$ and $S$ do not have a component along AdS$_4$ and so we use the same symbol to denote them and their component along $M^6$.   The warp factor $A$, $S$ and $X$ are functions of $M^6$, whereas $Y$, $H$ and $F$ are 4-form, 3-form and 2-form fluxes on $M^6$, respectively.  The  conditions imposed on the fields by the Bianchi identities and field equations after solving along the AdS$_4$ subspace can  be found in \cite{iiaads}. Relevant to the analysis that follows are the Bianchi identities
\begin{align}\label{iiabianchi}
dH&=0, \quad dS=S d\Phi~, \quad dY= d\Phi \wedge Y+ H\wedge F~,\notag\\
 dF&= d\Phi \wedge F + SH~, \quad d(A^4 X)= A^4 d\Phi~,
\end{align}
and the field equations for the fluxes
\begin{align}\label{ads4iiafieldeqs}
\nabla^2 \Phi &= -4A^{-1} \partial^i  A \, \partial_i \Phi + 2 (d\Phi)^2 + \frac{5}{4} S^2 + \frac{3}{8} F^2 -\frac{1}{12} H^2 + \frac{1}{96} Y^2 - \frac{1}{4} X^2~,\notag\\
\nabla^k H_{ijk} &= -4A^{-1} \partial^k  A \, H_{ijk} + 2 \partial^k\Phi H_{ijk} + S F_{ij} +\frac{1}{2} F^{k\ell} Y_{ijk\ell}~, \notag\\
\nabla^{j} F_{ij} &= -4A^{-1} \partial^j  A \, F_{ij} + \partial^j \Phi \, F_{ij} - \frac{1}{6} H^{jkl} Y_{ijkl}~, \notag\\
\nabla^{\ell} Y_{ijk\ell} &= -4A^{-1} \partial^\ell  A \, Y_{ijk\ell} + \partial^\ell \Phi Y_{ijk\ell}~,
\end{align}
along $M^6$. Moreover, we shall use  the field equation for the warp factor $A$ and the Einstein field
equation along $M^6$
\begin{align}\label{ads4einstiia}
\nabla^2 \log A &= -\frac{3}{\ell^2 A^2} - 4 (d \log A)^2 + 2 \, \partial_i \log A \partial^i \Phi + \frac{1}{96} Y^2 + \frac{1}{4} X^2 + \frac{1}{4} S^2 + \frac{1}{8} F^2~, \notag\\
R^{(6)}_{ij} &= 4 \nabla_i \partial_j \log A + 4 \partial_i \log A \, \partial_j \log A + \frac{1}{12} Y^2_{ij} - \frac{1}{96} Y^2 \delta_{ij} + \frac{1}{4} X^2 \delta_{ij} - \frac{1}{4} S^2 \delta_{ij} \notag\\
&\quad +\frac{1}{4} H_{ij}^2 + \frac{1}{2} F^2_{ij} - \frac{1}{8} F^2 \delta_{ij} - 2 \nabla_i \nabla_j \Phi~,
\end{align}
where $\nabla$ and $R^{(6)}_{ij}$ denote the Levi-Civita connection and  the Ricci tensor of $M^6$, respectively.

\subsection{The Killing spinor equations}

The solution to the (massive) IIA supergravity KSEs along the $AdS_4$ subspace   can again be written as \eqref{killingspinorsiib}, where now $\sigma_\pm$ and $\tau_\pm$ are $\mathfrak{spin}(9,1)$ Majorana  spinors  that satisfy the lightcone projections $\Gamma_\pm\sigma_\pm=\Gamma_\pm\tau_\pm=0$
 and   depend only on the coordinates of $M^6$. After the lightcone projections are imposed,   $\sigma_\pm$ and $\tau_\pm$ have 16 independent components. These satisfy   the gravitino KSEs
\begin{align}\label{ads4iiakse}
\nabla^{(\pm)}_i \sigma_\pm = 0~, \quad \nabla^{(\pm)}_i \tau_\pm = 0~,
\end{align}
the dilatino KSEs
\begin{align}\label{ads4iiadilat}
\mathcal{A}^{(\pm)} \sigma_{\pm} = 0~, \quad \mathcal{A}^{(\pm)} \tau_{\pm} = 0~,
\end{align}
and the algebraic KSEs
\begin{align}\label{ads4iiaalgkse}
\Xi^{(\pm)}\sigma_{\pm}=0~, \quad (\Xi^{(\pm)} \pm \frac{1}{\ell})\tau_{\pm}=0~,
\end{align}
where
\begin{align}
\nabla_i^{(\pm)} &= \nabla_i \pm \frac{1}{2} \partial_i\log A + \frac{1}{8} \sH_i \Gamma_{11} + \frac{1}{8} S \Gamma_i + \frac{1}{16} \sF \Gamma_i \Gamma_{11} + \frac{1}{192} \sY \Gamma_i \mp \frac{1}{8} X \Gamma_{zxi}~, \notag \\
\mathcal{A}^{(\pm)} &= \slashed{\partial}\Phi + \frac{1}{12} \sH \Gamma_{11} + \frac{5}{4} S + \frac{3}{8} \sF \Gamma_{11} + \frac{1}{96} \sY \mp \frac{1}{4} X \Gamma_{zx}~, \notag \\
\Xi^{(\pm)} &= -\frac{1}{2\ell} + \frac{1}{2} \slashed{\partial}A \Gamma_z - \frac{1}{8} A S \Gamma_z - \frac{1}{16} A \sF \Gamma_z \Gamma_{11} - \frac{1}{192} A \sY \Gamma_z \mp \frac{1}{8} A X \Gamma_x~.
\end{align}
The first two equations arise from the naive restriction of the gravitino and dilatino KSEs of the theory on $\sigma_\pm$ and $\tau_\pm$, respectively,  while the last algebraic equation
is  an integrability condition that arises  from the integration of the IIA KSEs  on AdS$_4$.
As in the IIB case, the solutions of the above KSEs are related as in (\ref{ksrel}) and so such
backgrounds preserve $4k$ supersymmetries.

\subsection{\texorpdfstring{$AdS_4$}{AdS4} solutions with \texorpdfstring{$N>16$}{N greater 16} in IIA}

\subsubsection{Conditions on spinor bilinears}

The methodology to establish conditions on the Killing spinor bilinears following from our assumption that either the solutions are smooth and the internal space is compact without boundary,
or that the even subalgebra of the Killing superalgebra decomposes as stated in section \ref{ads4introduction}, is the same as that presented for IIB. However, the formulae are somewhat different. Setting
$ \Lambda = \sigma_+ + \tau_+$ and
 using the gravitino KSE \eqref{ads4iiakse}, one finds
\bea\label{iiahopf1}
\nabla_i \parallel \Lambda \parallel^2 = -\nabla_i \log A \parallel \Lambda \parallel^2 - \frac{1}{4} S \langle \Lambda, \Gamma_i \Lambda \rangle - \frac{1}{8} \langle \Lambda, \sgF_i \Gamma_{11} \Lambda \rangle - \frac{1}{96} \langle \Lambda, \sgY_i \Lambda \rangle~.
\eea
After multiplying the algebraic KSE \eqref{ads4iiaalgkse} with  $\Gamma_{iz}$, one gets
\begin{align}
\frac{1}{2\ell} \langle \Lambda, \Gamma_{iz} (\sigma_+ - \tau_+) \rangle &= - \nabla_i A \parallel \Lambda \parallel^2 - \frac{A}{4} S \langle \Lambda, \Gamma_i \Lambda \rangle  - \frac{A}{8} \langle \Lambda, \sgF_i \Gamma_{11} \Lambda \rangle  \notag \\
 &\quad - \frac{A}{96} \langle \Lambda, \sgY_i \Lambda \rangle~.
\end{align}
Using this,  one can  rewrite \eqref{iiahopf1} as
\begin{align}\label{iiahopf2}
\nabla_i \parallel \Lambda \parallel^2 = \frac{2}{\ell A} \langle \tau_+, \Gamma_{iz} \sigma_+ \rangle~.
\end{align}
On the other hand, the gravitino KSE  \eqref{ads4iiakse} gives
\begin{align}
\nabla^i \left( A \langle \tau_+, \Gamma_{iz} \sigma_+ \rangle \right) = 0~.
\end{align}
Therefore, taking the divergence of \eqref{iiahopf2}, one finds
\begin{align}
\nabla^2 \parallel \Lambda \parallel^2 + 2 \nabla^i \log A \, \nabla_i \parallel \Lambda \parallel^2 = 0~.
\end{align}
Now, the Hopf maximum principle  implies that $\parallel \Lambda \parallel^2$ is constant, which when inserted back into \eqref{iiahopf1} and \eqref{iiahopf2} yields
\begin{align}
 -\nabla_i \log A \parallel \Lambda \parallel^2 - \frac{1}{4} S \langle \Lambda, \Gamma_i \Lambda \rangle - \frac{1}{8} \langle \Lambda, \sgF_i \Gamma_{11} \Lambda \rangle - \frac{1}{96} \langle \Lambda, \sgY_i \Lambda \rangle = 0~,
\end{align}
and
\begin{align}\label{iiahopf3}
\langle \tau_+, \Gamma_{iz} \sigma_+ \rangle = 0~,
\end{align}
 respectively. The above condition can also be expressed as $\langle \sigma^1_+, \Gamma_{ix}\sigma^2_+ \rangle = 0$ for any two solutions $\sigma^1_+$ and $\sigma^2_+$ of the KSEs.

As in  IIB,  the algebraic KSE (\ref{ads4iiaalgkse}) implies that  $\langle \tau_+, \Xi^{(+)} \sigma_+ \rangle -\langle \sigma_+, (\Xi^{(+)} + \ell^{-1}) \tau_+ \rangle = 0$.  This, together
with \eqref{iiahopf3}, gives that $\langle \sigma_+, \tau_+ \rangle = 0$ and so the $\tau_+$ and $\sigma_+$ Killing spinors are orthogonal.

\subsubsection{The warp factor is constant}

To begin with, for every pair of solutions $\sigma_+^1$ and $\sigma_+^2$ of the KSEs we define the 1-form bilinear

\begin{align}
W_i = A\, \mathrm{Im}\,\langle \sigma_+^1, \Gamma_{iz} \sigma_+^2 \rangle~.
\end{align}
Then, the gravitino KSE \eqref{ads4iiakse} implies that
\begin{align}
\nabla_{(i}W_{j)}=0~,
\end{align}
therefore $W$ is an Killing vector  on $M^6$.

Next, the difference  $\langle \sigma^1_+, \Xi^{(+)} \sigma^2_+ \rangle -\langle \sigma^2_+, \Xi^{(+)} \sigma^1_+ \rangle = 0$ implies that
\bea
W^i \, \nabla_i A =0~,
\label{wda}
\eea
where we have used (\ref{iiahopf3}).

So far we have not made use of the fact that the solutions preserve $N>16$ supersymmetries. However, assuming this, (\ref{wda})  implies that the warp factor $A$ is constant.
This is a consequence of an adaptation of the homogeneity theorem on $M^6$. The homogeneity theorem also implies that $\Phi$ and $S$ are constant.  $X$ is constant as well due to the Bianchi identity \eqref{iiabianchi}.
Therefore, we have established that if the backgrounds preserve $N>16$  supersymmetries, then
\bea
A=\mathrm{const}~,~~~\Phi=\mathrm{const}~,~~~S=\mathrm{const}~,~~~X=\mathrm{const}~.
\label{constsiia}
\eea
Since the warp factor is constant, all backgrounds that preserve $N>16$  supersymmetries are
 products of the form $AdS_4\times M^6$. In addition, as pointed out in section \ref{ads4introduction}, $M^6$ is  a homogeneous space admitting a transitive and effective
 action of a group $G$ with Lie algebra $\mathfrak{so}(N/4)$. These homogeneous spaces have been listed in (\ref{homo166}).  In what follows, we shall explore all these 6-dimensional homogeneous spaces searching for IIA solutions
that preserve $N>16$ supersymmetries.

\subsection{\texorpdfstring{$N=28$}{N=28}}

There are no maximally supersymmetric AdS$_4$ backgrounds in (massive) IIA supergravity \cite{maxsusy}.  So the next case  to be investigated is that with 28 supersymmetries.
In such a case  $M^6$ admits a transitive and effective action of a group with Lie algebra $\mathfrak{so}(7)$.
Amongst the homogeneous spaces presented in (\ref{homo166}), the only one with this  property is  $\mathrm{Spin}(7)/\mathrm{Spin}(6)=S^6$.

Since $\mathrm{Spin}(7)/\mathrm{Spin}(6)=S^6$ is a symmetric space, all left-invariant forms are parallel with respect to the Levi-Civita connection and thus represent classes in the de-Rham cohomology. As $H^2(S^6)=H^3(S^6)=H^4(S^6)=0$, one concludes
that $F=H=Y=0$.  Using this and (\ref{constsiia}),  the dilatino KSE \eqref{ads4iiadilat} implies that
\begin{align}
\left(\frac{5}{4} S - \frac{1}{4} X \Gamma_{zx}\right) \sigma_+ =0~.
\end{align}
As this is the sum of two commuting terms, one Hermitian and one anti-Hermitian, the existence of solutions requires that both vanish separately. Hence  $S=X=0$.
Therefore, all fluxes must vanish.  This in turn leads to a contradiction, because the warp factor field equation (\ref{ads4einstiia}) does not admit any such solutions. Thus there are no (massive) IIA AdS$_4$ backgrounds   preserving   28 supersymmetries.

\subsection{\texorpdfstring{$N=24$}{N=24}}\label{cp3}

The internal space of  AdS$_4$ backgrounds preserving 24 supersymmetries admits a  transitive and effective action of a group with Lie algebra $\mathfrak{so}(6)=\mathfrak{su}(4)$.
The only space  in (\ref{homo166}) compatible with  such an action is  ${SU(4)}/{S(U(1)\times U(3))}=\mathbb{CP}^3$. Again, this is a symmetric space and so
all invariant forms are parallel with respect to the Levi-Civita connection. In turn they represent classes in the de-Rham cohomology. As $H^{\mathrm{odd}}(\mathbb{CP}^3)=0$, this implies that $H=0$.

It is well-known that this homogeneous space is a K\"ahler manifold and the  left-invariant metric is given by the standard Fubini-Study metric on $\mathbb{CP}^3$. The even cohomology ring of $\mathbb{CP}^3$ is generated by the K\"ahler form $\omega$. As a result the 2- and 4-form fluxes can be written as
\begin{align}\label{iiafluxes}
F= \alpha \, \omega~,~~
Y=\frac{1}{2} \, \beta \,\omega\wedge \omega~,
\end{align}
for some real constants $\alpha$ and $\beta$ to be fixed.

To determine $\alpha$ and $\beta$, let us first consider  the dilatino KSE \eqref{ads4iiadilat} which after imposing (\ref{constsiia}) reads
\begin{align}
\left( \frac{5}{4} S + \frac{3}{8} \sF \Gamma_{11} + \frac{1}{96} \sY - \frac{1}{4} X \Gamma_{zx}\right) \sigma_ + = 0~.
\label{ksexxiia}
\end{align}
The Hermitian and anti-Hermitian terms in this equation commute and so we may impose these conditions separately. Notice that the only non-trivial commutator to check is $[\sF, \sY]$ which vanishes because $F$ is proportional to the K\"ahler form
while $Y$ is a (2,2)-form with respect to the associated complex structure. Thus, we have
\begin{align}\label{2-formintegr}
\left(\frac{3}{8} \sF \Gamma_{11} -  \frac{1}{4} X \Gamma_{zx}\right) \, \sigma_+ = 0~,
\end{align}
and
\begin{align}\label{Ykse}
\left(\frac{5}{4} S + \frac{1}{96} \sY \right)\, \sigma_+ = 0~.
\end{align}
Inserting these into the algebraic KSE \eqref{ads4iiaalgkse} simplifies to
\begin{align}
\left(3 S \Gamma_z -  X \Gamma_x\right) \, \sigma_+ = \frac{3}{\ell A}\sigma_+~.
\label{kseiiaa2x}
\end{align}
The integrability condition of this yields
\begin{align}\label{Xsqu}
X^2+9 S^2 = \frac{9}{\ell^2 A^2}~.
\end{align}

Next, let us focus on  (\ref{2-formintegr}) and \eqref{Ykse}. Without loss of generality choosing $\Gamma_{11}= \Gamma_{+-} \Gamma_{zx} \Gamma_{123456}$,
(\ref{2-formintegr}) can be rewritten as
\begin{align}
\alpha( \Gamma^{3456}+\Gamma^{1256}+\Gamma^{1234})\sigma_+= -\frac{X}{3}  \sigma_+~,
\label{kseiia3x}
\end{align}
and similarly \eqref{Ykse} as

\begin{align}\label{Ykse2}
\beta(\Gamma^{1234}+ \Gamma^{1256} + \Gamma^{3456}) \sigma_+ = -5 S \sigma_+~,
\end{align}
where we have chosen an orthonormal frame for which $\omega= {\bf e}^{12}+ {\bf e}^{34}+ {\bf e}^{56} $.

To solve (\ref{kseiia3x}) and \eqref{Ykse2}, we decompose $\sigma_+$ into eigenspaces of $J_1=\Gamma_{3456}$ and $J_2=\Gamma_{1256}$ and find that this leads to the relations
\begin{align}
\alpha = -\frac{1}{3} X, \quad \beta = -5S~,
\label{raa}
\end{align}
for the eigenspaces $|+,+\rangle$, $|+,-\rangle$, $|-,+\rangle$,  and
\begin{align}
\alpha = \frac{1}{9} X, \quad \beta = \frac{5}{3} S~,
\label{rbb}
\end{align}
for the eigenspace $|-,-\rangle$.

Before we proceed to investigate the  KSEs further, let us focus on the field equations for the fluxes and the warp factor. Observe that $\alpha\not=0$.  Indeed if $\alpha=0$,  then the KSEs would imply that  $X=0$.  As $H=X=0$,   the dilaton  field equation in (\ref{ads4iiafieldeqs})  implies  that
all fluxes vanish.  In such a case, the warp factor field equation in (\ref{ads4einstiia}) cannot be satisfied.

Thus  $\alpha\not=0$.  Then the   field equation for the 3-form flux in (\ref{ads4iiafieldeqs}) becomes   $\alpha (S+ 4 \beta)=0$ and so this implies that
 $\beta=-1/4\, S$.  This contradicts the  results from  KSEs in (\ref{raa}) and (\ref{rbb}) above unless  $\beta=S=0$.  Setting $S=Y=0$ in the dilaton field equation in (\ref{ads4iiafieldeqs}),
it is easy to see that this satisfied if and only if   $\alpha=-1/3 X$ and so $\sigma_+$ lies in the eigenspaces $|+,+\rangle$, $|+,-\rangle$ and  $|-,+\rangle$. As $S=0$,  (\ref{Xsqu}) implies that $X= \pm 3 \ell^{-1} A^{-1}$ and so  $\alpha=\mp \ell^{-1} A^{-1}$. The algebraic KSE (\ref{kseiiaa2x}) now reads $\Gamma_x \sigma_+ = \mp \sigma_+$.  As  $\alpha=-1/3 X$,  the common eigenspace  of $\Gamma_x$,   $\Gamma_{3456}$ and $\Gamma_{1256}$ on $\sigma_+$ spinors
has dimension 6. Thus, the number of supersymmetries that the background
\bea
&&ds^2=2 du (dr-2 \ell^{-1} r dz) + A^2 (dz^2 +e^{2z/\ell}dx^2) + ds^2(\mathbb{CP}^3)~,
\cr
&&G = \pm 3 \ell^{-1} A e^{z/\ell} du \wedge dr \wedge dz \wedge dx~,
\quad H = S=0~,
\cr
&& F=\mp \ell^{-1} A^{-1} \omega, \quad \Phi=\mathrm{const}~,
\label{24soliia}
\eea
with $R^{(6)}_{ij} \delta^{ij}=24 \ell^{-2}A^{-2}$, can preserve is  24.

To establish that (\ref{24soliia}) indeed preserves 24 supersymmetries, it remains to investigate the gravitino KSE \eqref{ads4iiakse}. Since $\mathbb{CP}^3$ is simply connected, it is sufficient to investigate the integrability condition
\begin{align}
\left( \frac{1}{4} R_{ijmn} \Gamma^{mn} - \frac{1}{8} F_{im} \, F_{jn} \Gamma^{mn} -\frac{1}{12} X\, F_{ij} \Gamma_{zx} \Gamma_{11} -\frac{1}{72} X^2 \Gamma_{ij} \right) \, \sigma_+ = 0~,
\label{intcp3}
\end{align}
of the gravitino KSE. The Riemann tensor of ${SU(4)}/{S(U(1)\times U(3))}$  is
\bea
R_{ij,kl}=\frac{1}{4 \ell^2 A^2} (\delta_{ik} \delta_{jl}-\delta_{il} \delta_{jk})+ \frac{3}{4 \ell^2 A^2} (\omega_{ij} \omega_{kl}- \omega_{i[j} \omega_{kl]})~.
\label{cp3curv}
\eea
Then a  substitution of this and the rest of the fluxes into the integrability condition reveals that it is satisfied without further conditions. In a similar manner, one can check that the Einstein equation along $M^6$ is also satisfied. This
is the IIA $N=24$  solution of  \cite{nillsonpope, dufflupope}.

\subsection{\texorpdfstring{$N=20$}{N=20}}\label{N=20}

The internal space of AdS$_4$ backgrounds preserving 20 supersymmetries admits an effective and transitive action of a group with Lie algebra $\mathfrak{so}(5)=\mathfrak{sp}(2)$. Inspecting the homogeneous spaces in table \ref{table:nonlin}, one finds that there are two candidate internal spaces namely the symmetric space ${Sp(2)}/{U(2)}$ and the homogeneous space ${Sp(2)}/{Sp(1)\times U(1)}$. The symmetric space is the space of complex structures on $\mathbb{H}^2$ which are compatible with the quaternionic inner product, while the homogeneous space
is identified with the coset space of the sphere $\bar x x+ \bar y y=1$, $x,y\in \mathbb{H}$, with respect to the action $(x, y)\rightarrow (a x, ay)$, $a\in U(1)$.
The latter is diffeomorphic to $\mathbb{CP}^3$.

 \subsubsection{\texorpdfstring{${Sp(2)}/{U(2)}$}{Sp(2) over U(2)}}\label{Sp2overU2}

The geometry and algebraic properties of this symmetric space are described in appendix \ref{so5=sp2}. The most general left-invariant metric is
\bea
ds^2=a\,\delta_{rs} \delta_{ab} \bbl^{ra} \bbl^{sb}=\delta_{rs} \delta_{ab} {\bf e}^{ra} {\bf e}^{sb}~,
\label{sp2u2metr}
\eea
where $a>0$ is a constant and $\bbl^{ra}$, and ${\bf e}^{ra}=\sqrt{a}\, \bbl^{ra}$ are the left-invariant and orthonormal frames, respectively,
and where $r,s=1,2,3$ and $a,b=4,5$.
The invariant forms are generated by the 2-form
\bea
\omega=\frac12 \delta_{rs} \epsilon_{ab}\, {\bf e}^{ra}\wedge {\bf e}^{sb}~.
\label{kahlersp2}
\eea
  ${Sp(2)}/{U(2)}$ is  a K\"ahler manifold with respect to the pair $(ds^2, \omega)$.

In order to proceed, we choose the metric on the internal manifold as (\ref{sp2u2metr}) and the fluxes as in the ${SU(4)}/{S(U(1)\times U(3))}$ case,  i.e.
 \bea
 F= \alpha \, \omega~,~~~Y= {\frac{1}{2}}\beta \, \omega\wedge \omega~,
 \eea
but now $\omega$ is given in (\ref{kahlersp2}), and where $\alpha$ and $\beta$ are constants. Since there are no invariant 3-forms on ${Sp(2)}/{U(2)}$, this implies $H = 0$. Performing a similar analysis to that in section \ref{cp3}, we find that  $\beta=S=0$, $\alpha={\mp} \ell^{-1} A^{-1}$ and $X= \pm 3 \ell^{-1} A^{-1}$,
 and $\sigma_+$ satisfies  the same Clifford algebra projections as in e.g. (\ref{kseiia3x}). This requires an appropriate re-labeling of the indices of the orthonormal frame $\bbe^{ra}$ so that the left-invariant tensors take the same canonical form as those of  $SU(4)/S(U(1)\times U(3))$ expressed in terms of the  orthonormal frame $\bbe^i$. As a result, there are 24 spinors that solve the  KSEs so far.

 It remains to investigate the solutions
 of the  gravitino KSE \eqref{ads4iiakse}. As in the  $SU(4)/S(U(1)\times U(3))$ case in section \ref{cp3}, we shall investigate the integrability condition instead. This   is again given as in (\ref{intcp3}).
 The curvature of the metric of this symmetric space is presented  in (\ref{curvsp2u2}).
 Using this, the integrability condition (\ref{intcp3}) may be written as
 \bea
&&\big[ \frac{1}{16 a} (\delta^{cd} \Gamma_{rcsd}- \delta^{cd} \Gamma_{scrd})\delta_{ab}+ \frac{1}{16 a} \delta^{tu}( \Gamma_{taub}-\Gamma_{tbua}) \delta_{rs}-
\cr &&
 \frac18 \ell^{-2} A^{-2} (\delta^{cd} \Gamma_{rcsd}\delta_{ab}-  \Gamma_{sb ra})+\frac14 \ell^{-2} A^{-2} \delta_{rs} \epsilon_{ab} \Gamma_{zx} \Gamma_{11}
 \cr &&
 -\frac18 \ell^{-2} A^{-2}
 \Gamma_{rasb}\big]\sigma_+=0~.
 \label{innxxx}
 \eea
 Contracting with $\delta_{ab}$, one finds that there are solutions preserving more than 8 supersymmetries provided that $a=\ell^{2} A^{2}$.
 Then, taking the trace of (\ref{innxxx}) with $\epsilon_{ab} \delta_{rs}$, we find that
 \bea
 \frac12\slashed{\omega}\sigma_+=-{12}\Gamma_{zx} \Gamma_{11} \sigma_+~,
 \eea
 which is in contradiction to the condition (\ref{2-formintegr})  arising from the dilatino KSE.   The symmetric space ${Sp(2)}/{U(2)}$ does not yield\footnote{${Sp(2)}/{U(2)}$ can also be excluded
 as a solution because it is not a spin manifold \cite{klausthesis}.}  AdS$_4$ solutions that preserve 20 supersymmetries.

\subsubsection{\texorpdfstring{${Sp(2)}/({Sp(1)\times U(1)})$}{Sp(2) over Sp(1)xU(1)}}\label{Sp2_over_Sp1xU1}

The ${Sp(2)}/({Sp(1)\times U(1)})$ homogeneous space is  described in appendix \ref{so5=sp2}. Introducing the  left-invariant frame $\bbl^A m_A= \bbl^a W_a+ \bbl^{\underline r} T^{(+)}_{\underline r}$, the most general left-invariant metric is
\bea
ds^2= a\, \delta_{ab} \bbl^a \bbl^b+ b \,\delta_{{\underline r}{\underline s}} \bbl^{\underline r}\bbl^{\underline s} = \delta_{ab} \bbe^a \bbe^{b} + \delta_{{\underline r}{\underline s}} \bbe^{{\underline r}} \bbe^{{\underline s}},
\eea
where we have introduced the orthonormal frame $\bbe^a=\sqrt{a}\,\bbl^a$ , $\bbe^{{\underline r}}=\sqrt{b}\,  \bbl^{\underline r}$, and where ${\underline r}=1,2$ and $a,b=1,\dots, 4$. The invariant forms are generated by
\bea
I^{(+)}_3={1\over2} (I^{(+)}_3)_{ab} \bbe^a\wedge \bbe^b ~,~~~\tilde{\omega} = \frac{1}{2}  \epsilon_{{\underline r}{\underline s}} \bbe^{\underline r}\wedge \bbe^{\underline s}~,~~~ \bbe^{\underline r}\wedge I^{(+)}_{\underline r}~,~~~
\eea
and their duals, where
\bea
I^{(+)}_{\underline r}={1\over2} (I^{(+)}_{\underline r})_{ab} \bbe^a\wedge \bbe^b~.
\eea
The matrices $\big((I^{(\pm)}_r)_{ab}\big)$ are a basis in the space of self-dual and anti-self dual 2-forms in $\bR^4$ and are defined in (\ref{Imatrpm}).  Imposing the  Bianchi identities \eqref{iiabianchi}, one finds   the relation
\begin{align}\label{sp2bianchi}
\frac{\alpha}{\sqrt{b}} - \frac{\beta\sqrt{b}}{2a} = S \, h~,
\end{align}
and that the fluxes can be written as
\begin{align}\label{iiasp2fluxes}
F &= \alpha I^{(+)}_3 + \beta \, \tilde{\omega}~, \quad H = h \, \epsilon_{{\underline r}{\underline s}} \, \bbe^{\underline{r}} \wedge I^{(+)}_{\underline{s}}~,\notag \\
Y&=\gamma \, \tilde{\omega}\wedge I^{(+)}_{3} +  \frac{1}{2} \, \delta \, I^{(+)}_{3} \wedge I^{(+)}_{3}~,
\end{align}
where $\alpha, \beta, h, \gamma$ and $\delta$ are constants.

The   dilatino KSE \eqref{ads4iiadilat} is the sum of Hermitian and anti-Hermitian Clifford algebra elements  which commute and thus lead to the two independent conditions
\begin{align} \label{splitkse}
\left(\frac{3}{8} \sF \Gamma_{11} -  \frac{1}{4} X \Gamma_{zx}\right) \, \sigma_+ &= 0~, \notag \\
\left(\frac{5}{4}\, S + \frac{1}{12} \sH \Gamma_{11} + \frac{1}{96} \sY   \right) \, \sigma_+ &=0~.
\end{align}
Using this to simplify the algebraic KSE \eqref{ads4iiaalgkse}, one finds

\begin{align}
\left( \frac{1}{12} \sH \, \Gamma_{11} \Gamma_z + S \Gamma_z - \frac{X}{3} \Gamma_{x}\right) \sigma_+ = \frac{1}{\ell A} \, \sigma_+~.
\label{kkkkk}
\end{align}
If we then insert the fluxes \eqref{iiasp2fluxes} into the above KSEs and set $J_1=\Gamma^{24\underline{1}}\Gamma_{11}$, $J_2=\Gamma^{13\underline{1}}\Gamma_{11}$ and $J_3=\Gamma^{23\underline{2}}\Gamma_{11}$, we obtain

\begin{align}
\left(\alpha (J_2 J_3 - J_1 J_3) + \beta J_1 J_2\right) \, \sigma_+ + \frac{X}{3} \sigma_+ &= 0~, \notag\\
\left( 5S + 2 h (J_1 -J_2-J_3 + J_1 J_2 J_3) + \gamma \, (J_2J_3 - J_1 J_3) + \delta \, J_1J_2   \right) \sigma_+ &=0~,\notag\\
\left( \frac{1}{2} h (J_1 -J_2-J_3 + J_1 J_2 J_3) \Gamma_z + S \Gamma_z - \frac{X}{3} \Gamma_{x} \right) \sigma_+ - \frac{1}{\ell A} \, \sigma_+ &= 0~.
\label{j1j2j3377}
\end{align}
As $J_1, J_2, J_3$  are commuting Hermitian Clifford algebra operators with eigenvalues
$\pm1$, the KSE (\ref{kkkkk}) can be decomposed along the common eigenspaces as described  in  table \ref{table2xx}.

\begin{table}[h]
\begin{center}
\vskip 0.3cm
 \caption{Decomposition of (\ref{j1j2j3377}) KSE into eigenspaces}
 \vskip 0.3cm
 \begin{tabular}{|c|c|c|}
		\hline
		&$|J_1,J_2,J_3\rangle$&  relations for the fluxes\\
		\hline
		(1)&$|+,+,+\rangle$, $|-,-,-\rangle$& $\beta=-\frac{X}{3}$, $5S + \delta = 0$ \\
		& $|+,+,-\rangle$, $|-,-,+\rangle$ &  $(S \, \Gamma_z - \frac{X}{3} \Gamma_{x} ) |\cdot\rangle = \frac{1}{\ell A} | \cdot \rangle$ \\
		\hline
		(2)&$|+,-,+\rangle$, $|-,+,-\rangle$& $2\alpha + \beta=\frac{X}{3}$, $5S - 2 \gamma - \delta = 0$ \\
		& & $(S \, \Gamma_z - \frac{X}{3} \Gamma_{x} ) |\cdot\rangle = \frac{1}{\ell A} | \cdot \rangle$ \\
		\hline
		(3)&$|+,-,-\rangle$	& $2 \alpha - \beta=-\frac{X}{3}$, $5S + 8h +2\gamma-\delta=0$ \\
		& & $((S+2h) \, \Gamma_z - \frac{X}{3} \Gamma_{x} ) |\cdot\rangle = \frac{1}{\ell A} | \cdot \rangle$ \\
		\hline
		(4)&$|-,+,+\rangle$	& $2 \alpha - \beta= -\frac{X}{3}$, $5S-8h + 2\gamma -\delta=0$ \\
		& & $((S-2h) \, \Gamma_z - \frac{X}{3} \Gamma_{x} ) |\cdot\rangle = \frac{1}{\ell A} | \cdot \rangle$ \\
		\hline
	\end{tabular}
 \vskip 0.2cm
  \label{table2xx}
 \end{center}
\end{table}

\vskip 0.5cm

From the results of table \ref{table2xx}, there are two possibilities to choose five $\sigma_+$  Killing spinors, namely those in eigenspaces (1) and (3) and those in eigenspaces (1) and (4). For both of these choices,  the Bianchi identity \eqref{sp2bianchi} and the dilaton field equation give
\begin{align}
\alpha=\beta=-\frac{X}{3}, \quad X= \pm \frac{3}{\ell A}, \quad b=2a, \quad S=h=\gamma=\delta=0~.
\label{4x4c}
\end{align}
In either case notice that  these conditions imply the existence of  six $\sigma_+$  Killing spinors, as the conditions
required for both $|+,-,-\rangle$ and $|-,+,+\rangle$ to be solutions are satisfied.  So potentially this background can preserve $N=24$ supersymmetries.
To summarize, the independent conditions
 on the Killing spinors arising from those  in  (\ref{splitkse}) and those in  table
 \ref{table2xx} are
\begin{align}\label{sp2ksproject}
\frac{1}{2} \left( \slashed{I^{(+)}_3} + \tilde{\slashed{\omega}}\right) \sigma_+ = \sigma_+~, \quad \Gamma_x \sigma_+ = -\frac{3}{\ell A X} \sigma_x~.
\end{align}
These are  the same conditions as those found in section \ref{cp3} for $M^6=\mathbb{CP}^3$.

\noindent It remains  to investigate the gravitino KSE \eqref{ads4iiakse} or, equivalently, as $Sp(2)/(Sp(1)\times U(1))$ is simply connected, the corresponding integrability condition is again of the form \eqref{intcp3}.  The curvature of the metric is given in (\ref{curvsp2sp1u1}). Moreover the Einstein equation \eqref{ads4einstiia} yields $a=\ell^2 A^2/2$. Using these and substituting  the  conditions  (\ref{4x4c}) into the integrability condition, one can show that this is automatically satisfied provided that  \eqref{sp2ksproject} holds. As a result, there are no AdS$_4$ backgrounds with
internal space ${Sp(2)}/({Sp(1)\times U(1)}$ which \emph{strictly} preserve  20 supersymmetries.  However as shown above,  there is a solution which preserves 24 supersymmetries for $b=2a$.  This   is locally isometric to the  $AdS_4\times \mathbb{CP}^3$ solution found in section \ref{cp3}.  Note that there are no $N>24$ solutions  as can be seen by a direct
computation or by observing that $\mathbb{CP}^3$ does not admit an effective and transitive action by the  $\mathfrak{so}(N/4)$ subalgebra of the Killing
superalgebra of such backgrounds. However, there are  $AdS_4\times {Sp(2)}/({Sp(1)\times U(1))}$ solutions which preserve 4 supersymmetries \cite{lust}.

\section{\texorpdfstring{$N>16$ $AdS_4\times_w M^7$}{AdS4} solutions in 11 dimensions}\label{ads411d}

\subsection{\texorpdfstring{$AdS_4$}{AdS4xM7} solutions in \texorpdfstring{$D=11$}{D=11}}

Let us first summarise some of the properties  of  $AdS_4\times_w M^7$ backgrounds  in 11-dimensional supergravity   as  described in  \cite{mads} that  we shall use later. The bosonic fields are given by
\begin{align}
ds^2&= 2 \bbe^+ \bbe^- + (\bbe^z)^2 + (\bbe^x)^2 + ds^2(M^7) ~, \notag\\
 F&= X\, \bbe^+\wedge \bbe^- \wedge \bbe^z \wedge \bbe^x + Y~,
\end{align}
where the null orthonormal  frame $(\bbe^+, \bbe^-, \bbe^z, \bbe^x, \bbe^i)$ is as  in (\ref{frameiib}), but now $i,j=1,\dots, 7$, and the metric on the internal space $M^7$ is $ ds^2(M^7)= \delta_{ij} \bbe^i \bbe^j$.  $X$ and $Y$ are a function and 4-form on $M^7$, respectively.

The Bianchi identities of 11-dimensional supergravity evaluated on an $AdS_4\times_w M^7$ background yield
\begin{align}\label{ads4mbianchi}
dY = 0, \quad d(A^4 X)=0~.
\end{align}
Similarly, the field equations give
\begin{align}\label{Xfieldeqn}
\nabla^k Y_{ki_1i_2i_3} + 4 \nabla^k A \, Y_{ki_1i_2i_3} = - \frac{1}{24} X \epsilon_{i_1i_2i_3}{}^{k_1k_2k_3k_4} Y_{k_1k_2k_3k_4}~,
\end{align}
\begin{align}\label{m-einst-ads}
\nabla^k \partial_k \log A = - \frac{3}{\ell^2 A^2} - 4 \partial_k \log A \, \partial^k \log A + \frac{1}{3} X^2 + \frac{1}{144} Y^2~,
\end{align}
and
\begin{align}\label{m-einst-trans}
R^{(7)}_{ij} - 4 \nabla_i \partial_j \log A - 4 \partial_i \log A \partial_j \log A = \frac{1}{12} Y^2_{ij} + \delta_{ij} \left(\frac{1}{6} X^2 - \frac{1}{144} Y^2 \right)~,
\end{align}
where $\nabla$ is the Levi-Civita connection on $M^7$.

\subsection{The Killing spinors}

The solution of the KSEs of $D=11$ supergravity  along the AdS$_4$ subspace of $AdS_4\times_w M^7$  given in \cite{mads} can be expressed as in (\ref{killingspinorsiib}), but now
$\sigma_\pm$ and $\tau_\pm$ are $\mathfrak{spin}(10,1)$ Majorana  spinors  that depend on the coordinates of $M^7$.  Again they satisfy the lightcone projections $\Gamma_\pm\sigma_\pm=\Gamma_\pm\tau_\pm=0$.
The remaining independent KSEs are
\begin{align} \label{mkse}
\nabla^{(\pm)}_i \sigma_\pm = 0~, \quad \nabla^{(\pm)}_i \tau_\pm = 0~,
\end{align}
and
\begin{align}\label{malgkse}
\Xi^{(\pm)}\sigma_{\pm}=0~, \quad (\Xi^{(\pm)} \pm \frac{1}{\ell})\tau_{\pm}=0~,
\end{align}
where
\begin{align}
\nabla_i^{(\pm)} &= \nabla_i \pm \frac{1}{2} \partial_i \log A - \frac{1}{288} \sgY_i + \frac{1}{36} \sY_i \pm \frac{1}{12} X \Gamma_{izx}~, \\
\Xi^{(\pm)} &= \mp \frac{1}{2\ell} - \frac{1}{2} \Gamma_z \slashed{\partial} A + \frac{1}{288} A \Gamma_z \sY \pm \frac{1}{6} A X \Gamma_x~.
\end{align}
The former KSE is the restriction of the gravitino KSE on $\sigma_\pm$ and $\tau_\pm$ while the latter arises as an integrability condition from integrating the gravitino KSE
of 11-dimensional supergravity
over the AdS$_4$ subspace of $AdS_4\times_w M^7$.

\subsection{\texorpdfstring{$AdS_4$}{AdS4} solutions with \texorpdfstring{$N>16$}{N greater 16} in 11 dimensions}

\subsubsection{Conditions on spinor bilinears}

The conditions that arise from the assumption that $M^7$ be compact without boundary and the solutions be smooth
are similar to those presented in the (massive) IIA case. In particular, one finds
\bea
\parallel \sigma_+\parallel=\mathrm{const}~,~~~\langle \tau_+, \Gamma_{iz} \sigma_+ \rangle = 0~,~~~  \langle \sigma_+, \tau_+ \rangle = 0~.
\eea
The proof follows along the same lines as in the (massive) IIA case, and so we shall not repeat it here.

\subsubsection{The warp factor is constant}

Using arguments  similar to those presented in the (massive) IIA case, one finds that
 $W_i = A\, \mathrm{Im} \,\langle \sigma_+^1, \Gamma_{iz} \sigma_+^2 \rangle$ are Killing vectors on $M^7$ for any pair of Killing spinors $\sigma_+^1$ and $\sigma_+^2$ and that $i_W dA=0$.

Next, let us suppose  that  the backgrounds preserve $N>16$  supersymmetries. Then, an argument similar to the one presented in the proof
  of the homogeneity conjecture implies that the vector fields $W$ span the tangent space of $M^7$ at every point and so $A$ is constant.
  From the Bianchi identity \eqref{ads4mbianchi} it then follows that $X$ is constant as well.
  Thus we have established that
\begin{align}\label{AXconst}
A=\mathrm{const}~,~~~X=\mathrm{const}~.
\end{align}
As a result, the spacetime is a product $AdS_4\times M^7$, where $M^7$ is a homogeneous space.
Further progress requires the investigation of individual homogeneous spaces of dimension 7 which have been classified in  \cite{niko7dim, bohmkerr} and
are listed in table~\ref{table:nonlin7}.  Requiring in addition that the homogeneous spaces which can occur as internal spaces of $N>16$ AdS$_4$ backgrounds  must admit an effective and transitive action of a group that has Lie algebra $\mathfrak{so}(N/4)$, one arrives at the homogeneous
spaces presented in (\ref{homo167}).  In what follows, we shall investigate the geometry of these homogeneous spaces in detail searching for $N>16$ AdS$_4$ backgrounds
in 11-dimensional supergravity.

\begin{table}\renewcommand{\arraystretch}{1.3}
	\caption{7-dimensional compact, simply connected,  homogeneous spaces}
	\centering
	\begin{tabular}{c l}
		\hline
		& $M^7=G/H$  \\  
		\hline
		(1)& $\frac{\mathrm{Spin}(8)}{\mathrm{Spin}(7)}= S^7$, symmetric space\\
		(2)&$\frac{\mathrm{Spin}(7)}{G_2}=S^7$ \\
		(3)& $\frac{SU(4)}{SU(3)}$ diffeomorphic to $S^7$ \\
        (4) & $\frac{Sp(2)}{Sp(1)}$ diffeomorphic to $S^7$ \\
		(5) & $\frac{Sp(2)}{Sp(1)_{max}}$, Berger space \\
		(6) & $ \frac{Sp(2)}{\Delta(Sp(1))}=V_2(\bR^5)$ \\
        (7) & $\frac{SU(3)}{\Delta_{k,l}(U(1))}=W^{k,l}$~~ $k, l$ coprime, Aloff-Wallach space\\
		(8)&$\frac{SU(2) \times SU(3) }{\Delta_{k,l}(U(1))\cdot (1\times SU(2))}=N^{k,l}$ ~$k,l$ coprime\\
		(9) & $\frac{SU(2)^3}{\Delta_{p,q,r}(U(1)^2)}=Q^{p,q,r}$ $p, q, r$ coprime\\
(10)&$M^4\times M^3$,~~$M^4=\frac{\mathrm{Spin}(5)}{\mathrm{Spin}(4)}, ~\frac{ SU(3)}{S(U(1)\times U(2))}, ~\frac{SU(2)}{U(1)}\times \frac{SU(2)}{U(1)}$\\
&~~~~~~~~~~~~~~~~$M^3= SU(2)~,~\frac{SU(2)\times SU(2)}{\Delta(SU(2))}$\\
(11)&$M^5\times \frac{SU(2)}{U(1)}$,~~$M^5=\frac{\mathrm{Spin}(6)}{\mathrm{Spin}(5)}, ~\frac{ SU(3)}{SU(2)}, ~\frac{SU(2)\times SU(2)}{\Delta_{k,l}(U(1))},~ \frac{ SU(3)}{SO(3)} $\\
[1ex]
		\hline
	\end{tabular}
	\label{table:nonlin7}
\end{table}

\subsection{\texorpdfstring{$N=28,~{Spin(7)}/{G_2}$}{N=28, Spin(7) over G2}}

The maximally supersymmetric solutions have been classified before \cite{maxsusy} where it has been shown that all are locally isometric to $AdS_4\times S^7$ with $S^7=\mathrm{Spin}(8)/\mathrm{Spin}(7)$. The only  solution that may preserve $N=28$ supersymmetries is associated with the homogeneous space ${Spin(7)}/{G_2}$, see (\ref{homo167}). The Lie algebra $\mathfrak{spin}(7)= \mathfrak{so}(7)$ is again spanned by matrices $M_{ij}$ as in \eqref{gen-sona} satisfying the commutation relations \eqref{son-commuta} where now $i, j=1,2,...,7$. Let us denote the generators of the $\mathfrak{g}_2$ subalgebra of $\mathfrak{spin}(7)$  and those of the module $\mathfrak{m}$, $\mathfrak{spin}(7)=\mathfrak{g}_2\oplus \mathfrak{m}$,   with $G$ and $A$, respectively. These are defined as

\begin{align}
G_{ij}= M_{ij} + \frac{1}{4} \ast_{{}_7}\!\varphi_{ij}{}{}^{kl}\, M_{kl}~, \quad A_i = \varphi_{i}{}^{jk} M_{jk}~,
\end{align}
where  $\varphi$ is the fundamental $G_2$ 3-form,  $\ast_{{}_7}\varphi$ is its dual and $\ast_{{}_7}$ is the duality operation along the 7-dimensional internal space.   The non-vanishing components of $\varphi$ and $\ast_{{}_7}\varphi$ can be chosen as
\begin{align}
\varphi_{123}&=\varphi_{147}=\varphi_{165}=\varphi_{246}=\varphi_{257}=\varphi_{354}=\varphi_{367}=1~, \notag\\
\ast_{{}_7}\varphi_{1276}&=\ast_{{}_7}\varphi_{1245}=\ast_{{}_7}\varphi_{1346}=\ast_{{}_7}\varphi_{1357}=\ast_{{}_7}\varphi_{2374}=\ast_{{}_7}\varphi_{2356}=\ast_{{}_7}\varphi_{4567}=1~,
\label{g2phi}
\end{align}
and we have raised the indices above using the flat metric.
We have used the conventions for  $\varphi$ and $\ast_{{}_7}\varphi$  of \cite{Gran:2016tqd}, where also several useful identities involving $\varphi$ and its Hodge dual are presented.
In particular, observe that $\varphi_{i}{}^{jk} G_{jk} = 0$.
The $\mathfrak{spin}(7)$ generators can be written as
\begin{align}
M_{ij} = \frac{2}{3} \, G_{ij} + \frac{1}{6} \varphi_{ij}{}^{k} \, A_k~,
\end{align}
and, using this, we obtain
\begin{align}
[G_{ij},G_{kl}] &= \frac{1}{2} (\delta_{il} G_{jk} + \delta_{jk} G_{il} - \delta_{ik} G_{jl} - \delta_{jl} G_{ik}) + \frac{1}{4} (\ast_{{}_7}\varphi_{ij [k}{}^{m} G_{\ell]m} - \ast_{{}_7}\varphi_{k\ell [i}{}^{m} G_{j]m})~, \notag \\
[A_i, G_{jk}] &= \frac{1}{2} (\delta_{ij} \, A_{k} - \delta_{ik} \, A_j) + \frac{1}{4} \ast_{{}_7}\varphi_{ijk}{}^{l} A_l~, \notag \\
[A_i, A_j] &= \varphi_{ij}{}^{k} \, A_k - 4 G_{ij}~.
\end{align}
Clearly, ${Spin(7)}/{G_2}$ is a homogeneous space.  As $G_2$ acts with the irreducible 7-dimensional representation on $\mathfrak{m}$,  the left-invariant metric on ${Spin(7)}/{G_2}$ is unique up to scale, therefore we may choose an ortho-normal frame $\bbe^i$ such that
\begin{align}
ds^2 = a\,  \delta_{ij} \bbl^i \bbl^j= \delta_{ij} \bbe^i \bbe^j~,
\end{align}
where $a>0$ is a constant. The left-invariant forms are
\begin{align}
\varphi = \frac{1}{3!} \, \varphi_{ijk} \, \bbe^{i}\wedge \bbe^{j}\wedge\bbe^{k}~,
\end{align}
and its dual $\ast_{{}_7}\varphi$. So the $Y$ flux can be chosen as
\begin{align} \label{Yspin7}
Y= \alpha \, \ast_{{}_7}\!\varphi~,~~ \alpha = \mathrm{const}~.
\end{align}
Using this, the algebraic KSE \eqref{malgkse} can be expressed as
\begin{align}\label{ads4Spin7G2_alg_KSE}
\bigg(\frac{1}{6} \alpha &\left( P_1 -P_2 +P_3 - P_1 \, P_2 \, P_3 - P_2 \, P_3 + P_1 \, P_3 -  P_1 \, P_2 \right) \Gamma_z + \notag\\
&+ \frac{1}{3} X \, \Gamma_x \bigg) \sigma_+ = \frac{1}{\ell A} \sigma_+~,
\end{align}
where $\{P_1, P_2, P_3\}= \{ \Gamma^{1245}, \Gamma^{1267}, \Gamma^{1346}\}$ are mutually commuting, hermitian Clifford algebra operators with eigenvalues $\pm 1$.
The solutions of the algebraic KSE on the eigenspaces of $\{P_1, P_2, P_3\}$ have been tabulated in table \ref{table3xxx}.
\vskip 0.5cm

\begin{table}[h]
\begin{center}
\vskip 0.3cm
 \caption{Decomposition of (\ref{ads4Spin7G2_alg_KSE}) KSE into eigenspaces }
 \vskip 0.3cm
\begin{tabular}{|c|c|}
		\hline
		$|P_1,P_2,P_3\rangle$&  relations for the fluxes\\
		\hline
		$|+,+,+\rangle$, $|+,+,-\rangle$, $|-,+,+\rangle$, $|+,-,-\rangle$& $(-\frac{1}{6} \alpha \, \Gamma_z + \frac{1}{3} X \Gamma_{x} ) |\cdot\rangle = \frac{1}{\ell A} | \cdot \rangle$ \\
		$|-,+,-\rangle$, $|-,-,+\rangle$, $|-,-,-\rangle$&   \\
		\hline
		$|+,-,+\rangle$ & $(\frac{7}{6} \alpha \, \Gamma_z + \frac{1}{3} X \Gamma_{x} ) |\cdot\rangle = \frac{1}{\ell A} | \cdot \rangle$ \\
		\hline
		
	\end{tabular}
 \vskip 0.2cm
  \label{table3xxx}
 \end{center}
\end{table}

For backgrounds preserving    $N>16$ supersymmetries, one has  to choose the first set of solutions in  table \ref{table3xxx} and so impose the condition
\begin{align}
\frac{1}{36} \alpha^2 + \frac{1}{9} X^2 = \frac{1}{\ell^2 A^2}~.
\label{algfluxspin7}
\end{align}
However, the field equation for the warp factor $A$  \eqref{m-einst-ads} gives
\begin{align}
\frac{3}{\ell^2 A^2} = \frac{1}{3} X^2 + \frac{7}{6} \alpha^2~.
\label{feqwarpfactorspin7}
\end{align}
These two equations imply that $\alpha=0$, hence $Y=0$.

As $Y=0$, the algebraic KSE  is simplified to
\begin{align}
\Gamma_x \sigma_+ =  \frac{3}{\ell A X} \, \sigma_+~,
\label{algspin7pro}
\end{align}
and so  $\sigma_+$  lies   in one of the 8-dimensional  eigenspaces of $\Gamma_x$ provided that
 $X= \pm \frac{3}{\ell A}$.  Thus,  instead of preserving 28 supersymmetries, the solution   can  be maximally supersymmetric.  Indeed, this is the case, as we shall now demonstrate.  The integrability
  condition of the gravitino KSE \eqref{mkse} becomes
\begin{align}
\left( R_{ijk \ell} \, \Gamma^{k \ell} -\frac{1}{18} \, X^2 \Gamma_{ij} \right) \sigma_+ = 0~.
\label{intgravspin7}
\end{align}
To investigate whether this can yield a new condition on $\sigma_+$, we find after a direct computation using the results of section \ref{homogeneousspaces} that the Riemann tensor in the orthonormal frame is given by
\begin{align}
R_{ijk\ell} = {9\over4} a^{-1}  ( \delta_{ik} \delta_{j\ell} - \delta_{i\ell} \delta_{jk} )~.
\end{align}
Therefore, $S^7={Spin(7)}/{G_2}$ is equipped with the  round metric. For supersymmetric solutions, one must set $a^{-1}={1\over 81} X^2={1\over 9\ell^2 A^2}$.
In such a case, the integrability condition of the gravitino KSE  is automatically satisfied and thus the solution preserves 32 supersymmetries.  This solution  is locally isometric to
the maximally supersymmetric AdS$_4\times S^7$ solution.

\subsection{\texorpdfstring{$N=24,~{SU(4)}/{SU(3)}$}{N=24, SU(4) over SU(3)}}

 As $\mathfrak{so}(6)= \mathfrak{su}(4)$, it follows from (\ref{homo167}) that the internal space  of  an AdS$_4$ solution with 24 supersymmetries is the 7-dimensional homogeneous manifold ${SU(4)}/{SU(3)}$. The geometry of this homogeneous space is described in appendix \ref{su(k)}.  The left-invariant metric can be rewritten as
\bea
ds^2= a\, \delta_{mn} \ell^m \ell^n+ b\, (\ell^7)^2= \delta_{mn}\bbe^m \bbe^n+ (\bbe^7)^2~,
\eea
where we have introduced an orthonormal frame $\bbe^m=\sqrt{a}\, \ell^m, \bbe^7=\sqrt{b}\, \ell^7$, and  $m,n=1,\dots, 6$.
The most general left-invariant  4-form flux $Y$ can be chosen as
\begin{align}
Y =  \frac{1}{2}\, \alpha\, \omega\wedge\omega + \beta \ast_{{}_7}\!(\mathrm{Re}\,\chi) + \gamma \ast_{{}_7}\!(\mathrm{Im}\, \chi)~,
\end{align}
where  $\alpha,\beta,\gamma $ are constants and  the left-invariant 4-forms are
\bea
 \ast_{{}_7}(\mathrm{Re}\,\chi)&=& \bbe^{1367} + \bbe^{1457} + \bbe^{2357}  - \bbe^{2467}~, ~~~\omega = \bbe^{12} + \bbe^{34} + \bbe^{56}~,~~
\cr
\ast_{{}_7}(\mathrm{Im}\, \chi)&=& -\bbe^{1357} + \bbe^{1467} + \bbe^{2367} + \bbe^{2457}~,
\eea
expressed in terms of the orthonormal frame.
Having specified the fields, it remains to solve the KSEs. For this define the mutually commuting Clifford algebra operators
\bea
&&J_1 = \cos\theta\,\Gamma^{1367} + \sin\theta\, \Gamma^{2457}~, \quad J_2 = \cos\theta\,\Gamma^{1457} + \sin\theta\,\Gamma^{2367}~, \quad
\cr
&&J_3 = \cos\theta\,\Gamma^{2357} + \sin\theta\,\Gamma^{1467}~,
\eea
with eigenvalues $\pm 1$,
where $\tan\theta=\gamma/\beta$. Then upon inserting $Y$ into the algebraic KSE \eqref{malgkse} and using the above Clifford algebra operators, we obtain
\bea
&&\Big[ -\frac{\alpha}{6} (J_1 J_2 + J_1 J_3 + J_2 J_3) ~ \Gamma_z + \frac{\sqrt{\beta^2+\gamma^2}}{6} (J_1 + J_2 + J_3 + J_1J_2J_3) ~ \Gamma_z
\cr
&&\qquad \qquad \qquad\qquad+ \frac{1}{3} X \Gamma_x\Big] \, \sigma_+ = \frac{1}{\ell A} \sigma_+~.
\label{429ccc}
\eea
The algebraic KSE \eqref{malgkse} can then be decomposed into the eigenspaces of $J_1, J_2$ and $J_3$.
 The different relations on the fluxes for all possible sets of eigenvalues of these operators are listed in   table \ref{table4xxxx}.

\begin{table}[h]
\begin{center}
\vskip 0.3cm
 \caption{Decomposition of (\ref{429ccc}) KSE into eigenspaces}
 \vskip 0.3cm

	\begin{tabular}{|c|c|}
		\hline
		$|J_1,J_2,J_3\rangle$&  relations for the fluxes\\
		\hline
		$|+,+,-\rangle$, $|+,-,+\rangle$, $|-,+,+\rangle$ & $(\frac{1}{6} \alpha \, \Gamma_z + \frac{1}{3} X \Gamma_{x} ) |\cdot\rangle = \frac{1}{\ell A} | \cdot \rangle$ \\
		$|+,-,-\rangle$, $|-,+,-\rangle$, $|-,-,+\rangle$&   \\
		\hline
		$|+,+,+\rangle$ & $[(- \frac{\alpha}{2} + \frac{2}{3} \sqrt{\beta^2+\gamma^2}) \, \Gamma_z + \frac{1}{3} X \Gamma_{x} ] |\cdot\rangle = \frac{1}{\ell A} | \cdot \rangle$ \\
		\hline
		$|-,-,-\rangle$ & $[(- \frac{\alpha}{2} - \frac{2}{3} \sqrt{\beta^2+\gamma^2}) \, \Gamma_z + \frac{1}{3} X \Gamma_{x} ] |\cdot\rangle = \frac{1}{\ell A} | \cdot \rangle$ \\
		\hline
		
	\end{tabular}
\vskip 0.2cm
  \label{table4xxxx}
 \end{center}
\end{table}

The only possibility  to obtain solutions with   $ N>16$  supersymmetries is to choose the first set of eigenspinors in table \ref{table4xxxx}. This leads to the integrability condition
\begin{align}
\frac{\alpha^2}{36} + \frac{1}{9} X^2 = \frac{1}{\ell^2 A^2}~,
\end{align}
from the remaining KSE. This together with  the warp factor field equation \eqref{m-einst-ads}
\begin{align}
\frac{1}{3} X^2 + \frac{1}{2} \alpha^2 + \frac{2}{3} (\beta^2 + \gamma^2) = \frac{3}{\ell^2 A^2}~,
\end{align}
implies
\begin{align}
\frac{5}{4} \alpha^2 + 2 (\beta^2 + \gamma^2) = 0~,
\end{align}
and so $\alpha=\beta=\gamma=0$.  Therefore   $Y=0$ and the solution is electric. As a result, the algebraic KSE \eqref{m-einst-ads} becomes
\begin{align}
\Gamma_x \sigma_+ = \frac{3}{\ell A X} \sigma_+~,
\end{align}
and so for $X=\pm 3 \ell^{-1}A^{-1}$  it admits 8 linearly independent  $\sigma_+$ solutions.  So potentially, the background is maximally supersymmetric.

It remains to investigate the gravitino KSE. First of all, we observe that for $Y=0$ the Einstein equation \eqref{m-einst-trans} along the internal space becomes
\begin{align}
R_{ij}=\frac{1}{6} X^2 \delta_{ij}~.
\end{align}
Therefore, the internal space is Einstein.  After some computation using the results in appendix \ref{su(k)},  one finds that the homogeneous space $SU(4)/SU(3)$ is Einstein, provided that    $b={9\over4}a$. In that case, the curvature of the metric in the orthonormal frame becomes
\begin{align}
R_{ij,mn} = \frac{1}{4a} (\delta_{im} \delta_{jn} - \delta_{in} \delta_{jm})~,
\end{align}
and so the internal space is locally isometric to the round 7-sphere.  As expected from this,
the integrability condition of  the gravitino KSE \eqref{mkse}
\begin{align}
(R_{ij,mn} \Gamma^{mn} - \frac{1}{18} X^2 \Gamma_{ij}) \sigma_+ = 0~,
\label{intads4grav}
\end{align}
 has non-trivial solutions for $X^2=9 a^{-1}$, i.e. $a=\ell^2 A^2$ and $b={9\over4}\ell^2 A^2$.   With this identification of parameters,  $AdS_4\times {SU(4)}/{SU(3)}$ is locally isometric to the maximally supersymmetric $AdS_4\times S^7$ background.

To summarise, there are no $AdS_4$ solutions with internal space ${SU(4)}/{SU(3)}$ which preserve  $16<N<32$ supersymmetries. However, for the choice of parameters for which
 ${SU(4)}/{SU(3)}$ is the round 7-sphere, the solution preserves 32 supersymmetries as expected.

\subsection{\texorpdfstring{$N=20$}{N=20}}
 As mentioned in section \ref{ads4introduction}, the internal space of AdS$_4$ backgrounds that preserve 20 supersymmetries admits an effective and  transitive action of a group which has Lie algebra $\mathfrak{so}(5)=\mathfrak{sp}(2)$.
The field equation for  $Y$~\eqref{Xfieldeqn} is
\begin{equation}
d\ast_{{}_7} Y = X\, Y \, .
\end{equation}
As $X$ is constant, note that for generic 4-forms $Y$ this defines a nearly-parallel $G_2$-structure on $M^7$, see e.g. \cite{Reidegeld2009} for homogeneous $G_2$ structures.  However, in what follows we shall not assume that $Y$ is generic.  In fact
in many cases, it vanishes.

Amongst the  7-dimensional compact homogeneous spaces of  (\ref{homo167}), there are three candidate internal spaces.
These are the Berger space $B^7 = Sp(2)/Sp(1)_{\mathrm{max}}$, $V_2(\bR^5) = Sp(2)/\Delta(Sp(1))$, and $J^7 = Sp(2)/Sp(1)$, corresponding to the three inequivalent embeddings of $Sp(1)$ into $Sp(2)$.
We will in the following examine each case separately, starting with the Berger space $Sp(2)/Sp(1)_{\mathrm{max}}$.

\subsubsection{\texorpdfstring{$Sp(2)/Sp(1)_{\mathrm{max}}$}{Sp(2) over Sp(1)max}}

The description of the Berger space  $B^7=Sp(2)/Sp(1)_{\mathrm{max}}$ as a   homogeneous manifold is summarized in  appendix \ref{bergerspace}.  $B^7$ is diffeomeorphic to the total space of an $S^3$ bundle over $S^4$ with Euler class $\mp 10$ and first Pontryagin class $\mp16$ \cite{difftype}.  As a result   $H^4(B^7,\mathbb{Z})=\mathbb{Z}_{10}$ and $B^7$ is a rational homology 7-sphere.
 Since $\mathfrak{sp}(2)= \mathfrak{so}(5)$ and $\mathfrak{sp}(1)= \mathfrak{so}(3)$, one writes $\mathfrak{so}(5)=\mathfrak{so}(3)\oplus \mathfrak{m}$ and the subalgebra $\mathfrak{so}(3)$ acts irreducibly on $\mathfrak{m}$ with the ${\bf 7}$ representation.  So  $B^7$ admits a unique
 invariant metric up to an overall scale that is Einstein. As the embedding of $\mathfrak{so}(3)$ into $\mathfrak{so}(7)$ factors through $\mathfrak{g}_2$, it also admits  an invariant 3-form $\varphi$ given in
 (\ref{g2phi}) which is unique up to a scale.  Because there is a unique invariant 3-form  $\varphi$,  $d\varphi\varpropto \ast_{{}_7}\varphi$ and $B^7$ is a nearly parallel $G_2$ manifold. Using these, we find that  the invariant fields of the theory are
\begin{equation}\label{SO5SO3_max_G_inv_metric}
ds^2= a \delta_{ij} \bbl^i \bbl^j= \delta_{ij} \bbe^i \bbe^j ~,~~~~Y={1\over 4!} \alpha  \ast_{{}_7}\varphi_{ijkm}\, \bbe^{i}\wedge \bbe^j\wedge \bbe^k\wedge \bbe^{m}~,
\end{equation}
where we have introduced the ortho-normal frame $\bbe^i=\sqrt{a}\, \bbl^i$, $\ast_{{}_7}\varphi$ is given in (\ref{g2phi}) and $a, \alpha$ are constants with $a>0$.

As the pair $(ds^2, Y)$ exhibits the same algebraic relations as that of the $Spin(7)/G_2$ case,  the algebraic KSE \eqref{ads4Spin7G2_alg_KSE} can be solved in the same way
yielding the results of table \ref{table3xxx}. To find $N>16$ AdS$_4$ solutions, one should consider the first set of eigenspinors  of the table
which in turn imply the relation (\ref{algfluxspin7}) amongst the fluxes. This together with the field equation of the warp factor (\ref{feqwarpfactorspin7}) leads again to the
conclusion that $\alpha=0$ and so $Y=0$.

As a result of the analysis  of the algebraic KSE, so far the background can admit up to 32 supersymmetries.  It remains to investigate the solutions of the gravitino KSE.
  The curvature of $B^7$ is given by
\bea
R_{ij, km}={1\over 10 \,a} \delta_{k[i}\, \delta_{j]m}-{1\over 5 a} \ast_{{}_7}\varphi_{ijkm}+{1\over a} \delta_{\alpha\beta} k^\alpha_{ij} k^\beta_{km}~,
\eea
where $k^\alpha$ is given in appendix \ref{bergerspace}.
The integrability condition of the gravitino KSE for $Y=0$ is given in (\ref{intgravspin7}).  To solve this condition, we decompose the expression into the ${\bf 7}$ and ${\bf 14}$
representations of $\mathfrak{g}_2$ using the projectors
\bea
(P^{\bf 7})^{ij}{}_{km}&={1\over 3} (\delta^i_{[k} \delta^j_{m]}-{1\over2} \ast_{{}_7}\varphi^{ij}{}_{km})~,\notag\\
(P^{\bf 14})^{ij}{}_{km}&={2\over 3} (\delta^i_{[k} \delta^j_{m]}+{1\over4} \ast_{{}_7}\varphi^{ij}{}_{km})~,
\eea
and noting that $k^\alpha$ as 2-forms are in the ${\bf 14}$ representation.  The integrability condition along the ${\bf 7}$ representation gives
$X^2={81\over 5} a^{-1}$ while along the  ${\bf 14}$ representation it implies that the Killing spinors must be invariant under  $\mathfrak{g}_2$.  It turns out that there are two such $\sigma_+$
spinors, however taking into account the remaining projection arising from the algebraic KSE, see (\ref{algspin7pro}),  we deduce that the solution preserves 4 supersymmetries in total.  This solution   has already been
derived in \cite{Castellani:1983yg}.

\subsubsection{\texorpdfstring{${Sp(2)}/{\Delta(Sp(1))}$}{Sp(2) over Delta Sp(1)}}

The decomposition  of the Lie algebra $\mathfrak{sp}(2)= \mathfrak{so}(5)$ suitable to describe this homogeneous space can be found in appendix \ref{so5=sp2}. Writing $\ell^A m_A= \ell^{ra} M_{ra}+ \ell^7 T_7$ for the left-invariant frame, $r=1,2,3$ and $a=4,5$, the most general left-invariant metric is
\bea
ds^2=\delta_{rs} g_{ab} \ell^{ra} \ell^{sb}+ a_4 (\ell^7)^2~,~~~
\eea
where $g_{ab}$ is a positive definite symmetric  $2\times 2$-matrix, $a>0$ a constant, and the left-invariant forms are generated by
\bea
 \ell^7=\ell^7~,~~~{1\over2} \delta_{rs} \epsilon_{ab} \ell^{ra}\wedge \ell^{sb}~,~~~{1\over3!} \epsilon_{rst} \ell^{ra}\wedge \ell^{sb}\wedge \ell^{tc}~.
\eea
In order to simplify the analysis of the geometry in what follows, we note  that without loss of generality the matrix $(g_{ab})$ can be chosen to be diagonal.  To see this, perform
an orthogonal transformation $O\in SO(2)$ to bring $(g_{ab})$ into a diagonal form.  Such a transformation can be compensated with a frame rotation
\bea
\ell^{ra}\rightarrow O^a{}_b \ell^{rb}~.
\eea
Demanding that $\ell^A m_A$ is invariant implies that $M_{ra}$ has to transform as $M_{ra}\rightarrow O^b{}_a M_{rb}$. Indeed, we observe that such a
transformation is an automorphism  of $\mathfrak{so}(5)$ that preserves the decomposition (\ref{comsp2so3a}), i.e.\ the structure constants of the Lie algebra
remain the same. As a result, we can diagonalise the metric and simultaneously use the same structure constants to calculate the geometric
quantities of the homogeneous space. Under these orthogonal transformations the first two left-invariant forms are invariant
while there is a change of basis in the space of left-invariant 3-forms.

To continue, take\footnote{We have performed the analysis that follows also without taking $(g_{ab})$ to be diagonal producing the same conclusions.}  $(g_{ab})=\mathrm{diag}(a_1, a_2)$.  Then  introduce the orthonormal frame $\bbe^7= \sqrt{a_4}\, \ell^7, \bbe^{r4}= \sqrt{a_1}\, \ell^{r4}$ and  $\bbe^{r5}= \sqrt{a_2}\, \ell^{r5}$.  In this
frame the most general left-invariant metric and $Y$ flux can be   written as
\begin{align}
ds^2&=\delta_{ab} \delta_{rs} \bbe^{ra} \bbe^{sb}+(\bbe^7)^2~, \notag\\
Y&=\beta_1\, \bbe^7\wedge \chi_{444}+ \beta_2\, \bbe^7\wedge\chi_{445}+\beta_3\, \bbe^7\wedge\chi_{455}+ \beta_4\, \bbe^7\wedge\chi_{555}+\beta_5\, \psi~,
\end{align}
where $\beta_1, \beta_2, \cdots, \beta_5$ are constants,
\begin{align}
\chi_{abc}= {1\over3!} \epsilon_{rst} \bbe^{ra}\wedge \bbe^{sb}\wedge \bbe^{tc}~,~~~\psi={1\over 2} \omega\wedge \omega~,
\end{align}
and
\bea
\omega={1\over2} \delta_{rs} \epsilon_{ab} \bbe^{ra}\wedge \bbe^{sb}~.
\eea
The Bianchi identity for $Y$ is automatically satisfied.  On the other hand, the field equation for $Y$ in (\ref{Xfieldeqn}) yields the conditions
\bea
&&{\beta_3\over2} \sqrt{{a_2\over a_4 a_1}} - \beta_1 X=0~,~~~
-\beta_2 \sqrt{{a_2\over a_4 a_1}}+{3\beta_4 \over 2} \sqrt{{a_1\over a_4 a_2}}- \beta_2 X=0~,
\cr
&&{3\beta_1\over 2} \sqrt{{a_2\over a_4 a_1}} -\beta_3 \sqrt{{a_1\over a_2 a_4}}- \beta_3 X=0~,~~~
{\beta_2\over 2} \sqrt{{a_1\over a_4 a_2}} - \beta_4 X=0~,
\cr
&&\beta_5 \Big(X+\sqrt{{a_4\over a_1 a_2}}\Big)=0~,
\label{sp2u2beta}
\eea
where we have chosen the top form on $M^7$ as $d\mathrm{vol}= \bbe^7\wedge \chi_{444}\wedge \chi_{555}$.

Before we proceed to investigate the various cases which arise from solving the linear system (\ref{sp2u2beta}), let us first consider the case in which $F$ is electric, i.e.\ proportional to the volume form of AdS$_4$.  In such a case $\beta_1=\cdots =\beta_5=0$.  The algebraic KSE then gives
\bea
{1\over3} X \Gamma_x\sigma_+={1\over \ell A} \sigma_+~,
\eea
and the field equations  along $M^7$ imply that
\bea
R_{ij}={1\over 6} X^2 \delta_{ij}~,
\eea
and so $M^7$ is Einstein. The Einstein condition on the metric of $M^7$ requires that
\bea
a_1=a_2~,~~~a_4={3\over2} a_1~.
\eea
To investigate whether there are solutions preserving 20 supersymmetries, it remains to consider the integrability condition of the gravitino KSE (\ref{intads4grav}).
Indeed, using the expressions (\ref{curvso5so3ab1}) and (\ref{curvso5so3ab2}) for the curvature of  this homogeneous space, the integrability condition along the directions $7$ and $ra$ gives $X^2=(27/8) a_1^{-1}$ while along the $ra$ and $sb$ directions additional
projections are required.  For example, after taking the trace with $\delta^{ab}$ and setting $r=1$ and $s=2$,  the condition is $\Gamma^{1245}\sigma_+=\sigma_+$ which  leads to solutions
that preserve 16 supersymmetries or less, where the gamma matrices are in the orthonormal frame and $\Gamma^{r4}=\Gamma^r, \Gamma^{r5}=\Gamma^{3+r}$.  Hence there are no $N>16$ AdS$_4$ solutions.

Next let us turn to investigate the solutions of the linear system (\ref{sp2u2beta}). The last condition implies that
\bea
\mathrm{either}\quad  \beta_5=0 \quad \mathrm{or}\quad X=-\sqrt{{a_4\over a_1 a_2}}~.
\label{sp2u2X}
\eea

To continue, first consider the case that $\beta_5\not=0$.
\vskip 0.3cm
$\underline {\beta_5\not=0}$

Substituting the second equation in (\ref{sp2u2X}) into the linear system (\ref{sp2u2beta}), one finds that
\bea
\beta_3 {a_2\over 2}+ \beta_1 a_4 =0~,~~~ (a_4-a_1) \beta_3+{3\over2} a_2 \beta_1=0~,
\cr
\beta_2 {a_1\over2}+ a_4 \beta_4=0~,~~~(a_4-a_2) \beta_2+{3\over2} a_1 \beta_4=0~.
\label{loonsyst}
\eea
Now there are several cases to consider.  First, suppose that the parameters of the metric $a_1,a_2,a_4$ are such that the only solutions of the
linear system above are $\beta_1=\beta_2=\beta_3=\beta_4=0$.  In such case $Y=\beta_5 \psi$ and $Y$ has the same algebraic properties as that of the $SU(4)/SU(3)$ case
with $\beta=\gamma=0$ and $\alpha=\beta_5$.  As a result, the algebraic KSE  together with the Einstein equation for the warp factor imply that $\beta_5=0$ as well and so $Y=0$.
This violates our assumption that $\beta_5\not=0$. In any case, the 4-form flux $F$ is electric which we have already investigated above and found that such a configuration  does not admit solutions with $N>16$ supersymmetries.

Next, suppose that the parameters of the metric are chosen such that
\bea
\mathrm{either} \quad \beta_1=\beta_3=0~,\quad \mathrm{or} \quad \beta_2=\beta_4=0~.
\eea
These two cases are symmetric, so it suffices to consider one of the two.  Suppose that $\beta_2=\beta_4=0$ and $\beta_1, \beta_3\not=0$.  In such  a case
\bea
{3\over4} a_2^2- a_4 (a_4-a_1)=0~,
\eea
with ${3\over4} a_1^2- a_4 (a_4-a_2)\not=0$.  Setting $P_1=\Gamma^{7156}, P_2=\Gamma^{7345}$ and $P_3=\Gamma^{7264}$, the algebraic KSE can be written as
\bea
&&\Big[{1\over18} \Big(-3\beta_1 P_1P_2P_3+\beta_3 (P_1+P_2+P_3)-3\beta_5 (P_1P_2+P_1 P_3+P_2P_3) \Big) \Gamma_z
\cr
&&\qquad +{1\over 3} X \Gamma_x\Big] \sigma_+={1\over \ell A} \sigma_+~.
\label{455cccv}
\eea
Since $P_1, P_2, P_3$ commute and have eigenvalues $\pm1$, the above algebraic equation decomposes into eigenspaces as tabulated in table \ref{table5xxxxxy}.

\begin{table}[h]
\begin{center}
\vskip 0.3cm
 \caption{Decomposition of (\ref{455cccv}) KSE into eigenspaces }
 \vskip 0.3cm
\begin{tabular}{|c|c|c|}\hline
$|P_1, P_2, P_3\rangle$ & relations for the fluxes \\ \hline
$|+,+,+\rangle$&$[ \frac{1}{6} (-\beta_1+\beta_3-3\beta_5) \Gamma_z + \frac13 X \Gamma_x ] |\cdot\rangle =  \frac{1}{\ell A} |\cdot\rangle$
\\ \hline
$|+,+,-\rangle$, $|+,-,+\rangle$, $|-,+,+\rangle$ & $[ \frac{1}{18} (3\beta_1+\beta_3+3\beta_5) \Gamma_z + \frac13 X \Gamma_x ] |\cdot\rangle =  \frac{1}{\ell A} |\cdot\rangle$ \\ \hline
$|-,-,+\rangle$, $|-,+,-\rangle$, $|+,-,-\rangle$ & $[ \frac{1}{18} (-3\beta_1-\beta_3+3\beta_5) \Gamma_z + \frac13 X \Gamma_x ] |\cdot\rangle =  \frac{1}{\ell A} |\cdot\rangle$ \\ \hline
$|-,-,-\rangle$ & $[ \frac{1}{6} (\beta_1-\beta_3-3\beta_5) \Gamma_z + \frac13 X \Gamma_x ] |\cdot\rangle =  \frac{1}{\ell A} |\cdot\rangle$ \\ \hline
	\end{tabular}
\vskip 0.2cm
  \label{table5xxxxxy}
 \end{center}
\end{table}

To find solutions with 20 supersymmetries or more, we can either choose one of the two eigenspaces with 3 linearly independent eigenspinors and both eigenspaces with a single
eigenspinor or both eigenspaces with 3 linearly independent eigenspinors. In the former case the algebraic KSE will admit 20 Killing spinors and in the latter 24 Killing spinors.

Let us first consider the case with 20 Killing spinors.  In such a case, we find that
\bea
\beta_1=\beta_3~,~~~\beta_1=3 \beta_5~,
\eea
and
\bea
{1\over36} \beta_1^2+{1\over9} X^2= {1\over\ell^2 A^2}~,
\label{algintsp2}
\eea
where we have considered the second eigenspace with 3 eigenspinors in table \ref{table5xxxxxy}. The  case where the first such eigenspace with 3 eigenspinors is chosen can be treated in a similar way.
The condition  (\ref{algintsp2}) follows as an integrability condition of the remaining algebraic KSE involving $\Gamma_z$ and $\Gamma_x$.
On the other hand, the field equation for the warp factor (\ref{m-einst-ads}) implies that
\bea
{7\over 54} \beta_1^2+{1\over 9} X^2= {1\over\ell^2 A^2}~,
\eea
which, together with (\ref{algintsp2}), gives $\beta_1=0$ and so $Y=0$. The solution cannot preserve $N>16$ supersymmetries.

Next consider the case with 24 Killing spinors.  In this case, we find that
\bea
3\beta_1=-\beta_3~,
\eea
and the integrability condition of the remaining algebraic KSE yields
\bea
{1\over36} \beta_5^2+ {1\over9} X^2={1\over\ell^2 A^2}~.
\label{algintsp2x}
\eea
On the other hand the field equation of the warp factor (\ref{m-einst-ads}) gives
\bea
{1\over 9} X^2+{2\over 9} \beta_1^2+{1\over 6} \beta_5^2={1\over\ell^2 A^2}~.
\eea
Comparing this with (\ref{algintsp2x}), one finds that the $\beta$'s vanish and so $Y=0$. Thus there are no solutions with $N>16$ for  either $\beta_1, \beta_3$ or $\beta_2, \beta_4$  non-vanishing.

It remains to investigate the case that all $\beta_1, \dots, \beta_5\not=0$. This requires that the determinant of the coefficients of the linear system (\ref{loonsyst})
must vanish, i.e.
\bea
{3\over4} a_2^2- a_4 (a_4-a_1)=0~,~~~{3\over4} a_1^2- a_4 (a_4-a_2)=0~.
\label{2detsp2}
\eea
Taking the difference of the two equations, we find that
\bea
\mathrm{either} \quad a_1=a_2~,\quad \mathrm{or}\quad a_4={3\over4} (a_1+a_2)~.
\eea
Substituting $a_4$ above into (\ref{2detsp2}), we find that $a_1=a_2$.  So without loss of generality, we set $a_1=a_2=a$.
Then the linear system (\ref{loonsyst}) can be solved to yield
\bea
\beta_3=-3\beta_1~,~~~\beta_2=-3\beta_4~.
\eea
Setting
\bea
&&P_1=\cos\theta \Gamma^{7156}+\sin\theta \Gamma^{7234}~,~~P_2=\cos\theta \Gamma^{7345}+\sin\theta \Gamma^{7126}~,~~
\cr
&&P_3=\cos\theta \Gamma^{7264}+\sin\theta \Gamma^{7315}~,
\label{procosx}
\eea
 the algebraic KSE (\ref{malgkse}) can be rewritten as
\bea
&&\big [{1\over18}\big(\alpha P_1P_2P_3+ \alpha (P_1+P_2+P_3)-3\beta_5 (P_1P_2+P_1P_3+P_2P_3)\big) \Gamma_z
\cr
&&\qquad \qquad\qquad\qquad +{1\over3} X\Gamma_x] \sigma_+={1\over\ell A} \sigma_+~,
\label{466cccv}
\eea
where $\tan\theta=\beta_3/\beta_2$ and $\alpha=\sqrt{\beta_2^2+\beta_3^2}$.
Again, these Clifford algebra operators commute and have eigenvalues $\pm1$, so we list the restrictions of this equation to the eigenspaces of $P_1, P_2$ and $P_3$ in  table \ref{table5xxxxxyy}.

\begin{table}[h]
\begin{center}
\vskip 0.3cm
 \caption{Decomposition of (\ref{466cccv}) KSE into eigenspaces }
 \vskip 0.3cm
\begin{tabular}{|c|c|c|}\hline
$|P_1, P_2, P_3\rangle$ & relations for the fluxes \\ \hline
$|+,+,+\rangle$&$[ \frac{1}{18} (4\alpha-9\beta_5) \Gamma_z + \frac13 X \Gamma_x ] |\cdot\rangle =  \frac{1}{\ell A} |\cdot\rangle$
\\ \hline
$|+,+,-\rangle$, $|+,-,+\rangle$, $|-,+,+\rangle$ &  \\
$|-,-,+\rangle$, $|-,+,-\rangle$, $|+,-,-\rangle$ & $[ \frac{1}{6} \beta_5 \Gamma_z + \frac13 X \Gamma_x ] |\cdot\rangle =  \frac{1}{\ell A} |\cdot\rangle$ \\ \hline
$|-,-,-\rangle$ & $[ \frac{1}{18} (-4\alpha-9\beta_5) \Gamma_z + \frac13 X \Gamma_x ] |\cdot\rangle =  \frac{1}{\ell A} |\cdot\rangle$ \\ \hline
	\end{tabular}
\vskip 0.2cm
  \label{table5xxxxxyy}
 \end{center}
\end{table}
To find solutions with 20 supersymmetries, one needs  to consider the eigenspace with  6 eigenspinors in table  \ref{table5xxxxxyy}.  In such a case the integrability
of the remaining KSE requires that
\bea
{1\over 36}\beta_5^2+{1\over9} X^2={1\over\ell^2A^2}~.
\eea
Comparing this with the field equation of the warp factor
\bea
{1\over 9}X^2 +{1\over 6} \beta^2_5+{1\over 18} (\beta_1^2+\beta_4^2)+{1\over 54} (\beta_2^2+\beta_3^2)={1\over\ell^2 A^2}~,
\eea
we find that all $\beta$'s must vanish and so $Y=0$.  Thus, the flux $F$ is electric and, as we have demonstrated, such a background does not admit  $N>16$ AdS$_4$ supersymmetries.

\vskip 0.3cm
$\underline {\beta_5=0}$

Since the backgrounds with electric flux $F$  cannot preserve $N>16$ supersymmetries, we have to assume that at least one of the pairs $(\beta_1, \beta_3)$ and $(\beta_2, \beta_4)$
do not vanish. If either the pair $(\beta_1, \beta_3)$ or  $(\beta_2, \beta_4)$ is non-vanishing, the investigation of the algebraic KSE proceeds as in the previous case
with $\beta_5\not=0$. In particular, we find that the algebraic KSE (\ref{malgkse}) together with the field equation for the warp factor imply that all $\beta$'s vanish
and the flux $F$ is electric.  So there are no solutions preserving $N>16$ supersymmetries.

It remains to investigate the case that $\beta_1, \beta_2, \beta_3, \beta_4\not=0$.  If this is the case, the determinant of the linear system (\ref{sp2u2beta}) must vanish which
in turn implies that
\bea
-{3\over4} {a_2\over a_1a_4}+X \Big(X+\sqrt{a_1\over a_2a_4}\Big)=0~,~~~-{3\over4} {a_1\over a_2a_4}+X \Big(X+\sqrt{a_2\over a_1a_4}\Big)=0~.
\label{det2ssp22}
\eea
The solution of these equations is
\bea
\mathrm{either}\quad  a_1=a_2~,\quad \mathrm{or} \quad X=-{3\over4} {a_1+a_2\over \sqrt{a_1a_2a_4}}~.
\label{2xsp2xy}
\eea
Substituting the latter equation into (\ref{det2ssp22}), one again finds that $a_1=a_2$.  So, without loss of generality, we take $a_1=a_2$   in which case
\bea
\mathrm{either}\quad X={1\over 2\sqrt{a_4}}~,\quad \mathrm{or} \quad X=-{3\over 2\sqrt{a_4}}~.
\label{2xsp2xy1}
\eea
For the latter case, the linear system (\ref{sp2u2beta}) gives
\bea
\beta_3=-3\beta_1~,~~~\beta_2=-3\beta_4~.
\eea
After setting   $\beta_5=0$,  the investigation  of the algebraic KSE can be carried out as  that described in table \ref{table5xxxxxyy}. As a result, after comparing with the
field equation for the warp factor, the $\beta$'s vanish and $F$ is electric.  Thus there are no solutions preserving $N>16$ supersymmetries.

\begin{table}[h]
\begin{center}
\vskip 0.3cm
 \caption{Decomposition of (\ref{474cccv}) KSE into eigenspaces }
 \vskip 0.3cm
\begin{tabular}{|c|c|c|}\hline
$|P_1, P_2, P_3\rangle$ & relations for the fluxes \\ \hline
$|+,+,+\rangle$, $|-,-,-\rangle$&$ \frac13 X \Gamma_x |\cdot\rangle =  \frac{1}{\ell A} |\cdot\rangle$
\\ \hline
$|+,+,-\rangle$, $|+,-,+\rangle$, $|-,+,+\rangle$ & $[ \frac{2}{9} \alpha \Gamma_z + \frac13 X \Gamma_x ] |\cdot\rangle =  \frac{1}{\ell A} |\cdot\rangle$ \\ \hline
$|-,-,+\rangle$, $|-,+,-\rangle$, $|+,-,-\rangle$ & $[ -\frac{2}{9} \alpha \Gamma_z + \frac13 X \Gamma_x ] |\cdot\rangle =  \frac{1}{\ell A} |\cdot\rangle$ \\ \hline
	\end{tabular}
\vskip 0.2cm
  \label{table5xxxyzz}
 \end{center}
\end{table}

It remains to investigate the case that $X=1/ (2\sqrt{a_4})$ in (\ref{2xsp2xy1}).  In this case, the linear system (\ref{sp2u2beta}) gives
\bea
\beta_1=\beta_3~,~~~\beta_2=\beta_4~.
\eea
Using the $P_1, P_2$ and $P_3$ as in (\ref{procosx}), the algebraic KSE (\ref{malgkse}) becomes
\bea
\big [{1\over18}\big(-3\alpha P_1P_2P_3+ \alpha (P_1+P_2+P_3)\big) \Gamma_z+{1\over3} X\Gamma_x] \sigma_+={1\over\ell A} \sigma_+~,
\label{474cccv}
\eea
and the solutions in the eigenspaces of $P_1,P_2$ and $P_3$ are described in table \ref{table5xxxyzz}.
To preserve $N>16$ supersymmetries, one has to consider either one of the eigenspaces with 3 eigenspinors and the eigenspace with 2 eigenspinors or both of the eigenspaces with 3
eigenspinors.  In either case, one finds that all $\beta$'s vanish and so $Y=0$.  Then $F$ is electric  and such solutions do not preserve $N>16$ supersymmetries.
Therefore we conclude that the homogenous space ${Sp(2)}/{\Delta(Sp(1))}$ does not give rise to AdS$_4$ backgrounds with $N>16$.

\subsubsection{\texorpdfstring{${Sp(2)}/{Sp(1)}$}{Sp(2) over Sp(1)}} \label{sp2modsp1}

The geometry of this homogeneous space is described in appendix \ref{so5=sp2} where the definition of the generators of the algebra and expressions
for the curvature and invariant forms can be found.  A left-invariant frame is given by  $\ell^A m_A= \ell^a W_a+ \ell^r T^{(+)}_r$, where $a=1,\dots, 4$ and $r=1,2,3$.  Then, the most general left-invariant metric is
\bea
ds^2= a \delta_{ab} \bbl^a \bbl^b+ g_{rs} \bbl^{r} \bbl^s~,
\label{mettrr}
\eea
where  $a>0$ is a constant and  $(g_{rs})$ is any constant $3\times 3$  positive definite symmetric matrix.

To simplify the computations that follow, it is convenient to use the covariant properties of the decomposition of $\mathfrak{sp}(2) = \mathfrak{so}(5)$ as described in (\ref{so5so3aa})  to restrict the number of parameters in the metric. In particular, observe that the decomposition  (\ref{so5so3aa}) remains invariant
under the transformation of the generators
\bea
T_r^{(+)}\rightarrow O_r{}^s T_s^{(+)}~,~~~W_a\rightarrow U_a{}^b W_b~,~~~T_r^{(-)}\rightarrow  T^{(-)}_r~,
\eea
where $O\in SO(3)$ and $U\in Spin(3)\subset SO(4)$  defined as
\bea
O_r{}^s I^{(+)}_s=  U I^{(+)}_r U^{-1}~,
\eea
since $I^{(+)}_r$ are  the gamma matrices of the Majorana spinor
representation of $\mathfrak{so}(3)$ on $\bR^4= \bC^2\oplus \bar\bC^2$. Furthermore notice that
$U I^{(-)}_r U^{-1}=I^{(-)}_r$ as $U$ is generated by the $I^{(+)}_r$ which commute with all $I^{(-)}_s$.
The orthogonal rotations $O$ act on the matrix $(g_{rs})$ as $g\rightarrow O g O^{-1}$.  As $(O, U)$
is an automorphism of $\mathfrak{so}(5)$ which leaves the decomposition (\ref{so5so3aa})
invariant,  we can use $O$ to put the matrix $(g_{rs})$ into diagonal form.  So from now on, without loss
of generality, we set $(g_{rs})=\mathrm{diag}(b_1, b_2, b_3)$ with $b_1, b_2, b_3>0$, see also \cite{ziller}.

The left-invariant 4-forms are generated by
\bea
\psi={1\over 4!}\epsilon_{abcd} \ell^a\wedge \ell^b\wedge \ell^c\wedge \ell^d~,~~~\rho_{rs}={1\over2} \epsilon_{rpq} \ell^p\wedge \ell^q \wedge I^{(+)}_s~,
\eea
where
\bea
I^{(+)}_s={1\over2} (I^{(+)}_s)_{ab}\, \ell^a\wedge \ell^b~.
\eea
Therefore, the 4-form flux $Y$ can  be chosen as
\bea
Y= \alpha \psi + \beta^{rs} \rho_{rs}~,
\eea
where $\alpha$ and $\beta^{rs}$ are constants.
Then it is straightforward to find  that the Bianchi identity $dY=0$ implies that
\bea
\beta^{rs}=\beta^{sr}~.
\eea
Furthermore, define $\sigma={1\over 3!}  \epsilon_{rst} \ell^r\wedge \ell^s\wedge \ell^t$ and choose as top form $d\mathrm{vol}=a^2 \sqrt{b_1 b_2b_3}\, \sigma\wedge \psi$. Then the field equation for $Y$, $d\ast_{{}_7}Y=XY$, gives the linear system
\bea
\sum_{r=1}^3 b_r \beta^{rr}= \sqrt{b_1b_2b_3}\, X \alpha~,~~~{\alpha\over2} {\sqrt{b_1b_2b_3}\over a^2}-{1\over3} {\sum_{r=1}^3 b_r \beta^{rr}\over \sqrt{b_1b_2b_3}}={X\over 3} \beta
\cr
\big( b_r \beta^{rs}+\beta^{rs} b_s-{2\over 3} \delta^{rs} \sum_{t=1}^3 b_t \beta^{tt}\big)=\sqrt{b_1b_2b_3} X (\beta^{rs}-{1\over3} \delta^{rs} \beta)~,
\label{lastls}
\eea
where there is no summation over the indices $r$ and $s$ on the left-hand side of the last equation and $\beta=\delta_{rs} \beta^{rs}$.

Before we proceed to investigate the solutions of the linear system, notice that if $\beta^{rs}=0$,
then $\alpha=0$ and so $F$ is electric.  The supersymmetry preserved by these
solutions will be investigated later.  We will show that such solutions cannot preserve more than 16
supersymmetries.

Furthermore writing $Y=\alpha \psi+ Y_\beta$, where $Y_\beta=\beta^{rs} \rho_{rs}$, the field equation
of the warp factor in (\ref{m-einst-ads}) can be written as
\bea
{1\over 9} X^2+{1\over18} \alpha^2 a^{-4} +{1\over 432} (Y_\beta)^2={1\over \ell^2 A^2}~.
\label{k1ller}
\eea
As we shall demonstrate, the compatibility of this field equation with the algebraic
KSE rules out the existence of $N>16$ backgrounds.

Returning to the solutions of (\ref{lastls}), let us focus on $\beta^{rs}$ with $r\not=s$. There are several cases
to consider.

\vskip0.3cm
\underline{Either $\beta^{rs}\not=0$ for all $r\not=s$ or $\beta^{rs}=0$ for all $r\not=s$ }

If $\beta^{rs}$,  $r\not=s$, are all non-vanishing, the last equation in  (\ref{lastls}) implies that

\bea
b_1=b_2=b_3~,~~~ X=2{b_1\over \sqrt{b_1b_2b_3}}~.
\eea
  As a result, the metric is invariant under  $SO(3)$ and this can be used  to
bring $\beta^{rs}$ into diagonal form.  Of course $(\beta^{rs})$ is also diagonal if
$\beta^{rs}=0$ for all $r\not=s$.

So, without loss of generality, we can assume   that  $(\beta^{rs})$
is diagonal. Setting
\bea
J_1= \Gamma^{6714}~,~~~J_2=\Gamma^{6723}~,~~~J_3=\Gamma^{7524}~,
\eea
where all gamma matrices are in the orthonormal basis and $\{\Gamma^i\}=\{\Gamma^a, \Gamma^{4+r}\}$,
the algebraic KSE can be written as
\bea
&&\Big({1\over6} \Big[-\alpha a^{-2} J_1 J_2+ {a^{-1} \over \sqrt{b_1 b_2b_3}}  \big(\sqrt{b_1}\beta^{11} (J_1+J_2)
+\sqrt{b_2} \beta^{22} J_3 (1+ J_1J_2)
\cr &&
+ \sqrt{b_3} \beta^{33}
J_3 (J_1+J_2)\big)\Big] \Gamma_z
+{1\over3} X \Gamma_x\Big) \sigma_+={1\over \ell A} \sigma_+~.
\label{493cccv}
\eea
The decomposition of the algebraic KSE into the eigenspaces of the commuting Clifford algebra
operators $J_1, J_2, J_3$ is  illustrated in table \ref{table5omega}.
\begin{table}[ht]
\begin{center}
\vskip 0.3cm
 \caption{Decomposition of (\ref{493cccv}) KSE into eigenspaces }
 \vskip 0.3cm
\begin{tabular}{|c|c|c|}\hline
$|J_1, J_2, J_3\rangle$ & relations for the fluxes \\ \hline
$|+,+,\pm\rangle$ & $\begin{array}{r@{}l@{}} \big( \frac{1}{6} [-\alpha a^{-2} & + 2 {a^{-1} \over \sqrt{b_1 b_2b_3}} (\sqrt{b_1}
 \beta^{11}\pm\sqrt{b_2} \beta^{22} \\ & \hspace*{1cm} \pm\sqrt{b_3} \beta^{33})] \Gamma_z + \frac13 X \Gamma_x \big) |\cdot\rangle =  \frac{1}{\ell A} |\cdot\rangle \end{array}$ \\ \hline
 $|+,-,\pm\rangle$ $|-,+,\pm\rangle$& $\left( \frac{1}{6} \alpha a^{-2} \Gamma_z + \frac13 X \Gamma_x \right) |\cdot\rangle =  \frac{1}{\ell A} |\cdot\rangle$
 \\
 \hline
$|-,-,\pm\rangle$ & $\begin{array}{r@{}l@{}} \big( \frac{1}{6} [-\alpha a^{-2} & + 2 {a^{-1} \over \sqrt{b_1 b_2b_3}} (-\sqrt{b_1}
 \beta^{11}\pm \sqrt{b_2} \beta^{22} \\ & \hspace*{1cm} \mp\sqrt{b_3} \beta^{33})] \Gamma_z + \frac13 X \Gamma_x \big) |\cdot\rangle =  \frac{1}{\ell A} |\cdot\rangle \end{array}$
 \\
 \hline
	\end{tabular}
\vskip 0.2cm
  \label{table5omega}
 \end{center}
\end{table}

To construct $N>16$ solutions, we have to include the eigenspace
with  four eigenspinors.  The integrability condition of the remaining KSE described in table
\ref{table5omega} gives
\bea
{1\over36} \alpha^2 a^{-4} +{1\over 9} X^2={1\over \ell^2 A^2}~.
\label{intconomega}
\eea
Comparing (\ref{intconomega}) with the field equation for the warp factor (\ref{k1ller}), we
find that $\alpha=\beta^{rs}=0$.  Therefore $Y=0$, so $F$ is electric.

\vskip 0.3cm

{\underline{ $\beta^{12}, \beta^{13}\not=0$ and  $\beta^{23}=0$}}

As the other two cases for which either $\beta^{13}=0$ or $\beta^{12}=0$ with  the rest of the components non-vanishing can be treated in a similar way,  we take, without loss of generality, that $\beta^{23}=0$ and $\beta^{12}, \beta^{13}\not=0$.  In such a case, the last condition in (\ref{lastls}) gives
\bea
X={b_1+b_2\over \sqrt{b_1 b_2b_3}}~,~~~b_2=b_3~.
\eea
The metric is invariant under an $SO(2)\subset SO(3)$ symmetry which acts with the vector
representation on the vector  $(\beta^{12}, \beta^{13})$ and leaves the form
of $(\beta^{rs})$ invariant.  As a result up to an $SO(2)$ transformation, we can set $\beta^{13}=0$ as well.
Furthermore, if $b_1\not= b_2$, the diagonal terms in the last condition in (\ref{lastls})  give
\bea
\beta^{11}=-\beta^{22}=-\beta^{33}~.
\eea
On the other hand if $b_1=b_2$ the analysis reduces to that of the previous case.
Therefore for $b_1\not= b_2$, $Y$ can be written as
\bea
Y=\alpha \psi+\beta^{11} (\rho_{11}-\rho_{22}-\rho_{33})+ \beta^{12} (\rho_{12}+\rho_{21})~.
\eea
Introducing  the  Clifford algebra operators
\begin{align}
J_1=\cos\theta \Gamma^{6714}+\sin\theta \Gamma^{6724}~,~~~J_2=\cos\theta \Gamma^{5724}-\sin\theta \Gamma^{5714}~,~~~J_3=\Gamma^{1234}~,
\label{SO5SO3Aprojectors1}
\end{align}
where $\tan\theta=\beta^{12}/\beta^{11}$,
the algebraic KSE can be written as
\begin{multline}
\Big({1\over6} \Big[\alpha a^{-2} J_3+\frac{a^{-1}}{\sqrt{b_1} b_2} (\sqrt{b_2}\beta^{11} J_1  J_2 (1 - J_3) \\
 + \sqrt{(\beta^{11})^2+(\beta^{12})^2} (\sqrt{b_1} J_1 + \sqrt{b_2}J_2)(1 - J_3) )\Big] \Gamma_z
+{1\over3} X \Gamma_x\Big) \sigma_+={1\over \ell A} \sigma_+ \; .
\label{493cccvi}
\end{multline}
The decomposition of the algebraic KSE into the eigenspaces of the commuting Clifford algebra
operators $J_1, J_2, J_3$ is  illustrated in table \ref{table5omega1}.

\begin{table}[ht]
\begin{center}
\vskip 0.3cm
 \caption{Decomposition of (\ref{493cccvi}) KSE into eigenspaces }
 \vskip 0.3cm
\begin{tabular}{|c|c|c|}\hline
$|J_1, J_2, J_3\rangle$ & relations for the fluxes \\ \hline
$|\pm,+,-\rangle$ & $\begin{array}{r@{}l@{}} \Big( \frac{1}{6} [-\alpha a^{-2} & +\frac{2 a^{-1}}{\sqrt{b_1} b_2} (\pm \sqrt{b_2} \beta^{11}  \\ & \hspace*{-1cm} +\sqrt{(\beta^{11})^2+(\beta^{12})^2} (\pm\sqrt{b_1}+\sqrt{b_2}))] \Gamma_z + \frac13 X \Gamma_x \Big) |\cdot\rangle =  \frac{1}{\ell A} |\cdot\rangle \end{array}$ \\ \hline
 $|+,\pm,+\rangle$ $|-,\pm,+\rangle$& $\left( \frac{1}{6} \alpha a^{-2} \Gamma_z + \frac13 X \Gamma_x \right) |\cdot\rangle =  \frac{1}{\ell A} |\cdot\rangle$
 \\
 \hline
$|\pm,-,-\rangle$ & $\begin{array}{r@{}l@{}} \Big( \frac{1}{6} [-\alpha a^{-2} & +\frac{2 a^{-1}}{\sqrt{b_1} b_2} (\mp \sqrt{b_2} \beta^{11} \\ & \hspace*{-1cm} +\sqrt{(\beta^{11})^2+(\beta^{12})^2} (\pm \sqrt{b_1}-\sqrt{b_2}))] \Gamma_z + \frac13 X \Gamma_x \Big) |\cdot\rangle =  \frac{1}{\ell A} |\cdot\rangle \end{array}$
 \\
 \hline
	\end{tabular}
\vskip 0.2cm
  \label{table5omega1}
 \end{center}
\end{table}

To construct solutions preserving more than 16 supersymmetries, we have to include the eigenspace
with  four eigenspinors leading again to the integrability condition (\ref{intconomega}).
Comparing again with the field equations of the warp factor (\ref{k1ller}), we deduce
that $F$ is electric.

\vskip 0.3cm
$\underline {\beta^{13}=\beta^{23}=0 \;\; \text{but $\beta^{12} \not=0$}}$

All three cases for which only  one of the three off-diagonal components of $(\beta^{rs})$ is non-zero  can be treated symmetrically. So, without loss of generality,  one can  take  $\beta^{13}=\beta^{23}=0$ but $\beta^{12} \not=0$.
In this case, the last equation in (\ref{lastls}) has four branches of solutions depending on the choice of the  $b_1$, $b_2$ and $b_3$ components of the metric.
\begin{enumerate}
\item $b_1=b_2=b_3=b$\newline The last equation in (\ref{lastls}) then implies $X=2/\sqrt{b}$ and the aforementioned residual $SO(3)$ symmetry can be used to set $\beta^{rs}$ to be diagonal.
\item $b_1=b_2$, $b_2 \not= b_3$\newline The last equation in (\ref{lastls}) then implies $X=2/\sqrt{b_3}$ and $\beta^{33}=0$. The aforementioned residual $SO(2)$ symmetry can be used to put $\beta^{rs}$ into diagonal form.
\item $b_2 \not= b_3$, $b_1 + b_2 = 2 b_3$\newline The last equation in (\ref{lastls}) then implies  $X=(b_1+b_2)/\sqrt{b_1 b_2 b_3}$ and $\beta^{11}=\beta^{22}=0$.
In such a case, $Y$ reads
\bea
Y=\alpha \psi + \beta^{33} \rho_{33} + \beta^{12} (\rho_{12}+\rho_{21}) \; .
\eea
Choosing
\bea\label{SO5SO3Acase43projectors}
J_1=\Gamma^{1457}~,~~~J_2=\Gamma^{2467}~,~~~J_3=\Gamma^{1234} \; ,
\eea
the algebraic KSE can be written as
\begin{multline}
\Big({1\over6} \Big[\alpha a^{-2} J_3+\frac{a^{-1}}{\sqrt{b_1 b_2 b_3}} \left(\sqrt{b_3} \beta^{33} J_1 J_2 (1-J_3) \right. \\ \left. - \beta^{12} (\sqrt{b_2}J_1-\sqrt{b_1}J_2) (1-J_3)\right)\Big] \Gamma_z
+{1\over3} X \Gamma_x\Big) \sigma_+={1\over \ell A} \sigma_+ \; .
\label{SO5SO3AKSEcase43}
\end{multline}
The decomposition of the algebraic KSE into the eigenspaces of $J_1, J_2, J_3$ is  illustrated in table \ref{SO5SO3AKSEcase43KSEtable}.
\begin{table}[ht]
\begin{center}
\vskip 0.3cm
 \caption{Decomposition of (\ref{SO5SO3AKSEcase43}) KSE into eigenspaces }
 \vskip 0.3cm
\begin{tabular}{|c|c|c|}\hline
$|J_1, J_2, J_3\rangle$ & relations for the fluxes \\ \hline
$|\pm,+,-\rangle$ & $\begin{array}{r@{}l@{}} \Big( \frac{1}{6} [-\alpha a^{-2} & +\frac{2 a^{-1}}{\sqrt{b_1 b_2 b_3}} (\pm \sqrt{b_3} \beta^{33}  \\ & -\beta^{12} (\pm\sqrt{b_2}-\sqrt{b_1}))] \Gamma_z + \frac13 X \Gamma_x \Big) |\cdot\rangle =  \frac{1}{\ell A} |\cdot\rangle \end{array}$ \\ \hline
 $|+,\pm,+\rangle$ $|-,\pm,+\rangle$& $\left( \frac{1}{6} \alpha a^{-2} \Gamma_z + \frac13 X \Gamma_x \right) |\cdot\rangle =  \frac{1}{\ell A} |\cdot\rangle$
 \\
 \hline
$|\pm,-,-\rangle$ & $\begin{array}{r@{}l@{}} \Big( \frac{1}{6} [-\alpha a^{-2} & +\frac{2 a^{-1}}{\sqrt{b_1 b_2 b_3}} (\mp \sqrt{b_3} \beta^{33}  \\ & -\beta^{12} (\pm\sqrt{b_2}+\sqrt{b_1}))] \Gamma_z + \frac13 X \Gamma_x \Big) |\cdot\rangle =  \frac{1}{\ell A} |\cdot\rangle \end{array}$
 \\
 \hline
	\end{tabular}
\vskip 0.2cm
  \label{SO5SO3AKSEcase43KSEtable}
 \end{center}
\end{table}
Again the eigenspace with four eigenspinors has to be included in the construction of $N>16$ backgrounds.  As a result, this leads to the integrability condition (\ref{intconomega}) which together with the warp factor field equation (\ref{k1ller}) imply that $F$ is electric.

\item $b_1 \not= b_2$, $b_1 + b_2 \not= 2 b_3$\newline The last equation in (\ref{lastls}) then implies
   \begin{align}
    X=\frac{b_1+b_2}{\sqrt{b_1 b_2 b_3}}~,~~~ \beta^{11}=-\beta^{22}=\beta^{33} \, \frac{2 b_3 - b_1 - b_2}{b_1 - b_2}~.
    \end{align}
     In such a case, $Y$ reads
\bea
Y=\alpha \psi + \beta^{11} (\rho_{11}-\rho_{22}+\frac{b_1 - b_2}{2 b_3 - b_1 - b_2} \rho_{33})+ \beta^{12} (\rho_{12}+\rho_{21}) \; .
\eea
With the choice of commuting Clifford algebra operators as in~\eqref{SO5SO3Aprojectors1}, the algebraic KSE can be written as
\begin{align}\label{SO5SO3AKSEcase44}
\Big({1\over6} &\Big[\alpha a^{-2} J_3+\frac{a^{-1}}{\sqrt{b_1 b_2 b_3}} \big(\frac{(b_1-b_2)\sqrt{b_3}}{b_1+b_2-2 b_3} \beta^{11} J_1 J_2 (1-J_3) \notag\\
 &+\sqrt{(\beta^{11})^2+(\beta^{12})^2} (\sqrt{b_1} J_1 + \sqrt{b_2}J_2) (1-J_3) \big)\Big] \Gamma_z \notag\\
&+{1\over3} X \Gamma_x\Big) \sigma_+={1\over \ell A} \sigma_+ \; .
\end{align}
The decomposition of the algebraic KSE into the eigenspaces of $J_1, J_2, J_3$ is  illustrated in table \ref{SO5SO3AKSEcase44KSEtable}.

To construct $N>16$ solutions, we again have to include the eigenspace
with  four eigenspinors which leads to the integrability condition (\ref{intconomega}).
Comparing with the warp factor field equation (\ref{k1ller}), we once again deduce
that $F$ is electric.

\end{enumerate}

\begin{table}[ht]
\begin{center}
\vskip 0.3cm
 \caption{Decomposition of (\ref{SO5SO3AKSEcase44}) KSE into eigenspaces }
 \vskip 0.3cm
\begin{tabular}{|c|c|c|}\hline
$|J_1, J_2, J_3\rangle$ & relations for the fluxes \\ \hline
$|\pm,+,-\rangle$ & $\begin{array}{r@{}l@{}} \Big( \frac{1}{6} [-\alpha a^{-2} & +\frac{2 a^{-1}}{\sqrt{b_1 b_2 b_3}} (\pm \frac{(b_1-b_2)\sqrt{b_3}}{b_1+b_2-2 b_3} \beta^{11}  \\ & \hspace*{-1cm} +\sqrt{(\beta^{11})^2+(\beta^{12})^2} (\pm\sqrt{b_1}+\sqrt{b_2}))] \Gamma_z + \frac13 X \Gamma_x \Big) |\cdot\rangle =  \frac{1}{\ell A} |\cdot\rangle \end{array}$ \\ \hline
 $|+,\pm,+\rangle$ $|-,\pm,+\rangle$& $\left( \frac{1}{6} \alpha a^{-2} \Gamma_z + \frac13 X \Gamma_x \right) |\cdot\rangle =  \frac{1}{\ell A} |\cdot\rangle$
 \\
 \hline
$|\pm,-,-\rangle$ & $\begin{array}{r@{}l@{}} \Big( \frac{1}{6} [-\alpha a^{-2} & +\frac{2 a^{-1}}{\sqrt{b_1} b_2} (\mp \frac{(b_1-b_2)\sqrt{b_3}}{b_1+b_2-2 b_3} \beta^{11} \\ & \hspace*{-1cm} +\sqrt{(\beta^{11})^2+(\beta^{12})^2} (\pm \sqrt{b_1}-\sqrt{b_2}))] \Gamma_z + \frac13 X \Gamma_x \Big) |\cdot\rangle =  \frac{1}{\ell A} |\cdot\rangle \end{array}$
 \\
 \hline
	\end{tabular}
\vskip 0.2cm
  \label{SO5SO3AKSEcase44KSEtable}
 \end{center}
\end{table}

It remains to investigate the number of supersymmetries preserved by the solutions for which $F$ is electric.
For this, one has to investigate the integrability condition of the gravitino KSE (\ref{intads4grav}).
Using the expression for the curvature in (\ref{curvsp2sp1aa})-(\ref{curvsp2sp1ad}) and requiring that the solution preserves $N>16$, we find that
\bea
\delta^{ca} \delta^{db} (I^{(-)}_r)_{ab}(R_{cd,mn} \Gamma^{mn} - \frac{1}{18} X^2 \Gamma_{cd}) \sigma_+ = 0~,
\eea
implies
\bea
a-{1\over 8} \delta^{rs} g_{rs}-{1\over18} a^2\, X^2=0~.
\label{ina}
\eea
Next requiring again that $N>16$, one finds that the condition
\bea
\delta^{ca} \delta^{db} (I^{(+)}_r)_{ab}(R_{cd,mn} \Gamma^{mn} - \frac{1}{18} X^2 \Gamma_{cd}) \sigma_+ = 0~.
\eea
gives
\bea
&&\delta^{pq} g_{pq} \epsilon_{rst}-{1\over2} a^{-1} \epsilon_t{}^{pq} g_{pr} g_{qs}- 2 g_{tp} \epsilon^p{}_{rs}=0~,
\cr
&&-{3\over4} g_{rs}+{1\over8} \delta^{pq} g_{pq} \delta_{rs}+ a\delta_{rs}- {1\over18} a^2\, X^2 \delta_{rs}=0~.
\label{inb}
\eea
Substituting (\ref{ina}) into the second equation in (\ref{inb}), one finds after a bit of analysis that
\bea
b_1=b_2=b_3~.
\eea
Setting  $b=b_1=b_2=b_3$ and substituting this back into (\ref{ina}) and (\ref{inb}), one deduces that
\bea
2a=b~,~~~X^2=9 b^{-1}~.
\eea
As $X^2=9 \ell^{-2} A^{-2}$, we have $b=\ell^2 A^2$ and $a=(1/2)\ell^2 A^2$.
The rest of the integrability condition is satisfied without further conditions. Therefore, every solutions that preserves $N>16$ supersymmetries is maximally supersymmetric
and so locally isometric to $AdS_4\times S^7$.

One can confirm this result  by investigating the Einstein equation (\ref{m-einst-trans}). As  all  solutions  with electric $F$  are Einstein $R^{(7)}_{ij}=(1/6) X^2 \delta_{ij}$, it suffices to identify
the left-invariant metrics on ${Sp(2)}/{Sp(1)}$ that are Einstein.  There are two Einstein metrics \cite{ziller, gibbons} on ${Sp(2)}/{Sp(1)}$ given by
\bea
X^2=9 b^{-1}~,~~~2a=b~,~~~b_1=b_2=b_3=b~,
\eea
and
\bea
X^2={81\over 25} b^{-1}~,~~~2a=5b~,~~~b_1=b_2=b_3=b~,
\eea
where the first one is the round metric on $S^7$, see also \cite{jensen}.  The second one does not give $N>16$ supersymmetric solutions.

\section{Summary}\label{ads4conclusions}

Up to local isometries, we have classified all warped AdS$_4$ backgrounds with the most general allowed fluxes in 10- and 11-dimensional supergravities that preserve $N>16$ supersymmetries.
We have demonstrated that up to an overall scale, the only solutions that arise are  the maximally supersymmetric solution $AdS_4\times S^7$ of 11-dimensional
supergravity \cite{fr, duffpopemax}  and the $N=24$ solution $AdS_4\times\mathbb{CP}^3$  of IIA supergravity \cite{nillsonpope}.  These two solutions are related via dimensional reduction  along
the  fibre of the Hopf fibration $S^1\rightarrow S^7\rightarrow \mathbb{CP}^3$.

The assumption we have made  to prove these results is that either the solutions are smooth and the internal space is compact without boundary or that the even part $\mathfrak{g}_0$ of
the Killing superalgebra of the backgrounds decomposes as $\mathfrak{g}_0=\mathfrak{so}(3,2)\oplus \mathfrak{t}_0$. In fact, these two assumptions are equivalent for $N>16$  AdS$_4$
backgrounds.  It may be possible to weaken these assumptions but they cannot be removed altogether.  This is because otherwise additional
solutions will exist. For example, the maximally supersymmetric $AdS_7\times S^4$ solution of 11-dimensional supergravity \cite{townsend} can be re-interpreted as a maximally supersymmetric  warped
AdS$_4$ solution. However, in such a case the ``internal'' 7-dimensional manifold $M^7$ is not compact and the even subalgebra of the Killing superalgebra $\mathfrak{g}_0$ does not decompose as $\mathfrak{so}(3,2)\oplus \mathfrak{t}_0$ (cf.\, sections \ref{sugrabackks} and \ref{ksaintro}).

We have identified all AdS$_4$ backgrounds up to a local isometry. Therefore, we have specified all the local geometries of the internal spaces $G/H$  of these solutions.  However, the  possibility remains that there are more solutions which arise via  additional discrete identifications   $Z\backslash G/H$, where $Z$ is a discrete subgroup of $Z\subset G$.  The  $AdS_4\times Z\backslash G/H$ solutions will preserve at most as many supersymmetries as the  $AdS_4\times G/H$ solutions.  As in IIB and massive IIA supergravities there are no $N>16$  $AdS_4\times G/H$ solutions, there are no $N>16$  $AdS_4\times Z\backslash G/H$ solutions either.   In  IIA theory, the possibility remains that there can be $AdS_4\times Z \backslash \mathbb{CP}^3$ solutions with 24 and 20 supersymmetries. In $D=11$ supergravity as $AdS_4\times S^7$ preserves 32 supersymmetries, there may be $AdS_4\times Z\backslash S^7$ solutions preserving 28, 24 and 20 supersymmetries.  Such solutions have been used in the context of AdS/CFT in \cite{abjm}. A systematic investigation of all possible $N>16$  $AdS_4\times Z\backslash G/H$ backgrounds
will involve the identification of all discrete subgroups of $G$. The relevant groups here are $SU(4)$
and $\mathrm{Spin}(8)$, see e.g. \cite{hanany} for an exposition of discrete subgroups of $SU(4)$ and references therein.

 It is clear from our results on AdS$_4$ backgrounds that supersymmetric AdS solutions which preserve $N>16$ supersymmetries in 10- and 11-dimensions are severely restricted.
Consequently there are few gravitational duals for superconformal theories with a large number of supersymmetries which have distinct local geometries.  For example, the
  superconformal theories of \cite{ferrara, aharony, garcia} have  gravitational duals which are locally isometric to  the $AdS_5\times S^5$ maximally supersymmetric background as there are no distinct local AdS$_5$ geometries that preserve strictly 24 supersymmetries \cite{ads5clas}. In general, our
results also suggest that there may not be a large number of backgrounds that preserve $N>16$ supersymmetries in 10- and 11-dimensional supergravities.

\chapter{A Non-existence Theorem for \texorpdfstring{$N > 16$}{N greater 16} Supersymmetric \texorpdfstring{AdS$_3$}{AdS3} Backgrounds}\label{ads3}

In this chapter, we show that there are no smooth warped   AdS$_3$ solutions with a compact without boundary internal manifold in  10- and 11-dimensional supergravities preserving strictly more than 16 supersymmetries. The following work was conducted in collaboration with Alexander Haupt and George Papadopoulos and published in \cite{ahslgp2}.

\section{Introduction}

The main purpose of this chapter is to complete the classification of AdS backgrounds that preserve strictly more than 16 supersymmetries in 10- and 11-dimensional supergravities. After our analysis of AdS$_4$ solutions in chapter \ref{ads4}, the only class of backgrounds that remains to be investigated are the warped AdS$_3$ backgrounds with the most general allowed fluxes in 10-dimensional type II and 11-dimensional supergravity theories preserving $N>16$ supersymmetries. For these, we shall demonstrate a non-existence theorem provided that the solutions are smooth
and their internal spaces are  compact manifolds without boundary. It suffices to establish the non-existence theorem up to  local isometries.  The more general
result follows, as there are no new geometries that can be constructed by taking quotients by discrete groups.

The method used to establish this result again relies on a number of previously mentioned recent developments, namely the integration of the KSEs along the AdS$_n$ subspace \cite{mads, iibads, iiaads}, the homogeneity theorem (cf.\ section \ref{homogentheorem}), as well as the classification of the Killing superalgebras (cf.\ section \ref{ksaintro}). 
Then the strategy of the proof is as follows.  First, one establishes that for all $\text{AdS}_3\times_w M^{D-3}$ backgrounds that preserve $N>16$ supersymmetry, the warp factor is constant. Therefore,
 the geometry is a product $\text{AdS}_3\times M^{D-3}$. To show this, either
one uses that the solutions are smooth and the internal space is compact without boundary as well as techniques from the proof of the homogeneity theorem or that the even subalgebra
$\mathfrak{g}_0$ of the Killing
superalgebra $\mathfrak{g}$ of $\text{AdS}_3\times_w M^{D-3}$ decomposes as $\mathfrak{g}_0=\mathfrak{iso}(\text{AdS}_3)\oplus \mathfrak{iso} (M^{D-3})$, where $\mathfrak{iso}(AdS_3)$
is  the isometry algebra of AdS$_3$
and   $\mathfrak{iso} (M^{D-3})=\mathfrak{t}_0$  is the isometry algebra of the internal space $M^{D-3}$.

Having established that the $N>16$ AdS$_3$ backgrounds are products, $\text{AdS}_3\times M^{D-3}$, and that $\mathfrak{g}_0=\mathfrak{iso}(\text{AdS}_3)\oplus \mathfrak{t}_0$,
where $\mathfrak{t}_0$ is an algebra of isometries on $M^{D-3}$, we obtain by virtue of the homogeneity theorem that the internal space is a homogeneous space $G/H$
with $\mathfrak{Lie} \,G=\mathfrak{t}_0$.  In addition, the theorem requires that all  fields are invariant under the left action of $G$ on $G/H$.

The final part of the proof involves the identification of all homogeneous spaces\footnote{As we are investigating supersymmetric backgrounds, we require that all the internal spaces are spin.}  in seven and eight dimensions that admit a transitive and an almost effective  action of a group $G$
with Lie algebra $\mathfrak{t}_0$.  For $\mathfrak{t}_0$  semisimple, one   can identify the relevant  homogeneous spaces     using  the classification results of (simply connected)  7- and 8-dimensional homogeneous
manifolds in  \cite{Castellani:1983yg,klausthesis,niko6dim,niko7dim,bohmkerr}; for a concise description see \cite{klausthesis}. We also need to find a procedure to help us determine whether a $\mathfrak{t}_0$ can act effectively on a given homogeneous space $G/H$.
 In section \ref{superxx},  we refer to this as ``modification'' of a homogeneous space. A similar approach can be used for the case that
 $\mathfrak{t}_0$  is not semisimple.   After identifying all the suitable homogeneous spaces,  a substitution of the geometric data into the field equations and KSEs of supergravity
theories in 10- and 11-dimensions establishes our non-existence theorem.

Before we proceed with the proof, let us investigate the necessity for the assumptions we have made. First,
 one can establish that if  $\text{AdS}_3\times_w M^{D-3}$ is smooth and $M^{D-3}$   is compact without boundary, then the even subalgebra of the Killing superalgebra
of  $\text{AdS}_3\times_w M^{D-3}$ will decompose as $\mathfrak{g}_0=\mathfrak{iso}(\text{AdS}_3)\oplus \mathfrak{t}_0$ \cite{superalgebra}.  The requirement that $M^{D-3}$   must be compact without boundary
may be weakened but not completely removed. If it is removed, then $\mathfrak{g}_0$ may  not decompose as stated above. In addition, the warp factor of $\text{AdS}_3\times_w M^{D-3}$
backgrounds with $N>16$ supersymmetries may not be constant and  there exist AdS$_3$ backgrounds that preserve $N>16$ supersymmetries, see \cite{desads} for a detailed exposition.  In particular the maximally supersymmetric $\text{AdS}_4\times S^7$ and  $\text{AdS}_7\times S^4$ backgrounds of 11-dimensional supergravity can be
viewed as warped AdS$_3$ backgrounds but the internal spaces are not compact without boundary.

This chapter is organised as follows. In section \ref{superxx}, we describe the Killing superalgebras of AdS$_3$ backgrounds and introduce the notion of a modification
of a homogeneous space, which will allow us to test whether an algebra can act effectively on a given space.  In sections \ref{ads311D}, \ref{ads3IIA} and \ref{ads3IIB} we prove the main result of this chapter
for 11-dimensional, IIA and IIB supergravities, respectively. Our conventions can be found in \ref{conventions}. In appendix \ref{modif}, we present some aspects of the structure
of homogeneous spaces admitting a transitive action by a compact but not semisimple Lie group which are useful in the proof of our results. In appendix \ref{appencx2},
we describe the geometry of the homogeneous space $N^{k,l}$.

\section{Symmetries of \texorpdfstring{AdS$_3$}{AdS3} backgrounds}\label{superxx}

\subsection{ Killing superalgebras of \texorpdfstring{AdS$_3$}{AdS3} backgrounds }

Since AdS$_3$ is locally a group manifold, the
Killing superalgebras of warped AdS$_3$ backgrounds with the most general allowed fluxes  decompose as $\mathfrak{g}=\mathfrak{g}_L\oplus \mathfrak{g}_R$, where $\mathfrak{g}_L$ and  $\mathfrak{g}_R$  are associated with the
left and right actions. The left and right Killing superalgebras $\mathfrak{g}_L$ and $\mathfrak{g}_R$ have been identified in \cite{superalgebra}.
This has been done under the assumption that either the internal space is compact without boundary or that the even  subalgebra decomposes as
$(\mathfrak{g}_L{})_0=\mathfrak{sl}(2, \bR)_L\oplus (\mathfrak{t}_L)_0$ and similarly for $ \mathfrak{g}_R$.  As $\mathfrak{g}_0=\mathfrak{iso}(\text{AdS}_3)\oplus \mathfrak{t}_0=(\mathfrak{g}_L{})_0\oplus (\mathfrak{g}_R{})_0$,
$\mathfrak{iso}(\text{AdS}_3)$
is isomorphic to either $\mathfrak{sl}(2, \bR)_L$ or $\mathfrak{sl}(2, \bR)_R$ if the background has only either left or right supersymmetries, respectively, or $\mathfrak{iso}(\text{AdS}_3)=\mathfrak{sl}(2, \bR)_L\oplus \mathfrak{sl}(2, \bR)_R$ if the background has both left and right supersymmetries. Furthermore $\mathfrak{t}_0= (\mathfrak{t}_L)_0\oplus (\mathfrak{t}_R)_0$.

 It has been shown in \cite{superalgebra} that for AdS$_3$ backgrounds $\mathfrak{t}_0$ may not be semisimple and in addition may admit   central terms $\mathfrak{c}$ which commute
 with all other generators of the superalgebra.  We shall show below that in all cases bar one $\mathfrak{c}_L=\{0\}$.  If $\mathfrak{c}_L\not=\{0\}$,  it will have at most dimension 3.  The left and right superalgebras are isomorphic
 and so it suffices to present only the left ones. These are tabulated in table\footnote{Throughout this thesis $\mathfrak{sp}(n)$, $n\geq1$, denotes the compact symplectic
 Lie algebras.  These have been denoted  with $\mathfrak{sp}^*(n)$ in \cite{superalgebra} to distinguish them from the non-compact ones.}  \ref{table:ads3ksa}.

 \subsection{Central terms}

We shall focus on $\mathfrak{g}_L$, as the description that follows  below also applies to $\mathfrak{g}_R$.
It has been observed in \cite{superalgebra} that the Killing superalgebras of AdS$_3$ backgrounds may exhibit central terms.  Such terms may occur in all cases apart from $\mathfrak{osp}(n|2)$ and
$\mathfrak{D}(2,1,\alpha)$.
However, it has been shown in \cite{superalgebra} that both $\mathfrak{f}(4)$ and $\mathfrak{g}(3)$ do not admit central extensions. The superalgebra $\mathfrak{sl}(2|2)/1_{4\times 4}$ can exhibit up to three central terms though.
This is because $\mathfrak{sl}(2|2)/1_{4\times 4}$ arises as a special case of  $\mathfrak{D}(2,1,\alpha)$ at certain values of the parameter $\alpha$.  For those values, three of the R-symmetry
generators of $\mathfrak{D}(2,1,\alpha)$ span  the R-symmetry algebra  $\mathfrak{so}(3)$ of  $\mathfrak{sl}(2|2)/1_{4\times 4}$ and the other three become central.

It can also be shown that $\mathfrak{sl}(n|2)$, $n>2$ and $\mathfrak{osp}(4|2n)$, $n>1$,  do not exhibit central terms either.  This can be seen after an analysis of the
condition
\bea
\alpha_{rsr'}{}^t \tilde V_{ts'}-
\alpha_{rss'}{}^t \tilde V_{tr'}+\alpha_{r's'r}{}^t \tilde V_{ts}-
\alpha_{r's's}{}^t \tilde V_{tr}=0~,
\label{consis}
\eea
of \cite{superalgebra}, where $\tilde V_{rs}=-\tilde V_{sr}$ are the generators of $(\mathfrak{t}_L)_0$ and $\alpha$ is described in \cite{superalgebra}.  For $\mathfrak{sl}(n|2)$, $n>2$, the central terms that can occur are  (2,0) and (0,2) components of the $\tilde V$.  However one can show that these do not satisfy (\ref{consis}) unless they vanish.  Thus $\mathfrak{c}=\{0\}$.

 It remains to investigate the superalgebra with $(\mathfrak{t}_L)_0/\mathfrak{c}_L= \mathfrak{sp}(n) \oplus \mathfrak{sp}(1)$. The central generators that may occur are the $\tilde V$
 which lie in the complement of $\mathfrak{sp}(n) \oplus \mathfrak{sp}(1)$ in $\mathfrak{so}(4n)$.
Taking the trace of (\ref{consis}) with one of the three complex structures that are associated with $\mathfrak{sp}(n) \oplus \mathfrak{sp}(1)$, one can  demonstrate that all such generators  $\tilde V$ must also vanish.  Thus again $\mathfrak{c}=\{0\}$. Therefore, apart from  $\mathfrak{sl}(2|2)/1_{4\times 4}$, all the other superalgebras in  table \ref{table:ads3ksa} do not admit central terms.

 \begin{table}
	\caption{$AdS_3$ Killing superalgebras in type II and 11D}
	\centering
	\begin{tabular}{|c|c|c|c|}
		\hline
		$N_L$ & $\mathfrak{g}_L/\mathfrak{c}_L$ & $(\mathfrak{t}_L)_0/\mathfrak{c}_L$&$\mathrm{dim}\, \mathfrak{c}_L $ \\
		\hline
		$2n$& $\mathfrak{osp}(n|2)$ & $\mathfrak{so}(n)$ & 0\\
		$4n,~n>2$ & $\mathfrak{sl}(n|2)$ & $\mathfrak{u}(n)$& 0 \\
		$8n, n>1$ & $\mathfrak{osp}(4|2n)$ & $\mathfrak{sp}(n) \oplus \mathfrak{sp}(1)$ &0\\
		16 & $\mathfrak{f}(4)$ & $\mathfrak{spin}(7)$ &0\\
		14 & $\mathfrak{g}(3)$ & $\mathfrak{g}_2$&0 \\
		8 & $\mathfrak{D}(2,1,\alpha)$ & $\mathfrak{so}(3) \oplus \mathfrak{so}(3)$&0 \\
		8 & $\mathfrak{sl}(2|2)/1_{4\times 4}$ & $\mathfrak{su}(2)$& $\leq 3 $\\ [1ex]
		\hline
	\end{tabular}
	\label{table:ads3ksa}
\end{table}

\subsection{On the \texorpdfstring{$G/H$}{G/H} structure of internal spaces} \label{modifx}

We shall demonstrate later that the spacetime of all AdS$_3$ backgrounds that preserve more than 16 supersymmetries in 10- and 11-dimensional supergravities  is a product $\text{AdS}_3\times M^{D-3}$
and that $M^{D-3}$ is a homogeneous space $M^{D-3}=G/H$ such that $\mathfrak{Lie} G=\mathfrak{t}_0$. Of course, $G$ acts transitively on $M^{D-3}$.  In addition, $G$ is required to
 act  ``almost effectively'' on $M^{D-3}$.  This means that the map of $\mathfrak{Lie}\, G$ into the space of Killing vector fields of $M^{D-3}$ is an inclusion, i.e.\ for every generator of
 $\mathfrak{Lie}\, G$ there is an associated {\it non-vanishing} Killing vector field on $M^{D-3}$.   We shall also refer to this property as $\mathfrak{Lie}\, G$ acting ``effectively'' on $M^{D-3}$.
This latter property is essential as otherwise  the super-Jacobi identities of the AdS  Killing superalgebra  will not be satisfied. It is also essential for the identification of the
manifolds that can arise as internal spaces of all AdS, and in particular AdS$_3$,  backgrounds preserving some supersymmetry.

For AdS$_3$ backgrounds, there are two cases to consider.  The first case arises,  whenever $\mathfrak{t}_0$ is a simple
Lie algebra.  Then the internal spaces can be identified, up to a factoring with a  finite group, using the classification of the simply connected 7- and 8-dimensional homogeneous spaces
 in  \cite{Castellani:1983yg,klausthesis,niko6dim,niko7dim,bohmkerr}.  This is sufficient to identify the internal spaces of all such AdS$_3$ backgrounds that preserve $N>16$ supersymmetries.

However for most AdS$_3$ backgrounds $\mathfrak{t}_0$ is not simple.  Typically, it is the sum of two Lie algebras, $\mathfrak{t}_0=(\mathfrak{t}_L)_0\oplus(\mathfrak{t}_R)_0$,  one arising from the left sector and another from the right sector. In addition, it may not be  semisimple.  For example, we have seen that $\mathfrak{t}_0=\mathfrak{u}(3)=\mathfrak{su}(3)\oplus \mathfrak{u}(1)$ for    the  $\mathfrak{sl}(3|2)$ Killing superalgebra  and    $\mathfrak{t}_0=\mathfrak{su}(2)\oplus \mathfrak{c}$ for the    $\mathfrak{sl}(2|2)/1_{4\times 4}$ superalgebra with a central term $\mathfrak{c}$.  Furthermore,  $\mathfrak{t}_0$ is not semisimple for all AdS$_3$ backgrounds that exhibit either $N_L=4$ or $N_R=4$ supersymmetries. Given that $\mathfrak{t}_0$ may not be simple, the question then arises how one can decide, given a $G'/H'$ space from the classification results of \cite{Castellani:1983yg,klausthesis,niko6dim,niko7dim,bohmkerr} whether $\mathfrak{t}_0$ acts both
transitively and effectively on $G'/H'$.

Let us illustrate this with some examples.  It is known that both $U(n)$ and $SU(n)$ act transitively and effectively on $S^{2n-1}$.  Thus $S^{2n-1}=U(n)/U(n-1)$ and $S^{2n-1}=SU(n)/SU(n-1)$.
However for $n>2$, it is  $\mathfrak{u}(n)$ which appears as a subalgebra of $\mathfrak{sl}(n|2)$ and so $\mathfrak{u}(n)$ is expected to act transitively and effectively
on the internal spaces instead of $\mathfrak{su}(n)$.   From this perspective  $U(n)/U(n-1)$ can arise as a potential internal space of an AdS$_3$ background whereas $SU(n)/SU(n-1)$ should be discarded.  Since it is not apparent which description is used for a given homogeneous space in the classification results but it is essential for the classification of AdS$_3$ backgrounds, let us investigate the above paradigm further. To see how $S^{2n-1}=SU(n)/SU(n-1)$ can be modified to be written as a $U(n)/U(n-1)$, consider the group homomorphism $i$ from $SU(n-1)\times U(1)$
into $SU(n)$ as
\bea
(A, z)\xrightarrow[]{i} \begin{pmatrix} A z &0\cr 0 & z^{1-n}\end{pmatrix}~.
\eea
In fact $i$ has kernel $\bZ_{n-1}$ and so factors to $U(n-1)$.
Next consider $SU(n)\times U(1)$ and the group homomorphism $j$ of  $SU(n-1)\times U(1)$ into $SU(n)\times U(1)$ as
\bea
(A, z)\xrightarrow{j} \left(\begin{pmatrix} A z &0\cr 0 &z^{1-n}\end{pmatrix}, z^{n-1}\right)~.
\eea
Again $j$  has kernel $\bZ_{n-1}$ and so factors to $U(n-1)$.
Then $SU(n)\times U(1)/j(SU(n-1)\times U(1))=S^{2n-1}$ with $SU(n)\times U(1)$ acting almost effectively on $S^{2n-1}$. Furthermore  one can verify that $U(n)=(SU(n)\times U(1))/\bZ_n$
acts effectively on $S^{2n-1}$, as expected.

The key point of the modification described above is the existence  of $U(1)\subset SU(n)$ such that  $SU(n-1)\times U(1)\subset SU(n)$ and that this $U(1)$ acts on both $SU(n)$ and
the $U(1)$ subgroup of $SU(n)\times U(1)$.  Observe that after the modification the isotropy group is larger and so the invariant geometry of $S^{2n-1}$ as a $U(n)/U(n-1)$
homogeneous space is more restrictive  than that of $S^{2n-1}=SU(n)/SU(n-1)$.

Another example that illustrates a similar point and which  will be used in the analysis that follows is $S^7=Sp(2)/Sp(1)$.
It is known that $S^7$ can also be described as $S^7=Sp(2)\cdot Sp(1)/Sp(1)\cdot Sp(1)$, where $Sp(2)\cdot Sp(1)=Sp(2)\times Sp(1)/\bZ_2$ and similarly for $Sp(1)\cdot Sp(1)$. The modification required to describe
$S^7$ as an $Sp(2)\cdot Sp(1)/Sp(1)\cdot Sp(1)$ coset starting from $Sp(2)/Sp(1)$ is as follows. View the elements of $Sp(2)$ as  $2\times 2$ matrices
with  quaternionic entries and consider the inclusion $i$ of $Sp(1)\times Sp(1)$ in $Sp(2)$ as
\bea
(x, y)\xrightarrow[]{i} \begin{pmatrix} x &0\cr 0 & y\end{pmatrix}~,
\eea
where $x$ and $y$ are quaternions of length one.
Then the map $j$ from $Sp(1)\times Sp(1)$ into $Sp(2)\times Sp(1)$ is constructed as
\bea
(x, y)\xrightarrow{j} \left(\begin{pmatrix} x &0\cr 0 &y\end{pmatrix}, y\right)~.
\label{modifsp2}
\eea
One finds that $Sp(2)\times Sp(1)/j(Sp(1)\times Sp(1))$ is diffeomorphic to $S^7$, with $Sp(2)\times Sp(1)$ acting almost effectively and descending to an effective action of $Sp(2)\cdot Sp(1)$.  Again the additional $Sp(1)$ introduced in the isotropy group acts both on $Sp(2)$ and the additional
$Sp(1)$ introduced in the transitive group.  The geometry of the homogeneous space $S^7=Sp(2)\cdot Sp(1)/Sp(1)\cdot Sp(1)$ is more restrictive than that
of $S^7=Sp(2)/Sp(1)$.  In fact, the former is a special case of the latter.  As a final example $SU(2)\times SU(2)/SU(2)$ can be seen as a modification of the homogeneous space $SU(2)/\{e\}$.  From now on we shall refer to  such  constructions as ``modifications'' of a homogeneous space.

On the level of Lie algebras the  modifications can be viewed as follows.  Suppose $\mathfrak{t}_0$ decomposes as $\mathfrak{t}_0=\mathfrak{k}\oplus \mathfrak{e}$, where $\mathfrak{k}$ and $\mathfrak{e}$ are Lie algebras, and that
there is a homogeneous space $K/L$ with $\mathfrak{Lie} (K) =\mathfrak{k}$.  To see whether $K/L$ can be modified to admit an effective
action of the whole $\mathfrak{t}_0$ algebra, it is first required that $\mathfrak{l}\oplus\mathfrak{e}$ is a subalgebra of $\mathfrak{k}$, where $\mathfrak{Lie} L=\mathfrak{l}$.
Then, up to possible discrete identifications, $K/L$ can be modified to $K\times E/L\times E$, where now $E$ with $\mathfrak{Lie}\, E=\mathfrak{e}$ acts on both the $K$ and $E$ subgroups of the transitive group.

All    7- and 8-dimensional  $K/L$  homogeneous spaces with $K$ semisimple are known up to possible modifications.  Because of this, for $\mathfrak{t}_0$ semisimple, one can systematically search for all  modifications of $K/L$ homogeneous spaces to determine whether a Lie algebra $\mathfrak{t}_0$ can act transitively and
effectively on a modified homogenous space.  If $\mathfrak{t}_0$ is not semisimple, we have argued in appendix \ref{modif} that up to discrete identifications one can construct all the homogeneous spaces $G/H$ with $\mathfrak{Lie}\,G=\mathfrak{t}_0$ as the product of a modification of $K/L$ with the abelian group $\times^k U(1)$, where $K$ is semisimple.

As we shall see the modifications of homogeneous spaces are necessary  to identify all possible internal spaces of AdS$_3$ backgrounds that can preserve some supersymmetry.  For such modifications of $K/L$ to exist,
a necessary condition is that the rank of $L$ must be strictly smaller than that of $K$.  It turns out that this is rather restrictive in the analysis that follows.

Let us now turn to investigate the homogeneous geometry of a modification $K\times E/L\times E$ of the homogenous $K/L$ space. One can show that this can be explored as a special case of that  of $K/L$.  Indeed
suppose that $\mathfrak{k}=\mathfrak{l}\oplus \mathfrak{m}$.  Then, we observe that we can choose  the generators of $\mathfrak{Lie}(K\times E)$ such that
$\mathfrak{Lie}(K\times E)= j(\mathfrak{l}\oplus \mathfrak{e})\oplus \mathfrak{m}$, where $j:~~\mathfrak{l}\oplus \mathfrak{e}\rightarrow \mathfrak{k}\oplus \mathfrak{e}$ is the
inclusion of the modification.  Therefore, the tangent space at the origin of the original $K/H$ space and that of the modification $K\times E/L\times E$ can be identified with the same vector space $\mathfrak{m}$.
The only difference is that $\mathfrak{m}$ as the tangent space at the origin of $K/L$ is the module of a representation of $\mathfrak{l}$ while after the modification $\mathfrak{m}$ is the module of a representation of $\mathfrak{l}\oplus \mathfrak{e}$.
Thus, all the local homogeneous geometry of the modification $K\times E/L\times E$ is that  of $K/L$ which in addition is invariant under the representation
of $\mathfrak{e}$ on $\mathfrak{m}$.

\section{\texorpdfstring{$N>16 ~ AdS_3\times_w M^8$}{N greater 16 AdS3xM8} solutions in 11 dimensions}\label{ads311D}

\subsection{Fields}

We consider  warped AdS$_3$ backgrounds with internal space $M^8$, $\text{AdS}_3\times_w M^8$, and the most general allowed fluxes invariant under the symmetries of the AdS$_3$ subspace.
 The bosonic fields of 11-dimensional supergravity are   a metric $ds^2$ and a 4-form field strength $F$. Following the description of $\text{AdS}_3\times_w M^8$ backgrounds presented in \cite{mads},  these can be written as
\begin{align}
ds^2 &= 2 du(dr+rh)+ A^2 dz^2 + ds^2(M^8)~, \notag\\
F&= du \wedge (dr+rh) \wedge dz \wedge Q + X~,
\end{align}
where $(u,r,z)$ are the coordinates of AdS$_3$,
\begin{align}
h= -\frac{2}{\ell}\, dz - 2 A^{-1} dA~,
\end{align}
$\ell$ is the AdS$_3$ radius, $A$ is the warp factor which is a function of  $M^8$, and $Q$ and $X$ are a 1-form and 4-form on $M^8$, respectively.  The dependence of the fields on the AdS$_3$ coordinates $(u,r,z)$
is explicit, while $ds^2(M^8), A, Q, X$ depend only on the coordinates $y^I$  of $M^8$.  Next we define a null orthonormal frame as
\begin{align}\label{nullframe}
\bbe^+ = du~,~~ \bbe^-=dr + rh~,~~ \bbe^z = A dz~, ~~ \bbe^i = \bbe^i_{I} dy^I~,
\end{align}
with $ds^2(M^8) = \delta_{ij} \bbe^i \bbe^j$. The Bianchi identity $dF=0$ of  $F$ implies that
\begin{align}\label{mbianchi}
d(A^2Q)=0~,\quad dX=0~.
\end{align}
The field equations for $F$ give that
\bea
d*_{{}_8}X=-3 d\log A\wedge *_{{}_8}X-A^{-1} Q\wedge X~,
\label{coX}
\eea
and
\bea
d(A^{-1}*_{{}_8}Q)=-{1\over2} X\wedge X~,
\label{fqxx}
\eea
where  our Hodge duality conventions can be found in appendix \ref{conventions}.
Similarly, the Einstein equation along $AdS_3$ gives rise to a field equation for the warp factor $A$
\begin{align}\label{warp}
A^{-1} \nabla^k \nabla_k A + 2 A^{-2} \nabla^k A \nabla_k A + \frac{2}{\ell^2 A^2} = \frac{1}{3A^2} Q^2 + \frac{1}{144} X^2~,
\end{align}
and the Einstein equation along $M^8$ reads
\begin{align}
R_{ij}^{(8)} = 3 A^{-1} \nabla_i\nabla_j A - \frac{1}{2} A^{-2} Q_i Q_j + \frac{1}{12} X^2_{ij} + \delta_{ij} \left(\frac{1}{6} A^{-2} Q^2 - \frac{1}{144} X^2 \right)~,
\end{align}
where $R^{(8)}_{ij}$ is the Ricci tensor of the internal manifold $M^8$. Note in particular that \eqref{warp} implies that $A$ is nowhere vanishing, provided that $A$ and all other fields are smooth.

\subsection{The Killing spinors}

Here, we summarise the solution of the gravitino KSE in 11-dimensional supergravity for warped $\text{AdS}_3\times_w M^8$  backgrounds in \cite{mads}. In this approach, the  KSE of 11-dimensional supergravity is first solved along the AdS$_3$ subspace and then the remaining independent KSEs
along the internal space $M^8$ are identified. The Killing spinors  can be expressed\footnote{The gamma matrices are always taken with respect to the null-orthonormal frame (\ref{nullframe}).}  as
\begin{align}\label{ks}
\epsilon  = \,&\sigma_+ +e^{-\frac{z}{\ell}} \tau_+ + \sigma_- + e^{\frac{z}{\ell}}  \tau_-  - \ell^{-1} u A^{-1} \Gamma_{+z} \sigma_- - \ell^{-1} r A^{-1} e^{-\frac{z}{\ell}} \Gamma_{-z} ~ \tau_+~,
\end{align}
where the dependence on the AdS$_3$ coordinates is explicit and $\sigma_\pm$ and $\tau_\pm$ are Majorana $\text{Spin}(10,1)$ spinors that depend only on the coordinates of
 $M^8$ and satisfy  the light-cone projections
\begin{align}
\Gamma_\pm \sigma_\pm = 0~, \quad \Gamma_\pm \tau_\pm = 0~.
\end{align}
The remaining independent KSEs on $M^8$ are
\begin{align}\label{graviKSE}
\nabla^{(\pm)}_i \sigma_\pm = 0~, \quad \nabla^{(\pm)}_i \tau_\pm = 0~,
\end{align}
and
\begin{align}\label{algKSE}
\Xi^{(\pm)}\sigma_{\pm}=0~, \quad (\Xi^{(\pm)} \pm \frac{1}{\ell})\tau_{\pm}=0~,
\end{align}
where
\begin{align}
\nabla_i^{(\pm)} &= \nabla_i \pm \frac{1}{2} \partial_i \log A - \frac{1}{288} \sgX_i + \frac{1}{36} \sX_i \mp \frac{1}{12} A^{-1} \Gamma_z \sgQ_i \pm \frac{1}{6} A^{-1} \Gamma_z Q_i~, \\
\Xi^{(\pm)} &= \mp \frac{1}{2\ell} - \frac{1}{2} \Gamma_z \slashed{\partial} A + \frac{1}{288} A \Gamma_z \sX \pm \frac{1}{6} \sQ~.
\end{align}
The conditions  \eqref{graviKSE} can be thought of as   the restriction of the gravitino KSE of 11-dimensional supergravity on  $M^8$ while  \eqref{algKSE} arises from integrating  the
gravitino KSE along the $AdS_3$ subspace.

To make a connection with the terminology used to describe the Killing superalgebras of AdS$_3$ backgrounds in section \ref{superxx},  the Killing spinors $\epsilon$ that depend only on the $\sigma_\pm$ type of spinors are in the left sector while those that depend on $\tau_\pm$ spinors are in the right sector.
The existence of  unrelated\footnote{In AdS$_n$, $n>3$, backgrounds the $\sigma_\pm$ and $\tau_\pm$ spinors are related by Clifford algebra operations.}
 $\sigma_\pm$ and $\tau_\pm$ types of spinors  is the reason that the Killing superalgebra $\mathfrak{g}$  of AdS$_3$ decomposes as $\mathfrak{g}=\mathfrak{g}_L\oplus \mathfrak{g}_R$.
Furthermore, it has been noted in \cite{mads} that if $\sigma_+$ and $\tau_+$ solve the KSEs (\ref{graviKSE}) and (\ref{algKSE}), so do
\begin{align}
\sigma_- = A \Gamma_{-z} \sigma_+~,\quad \tau_{-} = A \Gamma_{-z} \tau_+~.
\end{align}
Therefore the number of Killing spinors $N=N_L+N_R$ of AdS$_3$ backgrounds is always even, where $N_L$ and $N_R$  is the number of  Killing spinors of the left and right sector, respectively.

\subsection{For \texorpdfstring{$N>16$ $AdS_3$}{N greater 16 AdS3} solutions \texorpdfstring{$M^8$}{M8} is homogeneous}

\subsubsection{Factorization of Killing vectors}

It has been shown in \cite{superalgebra} that for  compact without boundary internal spaces $M^8$,  the even part of the Killing superalgebra $\mathfrak{g}_0$  decomposes into the algebra of symmetries of AdS$_3$ and those of the internal space $M^8$. This, together with the homogeneity theorem of \cite{homogen}, can be used to show that the internal space  $M^8$ is homogeneous for
$N>16$ backgrounds.

For AdS$_3$ backgrounds,  the condition \cite{superalgebra} for $\mathfrak{g}_0=\mathfrak{iso}(\text{AdS}_3)\oplus \mathfrak{t}_0$  is
\begin{align}\label{gamiz}
\langle \tau_+, \Gamma_{iz} \sigma_+ \rangle =0~,
\end{align}
for all $\sigma_+$ and $\tau_+$   spinors that satisfy (\ref{graviKSE}) and (\ref{algKSE}).  This can be derived using the compactness of $M^8$ as follows.
Setting $\Lambda= \sigma_+ + \tau_+$ and making use of the gravitino KSE \eqref{graviKSE}, one finds
\begin{align}
\nabla_i  \parallel \Lambda \parallel^2 = - \parallel \Lambda \parallel^2 A^{-1} \nabla_i A + \frac{1}{144} \langle \Lambda , \sgX_i \Lambda \rangle - \frac{1}{3} A^{-1} Q_i \langle \Lambda, \Gamma_z \Lambda \rangle~.
\label{middleqn}
\end{align}
Now, note that the algebraic KSE \eqref{algKSE} implies
\begin{align}
\frac{1}{\ell} (\sigma_+ -\tau_+) = (- \Gamma_{z}\slashed{d}A +\frac{A}{144} \Gamma_z \sX +\frac{1}{3} \sQ)~\Lambda~,
\end{align}
which, after multiplying by $A^{-1}\Gamma_{iz}$ and substituting back into (\ref{middleqn}), gives
\begin{align}\label{hopf1}
\nabla_i \parallel \Lambda \parallel^2=2 \ell^{-1} A^{-1}  \langle \tau_+, \Gamma_{iz} \sigma_+ \rangle~.
\end{align}
Furthermore, the gravitino KSE \eqref{graviKSE} also yields
\begin{align}
\nabla^i \left( A \langle \tau_+, \Gamma_{iz} \sigma_+ \rangle \right) = 0~.
\end{align}
Combining this with \eqref{hopf1}, one ends up with
\begin{align}
\nabla^2 \parallel \Lambda \parallel^2 + 2 A^{-1} \nabla^i A \nabla_i \parallel \Lambda \parallel^2 = 0~.
\end{align}
The Hopf maximum principle then implies that $\parallel \Lambda \parallel^2$ is constant, thus \eqref{hopf1} yields (\ref{gamiz}).

One consequence of (\ref{gamiz})  is that the linearly independent spinors  $\sigma_+$ and $\tau_+$,  on account of \eqref{algKSE},  are also orthogonal
\begin{align}
\langle \tau_+,  \sigma_+ \rangle =0~.
\end{align}
One can see this by taking $\langle \tau_+, \Xi^{(+)} \sigma_+ \rangle -\langle \sigma_+, (\Xi^{(+)} + \ell^{-1}) \tau_+ \rangle = 0$ and using \eqref{gamiz}.

\begin{table}
	\caption{8-dimensional compact, simply connected, homogeneous spaces}
	\centering
	\begin{tabular}{c l}
		\hline
		& $M^8=G/H$  \\  
		\hline
		(1)& $SU(3)$, group manifold\\
		(2)&$\frac{Sp(3)}{Sp(2)\times Sp(1)}=\mathbb{HP}^2$, symmetric space\\
		(3)& $\frac{SU(5)}{S(U(4)\times U(1))}=\mathbb{CP}^4$, symmetric space, not spin \\
		(4) & $\frac{Spin(9)}{Spin(8)}=S^8$, symmetric space\\
		(5) & $\frac{Sp(2)}{T^2}$, $T^2 \subset Sp(2)$ maximal torus\\
		(6) & $\frac{G_2}{SO(4)}$, symmetric space\\
		(7) & $\frac{SU(4)}{S(U(2)\times U(2))} = G_2(\mathbb{C}^4) =\frac{SO(6)}{SO(4)\times SO(2)} = G_2(\mathbb{R}^6)$, Grassmannian, symmetric space\\
		(8) & $\frac{SU(2)\times SU(2) \times SU(2)}{\Delta_{k,l,m}(U(1))}$\\
		(9) & $S^2\times S^6$ \\
		(10) & $S^2 \times \mathbb{CP}^3$ \\
		(11) & $S^2 \times \frac{SU(3)}{T^2}$\\
		(12) & $S^2 \times G_2(\mathbb{R}^5)$, not spin \\
		(13) & $S^3 \times S^5$ \\
		(14) & $S^3\times \frac{SU(3)}{SO(3)}$, not spin \\
		(15) & $S^4 \times S^4$ \\
		(16) & $S^4 \times \mathbb{CP}^2$, not spin \\
		(17) & $\mathbb{CP}^2 \times \mathbb{CP}^2$, not spin\\
		(18) & $S^2 \times S^2 \times S^4$ \\
		(19) & $ S^2 \times S^3 \times S^3$ \\
		(20) & $S^2\times S^2 \times S^2 \times S^2$ \\
		(21) & $S^2 \times S^2 \times \mathbb{CP}^2$, not spin \\[1ex]
		\hline
	\end{tabular}
	\label{table:8dim}
\end{table}

\subsubsection{\texorpdfstring{$A$}{A} is constant and  \texorpdfstring{$M^8$}{M8} is homogeneous}

Let us define the spinor bilinear
\begin{align}
W_i = A\, \text{Im} \langle \chi_1 , \Gamma_{iz} \chi_2 \rangle~,
\end{align}
where $\chi$ either stands for $\sigma_+$ or $\tau_+$~. The gravitino KSE \eqref{graviKSE} then implies
\begin{align}
\nabla_{(i} W_{j)} =0~,
\end{align}
i.e.\  $W$ is   a Killing vector\footnote{If the bilinear in (\ref{gamiz}) does not vanish, then the associated $W$ is not a Killing vector over the whole spacetime.} on $M^8$. From (\ref{gamiz}) it follows that the only non-vanishing Killing vector fields $W$  are those
that are constructed as bilinears of either $\sigma_+$ or  $\tau_+$ spinors.

As a consequence of the algebraic KSEs  \eqref{algKSE}, one has  $\text{Im} \langle \sigma_+^1, \Xi^{(+)} \sigma_+^2 \rangle =0$ and $\text{Im} \langle \tau_+^1, (\Xi^{(+)} +\ell^{-1}) \tau_+^2\rangle =0 $.  Expanding these,  one finds that
\begin{align}\label{iwda}
i_W dA =0~,
\end{align}
where $W$ is a bilinear of either $\sigma_+$ or $\tau_+$ spinors.

As we have mentioned,  (\ref{gamiz}) implies that the only non-vanishing Killing vectors $W$  on $M^8$ are  those constructed from either $\sigma_+$ or $\tau_+$ spinors.
Therefore (\ref{iwda}) will be  valid for all non-vanishing Killing vectors $W$ on $M^8$.
Suppose now that $N>16$. A similar argument to that used for the proof of the  homogeneity theorem in \cite{homogen} implies that the set of all Killing vectors  $W$ span the tangent space of $M^8$. Thus, $A$ is constant and $M^8$ is homogeneous.  It should be noted that if (\ref{gamiz}) is not valid, then the vector fields $W$ in (\ref{iwda}) may not span all Killing vectors on $M^8$.

Therefore we conclude that  all $N>16$ supersymmetric AdS$_3$  backgrounds are products AdS$_3\times M^8$, where $M^8$ is a homogeneous space.  In the analysis that follows, which includes that of AdS$_3$ backgrounds in type II 10-dimensional supergravities, we shall focus
only on such product spaces.

\subsection{Electric solutions do not preserve \texorpdfstring{$16<N<32$}{16 less N less 32} supersymmetries}

A consequence of the warp factor being constant is that it rules out the existence of electric solutions that preserve $16<N<32$ supersymmetries.  Indeed, for electric
solutions $X=0$.  The algebraic KSE (\ref{algKSE}) on $\sigma_+$ reduces to
\bea
{1\over 3} \slashed{Q} \sigma_+={1\over\ell} \sigma_+~,
\eea
which implies the integrability condition
\bea
{1\over9} Q^2={1\over\ell^2}~.
\eea
On the other hand the field equation for the warp factor (\ref{warp})  yields ${1\over6} Q^2={1\over\ell^2}$ which is a contradiction as the radius of AdS$_3$ is finite.

\subsection{\texorpdfstring{$N>16$}{N greater 16} solutions with left only supersymmetry}

Suppose first that the solutions only have  left-hand supersymmetry.  In such a case, the Lie algebras that must act transitively  and effectively on the
internal spaces are
\bea
&&\mathfrak{so}(n)_L~,~~~n=9,\cdots, 15~, ~~~(N=2n)~;~~~
\cr
&&\mathfrak{u}(n)_L~,~~~n=5,6,7~,~~~(N=4n)~;~~~
\cr
&&(\mathfrak{sp}(3)\oplus \mathfrak{sp}(1))_L~,~~~N=24~,
\label{nr0}
\eea
where $N<32$ as there are no AdS$_3$ solutions which preserve maximal supersymmetry.  Furthermore, solutions that preserve $N=30$ supersymmetries have already been excluded in \cite{11n30}.  An inspection of the list of homogeneous spaces reveals that the only possibility
that can occur is $S^8=\mathrm{Spin}(9)/\mathrm{Spin}(8)$ which can preserve 18 supersymmetries.  However $S^8$ is a symmetric space and there are no invariant 1- and 4-forms.
Thus $Q=X=0$ which in turn implies $F=0$.  This leads to a contradiction, as the field equation for the warp factor cannot be satisfied.

\subsection{\texorpdfstring{$N>16$}{N greater 16} solutions with \texorpdfstring{$N_R=2$}{NR=2}}

For $N_R=2$  there are no right isometries and so all the symmetries of the internal space are  generated by $(\mathfrak{t}_L)_0$. The Lie algebras $(\mathfrak{t}_L)_0$ that  act transitively  and effectively on the
internal spaces are
\bea
&&\mathfrak{so}(n)_L~,~~~n=8,\cdots, 14~,~~~(N=2n+2)~;~~~
\cr
&&\mathfrak{u}(n)_L~,~~~n=4, \cdots,7~,~~~(N=4n+2)~;~~~
\cr
&&(\mathfrak{sp}(n)\oplus \mathfrak{sp}(1))_L~,~~~n=2,3~,~~~(N=8n+2)~;~~~
\cr
&&
\mathfrak{spin}(7)_L~~(N=18)~,
\label{nr2}
\eea
where the last case is associated with the Killing superalgebra  $\mathfrak{f}(4)$.
An inspection of the 8-dimensional homogeneous spaces in table \ref{table:8dim} reveals that there are only two possibilities that can occur
\bea
&&S^8=\mathrm{Spin}(9)/\mathrm{Spin}(8)~~(N=20)~,~~~~
\cr
&&
\mathbb{CP}^3\times S^2=Sp(2)/(Sp(1)\times U(1))\times Sp(1)/U(1)~~~(N=18)~.
\eea
Observe that $G_2(\mathbb{C}^4) =SU(4)/S(U(2)\times U(2))$ could have been included as a potential internal space of an AdS$_3$ background with $N=18$ supersymmetries, provided that
it  admitted an effective $\mathfrak{u}(4)$ action.  However, this is not the case, since the rank of the isotropy group $S(U(2)\times U(2))$ is the same as that of $SU(4)$ and so it does not admit
a modification such that $U(4)$ acts almost effectively on $G_2(\mathbb{C}^4)$.
For confirmation, we have also excluded this case with an explicit calculation which we shall not present here.

In addition  $\text{AdS}_3\times S^8$ can also be excluded as a solution  with an identical argument to the one we produced in the previous case with no right-handed supersymmetries. The remaining
case is investigated below.

  \subsubsection{\texorpdfstring{$ \mathbb{CP}^3\times S^2=Sp(2)/(Sp(1)\times U(1))\times SU(2)/U(1)$}{CP3xS2}}

For the analysis that follows, we use  the description of the geometry of the homogeneous space $Sp(2)/(Sp(1)\times U(1))$ presented in  \cite{ads4Ngr16}, where more details can be found.
  The metric on the internal space $ \mathbb{CP}^3\times S^2$ is
\bea
ds^2(M^8)= ds^2(\mathbb{CP}^3)+ ds^2 (S^2)~,
\eea
where
\bea
 ds^2(\mathbb{CP}^3)= a\, \delta_{ij}\bbl^i\bbl^j+ b\, \delta_{{\underline r}{\underline s}} \bbl^{\underline r}\bbl^{\underline s}~,~~~ds^2(S^2)=c\big( (\bbl^7)^2+(\bbl^8)^2\big)~,
 \eea
 and $(\ell^i, \ell^{\underline r})$, $i=1,\dots, 4$, ${\underline r}=1,2$ is a  left-invariant frame\footnote{In \cite{ads4Ngr16}, the left-invariant frame on $\mathbb{CP}^3$
 has been denoted as $(\ell^a, \ell^{\underline r})$ $a=1,\dots, 4$, ${\underline r}=1,2$ instead.}
   on $\mathbb{CP}^3$ and $(\ell^7, \ell^8)$ is a left-invariant frame
 on $S^2$. Moreover $a,b,c>0$ are constants.
As there are no invariant 1-forms $Q=0$.  The most general invariant 4-form is
\bea
X&=\frac{1}{2} \, \alpha_1 \, I^{(+)}_{3} \wedge I^{(+)}_{3}+ \alpha_2 \, \tilde{\omega}\wedge I^{(+)}_{3}+\alpha_3\, \sigma\wedge \tilde{\omega} + \alpha_4\, \sigma\wedge I^{(+)}_{3}  ~,
\eea
where $\alpha_1, \dots, \alpha_4$ are constants, $I^{(+)}_{3}=\ell^{12}+\ell^{34}$ and $\tilde{\omega}=\ell^{\underline {12}}$ are invariant 2-forms on $\mathbb{CP}^3$ whose properties can be found in \cite{ads4Ngr16} and $\sigma=\ell^{78}$.

\begin{table}[h]
\begin{center}
\vskip 0.3cm
 \caption{Decomposition of (\ref{433ccv})  into eigenspaces}
 \vskip 0.3cm

	\begin{tabular}{|c|c|}
		\hline
		$|J_1,J_2, J_3\rangle$&  relations for the fluxes\\
		\hline
		$|\pm,\pm,+\rangle$ & ${1\over 6}  \Big( {\alpha_1\over a^2} \mp  {2\alpha_2\over ba}-{\alpha_3\over bc} \pm {2\alpha_4\over ac}\Big) ={1\over \ell A} $ \\
\hline
		$|+,-, \pm\rangle$, $|-,+, \pm\rangle$& ${1\over 6}  \Big( {\alpha_1\over a^2}\pm{\alpha_3\over bc} \Big) =\pm {1\over \ell A}$  \\
		\hline
		$|\pm,\pm, -\rangle$ & ${1\over 6}  \Big( {\alpha_1\over a^2} \mp  {2\alpha_2\over ba}+{\alpha_3\over bc} \mp {2\alpha_4\over ac}\Big) =-{1\over \ell A} $ \\
		\hline
		
	\end{tabular}
\vskip 0.2cm
  \label{table43xcv}
 \end{center}
\end{table}

The closure and co-closure of $X$ give a relation between $\alpha_1$ and $\alpha_2$, and between $\alpha_3$ and $\alpha_4$, but they are not essential here.  Also $X\wedge X=0$ implies
that $\alpha_1 \alpha_3=0$.

On the other hand the algebraic KSE (\ref{algKSE}) can be written as
\bea
{1\over6} \Big({\alpha_1\over a^2} J_1 J_2-{\alpha_2\over ba} (J_1+J_2)- {\alpha_3\over bc} J_3+ {\alpha_4\over ac} (J_1+J_2) J_3\Big) J_1 J_2 J_3\sigma_+={1\over \ell A} \sigma_+~,
\label{433ccv}
\eea
where $J_1=\Gamma^{12{\underline {12}}}$, $J_2= \Gamma^{34{\underline {12}}}$ and $J_3=\Gamma^{78{\underline {12}}}$.  We have chosen the orientation such that
$\Gamma_z \sigma_+=-J_1J_2J_3\sigma_+$.  The decomposition of the algebraic KSE into the eigenspaces of $J_1, J_2$ and $J_3$ as well as the relations implied amongst the fluxes for each eigenspace can be found in table \ref{table43xcv}.

As each common eigenspace  of $J_1$, $J_2$ and $J_3$ has dimension two for solutions with $N>16$ supersymmetries one has always to consider either one of the eigenspinors  $|+,-, \pm\rangle$
and  $|-,+, \pm\rangle$ or all the eigenspinors $|\pm,\pm,+\rangle$ and $|\pm,\pm, -\rangle$. In the former  case, we have  that
\bea
{1\over36}\Big( {\alpha_1\over a^2}+{\alpha_3\over bc} \Big)^2={1\over \ell^2 A^2}~,
\eea
where we have chosen, without loss of generality, the eigenvalue $+1$  of $J_3$.
Taking the difference of the equation above with the warp factor field equation
\bea
{1\over12} \Big({\alpha_1^2\over a^4}+{2 \alpha_2^2\over b^2 a^2}+{\alpha_3^2\over b^2 c^2}+{2\alpha^2_4\over a^2 c^2}\Big)={1\over \ell^2 A^2}~,
\eea
we find that $\alpha_1=\alpha_2=\alpha_3=\alpha_4=0$, and so $X=0$,  which is a contradiction.  In the latter case we have that $\alpha_1=\alpha_2=\alpha_4=0$ and
${1\over6} {\alpha_3\over bc}=-{1\over \ell A}$. Comparing this with the warp factor field equation above again leads to a contradiction. There are no solutions with
internal space $Sp(2)/(Sp(1)\times U(1))\times SU(2)/U(1)$ that preserve $N>16$ supersymmetries.

\subsection{\texorpdfstring{$N>16$}{N greater 16} solutions with \texorpdfstring{$N_R=4$}{NR=4}}

The only right superalgebra that gives rise to 4 supersymmetries is $\mathfrak{osp}(2|2)$ which in turn leads to an $\mathfrak{so}(2)_R$ right-handed symmetry.
Therefore the Lie algebras that act both transitively and effectively on the internal spaces $M^8$  are
\bea
&&
\mathfrak{so}(n)_L\oplus \mathfrak{so}(2)_R~,~~~n=7,\cdots, 13~,~~~(N=2n+4)~;~~~~
\cr
&&
\mathfrak{u}(n)_L\oplus \mathfrak{so}(2)_R~,~~~n=4, 5, 6~,~~~(N=4n+4)~;
\cr
&&(\mathfrak{sp}(n) \oplus \mathfrak{sp}(1))_L\oplus \mathfrak{so}(2)_R~,~~~n=2,3~,~~~(N=8n+4)~;~~~~
\cr
&&
\mathfrak{spin}(7)_L \oplus \mathfrak{so}(2)_R~,~~~(N=20)~;~~~~
\cr
&&(\mathfrak{g}_2)_L\oplus \mathfrak{so}(2)_R~,~~~(N=18)~.
\label{nr4}
\eea
Up to a finite cover, the allowed  homogeneous spaces are
\bea
&&\mathrm{Spin}(7)/G_2\times S^1~,~~~(N=18, 20)~;~~~\mathrm{Spin}(8)/\mathrm{Spin}(7)\times S^1~,~~~(N=20)~;~~~
\cr
&&S^7\times S^1=U(4)/U(3)\times S^1~,~~~(N=20)~;~~~
\cr
&& S^7\times S^1=(Sp(2)\times Sp(1))/(Sp(1)\times Sp(1))\times S^1~,~~~(N=20)~;
\cr
&&S^4\times S^3\times S^1=\mathrm{Spin}(5)/\mathrm{Spin}(4) \times SU(2)\times S^1~,~~~(N=20)~.
\eea
Observe that all the cases that arise, up to discrete identifications,  are products of 7-dimensional homogeneous spaces with $S^1$. This is because it is not possible
to modify the 8-dimensional homogeneous spaces which admit  an effective and transitive action of the $(\mathfrak{t}_L)_0$ Lie algebras in (\ref{nr4})
to homogeneous spaces on which $\mathfrak{t}_0=(\mathfrak{t}_L)_0\oplus \mathfrak{so}(2)_R$ acts transitively and effectively.  Note that this is due to the fact
that for all candidate homogeneous spaces which may occur the rank of the isotropy group is the same as the rank of  $(\mathfrak{t}_L)_0$.

However a modification
has been used to include the homogeneous space $S^7\times S^1=(Sp(2)\times Sp(1))/(Sp(1)\times Sp(1))\times S^1$.  An AdS$_3$  solution with internal space $Sp(2)/Sp(1)\times S^1$,
is expected to preserve $N=N_L+N_R=10+4=14<16$   supersymmetries as $\mathfrak{sp}(2)=\mathfrak{so}(5)$ and so should be discarded,  while with internal space $(Sp(2)\times Sp(1))/(Sp(1)\times Sp(1))\times S^1$ we expect it to preserve $N=20$ supersymmetries, as it is associated with the $(\mathfrak{sp}(n) \oplus \mathfrak{sp}(1))_L\oplus \mathfrak{so}(2)_R$ subalgebra in (\ref{nr4}) and therefore we need to include it.  We have also used a modification to include $S^7\times S^1=U(4)/U(3)\times S^1$ as a candidate space, since $S^7\times S^1=SU(4)/SU(3)\times S^1$ should have been discarded.

The coset space $\mathrm{Spin}(8)/\mathrm{Spin}(7)\times S^1$ can  immediately be excluded as the 4-form field strength $F$ is electric and we  have shown there are no electric
solutions which preserve $16<N\leq 32$ supersymmetries.  It remains to investigate the rest of the  cases.

\subsubsection{\texorpdfstring{$\mathrm{Spin}(7)/G_2\times S^1$}{Spin(7)/G2 x S1}}

The metric on the homogeneous space $\mathrm{Spin}(7)/G_2\times S^1$  can be chosen as
\begin{align}
ds^2(M^8)&= ds^2(\mathrm{Spin}(7)/G_2)+ ds^2(S^1)= a\,  \delta_{ij} \bbl^i \bbl^j+b\, (\bbl^8)^2\notag\\
&= \delta_{ij} \bbe^i \bbe^j+  (\bbe^8)^2~,
\end{align}
where the geometry of $\mathrm{Spin}(7)/G_2$ can be found in  \cite{ads4Ngr16} whose conventions we follow, $a,b>0$ are constants and $\ell^8$ is an invariant frame on $S^1$, $d\ell^8=0$.

The most general invariant fluxes are
\bea
Q=\gamma \bbe^8~,~~~X=\alpha\, *_{{}_7}\varphi+ \beta\, \bbe^8\wedge \varphi
\eea
where $*{}_7\varphi$ and $\varphi$ are the fundamental $G_2$ forms and $\alpha, \beta, \gamma$ are constants.  Furthermore, the Bianchi identity $dX=0$ implies that $ \beta=0$.

\vskip 0.5cm
\begin{table}[h]
\begin{center}
\vskip 0.3cm
 \caption{Decomposition of (\ref{Spin7G2_alg_KSE}) KSE into eigenspaces }
 \vskip 0.3cm
\begin{tabular}{|c|c|}
		\hline
		$|P_1,P_2,P_3\rangle$&  relations for the fluxes\\
		\hline
		$|+,+,+\rangle$, $|+,+,-\rangle$, $|-,+,+\rangle$, $|+,-,-\rangle$& $(-\frac{1}{6} \alpha \, \Gamma_z + \frac{1}{3} A^{-1} \gamma \Gamma_8  ) |\cdot\rangle = \frac{1}{\ell A} | \cdot \rangle$ \\
		$|-,+,-\rangle$, $|-,-,+\rangle$, $|-,-,-\rangle$&   \\
		\hline
		$|+,-,+\rangle$ & $(\frac{7}{6} \alpha \, \Gamma_z + \frac{1}{3} A^{-1} \gamma \Gamma_8  ) |\cdot\rangle = \frac{1}{\ell A} | \cdot \rangle$ \\
		\hline
		\end{tabular}
 \vskip 0.2cm
  \label{table3yyyx}
 \end{center}
\end{table}

It is straightforward to observe that the investigation of the number of supersymmetries preserved  by the algebraic KSE is exactly the same as that for the AdS$_4$ backgrounds
 with internal space $\mathrm{Spin}(7)/G_2$ in \cite{ads4Ngr16},
where instead of $\Gamma_x$ we have $\Gamma_8$.  In particular, the algebraic KSE can be written as
\begin{align}\label{Spin7G2_alg_KSE}
\bigg(\frac{1}{6} \alpha &\left( P_1 -P_2 +P_3 - P_1 \, P_2 \, P_3 - P_2 \, P_3 + P_1 \, P_3 -  P_1 \, P_2 \right) \Gamma_z \notag\\
&+ \frac{1}{3} \gamma \, A^{-1} \Gamma_8 \bigg) \sigma_+ = \frac{1}{\ell A} \sigma_+~,
\end{align}
where $\{P_1, P_2, P_3\}= \{ \Gamma^{1245}, \Gamma^{1267}, \Gamma^{1346}\}$ are mutually commuting, Hermitian Clifford algebra operators with eigenvalues $\pm 1$.
The solutions of the algebraic KSE on the eigenspaces of $\{P_1, P_2, P_3\}$ have been tabulated in table \ref{table3yyyx}.

To preserve $N>16$ supersymmetries, it is required to consider the subspace in table \ref{table3yyyx} with  7 eigenspinors. The integrability condition of the remaining  algebraic KSE gives
\bea
{1\over36} \alpha^2+{1\over 9} A^{-2} \gamma^2={1\over\ell^2 A^2}~,
\eea
while the warp factor field equation implies
\bea
 {7\over12} \alpha^2+{1\over 6} A^{-2} \gamma^2={1\over\ell^2 A^2}~.
\eea
Clearly, these are mutually inconsistent.  So there are no AdS$_3$  solutions that preserve $N>16$ supersymmetries with internal space $\mathrm{Spin}(7)/G_2\times S^1$.

\subsubsection {\texorpdfstring{$S^7\times S^1=U(4)/U(3)\times S^1$}{S7xS1}}\label{s7u4u3}

Let us briefly summarise  the homogeneous geometry of $S^7=U(4)/U(3)$ which is useful for our investigation of other cases below as well.
There is a left-invariant frame $(\ell^r, \ell^7)$, $r=1,\cdots, 6$,  on  $U(4)/U(3)$ such that the invariant metric can be written
as
\bea
ds^2(U(4)/U(3))= a\, (\ell^7)^2+b\, \delta_{rs} \ell^r \ell^s~,
\label{metru4u3}
\eea
where $a,b>0$ are constants.  The invariant forms on  $U(4)/U(3)$ are generated by the invariant 1-form $\ell^7$ and the
2-form $\omega$ which can be chosen as
\bea
\omega=\ell^{12}+\ell^{34}+\ell^{56}~.
\label{omegau4u3}
\eea
Furthermore
\bea
d\ell^7=\omega~.
\label{diffu4u3}
\eea
For more details see e.g.\ \cite{ads4Ngr16}, where the homogeneous geometry of $SU(4)/SU(3)$ is also described.

Turning to the investigation at hand, the metric on $U(4)/U(3)\times S^1$ can be written as
\bea
ds^2(M^8)=ds^2(U(4)/U(3)) + ds^2(S^1)~,~~~~ds^2(S^1)=c \, (\ell^8)^2~,
\eea
where $ds^2(U(4)/U(3))$ is as in (\ref{metru4u3}), $\ell^8$ is the invariant frame on $S^1$, $d\ell^8=0$,  and $c>0$ is constant.

The most general invariant fluxes $Q$ and $X$ that satisfy the Bianchi identities (\ref{mbianchi}),  $dX=dQ=0$, are
\begin{align}
X =  \frac{1}{2}\, \alpha\, \omega\wedge\omega ~,~~~Q=\beta\,  \ell^8~,
\end{align}
where  $\alpha,\beta $ are constants.

Next consider the Einstein equation along $S^1$.  As $X$ does not have non-vanishing components along $S^1$ and the metric factorises into that of  $U(4)/U(3)$ and $S^1$, we have
\begin{align}
R_{88}^{(8)} =  -  \frac{1}{3} A^{-2} Q^2 - \frac{1}{144} X^2 ~,
\label{r88s1}
\end{align}
where  $R_{88}^{(8)}$ is the Ricci tensor along $S^1$. This must vanish, $R_{88}^{(8)}=0$.  Thus  $Q=X=0$.  Then the warp factor field equation cannot be satisfied and so there are no AdS$_3$
solutions with internal space $U(4)/U(3)\times S^1$.

\subsubsection{\texorpdfstring{$S^7\times S^1=(Sp(2)\times Sp(1))/(Sp(1)\times Sp(1))\times S^1$}{S7xS1}} \label{sp2sp1sp1sp1s1}

The modification of $Sp(2)/Sp(1)$ to $(Sp(2)\times Sp(1))/(Sp(1)\times Sp(1))$ has already been described in section \ref{modifx} and in particular in (\ref{modifsp2}).
The geometry of this homogeneous space is a special case of that of  $Sp(2)/Sp(1)$.  In particular, the invariant forms on $(Sp(2)\times Sp(1))/(Sp(1)\times Sp(1))$
are those on $Sp(2)/Sp(1)$ which are invariant under both $Sp(1)$'s in the isotropy group.

Using the notation in \cite{ads4Ngr16}, we introduce a left-invariant frame $(\ell^a, \ell^r )$ on $(Sp(2)\times Sp(1))/(Sp(1)\times Sp(1))$, where $a=1,2,3,4$ and $r=5,6,7$.  Then imposing  invariance under both $Sp(1)$'s, one finds that
there are no invariant 1- and 2-forms on $(Sp(2)\times Sp(1))/(Sp(1)\times Sp(1))$.  However there are two invariant 3-forms and  two invariant 4-forms given by
\bea
\sigma={1\over 3!} \epsilon_{rst} \ell^{rst}~,~~~\tau=\ell^r\wedge I_r^{(+)}~,
\eea
\bea
\rho=\delta^{rs} \rho_{rs}={1\over2}\delta^{rs} \epsilon_{rpq} \ell^{pq}\wedge I_s^{(+)}~,~~~\psi={1\over4!}\epsilon_{abcd} \ell^{abcd}~,
\eea
respectively, where $I_r^{(+)}={1\over2} (I_r^{(+)})_{ab}\ell^{ab}$ and $\big( (I_r^{(+)})_{ab}\big)$ is a basis of self-dual 2-forms on $\bR^4$.  Moreover,
\bea
d\sigma={1\over2} \rho~,~~~d\tau=3 \psi- \rho~,~~~d\psi=d\rho=0~.~~~
\eea

After imposing the Bianchi identities $dQ=dX=0$, the  most general fluxes can be written as
\bea
X=\alpha_1 \psi+ \alpha_2 \rho~,~~~Q= \beta \ell^8~,
\eea
where $\ell^8$ is an invariant frame on $S^1$, $d\ell^8=0$.

The metric can be chosen as
\bea
ds^2 = f\, \delta_{ab} \ell^a \ell^b + h\, \delta_{rs} \ell^r \ell^s+ p\, (\ell^8)^2~,
\label{metrsp2sp1sp1sp1s1}
\eea
where $f,h,p>0$ are constants.  Substituting the metric and fluxes into the Einstein equation along the $S^1$ direction, we find again (\ref{r88s1}) which implies
 $Q=X=0$.  So there are no AdS$_3$ solutions with internal space $(Sp(2)\times Sp(1))/(Sp(1)\times Sp(1))\times S^1$.

\subsubsection{\texorpdfstring{$S^4\times S^3\times S^1=\mathrm{Spin}(5)/\mathrm{Spin}(4) \times SU(2)\times S^1$}{S4xS3xS1}}

The metric can be chosen as
\bea
ds^2(M^8)= a\, \delta_{ij} \ell^i \ell^j+ b_{rs} \ell^r \ell^s+ c\, (\ell^8)^2~,
\eea
where
\bea
ds^2(S^4)=a\, \delta_{ij} \ell^i \ell^j~,~~~ds^2(S^3)=b_{rs}\, \ell^r \ell^s~,~~~ds^2(S^1)=c\, (\ell^8)^2~,
\eea
and where $a, c>0$ are constants, $b=(b_{rs})$ is a constant symmetric positive definite matrix.  $(\ell^i)$, $i=1,2,3,4$, is a left-invariant frame on $S^4$ viewed
as a $\mathrm{Spin}(5)/\mathrm{Spin}(4)$ symmetric space and $(\ell^r)$, $r=5,6,7$, is a left-invariant frame on the group manifold $S^3$ with
\bea
d\ell^r={1\over2} \epsilon^r{}_{st} \ell^s\wedge \ell^t~,
\eea
and $\ell^8$ is an invariant  frame on $S^1$, $d\ell^8=0$.  Note that $\ell^r$ can be chosen up to an $SO(3)$ transformation.  This can be used to take $b$, without loss of generality,  to be
diagonal.

The most general invariant fluxes are
\bea
X&=&\alpha_1\, \ell^{1234}+ \alpha_2\, \ell^{5678}~,~~~
Q=\beta \ell^8+ \gamma_r \ell^r~.
\eea
As the Bianchi identities require that $dQ=0$, one finds that $\gamma_r=0$. Since $Q$ is also co-closed, we have that $X\wedge X=0$ which in turn gives
$\alpha_1\alpha_2=0$.

Suppose first that $\alpha_1=0$.  In that case, the algebraic KSE can be written as
\bea
\Big({1\over 6} {\alpha_2\over \sqrt {c b_1b_2b_3}} J_1+ {1\over3} {\beta \over A \sqrt{c}} J_2\Big)\sigma_+={1\over \ell A} \sigma_+~,
\eea
where $J_1=\Gamma^{5678} \Gamma_z$ and $J_2=\Gamma^8$ are commuting Hermitian Clifford algebra operators and $b=\mathrm{diag} (b_1, b_2,b_3)$. To find solutions with $N>16$ supersymmetries we have to consider
at least two of the common eigenspaces of $J_1$ and $J_2$, each of which has dimension 4.  This is possible if either $\alpha_2$ or  $\beta$ vanishes.  If $\alpha_2=0$,  then $X=0$ and the solution
is purely electric.  Such solutions cannot preserve $N>16$ supersymmetries.  On the other hand if $\beta=0$, the integrability condition of the KSE implies that
\bea
{1\over 36} {\alpha^2_2\over c\, b_1 b_2b_3}={1\over \ell^2 A^2}~.
\eea
Comparing this with the warp factor field equation, it leads to an inconsistency.  Thus, there are no such AdS$_3$ solutions which preserve $N>16$ supersymmetries with internal space $\mathrm{Spin}(5)/\mathrm{Spin}(4) \times SU(2)\times S^1$.

Suppose now that $\alpha_2=0$.  In such a case $X$ does not have components along $S^1$.  As a result the Einstein equations along $S^1$ can be written as in (\ref{r88s1})
and so $X=Q=0$.  There are no such AdS$_3$ solutions with internal space $\mathrm{Spin}(5)/\mathrm{Spin}(4) \times SU(2)\times S^1$.

\subsection{\texorpdfstring{$N>16$}{N greater 16} solutions with \texorpdfstring{$N_R=6$}{NR=6}}

The only right-handed superalgebra with 6 odd generators is $\mathfrak{osp}(3|2)$.  This   gives rise to an $\mathfrak{so}(3)_R$ action on the internal space.
Therefore the symmetry algebras that act transitively and effectively on  the internal spaces  are
\bea
&&\mathfrak{so}(n)_L\oplus \mathfrak{so}(3)_R~,~~~n=6, \cdots, 12~,~~~(N=2n+6)~;~~~~
\cr
&&\mathfrak{u}(n)_L\oplus \mathfrak{so}(3)_R~,~~~n=3, 4, 5, 6~,~~~(N=4n+6)~;
\cr
&&(\mathfrak{sp}(n) \oplus \mathfrak{sp}(1))_L\oplus \mathfrak{so}(3)_R~,~~~n=2,3~,~~~(N=8n+6)~;~
\cr
&&
\mathfrak{spin}(7)_L \oplus \mathfrak{so}(3)_R~,~~~(N=22)~;~~~~
\cr
&&(\mathfrak{g}_2)_L\oplus \mathfrak{so}(3)_R~,~~~(N=20)~.
\label{nr6}
\eea
An inspection of the homogeneous spaces in table \ref{table:8dim} reveals that, up to a finite covering, these are either $M^6\times S^2$ or $M^5\times S^3$, where $M^6$ and $M^5$ are
homogeneous 6- and 5-dimensional spaces.  So we have
\bea
&&S^6\times S^2=\mathrm{Spin}(7)/\mathrm{Spin}(6)\times SU(2)/U(1)~,~~~(N=20, 22)~;~~~
\cr
&&\mathbb{CP}^3\times S^2=SU(4)/S(U(1)\times U(3))\times SU(2)/U(1)~,~~~(N=18)~;
\cr
&&S^5\times S^3=\mathrm{Spin}(6)/\mathrm{Spin}(5)\times SU(2)~,~~~(N=18)~;~~~
\cr
&&S^5\times S^3=U(3)/U(2) \times SU(2)~,~~~(N=18)~;
\cr
&&S^4\times S^2\times S^2=\mathrm{Spin}(5)/\mathrm{Spin}(4)\times SU(2)/U(1)\times  SU(2)/U(1)~,~~~(N=22)~;
\cr
&&S^6\times S^2=G_2/SU(3)\times SU(2)/U(1)~,~~~(N=20)~.
\eea
The homogeneous space $SU(3)/T^2\times SU(2)/U(1)$ has been excluded as there is no modification that can be made such that  $U(3)$ can act almost  effectively on it.
Nevertheless we have performed the analysis to demonstrate that it cannot be the internal space of an AdS$_3$ solution that preserves $N>16$ supersymmetries.
 On the other hand, $SU(4)/S(U(1)\times U(3))\times SU(2)/U(1)$ has been included because $\mathfrak{su}(4)=\mathfrak{so}(6)$
and so $\mathbb{CP}^3$ admits an $\mathfrak{so}(6)$ effective and transitive action giving rise to  $N=18$ supersymmetries with $N_L=12$ and $N_R=6$.  $SU(4)/S(U(1)\times U(3))\times SU(2)/U(1)$ could have been considered as a background that preserves $20$ supersymmetries as well but it cannot be modified to admit an effective
$\mathfrak{u}(4)$ action.

The homogeneous spaces $S^6\times S^2=\mathrm{Spin}(7)/\mathrm{Spin}(6)\times SU(2)/U(1)$ and
$S^5\times S^3=\mathrm{Spin}(6)/\mathrm{Spin}(5)\times SU(2)$ can immediately be excluded as giving potential solutions. For $S^6\times S^2$, $X=Q=0$ and so the warp factor field equation cannot be satisfied.  The same is true for  $S^5\times S^3$ after applying the Bianchi identity $dQ=0$ to show that $Q=0$.

\subsubsection{\texorpdfstring{$\mathbb{CP}^3\times S^2=SU(4)/S(U(1)\times U(3))\times SU(2)/U(1)$}{CP3xS2}}

This homogeneous space is considered as an internal space because $\mathfrak{su}(4)=\mathfrak{so}(6)$ and so it may give rise to a solution which
preserves 18 supersymmetries.  The most general invariant  metric in the conventions of  \cite{ads4Ngr16} is
\begin{align}
ds^2(M^8)= ds^2(S^2)+ds^2(\mathbb{CP}^3)= a\, \delta_{ij} \ell^i \ell^j +b (\delta_{rs} \ell^r \ell^s + \delta_{\tilde{r}\tilde{s}} \ell^{\tilde{r}} \ell^{\tilde{s}})~,
\end{align}
where $(\ell^i)$, $i=7,8$, is a left-invariant frame on $S^2$ and $(\ell^r, \ell^{\tilde{r}})$  $r,\tilde{r}=1,2,3$, is a left-invariant frame on  $\mathbb{CP}^3$ and $a,b>0$ are constants. The invariant forms are generated by the volume form on $S^2$
\begin{align}
\sigma = \frac{1}{2} a\, \epsilon_{ij} \ell^i \wedge \ell^j~,
\end{align}
and the K\"ahler form on $\mathbb{CP}^3$
\begin{align}
\omega = b\, \delta_{r\tilde{s}} \ell^r \wedge \ell^{\tilde{s}}~.
\end{align}
Hence, the most general invariant  fluxes are
\begin{align}
Q=0~, \quad X= \alpha\, \frac{1}{2} \omega\wedge \omega + \beta\, \sigma\wedge\omega~.
\end{align}
The Bianchi identities are trivially satisfied but the field equation for $Q$ gives  the condition
\begin{align}
X\wedge X = \alpha \beta~ \sigma\wedge\omega\wedge\omega\wedge\omega = 0~.
\end{align}
Therefore, either $\alpha=0$ or $\beta=0$. It remains to investigate the KSEs.

\underline{$\beta=0$}

For $\beta=0$, the flux $X$ is simply $X= \frac{1}{2} \alpha\, \omega\wedge\omega$. Going to an  orthonormal frame, in which the K\"ahler form is $\omega= \bbe^{12} + \bbe^{34} +\bbe^{56}$, we find for the algebraic KSE \eqref{algKSE}
\begin{align}\label{alphakse}
\frac{\alpha}{6} (J_1+J_2 - J_1 J_2) \Gamma_z \sigma_+ = \frac{1}{\ell A} \sigma_+~,
\end{align}
where $J_1=\Gamma^{1234}$ and $J_2= \Gamma^{1256}$ are mutually commuting Clifford algebra operators with eigenvalues $\pm1$. The decomposition in terms of the common eigenspaces is summarised in table \ref{tablealphaksex}. A similar analysis applies to $\tau_+$, except that the right-hand side is $-1/(\ell A)$.

\begin{table}[h]
	\begin{center}
		\vskip 0.3cm
		\caption{Decomposition of \eqref{alphakse} KSE into eigenspaces}
		\vskip 0.3cm
		\begin{tabular}{|c|c|c|}
			\hline
			&$|J_1,J_2\rangle$&  relations for the fluxes\\
			\hline
			(1)&$|+,+\rangle$, $|+,-\rangle$, $|-,+\rangle$ & $\frac{\alpha}{6} \Gamma_z |\cdot\rangle = \frac{1}{\ell A} | \cdot \rangle$ \\
\hline
			(2)& $|-,-\rangle$ &  $-\frac{\alpha}{2} \Gamma_z |\cdot\rangle = \frac{1}{\ell A} | \cdot \rangle$ \\
			\hline
		\end{tabular}
		\vskip 0.2cm
		\label{tablealphaksex}
	\end{center}
\end{table}

To find solutions that preserve  $N>16$ supersymmetries, one has to choose  spinors from the  eigenspaces (1) in table \ref{tablealphaksex}. In such a case,  the integrability condition of the remaining $\Gamma_z$ projection on the spinors  is
\begin{align}
\frac{\alpha^2}{36} = \frac{1}{\ell^2 A^2}~,
\end{align}
whereas the field equation for the warp factor \eqref{warp} requires
\begin{align}
\frac{\alpha^2}{4} = \frac{1}{\ell^2 A^2}~.
\end{align}
Thus there is a contradiction and there are no AdS$_3$ solutions preserving $N>16$ supersymmetries.

\underline{$\alpha=0$}

For $\alpha=0$, the 4-form flux becomes $X=\beta \sigma\wedge\omega$. Going to an orthonormal frame, in which $\omega = \bbe^{12} + \bbe^{34} +\bbe^{56}$ and $\sigma = \bbe^{78}$, we find for the algebraic KSE \eqref{algKSE}
\begin{align}\label{betakse}
-\frac{\beta}{6} (J_1 + J_2 + J_3) J_1J_2J_3 \sigma_+ = \frac{1}{\ell A} \sigma_+~,
\end{align}
where the Clifford algebra operators $J$ are defined as
\begin{align}
J_1 = \Gamma^{1278}~,~~~J_2= \Gamma^{3478}~,~~~J_3= \Gamma^{5678}~,
\end{align}
and
\bea
\Gamma_z=-J_1J_2J_3~.
\eea
The decomposition of the algebraic KSE \eqref{betakse} into the eigenpaces of these mutually commuting Clifford algebra operators is illustrated in table \ref{tablebetaksexx}. A similar analysis applies to the $\tau_+$ spinors with the right-hand side replaced by $-\frac{1}{\ell A}$.

\begin{table}[h]
	\begin{center}
		\vskip 0.3cm
		\caption{Decomposition of \eqref{betakse} KSE into eigenspaces}
		\vskip 0.3cm
		\begin{tabular}{|c|c|c|}
			\hline
			&$|J_1,J_2,J_3\rangle$&  relations for the fluxes\\
			\hline
			(1)&$|\pm,\pm,\mp\rangle$, $|\pm,\mp,\pm\rangle$, $|\mp,\pm,\pm\rangle$ & $\frac{\beta}{6}  = \frac{1}{\ell A} $ \\
\hline
			(2)& $|\pm,\pm,\pm\rangle$ &  $-\frac{\beta}{2}  = \frac{1}{\ell A} $ \\
			\hline
		\end{tabular}
		\vskip 0.2cm
		\label{tablebetaksexx}
	\end{center}
\end{table}

For solutions to preserve  $N>16$ supersymmetries, we need to consider the eigenspinors given in row   (1) of table \ref{tablebetaksexx}.  This gives
\begin{align}
\frac{\beta^2}{36} = \frac{1}{\ell^2 A^2}~,
\end{align}
while the field equation for the warp factor \eqref{warp} leads to
\begin{align}
\frac{\beta^2}{4} = \frac{1}{\ell^2 A^2}~.
\end{align}
Clearly this is a  contradiction.  There are no AdS$_3$ backgrounds that preserve $N>16$ supersymmetries with internal space $SU(4)/S(U(1)\times U(3))\times SU(2)/U(1)$.

\subsubsection{\texorpdfstring{$S^5\times S^3=U(3)/U(2) \times SU(2)$}{S5xS3}} \label{u3u2su2}

The  geometry on $S^5$ as a $U(3)/U(2)$ homogeneous space can be described in a similar way as that for $S^7=U(4)/U(3)$ which can be found in section \ref{s7u4u3}. In particular the metric is
  \bea
ds^2(U(3)/U(2))= a\, (\ell^5)^2+b\, \delta_{rs} \ell^r \ell^s~,~~~r,s=1,2,3,4~,
\label{metru3u2}
\eea
where $a,b>0$ are constants.  The invariant forms on  $U(3)/U(2)$ are generated by the  1-form $\ell^5$ and the
2-form $\omega=\ell^{12}+\ell^{34}$.  Again  $d\ell^5=\omega$.

 The existence of AdS$_3$ solutions with internal space $S^5\times S^3$ can be ruled out with a cohomological argument.  Indeed, let $\ell^i$ be a left-invariant frame on $S^3$ such that
  \bea
  d\ell^i={1\over2} \epsilon^i{}_{jk} \ell^j\wedge \ell^k~,~~~i,j,k=1,2,3~.
  \eea
  The most general invariant 1-form $Q$ can be written as
\bea
Q= \alpha \bbl^5+ \beta_r \bbl^r~.
\eea
 The Bianchi identity, $dQ=0$, in \eqref{mbianchi} implies that $\alpha=\beta_r=0$.
So, we have  $Q=0$.

Furthermore, the Bianchi identities \eqref{mbianchi} also imply that $dX=0$, and as $Q=0$ the field equation \eqref{coX} also implies that $d*X=0$.  Thus $X$ is harmonic and  represents a class in $H^4(S^5\times S^3)$.
However $H^4(S^5\times S^3)=0$ and so $X=0$.  This leads to a contradiction as the field equation for the warp factor \eqref{warp} cannot be satisfied.

Note that the above calculation rules out the existence of AdS$_3$ solutions with internal space $S^5\times S^3=\mathrm{Spin}(6)/\mathrm{Spin}(5) \times SU(2)$ as this is a special case of the
background examined above.

\subsubsection{\texorpdfstring{$S^4\times S^2\times S^2=\mathrm{Spin}(5)/\mathrm{Spin}(4)\times SU(2)/U(1)\times  SU(2)/U(1)$}{S4xS2xS2}}

The most general invariant metric is
\bea
ds^2(M^8)&=& ds^2(S^4)+ ds^2(S^2)+ ds^2(S^2)
\cr
&=& a\, \delta_{ij} \ell^i \ell^j+ b\, \left((\ell^5)^2+(\ell^6)^2\right)+ c\,\left(\ell^7)^2+(\ell^8)^2\right)~,
\eea
where $a,b,c>0$ are constants, $\ell^i$, $i=1,2,3,4$, is a left-invariant frame on $S^4$ viewed as the symmetric space $\mathrm{Spin}(5)/\mathrm{Spin}(4)$, and $(\ell^5, \ell^6)$ and $(\ell^7, \ell^8)$ are left-invariant frames on the two $S^2$'s, respectively.

As there are no invariant 1-forms, $Q=0$.  Moreover, $X$ can be written as
\bea
X=\alpha_1 \ell^{1234}+\alpha_2 \ell^{5678}~.
\eea
Since $X\wedge X=0$, which follows from the field equation of $Q$, we have that $\alpha_1\alpha_2=0$.  If $\alpha_2=0$, then the integrability condition of the algebraic KSE will
give
\bea
{1\over 36} {\alpha_1^2\over a^4}={1\over \ell^2 A^2}~.
\eea
Comparing this with the field equation of the warp factor leads to a contradiction.  This is also the case if instead $\alpha_1=0$.  There are no supersymmetric AdS$_3$ solutions with internal space  $S^4\times S^2\times S^2$.

\subsubsection{\texorpdfstring{$S^6\times S^2=G_2/SU(3)\times SU(2)/U(1)$}{S6xS2}}

The existence of AdS$_3$ solutions with  $G_2/SU(3)\times SU(2)/U(1)$ internal space can be ruled out by a cohomological argument.  Observe that $\mathfrak{su}(3)$ acts on $\mathfrak{m}$ with the $[\bf 3]_R={\bf 3}\oplus \bar{\bf{3}}$ representation. Using this, one concludes that there are no  invariant 1-forms on $M^8$ and so  $Q=0$.  In such a case $X$ is both closed and co-closed and so harmonic.   However,  $H^4(M^8)=0$ as $M^8=S^6\times S^2$ and so $X=0$.  This  in turn leads to a  contradiction as the field equation
for the warp factor cannot be satisfied.

\subsection{\texorpdfstring{$N>16$}{N greater 16} solutions with \texorpdfstring{$N_R=8$}{NR=8}}

The right-handed superalgebras with 8 supercharges are $\mathfrak{osp}(4|2)$, $\mathfrak{D}(2,1, \alpha)$ and $\mathfrak{sl}(2|2)/1_{4\times 4}$.
These give rise to right-handed isometries with Lie algebras $\mathfrak{so}(4)_R$, $(\mathfrak{so}(3)\oplus \mathfrak{so}(3))_R$ and $\mathfrak{su}(2)_R$,
respectively.  In the latter case there can also be up to three additional central generators. As $\mathfrak{so}(4)_R=(\mathfrak{so}(3)\oplus \mathfrak{so}(3))_R$, it suffices to consider  $(\mathfrak{so}(3)\oplus \mathfrak{so}(3))_R$ and $\mathfrak{su}(2)_R$, and, for the latter, include up to 3 central generators.   Furthermore, since $N>16$, one has
 $N_L> 8$. Collecting the above and using the results of table \ref{table:ads3ksa},  the allowed algebras that act transitively and effectively on the internal space are the following.
 \bea
&& \mathfrak{so}(n)_L\oplus (\mathfrak{t}_R)_0~,~~~n=5,\dots,11~,~~~(N=2n+8)~;
\cr
&&\mathfrak{u}(n)_L\oplus (\mathfrak{t}_R)_0~,~~~n=3,4,5~,~~~(N=4n+8)~;
\cr
&&(\mathfrak{sp}(2)\oplus\mathfrak{sp}(1))_L\oplus (\mathfrak{t}_R)_0~,~~~(N=24)~;
\cr
&&\mathfrak{spin}(7)_L\oplus (\mathfrak{t}_R)_0~,~~~(N=24)~;~~~(\mathfrak{g}_2)_L\oplus (\mathfrak{t}_R)_0~,~~~(N=22)~,
\label{nr8}
\eea
where $(\mathfrak{t}_R)_0$ is either $(\mathfrak{so}(3)\oplus \mathfrak{so}(3))_R$ or $\mathfrak{su}(2)_R\oplus \mathfrak{c}_R$ with  $\mathfrak{c}_R$ spanned by up to 3 central generators.
The homogeneous spaces that can admit a transitive and an effective action  by the above Lie algebras  have been tabulated in table \ref{nr8table}.

\begin{table}[h]
	\begin{center}
		\vskip 0.3cm
		\caption{Homogeneous spaces for $N_R=8$}
		\vskip 0.3cm
\scalebox{0.8}{%
		\begin{tabular}{|c|c|c|}
			\hline
$\mathfrak{t}_0$& Homogeneous spaces& $N$\\
\hline
			$\mathfrak{so}(5)_L\oplus (\mathfrak{t}_R)_0$&  $ S^4\times S^2\times S^2=\mathrm{Spin}(5)/\mathrm{Spin}(4)\times SU(2)/U(1)\times  SU(2)/U(1)$& 18\\
			& $S^4\times S^2\times
 T^2=\mathrm{Spin}(5)/\mathrm{Spin}(4)\times SU(2)/U(1)\times  T^2 $ & 18\\
			  &  $
 \mathbb{CP}^3\times S^2=Sp(2)/(Sp(1)\times U(1))\times SU(2)/U(1) $&18 \\
  &$S^7\times S^1=(Sp(2)\times Sp(1))/(Sp(1)\times Sp(1))\times S^1$&18\\
			\hline
$\mathfrak{so}(6)_L\oplus (\mathfrak{t}_R)_0$&$S^5\times S^3=\mathrm{Spin}(6)/\mathrm{Spin}(5)\times SU(2)$&20\\
&$
 S^5\times S^2\times
 S^1=\mathrm{Spin}(6)/\mathrm{Spin}(5)\times SU(2)/U(1)\times  S^1$&20\\
 &$\mathbb{CP}^3\times S^2=SU(4)/S(U(3)\times U(1))\times SU(2)/U(1)$ &20\\
 \hline
 $\mathfrak{so}(7)_L\oplus (\mathfrak{t}_R)_0$&$S^6\times S^2=\mathrm{Spin}(7)/\mathrm{Spin}(6)\times SU(2)/U(1)$&22, 24\\
 \hline
 $\mathfrak{u}(3)_L\oplus (\mathfrak{t}_R)_0$&$SU(3)^{k,l}=(SU(3)\times SU(2)\times U(1)/(SU(2)\times \Delta_{k,l}U(1))$&20\\
 &$S^5\times S^3=U(3)/U(2)\times SU(2)$&20\\
 &$S^5\times S^2\times S^1=U(3)/U(2)\times SU(2)/U(1)\times S^1$&20\\
 &$N^{k,l,m}\times S^1=\frac{U(1)\times SU(2) \times SU(3) }{\Delta_{k,l,m}((U(1)^2)\cdot (1\times SU(2))}\times S^1$&20\\
 \hline
 $\mathfrak{sp}(2)\oplus \mathfrak{sp}(1))_L\oplus (\mathfrak{t}_R)_0$&$S^4\times S^2\times S^2=\mathrm{Spin}(5)/\mathrm{Spin}(4)\times SU(2)/U(1)\times SU(2)/U(1)$&24\\
 &$S^4\times S^3\times S^1=\mathrm{Spin}(5)/\mathrm{Spin}(4)\times (SU(2)\times SU(2))/SU(2)\times S^1$&24\\

 \hline
 $( \mathfrak{g}_2)_L\oplus (\mathfrak{t}_R)_0$&$S^6\times S^2=G_2/SU(3)\times SU(2)/U(1)$&22\\
 \hline
 \end{tabular}}
		\vskip 0.2cm
		\label{nr8table}
	\end{center}
\end{table}

A detailed examination of the homogeneous spaces that may give rise to supersymmetric AdS$_3$ solutions with $N_R=8$ reveals that the only cases that have not
been investigated so far are $S^4\times S^2\times T^2$, $S^5\times S^2\times S^1$ with $S^5$ either $\mathrm{Spin}(6)/\mathrm{Spin}(5)$ or $U(3)/U(2)$, $SU(3)$ and
$N^{k,l,m}\times S^1$.  The remaining homogeneous spaces have already been excluded as internal spaces  in the analysis of AdS$_3$ backgrounds with  $N_R<8$ backgrounds.    The presence of additional right-handed supersymmetries here for $N_R=8$
are not sufficient to bring these backgrounds into the range  of $N>16$ supersymmetries.  So again they are excluded as solutions.

 \subsubsection{\texorpdfstring{$ S^4\times S^2\times
 T^2=\mathrm{Spin}(5)/\mathrm{Spin}(4)\times SU(2)/U(1)\times  T^2$}{S4xS2xT2}}

  The most general invariant metric is
\begin{align}
ds^2 (M^8)=ds^2(S^4)+ ds^2(S^2)+ds^2(T^2)=a~ \delta_{rs} \ell^r \ell^s+ b~ \delta_{\hat a\hat b} \ell^{\hat a} \ell^{\hat b} + c_{\tilde{a}\tilde{b}} \ell^{\tilde{a}} \ell^{\tilde{b}} ~,
\end{align}
where $\ell^r$, $r=1,...,4$, is a left-invariant frame on $S^4$, $ \ell^{\hat a}$,  $\hat a = 5,6$ is a left-invariant frame on $S^2$ and  $\ell^{\tilde{a}}$,  $\tilde{a}=7,8$,  is a left
 invariant frame on $T^2$, $d\ell^{\tilde{a}}=0$, and $a,b>0$ are constants and $(c_{\tilde a \tilde b})$ is a positive definite matrix. The invariant forms on this $M^8$ are generated by
   $\ell^{\tilde{a}}$ and the top forms on $S^4$ and $S^2$.  Hence, the 4-form flux $X$ is
\begin{align}
X= \alpha \,\sigma \wedge \rho + \beta\, \psi~,
\end{align}
where $\alpha$ and $\beta$ are constant parameters and $\psi=\ell^{1234}$, $\sigma=\ell^{56}$ and $\rho=\ell^{78}$. Furthermore
\bea
Q=\gamma_{\tilde a} \bbl^{\tilde a}~,
\eea
where $\gamma$ are constants.  As $Q$ is parallel,
 the field equation for $Q$, \eqref{fqxx}, gives $X\wedge X=0$ and so we obtain the condition that either $\alpha=0$ or $\beta=0$. Let us proceed to investigate $\alpha=0$, as the case for $\beta=0$ can be dealt with  analogously. As $X=\beta \psi$, the algebraic KSE \eqref{algKSE} becomes
\begin{align}
({1\over 3A} \slashed {Q}+\frac{\beta}{6a^2} \Gamma^{1234} \Gamma_z )\sigma_+ = \frac{1}{\ell A} \sigma_+~.
\end{align}
The integrability condition of this is
\bea
{1\over 9 A^2} Q^2+\frac{\beta^2}{36a^4}=\frac{1}{\ell^2 A^2}~.
\eea
On the other hand  the warp factor field equation \eqref{warp} gives
\begin{align}
{1\over 6 A^2} Q^2+\frac{\beta^2}{12a^4} = \frac{1}{\ell^2 A^2}~.
\end{align}
The last two equations are incompatible and so there are no supersymmetric solutions.

 \subsubsection{\texorpdfstring{$M^8=S^5\times S^2\times S^1$}{S5xS2xS1}}

Here we shall consider two cases that  with $S^5=U(3)/U(2), SU(3)/SU(2)$ and that with $S^5=\mathrm{Spin}(6)/\mathrm{Spin}(5)$.  The latter can be excluded immediately.
 As $M^8$ is a product of symmetric spaces all
 left-invariant forms are parallel and represent classes in the de-Rham cohomology of $M^8$.  As  $H^4(S^5\times S^2\times S^1)=0$, we have that $X=0$.
  The solution becomes electric and as we have seen such solutions cannot preserve $N>16$ supersymmetries.

 Next suppose that $S^5=U(3)/U(2)$.
  The metric on $M^8$ can be chosen as
  \bea
  ds^2(M^8)= ds^2(S^5)+ ds^2(S^2)+ ds^2(S^1)~,
  \eea
  where
  \begin{align}
  ds^2(S^5)&= b\sum_{r=1}^4 (\bbl^r)^2+ a (\bbl^5)^2~,~~~ds^2(S^2)= c \left((\bbl^6)^2+(\bbl^7)^2\right)~,\notag\\
  ds^2(S^1)&= f (\bbl^8)^2~,
  \end{align}
  and where  $a,b,c,f>0$ are constants.
  The invariant forms are generated by $\bbl^5$, $\bbl^8$, $\omega=\bbl^{12}+ \bbl^{34}$ and  $\sigma= \bbl^{67}$.  The independent differential relations between the invariant forms are
  \bea
  d\bbl^5= \omega~,~~~d\bbl^8=0~,~~~d\sigma=0~,~~~
  \eea
  where we have used the description of the geometry on $S^5$ as in section \ref{u3u2su2}.
  Since $dQ=0$, we have that $Q=\gamma \bbl^8$. Furthermore after imposing $dX=0$ the most general flux $X$  is
  \bea
   X={1\over 2} \alpha\, \omega\wedge \omega+\beta\, \omega\wedge \sigma~,
   \label{xu4u3s2s1}
   \eea
   where  $\alpha, \beta$ are constants.
   
   The algebraic KSE gives
   \bea
  \Big[ {1\over6}\Big( {\alpha\over b^2} \Gamma^{1234}+{\beta\over bc} (\Gamma^{1267}+\Gamma^{3467})\Big)\Gamma_z+{1\over3} {\gamma\over \sqrt f\, A} \Gamma^8\Big]\sigma_+={1\over \ell A} \sigma_+~.
  \label{ksenr8}
  \eea
Squaring this, we find
\bea
\Big[{1\over 36}  \Big( {\alpha^2\over b^4} +{2\beta^2\over b^2c^2} -2{\alpha\beta\over b^3c} (J_1+J_2)+ {2\beta^2\over b^2 c^2} J_1 J_2 \Big)+{1\over9} {\gamma^2\over  f\, A^2} \Big]\sigma_+={1\over \ell^2 A^2} \sigma_+~,
\label{44433c}
\eea
where $J_1=\Gamma^{1267}$ and $J_2=\Gamma^{3467}$.  The decomposition of this condition on $\sigma_+$ into eigenspaces of $J_1$ and $J_2$ is given in table \ref{table43x}.

Each common eigenspace of $J_1$ and $J_2$ has dimension 4.  So, to find solutions with $N>16$ supersymmetries, we have to consider at least two of these eigenspaces.  Hence, this would necessarily involve either one of the eigenspinors  $|+,-\rangle$ and $|-,+\rangle$ or both eigenspinors $|\pm,\pm\rangle$.  In the former case taking the difference of the condition
that arises on the fluxes
with  the warp factor field equation
\bea
{1\over12} \Big( {\alpha^2\over b^4} +{2\beta^2\over b^2c^2}\Big)+ {\gamma^2\over 6 f A^2} ={1\over \ell^2 A^2} ~,
\eea
one finds that $\alpha=\beta=\gamma=0$ which is a contradiction.  In the latter case, we find that $\alpha\beta=0$.  Using this and comparing the condition on the fluxes
in table \ref{table43x} with the warp factor field equation above, again leads to a contradiction.
There are no AdS$_3$ solutions that preserve $N>16$ supersymmetries with internal space $S^5\times S^2\times S^1$.

\begin{table}[h]
\begin{center}
\vskip 0.3cm
 \caption{Decomposition of (\ref{44433c})  into eigenspaces}
 \vskip 0.3cm

	\begin{tabular}{|c|c|}
		\hline
		$|J_1,J_2\rangle$&  relations for the fluxes\\
		\hline
		$|+,-\rangle$, $|-,+\rangle$& ${1\over 36}   {\alpha^2\over b^4}  +{1\over9} {\gamma^2\over  f\, A^2} ={1\over \ell^2 A^2} $ \\
\hline
		$|\pm,\pm\rangle$& ${1\over 36}  \Big( {\alpha^2\over b^4} +{4\beta^2\over b^2c^2} \mp {4\alpha\beta\over b^3c} \Big)+{1\over9} {\gamma^2\over  f\, A^2} ={1\over \ell^2 A^2}$  \\
		\hline
		\end{tabular}
\vskip 0.2cm
  \label{table43x}
 \end{center}
\end{table}

We have also performed the calculation for $S^5=SU(3)/SU(2)$ which gives rise to an $X$ flux with additional  terms to those in (\ref{xu4u3s2s1}) because of the presence
 of an invariant complex (2,0) form. After some investigation, we find
that again there are no solutions with $N>16$ supersymmetry.

\subsubsection{\texorpdfstring{$SU(3)^{k,l}$}{SU(3)k,l}}

In this context $SU(3)$ is viewed, up to a discrete identification, as a homogeneous space with isotropy group $SU(2)\times U(1)$ and almost effective transitive group $SU(3)\times SU(2)\times U(1)$, where the inclusion map of $SU(2)\times U(1)$ in $SU(3)\times SU(2)\times U(1)$ is
\bea
(a, z)\rightarrow \left(\begin{pmatrix}a z^k & 0\cr 0 & z^{-2k}\end{pmatrix}, a, z^l\right)
\eea
As we have mentioned the geometry of such cosets is more restrictive than that of $SU(3)$ viewed as the homogeneous space $SU(3)/\{e\}$.
Thus it suffices to investigate whether $SU(3)$ is a solution.  As $SU(3)$ does not admit closed 1-forms, $Q=0$.  In such case $X$ is harmonic. However,
$H^4(SU(3), \bR)=0$ and so $X=0$.  This leads to a contradiction, since the warp factor field equation cannot be satisfied.

 \subsubsection{\texorpdfstring{$N^{k,l, m}\times S^1= \frac{SU(2) \times SU(3)\times U(1) }{\Delta_{k,l, m}(U(1)^2)\cdot (1\times SU(2))}\times S^1$}{N(k,l,m)xS1xS1}}

 Let us denote the left-invariant frame along $S^1$ with $\ell^8$, $d\ell^8=0$.  $N^{k,l, m}$ can be thought of as a modification of $N^{k,l}$ and so for the analysis that follows we can use
 the description of the geometry of  $N^{k,l}$ in appendix \ref{appencx2}.  In particular, the most general $Q$ flux is
 \bea
 Q=\gamma_1\, \ell^8+\gamma_2\, \ell^7~.
 \eea
 As $dQ=0$, we deduce that $\gamma_2=0$ and set $\gamma_1=\gamma$.  The most general invariant metric is
 \bea
 ds^2(M^8)=  a\, \left((\ell^5)^2+(\ell^6)^2\right)+ b\, \delta_{rs} ( \ell^r \ell^s+ \hat\ell^r \hat\ell^s)+ c\, (\ell^7)^2+ f\, (\ell^8)^2~,
 \eea
 where $(\ell^r, \hat\ell^r, \ell^5, \ell^6, \ell^7)$, $r,s=1,2$, is a left-invariant frame on $N^{k,l}$, $\ell^8$ is a left-invariant frame on $S^1$ and   $a,b,c,f>0$ are constants.
 Next $X$ can be chosen as
 \bea
 X={1\over2} \alpha_1\, \omega_1\wedge \omega_1+ \alpha_2\, \omega_1\wedge \omega_2+ \alpha_3\, \omega_1\wedge\ell^7\wedge\ell^8+\alpha_4\, \omega_2\wedge\ell^7\wedge\ell^8~,
 \eea
 where $\alpha_1, \alpha_2, \alpha_3, \alpha_4$ are constants. Since $dX=0$, one deduces that $\alpha_3=\alpha_4=0$.  Choosing an  orthonormal frame as
 \bea
 &&\bbe^1=\sqrt{a}\, \ell^5~,~~~\bbe^2=\sqrt{a}\, \ell^6~,~~~\bbe^{2r+1}=\sqrt{b}\, \ell^r~,~~~\bbe^{2r+2}=\sqrt{b}\, \hat \ell^r~,~~~
 \cr
 &&\bbe^7=\sqrt{c}\,\ell^7~,~~~\bbe^8=\sqrt{f}\,\ell^8~,
 \eea
 the algebraic KSE can be written as
 \bea
  \Big[ {1\over6}\Big( {\alpha_1\over b^2} \Gamma^{3456}+{\alpha_2\over ab} (\Gamma^{1234}+\Gamma^{1256})\Big)\Gamma_z+{1\over3} {\gamma\over \sqrt f\, A} \Gamma^8\Big]\sigma_+={1\over \ell A} \sigma_+~.
  \label{ksenr8-2}
  \eea
  The form of this KSE is the same as that in (\ref{ksenr8}).  A similar analysis again shows that there are no solutions preserving $N>16$ supersymmetries.

\subsection{\texorpdfstring{$N>16$}{N greater 16} solutions with \texorpdfstring{$N_R=10$}{NR=10}}

The only  superalgebra that gives rise to ten right-handed supersymmetries   is $\mathfrak{osp}(5|2)$ with $(\mathfrak{t}_R)_0=\mathfrak{so}(5)$.
As we are investigating  backgrounds with $N>16$ and we have chosen that $N_L\geq N_R$, we conclude that $10\leq N_L<22$.  Using this and the results of table \ref{table:ads3ksa}, the allowed
 algebras that can act transitively and effectively on the internal spaces are
\bea
&&\mathfrak{so}(n)_L\oplus \mathfrak{so}(5)_R~,~~n=5,6,7,8,9,10~, ~~(N=2n+10)~;~~~
\cr
&&
 \mathfrak{u}(n)_L\oplus \mathfrak{so}(5)_R~,~~n=3,4,5~,~~(N=4n+10)~;~~~
\cr
&&(\mathfrak{sp}(2) \oplus \mathfrak{sp}(1))_L\oplus \mathfrak{so}(5)_R~,~~~(N=26)~;~~~
\cr
&&
\mathfrak{spin}(7)_L\oplus \mathfrak{so}(5)_R~,~~~~(N=26)~;~~~
\cr
&&
(\mathfrak{g}_2)_L\oplus \mathfrak{so}(5)_R~,~~~~(N=24)~.
\label{nr10}
\eea
The only 8-dimensional homogeneous space  that admits such an action by the algebras presented above is
\bea
S^4\times S^4=\mathrm{Spin}(5)/\mathrm{Spin}(4) \times \mathrm{Spin}(5)/\mathrm{Spin}(4)~.
\eea
It remains to examine whether such a background solves the KSE and field equations of 11-dimensional supergravity.

\subsubsection{\texorpdfstring{$S^4 \times S^4$}{S4xS4}}

The most general invariant metric on $S^4 \times S^4$ is
\begin{align}
ds^2(M^8)= ds^2(S^4)+ds^2(S^4)=a\, \delta_{ij} \ell^i \ell^j + b\, \delta_{rs} \ell^r \ell^s= \delta_{ab} \bbe^a \bbe^b+ \delta_{rs} \bbe^r \bbe^s~,
\end{align}
where $a,b>0$ are constants and $\ell^i$ ($\bbe^i$), $i=1,\dots4$,  and $\ell^r$ ($\bbe^r$), $r=5,\dots, 8$,  are the left-invariant (orthonormal) frames of the two $S^4$'s, respectively. There are no invariant 1-forms on $M^8$, and therefore $Q=0$.
The invariant 4-forms are just the volume forms on the two spheres, hence the most general 4-form flux is
\begin{align}
X = \alpha \bbe^{1234} + \beta \bbe^{5678}~,
\end{align}
where $\alpha,\beta$ are constants.
The field equation  for $Q$, \eqref{fqxx}, yields the condition that either $\alpha =0$ or $\beta=0$. Without loss of generality, we take $\beta=0$. Substituting $X$ into the algebraic KSE \eqref{algKSE}, one finds
\begin{align}
\frac{\alpha}{6} \Gamma^{1234} \Gamma_z \sigma_+ = \frac{1}{\ell A} \sigma_+~,
\end{align}
and  hence obtains
\begin{align}
\frac{\alpha^2}{36 } = \frac{1}{\ell^2 A^2}~,
\end{align}
as an integrability condition. However, the warp factor field equation  \eqref{warp} implies that
\begin{align}
\frac{\alpha^2}{12 }=\frac{1}{\ell^2 A^2}~.
\end{align}
Thus, we get a contradiction and there are no such supersymmetric solutions.

\subsection{\texorpdfstring{$N>16$}{N greater 16} solutions with \texorpdfstring{$N_R\geq 12$}{NR greater equal 12}}

Imposing the restriction that  $N_L\geq N_R$, it is easy to see that there are no homogeneous spaces that admit a transitive and effective $\mathfrak{t}_0$ action. This follows from a detailed
examination of the classification results of \cite{Castellani:1983yg,klausthesis,niko6dim,niko7dim,bohmkerr}, as well as their modifications.

\section{\texorpdfstring{$N>16 ~ AdS_3\times_w M^7$}{N greater 16 AdS3xM7} solutions in (massive) IIA}\label{ads3IIA}

\subsection{Field equations and Bianchi identities for \texorpdfstring{$N>16$}{N greater 16}}

 The bosonic fields of (massive) IIA supergravity are the metric $ds^2$, a 4-form field strength $G$, a 3-form field strength $H$, a 2-form field strength $F$, the dilaton $\Phi$  and    the cosmological constant  dressed with the dilaton $S$. Following the description of warped AdS$_3$ backgrounds  in \cite{iiaads}, we write  the fields as
\begin{align}
ds^2 &=2 \bbe^+ \bbe^- + (\bbe^z)^2 +ds^2(M^7)~,\notag\\
G&=\bbe^+\wedge\bbe^-\wedge\bbe^z\wedge Y + X~,\quad H = W \bbe^+\wedge\bbe^-\wedge\bbe^z + Z~,\notag\\
&F~,\quad \Phi~,\quad S~,
\end{align}
where we have used a  null-orthonormal frame $(\bbe^+, \bbe^-, \bbe^i)$, $i=1,\dots, 7$, defined as in \eqref{nullframe} and  $ds^2(M^7) = \delta_{ij} \bbe^i \bbe^j$. The fields $\Phi, W,  S$ and the warp factor $A$ are functions,
$Y$ is a 1-form, $F$ is a 2-form, $Z$ is a 3-form and $X$ is a 4-form on $M^7$.  As the 2-form field strength $F$ is purely magnetic we have denoted the field
and its component on $M^7$ by the same symbol.  This is also the case for $\Phi$ and $S$.  The dependence of the fields on the AdS$_3$ coordinates  is hidden in the definition of the frame $\bbe^+, \bbe^-$ and $\bbe^z$.
The components of the fields in this frame depend only on the coordinates of $M^7$.

As we have demonstrated in  11-dimensional supergravity,  the description for the fields simplifies considerably for AdS$_3$ backgrounds preserving $N>16$ supersymmetries.
In particular, a similar argument to the one presented for 11-dimensional backgrounds gives that the warp factor $A$ is constant. The proof of this is very similar to that given in eleven dimensions
and so we shall not repeat the analysis.  Furthermore, it is a consequence of the homogeneity
theorem and the Bianchi identities of the theory  that the scalars $\Phi, S$ and $W$ are constant.

Focusing on the analysis of the IIA AdS$_3$ backgrounds that preserve $N>16$ supersymmetries, we shall impose these conditions on the Bianchi identities, field equations and Killing spinor
equations.  The general formulae can be found in \cite{iiaads}.  In particular taking $A,\,W,\, \Phi$ and $S$ to be constant the Bianchi identities can be simplified as
\begin{align}
dZ&=0~,~~~
dF=  SZ~,~~~ S W =0, \quad dX=  Z\wedge F~,\notag\\
dY&=  - W F~.
\label{iiab}
\end{align}
A consequence of this is that either $S=0$ or $W=0$.
Furthermore, the field equations of the form fluxes can be written as
\begin{align}\label{iiaf}
 d*_{{}_7}Z&=*_{{}_7}X\wedge F+ S *_{{}_7}F~,~~~d*_{{}_7}F=-W *_{{}_7}Y+ *_{{}_7}X\wedge Z~,\notag\\
 d *_{{}_7}Y&=-Z\wedge X~,~~~ d*_{{}_7}X=Z\wedge Y- W X~,
 \end{align}
 respectively.  As $M^7$ is compact without boundary observe that $d *_{{}_7}Y=-Z\wedge X$ implies that
 \bea
 Z\wedge X=0~.
 \label{zxz}
 \eea
To see this, first observe  that homogeneity  implies that $*_{{}_7}(Z\wedge X)$  is constant. On the other hand,   the integral of $Z\wedge X$ over $M^8$ is the constant $*_{{}_7}(Z\wedge X)$ times the volume of $M^8$.  As the integral of $Z\wedge X$ is zero, this constant must vanish giving (\ref{zxz}).

The dilaton field equation is
\begin{align}\label{iiafieldeqs}
 -\frac{1}{12} Z^2+ \frac{1}{2} W^2 +\frac{5}{4} S^2+ \frac{3}{8}F^2 + \frac{1}{96} X^2 - \frac{1}{4}Y^2=0~.
\end{align}
The Einstein equation along $AdS_3$ and $M^7$ implies
\begin{align}\label{einstiia}
& \frac{1}{2} W^2 + \frac{1}{96} X^2 + \frac{1}{4} Y^2 + \frac{1}{4} S^2 + \frac{1}{8} F^2=\frac{2}{\ell^2 A^2}~, \notag\\
&R^{(7)}_{ij} = \frac{1}{12} X^2_{ij} -\frac{1}{2} Y_i Y_j - \frac{1}{96} X^2 \delta_{ij} + \frac{1}{4} Y^2 \delta_{ij} \notag\\
&\quad~~~~~~~~- \frac{1}{4} S^2 \delta_{ij} +\frac{1}{4} Z_{ij}^2 + \frac{1}{2} F^2_{ij} - \frac{1}{8} F^2 \delta_{ij} ~,
\end{align}
where $\nabla$ and $R^{(7)}_{ij}$ denote the Levi-Civita connection and the Ricci tensor of $M^7$, respectively.  The former condition is the warp factor field equation.

\subsection{The Killing spinor equations}

The solutions to the KSEs of (massive) IIA along $AdS_3$ may be written as in  \eqref{ks}, although now $\sigma_\pm$ and $\tau_\pm$ are $\text{Spin}(9,1)$ Majorana spinors which satisfy the lightcone projections $\Gamma_\pm\sigma_\pm=\Gamma_\pm\tau_\pm=0$ and only depend on the coordinates of $M^7$. These are subject to the gravitino KSEs
\begin{align}\label{iiakse}
\nabla^{(\pm)}_i \sigma_\pm = 0~, \quad \nabla^{(\pm)}_i \tau_\pm = 0~,
\end{align}
the dilatino KSEs
\begin{align}\label{iiadilat}
\mathcal{A}^{(\pm)} \sigma_{\pm} = 0~, \quad \mathcal{A}^{(\pm)} \tau_{\pm} = 0~,
\end{align}
and the algebraic KSEs
\begin{align}\label{iiaalgkse}
\Xi^{(\pm)}\sigma_{\pm}=0~, \quad (\Xi^{(\pm)} \pm \frac{1}{\ell})\tau_{\pm}=0~,
\end{align}
where
\begin{align}
\nabla_i^{(\pm)} &= \nabla_i  + \frac{1}{8} \slashed{Z}_i \Gamma_{11} + \frac{1}{8} S \Gamma_i + \frac{1}{16} \sF \Gamma_i \Gamma_{11} + \frac{1}{192} \sX \Gamma_i \pm \frac{1}{8} \sY \Gamma_{zi}~, \notag \\
\mathcal{A}^{(\pm)} &=  \frac{1}{12} \slashed{Z} \Gamma_{11} \mp \frac{1}{2}W \Gamma_z \Gamma_{11} + \frac{5}{4} S + \frac{3}{8} \sF \Gamma_{11} + \frac{1}{96} \sX \pm \frac{1}{4} \sY \Gamma_{z}~, \notag \\
\Xi^{(\pm)} &= \mp\frac{1}{2\ell}  \pm \frac{1}{4} A W \Gamma_{11} - \frac{1}{8} A S \Gamma_z - \frac{1}{16} A \sF \Gamma_z \Gamma_{11} - \frac{1}{192} A \sX \Gamma_z \mp \frac{1}{8} A \sY~.
\end{align}
  If $M^7$ is compact without boundary, one can demonstrate that
\begin{align}
\parallel \sigma_+ \parallel = \text{const}~, \quad \parallel \tau_+ \parallel = \text{const}~, \quad \langle \sigma_+, \tau_+ \rangle =0~,~~~\langle \tau_+, \Gamma_{iz} \sigma_+\rangle=0~.
\end{align}
As in eleven dimensions, the last condition is essential to establish that the warp factor $A$ is constant for IIA AdS$_3$ backgrounds preserving $N>16$ supersymmetries
with compact without boundary internal space $M^7$.

\begin{table}\renewcommand{\arraystretch}{1.3}
	\caption{7-dimensional compact, simply connected,  homogeneous spaces}
	\centering
	\begin{tabular}{c l}
		\hline
		& $M^7=G/H$  \\  
		\hline
		(1)& $\frac{\mathrm{Spin}(8)}{\mathrm{Spin}(7)}= S^7$, symmetric space\\
		(2)&$\frac{\mathrm{Spin}(7)}{G_2}=S^7$ \\
		(3)& $\frac{SU(4)}{SU(3)}$ diffeomorphic to $S^7$ \\
        (4) & $\frac{Sp(2)}{Sp(1)}$ diffeomorphic to $S^7$ \\
		(5) & $\frac{Sp(2)}{Sp(1)_{max}}$, Berger space \\
		(6) & $ \frac{Sp(2)}{\Delta(Sp(1))}=V_2(\bR^5)$  \\
        (7) & $\frac{SU(3)}{\Delta_{k,l}(U(1))}=W^{k,l}$~~ $k, l$ coprime, Aloff-Wallach space\\
		(8)&$\frac{SU(2) \times SU(3) }{\Delta_{k,l}(U(1))\cdot (1\times SU(2))}=N^{k,l}$ ~$k,l$ coprime\\
		(9) & $\frac{SU(2)^3}{\Delta_{p,q,r}(U(1)^2)}=Q^{p,q,r}$ $p, q, r$ coprime\\
(10)&$M^4\times M^3$,~~$M^4=\frac{\mathrm{Spin}(5)}{\mathrm{Spin}(4)}, ~\frac{ SU(3)}{S(U(1)\times U(2))}, ~\frac{SU(2)}{U(1)}\times \frac{SU(2)}{U(1)}$\\
&~~~~~~~~~~~~~~~~$M^3= SU(2)~,~\frac{SU(2)\times SU(2)}{\Delta(SU(2))}$\\
(11)&$M^5\times \frac{SU(2)}{U(1)}$,~~$M^5=\frac{\mathrm{Spin}(6)}{\mathrm{Spin}(5)}, ~\frac{ SU(3)}{SU(2)}, ~\frac{SU(2)\times SU(2)}{\Delta_{k,l}(U(1))},~ \frac{ SU(3)}{SO(3)} $\\
[1ex]
		\hline
	\end{tabular}
	\label{table:7dhom}
\end{table}

\subsection{\texorpdfstring{$N>16$}{N greater 16} solutions with left only supersymmetry}

AdS$_3$ backgrounds admit the same Killing  superalgebras in  11-dimensional, IIA and IIB supergravities.  As a result the
  Lie algebras $\mathfrak{t}_0$  that must act transitively  and effectively on the
internal spaces of IIA and IIB  AdS$_3$ backgrounds can be read off  those  found in the  11-dimensional  analysis.  So for $N_R=0$, these are given in (\ref{nr0}).  An inspection of the 7-dimensional homogeneous spaces in table \ref{table:7dhom}   reveals that there are no $N>16$ supersymmetric AdS$_3$ backgrounds with $N_R=0$.

\subsection{\texorpdfstring{$N>16$}{N greater 16} solutions with \texorpdfstring{$N_R=2$}{NR=2}}

The 7-dimensional homogeneous spaces\footnote{There are several embeddings of $Sp(1)$ in $Sp(2)$ however only one of them admits a modification such that
 the internal space is associated to a background that can preserve $N>16$ supersymmetries.} that admit an effective and transitive action of the Lie algebras in (\ref{nr2}) are
\bea
&&S^7=\mathrm{Spin}(8)/\mathrm{Spin}(7)~~(N=18)~,~~~
\cr
&&S^7=U(4)/U(3)~,~~~(N=18)~,
\cr
&&S^7=(Sp(2)\times Sp(1))/Sp(1)\times Sp(1)~,~~~(N=18)~,
\cr
&& S^4\times S^3=\mathrm{Spin}(5)/\mathrm{Spin}(4)\times SU(2)~~(N=18)~,~~~
\cr
&& S^7=\mathrm{Spin}(7)/G_2~~(N=18)~.
\label{cnr2}
\eea
Solutions with internal space $\mathrm{Spin}(8)/\mathrm{Spin}(7)$ can immediately be excluded.  This is a symmetric space and so all fluxes are parallel.  On the other hand,
  the only parallel forms on $S^7$ are the constant
functions and the volume form.  Therefore all k-form fluxes for $k>0$ must vanish.  In such a case the dilaton field equation in (\ref{iiafieldeqs}) implies that $W=S=0$.  In turn, the warp
factor field equation in (\ref{einstiia})  becomes inconsistent.  The remaining cases are investigated below.

\subsubsection{\texorpdfstring{$S^7=U(4)/U(3)$}{S7=U(4)/U(3)}} \label{u4u3x}

The geometry of $S^7=U(4)/U(3)$ has been summarised in the beginning of section \ref{s7u4u3}. The metric is given in (\ref{metru4u3}).  The invariant forms
are generated by the 1-form $\ell^7$ and 2-form $\omega$ as in (\ref{omegau4u3}), $d\ell^7=\omega$.
Given these data, the most general invariant  fluxes can be chosen as
\bea
X={\alpha\over2}\, \omega^2~,~~~Z=\beta\, \ell^7\wedge \omega~,~~~F=\gamma\, \omega~,~~~Y=\delta\, \ell^7~.
\eea
As the Bianchi identities require that $dZ=0$, we have $\beta=0$.  Furthermore, the remaining Bianchi identities imply
\bea
SW=0~,~~~\delta=-W\gamma~,
\eea
and the field equations for the fluxes give
\bea
{\alpha\,\gamma \over b}+{1\over2} \gamma\, S\, b=0~,~~~\gamma\, \sqrt{a}=-{1\over3}W{b^2\delta\over \sqrt{a}}~,~~~{\alpha\sqrt{a}\over b}=-{1\over2} W\,\alpha~.
\eea

Suppose first that $S\not=0$. Then $W=0$ which in turn gives $\alpha=\gamma=\delta=0$.  As both $Z=Y=0$, the dilaton field equation in (\ref{iiafieldeqs}) implies that the rest of the fluxes vanish which in turn
leads to a contradiction, since the warp factor field equation in (\ref{einstiia})  cannot be satisfied.

Next suppose that $S=0$. Then $\alpha\,\gamma=0$.  Take that $W\not=0$ otherwise there will be a contradiction as described for $S\not=0$ above.  If $\gamma=0$, this will imply that $\delta=0$
and so again the dilaton field equation in (\ref{iiafieldeqs}) will imply that the rest of the fluxes must vanish.

It remains to investigate the case  $\alpha=0$.  The dilatino KSE (\ref{iiadilat}) and algebraic KSE (\ref{iiaalgkse}) become
\bea
&&(-{1\over2} W \Gamma_{11}+{3\over8} \slashed{F} \Gamma_z \Gamma_{11}-{1\over4} \slashed{Y})\sigma_+=0~,
\cr
&&({1\over2} W \Gamma_{11}-{1\over8} \slashed{F} \Gamma_z \Gamma_{11}-{1\over4} \slashed{Y})\sigma_+={1\over \ell A} \sigma_+~.
\label{wfyalg}
\eea
Eliminating the flux $F$, one finds
\bea
(W \Gamma_{11}-\slashed{Y})\sigma_+={3\over\ell A}\sigma_+~.
\label{wyalg}
\eea
The integrability condition gives
\bea
W^2+ Y^2={9\over\ell^2 A^2}~.
\label{wfyint}
\eea
Comparing this with the field equation for the warp factor (\ref{einstiia}) leads to a contradiction.  There are no supersymmetric solutions.

\subsubsection{\texorpdfstring{$S^7=(Sp(2)\times Sp(1))/Sp(1)\times Sp(1)$}{S7 = Sp(2)xSp(1)/Sp(1)xSp(1)}}

The geometry of $S^7=(Sp(2)\times Sp(1))/Sp(1)\times Sp(1)$  has been described in section \ref{sp2sp1sp1sp1s1}. As we have pointed out, there are no invariant 1- and 2-forms, and no invariant closed 3-forms on this homogeneous space.  As a result $Y=F=Z=0$.  Then, the dilaton field equation in (\ref{iiafieldeqs}) implies that
$W=S=X=0$ and therefore the warp factor field equation  in (\ref{einstiia}) cannot be satisfied.  There are no AdS$_3$ solutions with internal space $S^7=(Sp(2)\times Sp(1))/Sp(1)\times Sp(1)$.

\subsubsection{\texorpdfstring{$M^7=S^4\times S^3=\mathrm{Spin}(5)/\mathrm{Spin}(4)\times SU(2)$}{S4xS3}} \label{o5o4su2}

The metric on the internal space can be chosen as
\bea
ds^2(M^7)=  ds^2(S^4)+ ds^2(S^3)= a\, \delta_{ij} \bbl^i\bbl^j+ b_{rs} \bbl^r \bbl^s~,~~~
\eea
where $(\bbl^i)$, $i=4,\dots, 7$,  is a left-invariant frame on $S^4$ and $(\bbl^r)$, $r=1,2,3$ is a left-invariant frame on $S^3=SU(2)$,
$a>0$ is a constant and $(b_{rs})$ a positive definite $3\times 3$ symmetric matrix. Note that
\bea
d\bbl^r={1\over2} \epsilon^r{}_{st} \bbl^s\wedge \bbl^t~.
\eea
Before we proceed, observe that without loss of generality $b=(b_{rs})$ can be chosen  to be
diagonal.  This is because  any transformation $\bbl^r\rightarrow O^r{}_s \bbl^s$ of the left-invariant frame with $O\in SO(3)$ leaves the structure constants of $\mathfrak{su}(2)$
invariant and acts on $b$ as $O^\text{T}\, b \,O$.  So, there is a choice of frame such that $b=\mathrm{diag}(b_1, b_2, b_3)$ with $b_1,b_2,b_3>0$ constants.  From here on, we shall take
$b$ to be diagonal.

The most general invariant fluxes are
\bea
X=\alpha\, \bbl^{4567}~,~~~Z=\beta\, \bbl^{123}~,~~~F={1\over2}\gamma_r\, \epsilon^r{}_{st} \bbl^s\wedge \bbl^t~,~~~Y=\delta_r \bbl^r~,
\eea
where $\alpha$, $\beta$, $\gamma_r$ and $\delta_r$, $r=1,2,3$, are constants.

First observe that $Z\wedge X=0$ implies that $\alpha \beta=0$.   Next suppose that $S\not=0$.  It follows from the Bianchi identities (\ref{iiab}) that  $Z=0$,  as $dF=0$.  In addition,    the Bianchi identities (\ref{iiab}) give
\bea
W=Y=0~.
\eea
Next, the dilaton field equation in (\ref{iiafieldeqs}) implies that $S=X=F=0$ which is a contradiction to the assumption that $S\not=0$.

So, let us now consider $S=0$.  Again $\alpha \beta=0$ and so either $Z=0$ or $X=0$.  Let us first take $Z=0$ and $X\not=0$.  In such a case the field equation for $X$ in (\ref{iiaf})
gives $W=0$.  If $W=0$, the Bianchi identities (\ref{iiab}) will imply that $Y=0$.  This in turn leads to a contradiction as the field equation for the dilaton in (\ref{iiafieldeqs}) implies that
$X=0$.

Suppose now that both $Z=X=0$. As $S=0$ as well,  the dilatino and algebraic KSEs can be re-written as in (\ref{wfyalg}).  This in turn gives (\ref{wyalg}) which leads to
the integrability condition (\ref{wfyint}).
Substituting this into the field equation for the warp factor in (\ref{einstiia})  and after  eliminating $Y^2$, one finds a contradiction.

It remains to investigate the case that $Z\not=0$ and $X=0$.  First the Bianchi identity for $Y$  (\ref{iiab}) implies that
\bea
\delta_r=-W \gamma_r~.
\label{dwg}
\eea
Then field equation for $F$, $d*_{{}_7} F=-W *_{{}_7} Y$, together with (\ref{dwg}) imply that
\bea
W^2 \,\gamma_r={b_r^2\over b_1 b_2 b_3}\, \gamma_r~,~~~\text{no summation over $r$}~.
\label{w2b3}
\eea
Next turn to the Einstein equation along  $S^4$.  As $X=0$ and the fields $Z$, $F$ and $Y$ have non-vanishing components only along $S^3$, we find that
\bea
R^{(7)}_{ij}= ({1\over 4} Y^2-{1\over 8} F^2) \delta_{ij}~.
\eea
Using (\ref{w2b3}), one can show that $R^{(7)}_{ij}=0$.  This   is a contradiction as $R^{(7)}_{ij}$ is the Ricci tensor of the  $S^4$ subspace which is required to be  strictly positive.  Therefore, we conclude
that there are no supersymmetric IIA AdS$_3$ solutions with internal space $S^4\times S^3$.

\subsubsection{\texorpdfstring{$M^7=S^7=\mathrm{Spin}(7)/G_2$}{S7 = Spin(7)/G2}}
This homogeneous space admits invariant 3- and 4-forms which are the fundamental  $G_2$ forms  $\varphi$ and $*{{}_7}\varphi$.  However the 3-form $\varphi$ is not closed and so $Z=0$.
As there are no invariant 1-forms and 2-forms $Y=F=0$.  In such a case the dilaton field equation in (\ref{iiafieldeqs})  implies that $W=S=X=0$.  In turn, the warp factor field equation in
(\ref{einstiia}) becomes inconsistent.

\subsection{\texorpdfstring{$N>16$}{N greater 16} solutions with \texorpdfstring{$N_R=4$}{NR = 4}}

The Lie algebras that must act both effectively and transitively on the internal space $M^7$ are the same as those found in $D=11$ supergravity and  given in (\ref{nr4}).  With regard to the 7-dimensional homogeneous
spaces, those admitting an effective and transitive action by these Lie algebras are the following
\begin{align}
&S^6\times S^1= \mathrm{Spin}(7)/\mathrm{Spin}(6)\times S^1~,~~(N=18)~;\notag\\
 ~&G_2/SU(3)\times S^1~,~~(N=18)~.
\label{cnr4}
\end{align}
Both are products of 6-dimensional homogeneous spaces with $S^1$.

\subsubsection{\texorpdfstring{$M^7=\mathrm{Spin}(7)/\mathrm{Spin}(6)\times S^1$}{Spin(7)/Spin(6)xS1}}

The only non-vanishing k-form flux, $k>0$, allowed is $Y=\alpha \bbl^7$, where $\bbl^7$ is a left-invariant frame along $S^1$ and $\alpha$ a constant.  The dilatino
KSE (\ref{iiadilat}) can be re-written as
\bea
\Big(-{1\over2} W\Gamma_{11}+{5\over4} S\Gamma_z-{1\over4} \slashed{Y}\Big)\sigma_+=0~,
\eea
which leads to the integrability condition
\bea
W^2+{25\over 4}S^2+{1\over 4} Y^2=0~.
\eea
As a result $W=S=Y=0$.  This leads to an inconsistency, since the warp field equation (\ref{einstiia}) cannot be satisfied.
There are no supersymmetric IIA AdS$_3$ backgrounds with $\mathrm{Spin}(7)/\mathrm{Spin}(6)\times S^1$ internal space.

\subsubsection{\texorpdfstring{$M^7=S^6\times S^1=G_2/SU(3)\times S^1$}{G2/SU(3)xS1}} \label{g2su3s1}

The differential algebra of a  left-invariant frame on $M^7$ modulo terms in $\mathfrak{su}(3)\wedge \mathfrak{m}$ which involve the canonical connection is
\bea
d\lambda^{\bar r}={1\over2} \epsilon^{\bar r}{}_{st} \lambda^r\wedge \lambda^t~,~~~d\bbl^7=0~,~~~r=1,2,3~,
\eea
where $\lambda^r$ is a complex frame, $\bar{\lambda}^r= \lambda^{\bar r}$, on $S^6$ and $\bbl^7$ is a left-invariant frame on $S^1$.  The invariant forms on $S^6$ are the 2-form
\bea
\omega={i\over 2} \delta_{r\bar s} \lambda^r\wedge \lambda^{\bar s}~,
\eea
and the holomorphic 3-form
\bea
\chi={1\over 6} \epsilon_{rst} \lambda^r\wedge \lambda^s\wedge \lambda^t~.
\eea
Clearly,
\bea
d\omega=3\,\mathrm{Im} \chi~,~~~d\,\mathrm{Re}\chi=2\omega\wedge \omega~.
\eea
The most general invariant metric  on $M^7$ is
\bea
ds^2(M^7)=a\, \delta_{r\bar s} \lambda^r \lambda^{\bar s}+ b\, (\bbl^7)^2~,
\eea
where $a,b>0$ are constants. Moreover, the most general invariant fluxes are
\begin{align}
X&= {1\over 2}\alpha_1\,\omega^2+\alpha_2 \bbl^7\wedge \mathrm{Re}\,\chi+\alpha_3 \bbl^7\wedge \mathrm{Im}\,\chi~,~~~Z=\beta\, \mathrm{Im} \chi~,\notag\\
F&=\gamma\, \omega~,~~~Y=\delta\, \bbl^7~,
\end{align}
where the $\alpha$'s, $\beta, \gamma$ and $\delta$ are constants and we have used that $dZ=0$.  As $Z\wedge X=0$, we have that $\alpha_2 \beta=0$.
Furthermore, $dF=SZ$ yields
\bea
3\,\gamma=S\beta~.
\eea

Let us first consider the case that $S\not=0$.  This implies that $W=0$.  As either $\alpha_2=0$ or $Z=0$, let us first investigate $Z=0$.  In such a case  the Bianchi identities
(\ref{iiab}) and the field equations (\ref{iiaf}) imply that $X$ is harmonic and since $H^4(S^6\times S^1)=0$, we have $X=0$. Using this, we also find  that  $F$ is harmonic and so because $H^2(S^6\times S^1)=0$, $F=0$.  Next the dilatino KSE (\ref{iiadilat}) becomes $(5S+\slashed{Y}\Gamma_z)\sigma_+=0$, which in turn implies that $25S^2+Y^2=0$.  This is a contradiction, as that would mean $S=0$.
Thus, there are no such supersymmetric AdS$_3$ backgrounds.

Next suppose that $\alpha_2=0$. The field equation $d*_{{}_7}X=Z\wedge Y$ gives
\bea
\alpha_3=0~,
\eea
and so $X$ does not have a component along $S^1$.  Then the Einstein equation along $S^1$ gives
\bea
R^{(7)}_{77}=-{1\over4} Y^2-{1\over 4} S^2-{1\over 8} F^2-{1\over 96} X^2=0~.
\eea
Thus, again $S=0$ which is a contradiction.

So to find solutions, we have to set $S=0$.  The Bianchi identity $dF=0$ gives $F=0$.   Furthermore, the field equation $d*_{{}_7}Z=0$ implies $Z=0$.
We also have from the field equations (\ref{iiaf}) that $W *_{{}_7}Y=0$.  If we choose $Y=0$ and since $Z$  vanishes as well, $Z=0$ ,  the dilaton field equation in (\ref{iiafieldeqs}) implies that the rest
of the fields vanish which contradicts the warp factor field equation. So, let us take $W=0$.  In that case $X$ is harmonic and hence $X=0$. This is also the case for $Z$ and so $Z=0$. The dilatino KSE in turn
implies that $\slashed{Y}\sigma_+=0$, which gives $Y=0$.  Thus, all the fields vanish, leading to a contradiction with the warp factor field equation.  There are no  supersymmetric AdS$_3$ solutions
with internal space $G_2/SU(3)\times S^1$.

\subsection{\texorpdfstring{$N>16$}{N greater 16} solutions with \texorpdfstring{$N_R=6$}{NR=6}}
The 7-dimensional homogeneous spaces that admit an effective and transitive action of one of the Lie algebras in (\ref{nr6}) are
\begin{align}
&S^5\times S^2=\mathrm{Spin}(6)/\mathrm{Spin}(5)\times SU(2)/U(1)~,~~(N=18)~;\notag\\
&S^5\times S^2=U(3)/U(2)\times SU(2)/U(1)~,~~(N=18)~;\notag\\
&S^4\times S^3=\mathrm{Spin}(5)/\mathrm{Spin}(4)\times \big(SU(2)\times SU(2)\big)/SU(2)~,~~(N=22)~.
\label{cnr6}
\end{align}
The $\mathrm{Spin}(6)/\mathrm{Spin}(5)\times SU(2)/U(1)$ case can be easily ruled out, as $Y=Z=0$.  Then, the dilaton field equation implies that $X=W=S=F=0$,
which in turn leads to a contradiction, since the warp factor field equation cannot be satisfied. Moreover, the $\mathrm{Spin}(5)/\mathrm{Spin}(4)\times \big(SU(2)\times SU(2)\big)/SU(2)$
internal space has been investigated already, as it is a special case of $\mathrm{Spin}(5)/\mathrm{Spin}(4)\times SU(2)$.

\subsubsection{\texorpdfstring{$S^5\times S^2=U(3)/U(2)\times SU(2)/U(1)$}{U(3)/U(2)xSU(2)/U(1)}}\label{u3u2su2u1}

The geometry of the homogeneous space  $S^5=U(3)/U(2)$ has been described in section \ref{u3u2su2}.  Using this, the most general invariant metric on $M^7$ can be written as
\bea
ds^2(M^7)= ds^2(S^5)+ds^2(S^2)= a\, (\ell^5)^2+ b\, \delta_{rs} \ell^r \ell^s+ c\,\left( (\ell^6)^2+(\ell^7)^2\right)~,
\label{metru3u2su2u1}
\eea
where $a,b,c>0$ are constants $(\ell^r, \ell^5)$, $r=1,2,3,4$,  is a left-invariant frame on $S^5$ and $(\ell^6, \ell^7)$ is a left-invariant frame
on $S^2$.  The invariant forms on the homogeneous space are generated by
\bea
\bbl^5~,~~~\omega=\ell^{12}+\ell^{34}~,~~~\sigma=\ell^{67}~,
\eea
with
\bea
d\bbl^5=\omega~,~~~d\sigma=0~.
\eea
The most general invariant fluxes are
\bea
&&X={1\over2} \alpha_1\, \omega^2+\alpha_2\, \omega\wedge \sigma~,~~~Z=\beta_1\, \ell^5\wedge \omega+ \beta_2\, \ell^5\wedge \sigma~,
\cr
&& F=\gamma_1\, \omega+\gamma_2 \sigma~,~~~Y=\delta\, \ell^5~.
\eea
The Bianchi identity $dZ=0$ implies that $\beta_1=\beta_2=0$.  So $Z=0$.  The remaining Bianchi identities imply that
\bea
SW=0~,~~~\delta=-W\gamma_1~,~~~W \gamma_2=0~.
\label{xxe}
\eea

To continue first take $S\not=0$.  In such a case $W=0$ and so $\delta=0$.  As both $Y=Z=0$, the dilaton field equation implies that $S=X=F=0$.  This is a contradiction to the assumption that $S\not=0$.

Therefore we have to set $S=0$. Furthermore $W\not=0$ as otherwise $Y=Z=0$ and the dilaton field equation will imply that all other fluxes must vanish.  This in turn leads to a contradiction as the warp factor field equation cannot be satisfied. As $W\not=0$, we have $\gamma_2=0$.  Then,  the field equation for the fluxes (\ref{iiaf})  give $*_{{}_7}X\wedge F=0$, which implies that
\bea
\alpha_2 \gamma_1=0~,~~~\gamma_1\alpha_1=0~.
\eea
Notice that $\gamma_1\not=0$ as otherwise $\delta=0$ and so $Y=Z=0$ leading again to a contradiction. Thus, we find $\alpha_1=\alpha_2=0$ and so $X=0$.

As we have established that $Z=X=0$, we can follow the analysis of the KSEs in section \ref{u4u3x} that leads to the conclusion
that there are no supersymmetric AdS$_3$ solutions with internal space $U(3)/U(2)\times SU(2)/U(1)$.

\subsection{\texorpdfstring{$N>16$}{N greater 16} solutions with \texorpdfstring{$N_R=8$}{NR=8}}
The 7-dimensional homogeneous spaces that admit an effective and transitive action of one of the Lie algebras in (\ref{nr8}) are
\begin{align}
&S^7=(Sp(2)\times Sp(1))/Sp(1)\times Sp(1)~,~~(N=18)~;\notag\\
&S^4\times S^3=\mathrm{Spin}(5)/\mathrm{Spin}(4)\times SU(2)~,~~(N=18)~;\notag\\
&S^4\times S^2\times S^1=\mathrm{Spin}(5)/\mathrm{Spin}(4)\times SU(2)/U(1)\times S^1~,~~(N=18)~;\notag\\
&S^5\times S^2=\mathrm{Spin}(6)/\mathrm{Spin}(5)\times SU(2)/U(1)~,~~(N=20)~;\notag\\
&S^5\times S^2=U(3)/U(2)\times SU(2)/U(1)~,~~(N=20)~;\notag\\
& N^{k,l,m}=(SU(2)\times SU(3)\times U(1))/\Delta_{k,l,m} (U(1)\times U(1))\cdot (1\times SU(2))~,~~(N=20)~;\notag\\
&S^4\times S^3=\mathrm{Spin}(5)/\mathrm{Spin}(4)\times \big(SU(2)\times SU(2)\big)/SU(2)~,~~(N=18)~.
\label{cnr8}
\end{align}
The only new cases that arise and have not already been investigated are those with internal space $S^4\times S^2\times S^1=\mathrm{Spin}(5)/\mathrm{Spin}(4)\times SU(2)/U(1)\times S^1$
and $N^{k,l,m}$.
All the remaining ones do not give supersymmetric solutions with $N>16$ and $N_R=8$.

\subsubsection{\texorpdfstring{$S^4\times S^2\times S^1=\mathrm{Spin}(5)/\mathrm{Spin}(4)\times SU(2)/U(1)\times S^1$}{S4xS2xS1}} \label{so5so4su2u1s1}

The most general invariant  metric on this homogeneous space can be written as
\begin{align}
ds^2(M^7)&= ds^2(S^4)+  ds^2(S^2)+ ds^2(S^1)\notag\\
&=a\, \delta_{rs} \ell^r \ell^s+ b\, \big((\ell^5)^2+ (\ell^6)^2\big)+ c\, (\ell^7)^2~,
\label{metrso5so4su2u1s1}
\end{align}
where $a,b,c>0$ are constants and $\ell^r$, $r=1,2,3,4$, is a left-invariant frame on $S^4$, $(\ell^5, \ell^6)$ is a left-invariant frame on $S^2$ and $\ell^7$
is a left-invariant frame on $S^1$.  The most general invariant form fluxes can be chosen as
\bea
X=\alpha\, \ell^{1234}~,~~~Z= \gamma\, \ell^7\wedge \ell^{56}~,~~~F=\beta\, \ell^{56}~,~~~Y=\delta\, \ell^7~,
\eea
where $\alpha, \beta, \gamma, \delta$ are constants.

From the Bianchi identities (\ref{iiab}) and  the field equation (\ref{zxz}), we find that
\bea
S\gamma=0~,~~~\alpha\gamma=0~,~~~SW=0~,~~~W\beta=0~.
\eea
First, suppose that $S\not=0$.  It follows that $W=Z=0$.  Moreover, from the field equation of $Z$ (\ref{iiaf}) follows that $F=0$.  Next, consider the Einstein field equation
to find that
\bea
R^{(7)}_{77} &= -\frac{1}{4 } Y^2 - \frac{1}{96} X^2 - \frac{1}{4} S^2~.
\eea
However, this is the Ricci tensor of $S^1$ and hence vanishes. This in turn gives $S=0$ which is a contradiction to our assumption that $S\not=0$.

Thus, we have to set $S=0$. The Bianchi identities (\ref{iiab}) and  the field equation (\ref{zxz}) give that
\bea
\alpha \gamma=0~,~~~W\beta=0~,
\eea
and the field equations (\ref{iiaf}) of the form field strengths imply that
\bea
W \delta=0~,~~~W\alpha=0~.
\eea
Therefore, if $W\not=0$, we will have $F=Y=X=0$.  Furthermore as $R^{(7)}_{77}=0$, the Einstein equation reveals that $Z=0$.
 Then the dilaton field equation implies that $W=0$ which is a contradiction to our assumption that $W\not=0$.

It remains to investigate solutions with $W=S=0$.  Notice that we should take $Z\not=0$, or equivalently $\gamma\not=0$, as otherwise the Einstein equation
$R^{(7)}_{77}=0$  will imply that $X=Y=F=0$ and so the warp factor field equation cannot be satisfied leading to a contradiction.
Thus $Z\not=0$ and since $\alpha\gamma=0$, we have that $X=0$.  Inserting $X=S=W=0$ into the dilatino and algebraic KSEs we find that they can be rewritten as
\bea
&&\Big(-{\gamma\over b\sqrt c} J_1 J_2+{3\over2} {\beta\over b} J_1-{1\over2} {\delta\over \sqrt c} J_2\Big)\sigma_+=0~,
\cr
&&\Big({\beta\over b} J_1+{\delta\over \sqrt c} J_2\Big)\sigma_+=-{4\over \ell A} \sigma_+~,
\label{systkse}
\eea
where $J_1=\Gamma^{56} \Gamma_z \Gamma_{11}$ and $J_2=\Gamma^7$.  As each common eigenspace of  $J_1$ and $J_2$ has dimension 4 to find solutions with $N>16$ supersymmetries we have to choose
at least two of these eigenspaces.  One can verify after some calculation that for all possible pairs of eigenspaces the resulting system of equations arising from (\ref{systkse})
does not have solutions.  Therefore, there are no AdS$_3$ solutions that have internal space $\mathrm{Spin}(5)/\mathrm{Spin}(4)\times SU(2)/U(1)\times S^1$ and preserve
$N>16$ supersymmetries.

\subsubsection{\texorpdfstring{$N^{k,l,m}=(SU(2)\times SU(3)\times U(1))/\Delta_{k,l,m} (U(1)\times U(1))\cdot (1\times SU(2))$}{N(k,l,m)}}
As $N^{k,l,m}$ is a modification of $N^{k,l}$, see \cite{Witten:1981me, fre}, we can use the local description of the geometry of the latter in appendix \ref{appencx2} to describe this space.
In particular, the metric can be written as
\begin{align}
ds^2(M^7)= a\, (\ell^7)^2+ b\, (\delta_{rs} \ell^r \ell^s+\delta_{rs} \hat\ell^r \hat\ell^s)+ c\, ((\ell^5)^2+(\ell^6)^2)~,~~~r,s=1,2~,
\label{metrnklm}
\end{align}
where $(\ell^r, \hat\ell^r, \ell^5, \ell^6, \ell^7)$ is a left-invariant frame and $a,b,c>0$ constants. From the results of appendix \ref{appencx2}, one can deduce that there
are no closed 3-forms and so $Z=0$. The remaining invariant form field strengths are
\bea
X={1\over2}\alpha_1\omega_1^2+\alpha_2 \omega_1\wedge \omega_2~,~~~F=\gamma_1\omega_1+\gamma_2 \omega_2~,~~~Y=\delta \ell^7~,
\eea
where $\alpha_1, \alpha_2, \gamma_1, \gamma_2, \delta$ are constants.
The Bianchi identities (\ref{iiab})  imply that
\bea
SW=0~,~~~-{\delta\over 8 l}=\gamma_1 W~,~~~{\delta\over 4k}=\gamma_2 W~.
\label{bbb5}
\eea
Furthermore, the field equation for $Z$ in (\ref{iiaf}) gives
\bea
&&{c\over b^2} \alpha_1 \gamma_1+ {1\over c} \alpha_2 \gamma_2+ S c \gamma_1=0
\cr
&& \alpha_2 \gamma_1+{1\over2} S \gamma_2 b^2=0~.
\label{fff5}
\eea
Clearly, from (\ref{bbb5}) either $S=0$ or $W=0$.  Suppose that $S\not=0$.  Then $W=0$ and from the rest of the conditions arising in the Bianchi identities, $Y=0$.
As both $Y=Z=0$, the dilaton field equation implies that $S=F=X=W=0$ which is  a contradiction to our assumption that $S\not=0$.

Therefore, we set $S=0$.  We also take $W\not=0$ as otherwise the same argument presented above leads to a contradiction again.
As $S=0$, the last condition in (\ref{fff5}) implies that $\alpha_2 \gamma_1=0$.  However, $\gamma_1$ cannot vanish. Indeed, if $\gamma_1=0$, then (\ref{bbb5}) will  lead to $Y=0$. Since $Y=Z=0$, the dilaton field equation in (\ref{iiafieldeqs})  will imply that the rest of the fields vanish.  In turn the warp factor field equation (\ref{einstiia}) cannot be satisfied.  Thus we have to set $\gamma_1\not=0$.  In that case  $\alpha_2=0$ and the first equation in (\ref{fff5}) gives $\alpha_1=0$.  As both $\alpha_1=\alpha_2=0$, $X=0$.

We have shown that the remaining non-vanishing fields are $W$, $Y$ and $F$.  To continue, consider the dilatino and algebraic KSEs. These can be written as in (\ref{wfyalg}).
Then  a similar argument as that presented in section \ref{u4u3x} leads to a contradiction.  There are no supersymmetric AdS$_3$ solutions with internal space $N^{k,l,m}$.

\section{\texorpdfstring{$N>16 ~ AdS_3\times_w M^7$}{N greater 16 AdS3xM7} solutions in IIB}\label{ads3IIB}

\subsection{Field equations and Bianchi identities for \texorpdfstring{$N>16$}{N greater 16}}

The bosonic fields of IIB supergravity are a metric $ds^2$, a complex 1-form field strength $P$, a complex 3-form field strength $G$ and a real self-dual 5-form field strength $F$.
For the investigation of IIB $AdS_3\times_w M^7$ backgrounds that  follows, we shall employ the analysis presented  in \cite{iibads}, where all the necessary  formulae can be found.  Since we are focusing on backgrounds that preserve $N>16$ supersymmetries,
the homogeneity theorem implies that the scalars are constant and so $P=0$.  We shall use this from the beginning to simplify the relevant field equations, Bianchi identities and KSEs.
Imposing the symmetries of the AdS$_3$ subspace on the fields, one finds that the non-vanishing fields are
\begin{align}
ds^2 &= 2 \bbe^+ \bbe^- + (\bbe^z)^2 +ds^2(M^7)~, \quad F= \bbe^+\wedge \bbe^- \wedge \bbe^z\wedge Y - *_{{}_7} Y \notag\\
G &= X\, \bbe^+\wedge\bbe^-\wedge \bbe^z + H~, \
\end{align}
where a null orthonormal frame $(\bbe^+, \bbe^-, \bbe^z, \bbe^i)$, $i=1,\dots, 7$,  is defined as \eqref{nullframe} and  $ds^2(M^7) = \delta_{ij} \bbe^i \bbe^j$. $Y$ is a real 2-form, $X$ is a complex function and $H$ a complex 3-form on $M^7$. The dependence of the fields on AdS$_3$ coordinates is hidden in the definition of the frame $(\bbe^+, \bbe^-, \bbe^z)$.  All the components
of the fields in this frame depend on the coordinates of $M^7$.

The Bianchi identities of the k-form  field strengths can be written as
\begin{align}
dY&= \frac{i}{8} (\overline{X} H- X \overline{H})~,\quad dX= 0 \notag\\
d*_7Y&= -\frac{i}{8} H\wedge\overline{H}~, \quad dH= 0~, \label{iibb}
\end{align}
while their field equations are
\bea\label{iibf}
 \frac{1}{6} H^2 +  X^2=0~,~~~
 d*_{{}_7} H=4i X *_{{}_7}Y+ 4i Y\wedge H~.
\eea
Note that the Bianchi identities imply that $X$ is constant.  We have also used that the warp factor $A$ is constant.  This is proved as in eleven dimensions upon making use
of the compactness of $M^7$ and the homogeneity theorem.

The Einstein equation along AdS$_3$ and $M^7$ becomes
\begin{align}\label{iibeinst}
 &2Y^2 +\frac{3}{8} {X \overline{X}} + \frac{1}{48} {H_{ijk} \overline{H}^{ijk}} = \frac{2}{\ell^2} A^{-2}~, \notag\\
R_{ij}^{(7)}&=  2 Y^2 \delta_{ij} - 8 Y^2_{ij} + \frac{1}{4} H_{(i}{}^{kl} \overline{H}_{j)kl} \notag\\
&\quad + \frac{1}{8} {X \overline{X}} \delta_{ij} - \frac{1}{48} {H_{klm} \overline{H}^{klm}} \delta_{ij}~,
\end{align}
respectively. Here, $\nabla$ denotes the Levi-Civita connection on $M^7$ and $R^{(7)}$ is the Ricci tensor on the transverse space.  The first condition above is
the field equation for the warp factor.

\subsection{The Killing spinor equations}

The solution of the KSEs of IIB supergravity along the $AdS_3$-subspace can be expressed as in \eqref{ks}, only that now $\sigma_\pm$ and $\tau_\pm$ are $\text{Spin}(9,1)$  Weyl spinors which depend only on the coordinates of $M^7$ and satisfy the lightcone projections $\Gamma_\pm \sigma_{\pm}=\Gamma_\pm \tau_\pm =0$. The remaining independent KSEs are the gravitino
\begin{align} \label{iibgravkse}
\nabla^{(\pm)}_i \sigma_\pm = 0~, \quad \nabla^{(\pm)}_i \tau_\pm = 0~,
\end{align}
dilatino
\begin{align}\label{algkseiib2}
\mathcal{A}^{(\pm)} \sigma_\pm = 0~, \quad \mathcal{A}^{(\pm)} \tau_\pm = 0~,
\end{align}
and algebraic
\begin{align} \label{algkseiib}
\Xi^{(\pm)} \sigma_\pm = 0~, \quad \left(\Xi^{(\pm)} \pm \frac{1}{\ell} \right) \tau_\pm = 0~,
\end{align}
KSEs, where
\begin{align}
\nabla^{(\pm)}_i &= \nabla_i \pm \frac{i}{4} \sgY_i \Gamma_z \mp \frac{i}{2} \sY_i \Gamma_z \notag\\
&\quad +\left(-\frac{1}{96} \sgH_i + \frac{3}{32} \sH_i \mp \frac{1}{16} X \Gamma_{zi}\right)C*~,\notag\\
\mathcal{A}^{(\pm)} &= \mp\frac{1}{4} X \Gamma_z + \frac{1}{24} \sH~,\notag\\
\Xi^{(\pm)} &= \mp\frac{1}{2\ell}  \pm \frac{i}{4} A \sY + \left(\frac{1}{96} A \Gamma_z \sH \pm \frac{3}{16} A X\right)C*~,
\end{align}
and $C$ is the charge conjugation matrix followed by complex conjugation. In the expressions above we have used that $P=0$ and that $A$ is constant.
As in the 11-dimensional and IIA supergravities, the IIB AdS$_3$  backgrounds preserve an even number of supersymmetries.

\subsection{\texorpdfstring{$N>16$}{N greater 16} solutions with \texorpdfstring{$N_R=0$ and  $N_R=2$}{Nr=0 and NR=2}}

The existence of solutions that preserve strictly 28 and 30 supersymmetries has already been excluded in \cite{Gran:2009cz}.
As in the IIA case,  IIB $N>16$ supersymmetric AdS$_3$ solutions with $N_R=0$ can also be ruled out because there are no 7-dimensional homogeneous manifolds that admit a transitive and effective
action of the $\mathfrak{t}_0$ subalgebra of the expected symmetry superalgebra of such backgrounds.  So we shall begin with backgrounds with $N_R=2$.  The homogeneous spaces are as those
in IIA and are given in (\ref{cnr2}).

The homogeneous space $S^7=\mathrm{Spin}(8)/\mathrm{Spin}(7)$ can be ruled out immediately.  This symmetric space does not admit invariant 2- and 3-forms.  Therefore $Y=H=0$.
Then, a field equation in (\ref{iibf}) implies that $X=0$ as well and so the warp factor field equation in (\ref{iibeinst}) cannot be satisfied.

Similarly, $S^7=\mathrm{Spin}(7)/G_2$ can also be ruled out, as it does not admit an invariant closed 3-form and so $H=0$. Also,  it does not admit an invariant 2-form either, i.e.\ $Y=0$.  Then because of the field equations in (\ref{iibf}), one deduces that $X=0$ and so the warp factor field equation in (\ref{iibeinst}) becomes inconsistent.

\subsubsection{\texorpdfstring{$S^7=U(4)/U(3)$}{S7 = U(4)/U(3)}}

Following the description of the geometry of the homogeneous space $U(4)/U(3)$ as in section \ref{s7u4u3}, the most general allowed fluxes are
\bea
Y=\alpha\, \omega~,~~~H=\beta\, \ell^7\wedge \omega~.
\eea
The Bianchi identity $dH=0$ requires that $\beta=0$. In turn a field equation in (\ref{iibf}) implies that $X=0$.  Substituting this back into the Bianchi identities
(\ref{iibb}), one finds that $Y$ is harmonic and so it must vanish.  As all fluxes vanish, the warp factor field equation in (\ref{iibeinst}) becomes inconsistent. There are no AdS$_3$ solutions with
internal space $U(4)/U(3)$.

\subsubsection{\texorpdfstring{$S^7=(Sp(2)\times Sp(1))/Sp(1)\times Sp(1)$}{S7 = Sp(2)xSp(1)/Sp(1)xSp(1)}}

The geometry of this homogeneous space described in section \ref{sp2sp1sp1sp1s1} reveals that there are no invariant 2-forms and closed 3-forms.  As a result $Y=H=0$.
The field equations (\ref{iibf}) imply that $X=0$ as well.  Therefore, there are no solutions as the warp factor field equation cannot be satisfied.

\subsubsection{\texorpdfstring{$S^4\times S^3=\mathrm{Spin}(5)/\mathrm{Spin}(4)\times SU(2)$}{S4xS3}} \label{s533}

The geometry of this homogeneous space space has been described in section \ref{o5o4su2}.  The most general fluxes can be chosen as
\bea
Y={1\over2} \alpha_r \epsilon^r{}_{st} \ell^s\wedge \ell^t~,~~~H=\beta\, \ell^{123}~,
\eea
where $r,s,t=1,2,3$.
As $Y$ is both closed and co-closed and $H^2(S^4\times S^3)=0$, we deduce that $Y=0$ and  so a Bianchi identity in (\ref{iibb}) implies that
\bea
\bar X \beta- X \bar \beta=0~.
\eea
This together with  a field equation in (\ref{iibf}) imply $|\beta|^2+|X|^2=0$ and so $X=H=0$.
Then the warp factor field equation in (\ref{iibeinst}) cannot be satisfied. There are no AdS$_3$ solutions with internal space $\mathrm{Spin}(5)/\mathrm{Spin}(4)\times SU(2)$.

\subsection{\texorpdfstring{$N>16$}{N greater 16} solutions with \texorpdfstring{$N_R=4$}{NR=4}}

The homogeneous internal spaces are given in (\ref{cnr4}).  It is straightforward to show that $S^6\times S^1=\mathrm{Spin}(7)\mathrm{Spin}(6)\times S^1$ is not a solution
as $Y=H=X=0$ which contradicts the warp factor field equation.

\subsubsection{\texorpdfstring{$G_2/SU(3)\times S^1$}{G2/SU(3)xS1}}

In the notation of section \ref{g2su3s1} the metric can be chosen as $ds^2=a \delta_{r\bar s} \lambda^r \lambda^{\bar s}+ b (\bbl^7)^2$ and the most general
invariant $Y$ and $H$ forms are
\bea
Y=\alpha\, \omega~,~~~H=\beta_1\, \mathrm{Im}\chi+ \beta_2\, \ell^7\wedge \omega+\beta_3\, \mathrm{Re}\chi~.
\eea
The Bianchi identity $dH=0$ implies that $\beta_2=\beta_3=0$.  Set $\beta_1=\beta$.  It follows from the Bianchi identity for $Y$ in (\ref{iibb}) that
\bea
3\alpha={i\over8} (\bar X \beta- \bar\beta X)~.
\label{axxbb}
\eea
Furthermore the field equation for $H$ in (\ref{iibf}) gives
\bea
\beta=-i X\, a\, \alpha~.
\label{axxbb2}
\eea
Next turn to the dilatino KSE. Setting $\lambda^r=\ell^{2r-1}+i \ell^{2r}$, it can be written as
\bea
{\beta\over a^{{3\over2}}}(J_1+J_2-J_3-J_1J_2J_3) \sigma_+=X \sigma_+
\eea
where $J_1=\Gamma_z \Gamma_{136}$, $J_2=\Gamma_z \Gamma_{235}$ and $J_3=\Gamma_z \Gamma_{246}$.  These are commuting Hermitian Clifford algebra operators with eigenvalues
$\pm 1$.  For all choices of eigenspaces either $X=0$ or $X=\pm 4\beta/a^{{3\over2}}$.  Substituting this into the first field equation in (\ref{iibf}), we find that $\beta=0$.
Therefore, $X=0$ as well. Then (\ref{axxbb}) and (\ref{axxbb2}) imply that $Y=H=0$.  Thus  the warp factor field equation in (\ref{iibeinst})  cannot be satisfied.  Hence, there are no supersymmetric AdS$_3$ solutions with internal space
$G_2/SU(3)\times S^1$.

\subsection{\texorpdfstring{$N>16$}{N greater 16} solutions with \texorpdfstring{$N_R=6$}{NR=6}}

The allowed homogeneous internal spaces are given in (\ref{cnr6}).  We have already investigated the AdS$_3$ backgrounds with internal space $S^4\times S^3=\mathrm{Spin}(5)/\mathrm{Spin}(4)\times (SU(2)\times SU(2))/SU(2)$ as they are a special case of those explored in section \ref{s533} and  we have found that there are no solutions. Next, we shall  examine the remaining two cases.

\subsubsection{\texorpdfstring{$S^5\times S^2=\mathrm{Spin}(6)/\mathrm{Spin}(5)\times SU(2)/U(1)$}{S5xS2}}
The metric can be chosen as
\bea
ds^2(M^7)=  ds^2(S^5)+  ds^2(S^2)=a\, \delta_{rs} \ell^r \ell^s+ b\,\big((\ell^6)^2+(\ell^7)^2\big)~,
\eea
where $a,b>0$ are constants, $\ell^r$, $r=1,\dots, 5$ is a left-invariant frame on $S^5$ and $(\ell^6, \ell^7)$ is a left-invariant frame on $S^2$.
As this symmetric space does not admit  invariant 3-forms, we have  $H=0$.  Then a field equation in (\ref{iibf}) implies that $X=0$.  Setting $Y=\alpha\,\ell^{67}$,
the Einstein equation along $S^2$ gives
\bea
R_{pq}^{(7)}=-4{\alpha^2\over b^2} \delta_{pq}~,~~~p,q=6,7~.
\eea
However the Ricci tensor of $S^2$ is strictly positive.  Thus there are no AdS$_3$ solutions with internal space $\mathrm{Spin}(6)/\mathrm{Spin}(5)\times SU(2)/U(1)$.


\subsubsection{\texorpdfstring{$S^5\times S^2=U(3)/U(2)\times SU(2)/U(1)$}{U(3)/U(2)xS2}}

The geometry of this homogeneous space has already been described in section \ref{u3u2su2u1} and the metric is given in (\ref{metru3u2su2u1}).
The most general fluxes can be chosen as
\bea
Y=\alpha_1\,\omega+ \alpha_2\, \sigma~,~~~H=\beta_1\, \ell^5 \wedge \omega+\beta_2\, \ell^5\wedge \sigma~.
\eea
The Bianchi identity $dH=0$ implies that $\beta_1=\beta_2=0$ and so $H=0$.  Furthermore, a field equation in (\ref{iibf})  yields $X=0$, and $d*_7Y=0$ implies $\alpha_1=0$.

Next, consider the Einstein equation (\ref{iibeinst}) along $S^2$.  A direct calculation reveals that
\bea
R_{pq}^{(7)}=-4{\alpha_2^2\over b^2}\, \delta_{pq}~,~~~p,q=6,7~.
\label{s2ein}
\eea
However, the Ricci tensor of $S^2$ is strictly positive.  There are no AdS$_3$ solutions with internal space $U(3)/U(2)\times SU(2)/U(1)$.

\subsection{\texorpdfstring{$N>16$}{N greater 16} solutions with \texorpdfstring{$N_R=8$}{NR=8}}

The allowed homogeneous internal spaces are given in (\ref{cnr8}).  All these cases have already been investigated apart from those with internal space
$S^4\times S^2 \times S^1=\mathrm{Spin}(5)/\mathrm{Spin}(4)\times SU(2)/U(1)\times S^1$ and $N^{k,l,m}$ which we shall  examine next.

\subsubsection{\texorpdfstring{$S^4\times S^2 \times S^1=\mathrm{Spin}(5)/\mathrm{Spin}(4)\times SU(2)/U(1)\times S^1$}{S4xS2xS1}}

The geometry of this symmetric space has been described in section \ref{so5so4su2u1s1}. The metric can be chosen as
in (\ref{metrso5so4su2u1s1}) and the most general invariant fluxes are
\bea
Y=\alpha\, \ell^{56}~,~~~H=\beta\, \ell^{567}~.
\eea
As $dY=0$, the Bianchi identities (\ref{iibb}) give that
\bea
\bar X \beta- \bar\beta X=0~,
\eea
which, together with a field equations in (\ref{iibf}), imply that $H=X=0$.  The only non-vanishing field is $Y$.  However as in the previous case after evaluating the
Einstein equation along $S^2$, one finds a similar relation to (\ref{s2ein}).  This is a contradiction as the Ricci tensor of $S^2$ is strictly positive, i.e.\ there are
no AdS$_3$ solutions with internal space $\mathrm{Spin}(5)/\mathrm{Spin}(4)\times SU(2)/U(1)\times S^1$.

\subsubsection{\texorpdfstring{$N^{k,l,m}=(SU(2)\times SU(3)\times U(1))/\Delta_{k,l,m} (U(1)\times U(1))\cdot (1\times SU(2))$}{N(k,l,m)}}

The metric can be chosen as in (\ref{metrnklm}).
 From the results of appendix \ref{appencx2}, one can deduce that there
are no closed invariant 3-forms and so $H=0$. The field equations (\ref{iibf}) imply that $X=0$ as well. The most general 2-form $Y$ is
\bea
Y=\alpha_1\, \omega_1+ \alpha_2\, \omega_2~.
\eea
The Bianchi identities imply that $Y$ must be harmonic.  Observe that $dY=0$.  The co-closure condition implies that
\bea
{\alpha_1 \over l\, b^2}-{\alpha_2 \over k\, c^2} =0~,
\label{coco}
\eea
where $d\text{vol}={1\over2}\omega_1^2\wedge \omega_2\wedge \ell^7$.

Next the algebraic KSE (\ref{algkseiib}) can be written as
\bea
\left({\alpha_1\over b} (J_1+J_2)+{\alpha_2\over c} J_3\right)\sigma_+={1\over\ell A} \sigma_+~,
\label{nkij}
\eea
where $J_1=i\Gamma^{12}$, $J_2=i \Gamma^{34}$ and $J_3=i\Gamma^{56}$. The relations amongst the fluxes for each of the eigenspaces can be found in table \ref{tableklx}.  The warp factor field equation in (\ref{iibeinst}) also gives
\bea
4{\alpha_1^2\over b^2}+2 {\alpha_2^2\over c^2}={1\over \ell^2 A^2}~.
\eea
Since the common eigenspaces of $J_1, J_2, J_3$ have dimension 2   to find solutions preserving $N>16$ supersymmetries, one needs to choose at least three such  eigenspaces.

\begin{table}[h]
\begin{center}
\vskip 0.3cm
 \caption{Decomposition of (\ref{nkij})  into eigenspaces}
 \vskip 0.3cm

	\begin{tabular}{|c|c|}
		\hline
		$|J_1,J_2, J_3\rangle$&  relations for the fluxes\\
		\hline
		$|\pm,\mp,+\rangle$& ${\alpha_2\over c} ={1\over \ell A} $ \\
\hline
		$|\pm,\mp,-\rangle$& ${\alpha_2\over c} =-{1\over \ell A} $ \\
		\hline
		$|\pm,\pm,\pm\rangle$& ${2\alpha_1\over b}+{\alpha_2\over c} =\pm{1\over \ell A} $ \\
		\hline
$|\pm,\pm,\mp\rangle$& ${2\alpha_1\over b}-{\alpha_2\over c} =\pm{1\over \ell A} $ \\
		\hline
		\end{tabular}
\vskip 0.2cm
  \label{tableklx}
 \end{center}
\end{table}

The eigenspaces that lead to the relation ${\alpha_2\over c} =\pm{1\over \ell A} $  for the fluxes can be ruled out because of the warp factor field equation.  Therefore, we have to choose
three eigenspaces from the remaining cases in table \ref{tableklx}.  For every choice of a pair of relations either $\alpha_1$ or $\alpha_2$ vanishes. Then, the co-closure condition (\ref{coco}) implies that
$Y=0$. There are no AdS$_3$ supersymmetric solutions preserving $N>16$ supersymmetries.

\chapter{A Uniqueness Theorem for warped \texorpdfstring{$N>16$}{N greater 16} Minkowski Backgrounds with Fluxes}\label{minkngr16}

In this chapter, we demonstrate that warped Minkowski space backgrounds, $\bR^{n-1,1}\times_w M^{d-n}$, $n\geq3$, preserving strictly more than 16 supersymmetries in 11-dimensional and type II supergravities, and with fields, which may not be smooth everywhere, are locally isometric to the $\bR^{d-1,1}$ Minkowski vacuum. In particular, all such flux compactification  vacua of  these theories  have the same local geometry as the  maximally supersymmetric vacuum $\bR^{n-1,1}\times T^{d-n}$. The following work was conducted in collaboration with George Papadopoulos and published in \cite{Lautz:2018qcb}.

\section{Proof of the main statement}

Recently,  all  warped anti-de-Sitter  (AdS) backgrounds with fluxes that preserve $N>16$ supersymmetries in $d=11$ and $d=10$ supergravities  have been classified up to a local isometry
in \cite{ads5clas, ads4Ngr16, ahslgp2}.  In this chapter, we extend this result to include all warped $\bR^{n-1,1}\times_w M^{d-n}$   backgrounds of these theories.  In particular, we demonstrate that all warped $\bR^{n-1,1}\times_w M^{d-n}$, $n\geq3$, solutions with fluxes of $d=11$,  IIA $d=10$  and IIB $d=10$  supergravities that preserve $N>16$ supersymmetries are locally isometric to the $\bR^{d-1,1}$ maximally supersymmetric vacuum of these theories. Massive IIA supergravity does not admit such solutions.   A consequence of this
is that all  $N>16$ flux compactification vacua, $\bR^{n-1,1}\times_w M^{d-n}$,  of these theories   are locally isometric to the maximally supersymmetric toroidal vacuum $\bR^{n-1,1}\times T^{d-n}$. To prove these results we have made an  assumption that  the translation isometries along the  $\bR^{n-1,1}$ subspace of these backgrounds commute with all the odd generators of their Killing superalgebra.  The necessity and justification of this assumption will be made clear below.

To begin, we shall first describe the steps of the proof common   to  all $d=11$ and $d=10$ theories and then at the end specialise to present the theory-specific features  in each case.  Schematically, the fields of $\bR^{n-1,1}\times_w M^{d-n}$ backgrounds are
\bea
ds^2&=&A^2 ds^2(\bR^{n-1,1})+ds^2 (M^{d-n})~,
\cr
F&=& W\wedge \mathrm{dvol}_A(\bR^{n-1,1})+ Z~,
 \eea
 where $A$ is the warp factor that depends only on the coordinates of the internal space $M^{d-n}$, $\mathrm{dvol}_A(\bR^{n-1,1})$ denotes the volume form of $\bR^{n-1,1}$ evaluated in the warped metric and $F$ denotes collectively all the k-form fluxes of the
 supergravity theories. We take $ds^2(\bR^{n-1,1})=2dudv+dz^2+\delta_{ab} dx^a dx^b$, where we have singled out a spatial coordinate $z$ which will be useful later. $W$ and $Z$ are $(k-n)$- and $k$-forms on $M^{d-n}$ which depend only on the coordinates of $M^{d-n}$.    Clearly if $n>k$,  $W=0$. Therefore, these backgrounds are invariant under the Poincar\'e isometries of the $\bR^{n-1,1}$ subspace. It is known that there are no  smooth compactifications with non-trivial fluxes  of $d=10$ and $d=11$ supergravities \cite{Gibbons:1984kp, nunez}, i.e.\ solutions for which all fields are smooth including the warp factor and $M^{d-n}$ is compact without boundary.  However, here we do not make these assumptions. $M^{d-k}$ is allowed to be non-compact and the fields may not be smooth.

 To continue following  the description of $\bR^{n-1,1}\times_w M^{d-n}$ backgrounds in \cite{mads,iibads, iiaads}, where one can also find more details about our notation, we  introduce a light-cone orthonormal frame
 \bea
 \bbe^+= du~,~~~ \bbe^-= (dr-2 r A^{-1} dA)~,~~~ \bbe^m=A\, dx^m~,~~~ \bbe^i=e^i_\tI dy^\tI~,
 \label{orthonorm}
  \eea
 on the spacetime with $ds^2(M^{d-n})=\delta_{ij} \bbe^i \bbe^j$.   Then the Killing spinors of the $\bR^{n-1,1}\times_w M^{d-n}$ backgrounds can be written  as
\begin{eqnarray}
\epsilon&=&\sigma_++ u \Gamma_+\Gamma_z\Xi^{(-)}\sigma_-+A\sum_m x^m\Gamma_m  \Gamma_z\Xi^{(+)}\sigma_+
\cr
&&~~+\sigma_-+r \Gamma_-\Gamma_z\Xi^{(+)} \sigma_++ A\sum_m x^m\Gamma_m  \Gamma_z\Xi^{(-)}\sigma_-~,
\label{minksp1}
\end{eqnarray}
where $x^m=(z, x^a)$,  all the gamma matrices are in the frame basis (\ref{orthonorm}) and the spinors $\sigma_\pm$, $\Gamma_\pm\sigma_\pm=0$, only depend on the coordinates of $M^{d-n}$.  The remaining independent  KSEs are
a restriction of the gravitino and algebraic KSEs of the supergravity theories on $\sigma_\pm$ which schematically can be written as
\bea
D^{(\pm)}_i\sigma_\pm=0~,~~~{\cal A}^{(\pm)}\sigma_\pm=0~,
\label{rkse}
\eea
respectively,
and, in addition, an integrability condition
\bea
(\Xi^{(\pm)})^2\sigma_\pm=0~,
\label{minkremkses}
\eea
where $\Xi^{(\pm)}$ is a Clifford algebra element that depends on the fields. $\Xi^{(\pm)}$ will not be given here and can be found in the references above. The latter arises as a consequence of integrating the gravitino KSE of the theories along the $\bR^{n-1,1}$ subspace.

Notice that the (spacetime) Killing spinors (\ref{minksp1}) may depend on the coordinates of the $\bR^{n-1,1}$ subspace. Such a dependence  arises, whenever $\sigma_\pm$ is not in the kernel of $\Xi^{(\pm)}$. Of course $\sigma_\pm$ is required to lie in the kernel of
of $(\Xi^{(\pm)})^2$.  To see why this dependence can arise for $\bR^{n-1,1}\times_w M^{d-n}$ backgrounds, notice that $AdS_{n+1}$ in Poincar\'e coordinates
can be written as a warped product of $\bR^{n-1,1}\times_w \bR$.  Therefore all  AdS backgrounds, warped or otherwise, can be interpreted as warped
Minkowski space backgrounds.  It is also known that the former admit Killing spinors that depend on all AdS coordinates including those of
the Minkowski subspace. Therefore $\bR^{n-1,1}\times_w M^{d-n}$ may also admit  Killing spinors that depend on the coordinates of $\bR^{n-1,1}$, see also \cite{desads} for a more detailed explanation.

The assumption we have made that the commutator  of the translations $P$ along $\bR^{n-1,1}$ and the odd generators $Q$ of the
Killing superalgebra \cite{Gauntlett:1998kc, Figueroa-OFarrill:1999klq} must vanish, $[P,Q]=0$, is required for  Killing spinors $\epsilon$  not to exhibit a dependence on the coordinates of $\bR^{n-1,1}$. Indeed, if the Killing spinors have a dependence on the Minkowski subspace coordinates, then the commutator $[P,Q]$ of the Killing superalgebra  will not vanish. This can be verified with an explicit computation of the spinorial Lie
derivative of $\epsilon$ in (\ref{minksp1})  along the translations of $\bR^{n-1,1}$.   Although this may seem to be a technical assumption, it also has a physical significance in the context of flux compactifications.  Typically, the reduced theory is invariant under the Killing superalgebra of the  compactification vacuum. So, for the reduced theory to exhibit at most super-Poincar\'e invariance,  one must set $[P,Q]=0$ for all $P$ and $Q$ generators.
 This physical justification applies only to compactification vacua, but we shall take it to be valid for all backgrounds that we are investigating below. Of course such an assumption excludes all AdS solutions of supergravity theories re-interpreted as  warped Minkowski backgrounds.  Therefore from now on we shall take $\Xi^{(\pm)}\sigma_\pm=0$ and so {\it all} Killing spinors $\epsilon$ will not depend on
the coordinates of the $\bR^{n-1,1}$ subspace.

Before we proceed further, let us describe the Killing spinors of $\bR^{n-1,1}\times_w M^{d-n}$  backgrounds in more detail.  It turns out that if $\sigma_+$ is a Killing spinor, then $\sigma_-\defeq A\Gamma_{-z} \sigma_+$ is also a Killing spinor.  Similarly, if $\sigma_-$ is a Killing spinor,  then $\sigma_+ \defeq A^{-1} \Gamma_{+z} \sigma_-$ is  also a Killing spinor.  Furthermore, if $\sigma_+$ is a Killing spinor, then  $\sigma'_+\defeq\Gamma_{mn} \sigma_+$ are also Killing spinors for every $m,n$. Therefore, the Killing spinors form multiplets under these Clifford algebra operations.  The counting of Killing spinors of a background  proceeds by identifying the linearly  independent Killing spinors in each multiplet and then counting the number of different multiplets that can occur \cite{mads,iibads, iiaads}. As all Killing spinors are generated from $\sigma_+$ Killing spinors, we shall express all key formulae in terms of the latter.

The 1-form bilinears of the Killing spinors $\epsilon^r$, $r=1,\dots, N$, that are associated with spacetime Killing vectors, and also leave all other fields invariant are
\bea
X(\epsilon^r, \epsilon^s)\defeq \langle (\Gamma_+-\Gamma_-) \epsilon^r, \Gamma_\tA \epsilon^s\rangle_s\, \bbe^\tA~,
\eea
where in $d=11$ and IIA supergravity theories  $\langle(\Gamma_+-\Gamma_-)\cdot, \cdot\rangle_s$ is the Dirac inner product restricted on the Majorana representation of $Spin(10,1)$ and $Spin(9,1)$, respectively, while in IIB it is the real part of the Dirac inner product.
Note that $X(\epsilon^r, \epsilon^s)=X(\epsilon^s, \epsilon^r)$.
In particular, one finds that
\bea
X(\sigma_-^r, \sigma_-^s)&=& 2 A^2\langle \sigma_+^r, \Gamma_z\Gamma_\tA \Gamma_-\Gamma_z\sigma_+^s\rangle_s \bbe^\tA~,
\cr
X(\sigma^r_-, \sigma_+^s) &=&2 A \langle \sigma_+^r, \Gamma_z\Gamma_\tA \sigma_+^s\rangle_s \bbe^\tA~,
\cr
X(\sigma_+^r, \sigma_+^s)&=& -\langle \sigma_+^r, \Gamma_+\Gamma_\tA \sigma_+^s\rangle_s \bbe^\tA~.
\label{bilin}
\eea
Clearly, the last 1-form bilinear  is $X(\sigma_+^r, \sigma_+^s)= -2 \langle \sigma_+^r, \sigma_+^s\rangle_s \bbe^-$.  The requirement that $X(\sigma_+^r, \sigma_+^s)$  is Killing,  implies that $\langle \sigma_+^r, \sigma_+^s\rangle_s$ are constants. In particular, one can choose, without loss of generality, that $\langle \sigma_+^r, \sigma_+^s\rangle_s=(1/2) \delta^{rs}$.  Then, the first 1-form bilinear in (\ref{bilin}) is $X(\sigma_-^r, \sigma_-^s)=2 A^2 \delta^{rs} \bbe^+$.

Next, consider the middle 1-form bilinear in (\ref{bilin}). If $\sigma_+^r$ is in the same multiplet as $\sigma_+^s$, i.e. $\sigma_+^s=\Gamma_{za} \sigma_+^r$, then $X(\sigma^r_-, \sigma_+^s)=-\delta^{rs} A\, \bbe^a$. On the other hand, if   $\sigma_+^s=\sigma_+^r$, then
$X(\sigma^r_-, \sigma_+^r)=A\, \bbe^z$.  Thus all these bilinears generate the translations in $\bR^{n-1,1}$.  However, if $\sigma_+^r$ and $\sigma_+^s$ are not in the same multiplet, then the bilinear
\bea
\tilde X_{rs}\defeq X(\sigma^r_-, \sigma_+^s) =2 A \langle \sigma_+^r, \Gamma_z\Gamma_i \sigma_+^s\rangle_s\, \bbe^i~,
\label{inkill}
\eea
will generate the isometries of the internal space.  The Killing condition of $\tilde X$ implies that
\bea
\tilde X_{rs}^i\partial_i A=0~.
\label{txia}
\eea
As the $\tilde X$ isometries commute with the translations,  the even part, $\mathfrak{g}_0$, of the Killing superalgebra decomposes as $\mathfrak{g}_0=\mathfrak{p}_0\oplus \mathfrak{t}_0$, where $\mathfrak{p}_0$ is the Lie algebra of translations in $\bR^{n-1,1}$ and $\mathfrak{t}_0$ is the Lie algebra of isometries in the internal space $M^{d-n}$.

So far we have not used the assumption that the backgrounds preserve $N>16$ supersymmetries. If this is the case, the Killing vectors generated by $\mathfrak{g}_0$ span the tangent space of the spacetime at each point.  This is a consequence of the homogeneity theorem proven for $d=11$ and $d=10$ supergravity backgrounds in \cite{homogen, figueroab}.  This states that all solutions of these theories that preserve more than 16 supersymmetries must be locally homogeneous. In this particular case, because of the decomposition of $\mathfrak{g}_0$, the Killing vector fields generated by $\mathfrak{t}_0$ span the tangent space of $M^{d-n}$ at every point. As a result,   the condition (\ref{txia}) implies that $A$ is constant.    The main result of this chapter then follows as a consequence of the field equation of the warp factor and those of the rest of the scalar fields of these theories.  So to complete the proof we shall state the relevant equations on a case by case basis.

In 11-dimensional supergravity, the 4-form field strength of the theory for   $\bR^{n-1,1}\times_w M^{11-n}$, $n\geq 3$, backgrounds can be expressed as
\bea
F= d\mathrm{vol}_A(\bR^{n-1,1})\wedge W^{4-n}+ Z~,
\eea
with $W^{4-n}=0$ for $n>4$, and the warp factor field equation is
\bea
\tilde \nabla^2 \log A=-n (\partial\log A)^2+{1\over 3\cdot (4-n)!} (W^{4-n})^2+{1\over 144} Z^2~,
\eea
where $\tilde \nabla$ is the Levi-Civita connection on $M^{11-n}$. The superscripts on the forms denote their degree whenever it is required for clarity. Clearly if $A$ is constant, as it has been demonstrated above for $N>16$ backgrounds, then $W^{n-4}=Z=0$.  So $F=0$ and thus all the fluxes vanish.  In fact, this is also the case for $n=2$ provided that $A$ is taken to be constant.
As $F=0$ and $A$ is constant, the gravitino KSE  in (\ref{rkse}) implies that all the Killing spinors $\sigma_\pm$ are  parallel with respect to the Levi-Civita connection, $\tilde \nabla$, on $M^{11-n}$. This in turn means that all the Killing vector fields $\tilde X$ in (\ref{inkill}),  which span the tangent space of $M^{11-n}$,  are also parallel with respect to $\tilde \nabla$.  Thus  $M^{11-n}$  is locally isometric to  $\bR^{11-n}$.  Therefore, the backgrounds
$\bR^{n-1,1}\times_w M^{11-n}$ are locally isometric to the maximally supersymmetric vacuum $\bR^{10,1}$. Notice that the last step of the proof requires the use of
the homogeneity theorem.

In (massive) IIA supergravity, the   4-form $F$, 3-form $H$ and  2-form $G$ field strengths of the theory for  $\bR^{n-1,1}\times_w M^{10-n}$, $n\geq 3$, backgrounds  can be written as
\bea
F&=&\mathrm{dvol}_A(\bR^{n-1,1})\wedge W^{4-n}+ Z~,
\cr
H&=&\mathrm{dvol}_A(\bR^{n-1,1})\wedge P^{3-n}+ Q~,
\cr
G&=& L~,
\eea
where $W^{4-n}$ vanishes for $n>4$ and similarly $P^{3-n}$ vanishes for $n>3$. The
 field equations for  the warp factor $A$ and dilaton field $\Phi$, $n>2$, are
\begin{align}
\tilde\nabla^2\log\, A&=- n (\partial\log\, A)^2+2 \partial_i\log\, A \partial^i\Phi+{1\over 2} (P^{3-n})^2+{1\over 4} S^2+{1\over 8} L^2\notag\\
&~~~+{1\over 96} Z^2+{1\over 4} (W^{4-n})^2~,\notag\\
\tilde\nabla^2\Phi &=-n \partial_i\log A \partial^i\Phi+ 2 (d\Phi)^2-{1\over12} Q^2+{1\over2} (P^{3-n})^2+{5\over 4} S^2+{3\over 8} L^2 \notag\\
 & ~~~ +{1\over 96} Z^2-{1\over 4} (W^{4-n})^2~,
\end{align}
where $S=e^\Phi m$ and $m$ is the cosmological constant of (massive) IIA supergravity.
Clearly, if both $A$ and $\Phi$ are constant, which is the case for all $N>16$ backgrounds,  then the above two field equations imply that all the form fluxes will vanish.  Significantly, the cosmological constant must vanish as well. There are no $\bR^{n-1,1}\times_w M^{10-n}$ solutions in massive IIA supergravity that preserve $N>16$ supersymmetries.  In IIA supergravity, an argument similar to the one presented above in $d=11$ supergravity reveals
that $M^{10-n}$ is locally isometric to $\bR^{10-n}$ and so all $N>16$  $\bR^{n-1,1}\times_w M^{10-n}$ backgrounds are locally isometric to the maximally
supersymmetric vacuum $\bR^{10,1}$.   It is not apparent that the theorem holds for $n=2$ even if $A$ and $\Phi$ are taken to be constant.

In IIB supergravity  the self-dual real  5-form $F$ and complex  3-form $H$  field strengths of the theory for  $\bR^{n-1,1}\times_w M^{10-n}$, $n\geq 3$, backgrounds  can be expressed as
\bea
F&=&\mathrm{dvol}_A(\bR^{n-1,1})\wedge W^{5-n}+ * W^{5-n}~,
\cr
H&=&\mathrm{dvol}_A(\bR^{n-1,1})\wedge P^{3-n}+ Q~,
\eea
where $W^{5-n}$ vanishes for $n>5$ and $P^{3-n}$ vanishes for $n>3$.
The field equation of the warp factor is
\bea
\tilde\nabla^2\log A=-n (\partial\log A)^2+{3\over 8} |P^{3-n}|^2+{1\over 48} |Q|^2+{4\over (5-n)!} (W^{5-n})^2~.
\eea
Clearly if $A$ is constant, which we have demonstrated is the case for $N>16$ backgrounds, then $F=H=0$.  Moreover the homogeneity theorem implies that the two scalar fields of
IIB supergravity, the axion and the dilaton, are also constant. A similar argument to that used in $d=11$ supergravity
implies  that $M^{10-n}$ is locally isometric to $\bR^{10-n}$.  Thus the $\bR^{n-1,1}\times_w M^{10-n}$ are locally isometric
to the $\bR^{9,1}$ maximally supersymmetric vacuum of the theory. The same conclusion holds for $n=2$ as well, provided that $A$ is taken to be constant.


\chapter{Conclusions}\label{concluions}

In this thesis, we investigated all warped AdS$_4$ and AdS$_3$ backgrounds preserving more than 16 real supercharges in 10- dimensional type II and 11-dimensional supergravities. We made the assumption that either the internal manifold is compact without boundary or equivalently that the isometry algebra of the background decomposes into that of AdS and that of the transverse space. As we pointed out in section \ref{sugrabackks} and \ref{ksaintro}, this is crucial for performing any sort of meaningful and unambiguous classification. Otherwise, we could, for example, rewrite the maximally supersymmetric $AdS_7\times S^4$ solution as a warped product of some lower dimensional AdS$_k$ space ($k<7$) with a non-compact manifold. Assuming a closed internal space or, alternatively, the aforementioned decomposition of the bosonic KSA, culls such ``fake'' vacua.

The key developments that enabled us to obtain the results in chapters \ref{ads4}, \ref{ads3} and \ref{minkngr16} are the proof of the homogeneity theorem \cite{homogen}, the integration of the Killing spinor and field equations along the AdS subspace \cite{mads, iibads, iiaads}, as well as the classification of the Killing superalgebras \cite{superalgebra,ads2}. With these tools at our disposal, we proved in chapter \ref{ads4} that there are no $N>16$ AdS$_4$ backgrounds in type IIB supergravity. Similarly, we showed that all 11-dimensional $N>16$ AdS$_4$ backgrounds are locally isometric to the maximally supersymmetric $AdS_4\times S^7$  solution. In type IIA the unique such background is its reduction on the circle fibre, namely $AdS_4\times \mathbb{CP}^3$, which preserves $N=24$ supersymmetries. Furthermore in chapter \ref{ads3}, we proved a non-existence theorem for AdS$_3$ solutions preserving strictly more than 16 supersymmetries. Finally in chapter \ref{minkngr16}, we showed that all warped 10- and 11-dimensional $N>16$ Minkowski backgrounds $\mathbb{R}^{n-1,1}\times_w M^{D-n}$ ($n\geq 3, D=10,11$) with fields which may not be smooth everywhere and possibly non-compact internal spaces are locally isometric to the Minkowski vacuum $\mathbb{R}^{D-1,1}$. Particularly in the context of flux compactifications, this means that all vacua of this type have the same local geometry as the maximally supersymmetric toroidal vacuum $\mathbb{R}^{n-1,1}\times T^{D-n}$.

\begin{table}	
\renewcommand{\arraystretch}{1.3}
\centering
\caption{$AdS$ backgrounds preserving more than half maximal supersymmetry.}

\begin{tabular}{cccc}
	\toprule
   & $\mathbf{D=11}$ & \textbf{IIA} & \textbf{IIB} \\
  \midrule
  $\mathbf{AdS_7}$ & $AdS_7\times S^4~(N=32)$ & \textcolor{red}{\ding{55}} & \textcolor{red}{\ding{55}}  \\
  $\mathbf{AdS_6}$ & \textcolor{red}{\ding{55}} & \textcolor{red}{\ding{55}} &\textcolor{red}{\ding{55}} \\
  $\mathbf{AdS_5}$ & \textcolor{red}{\ding{55}} & \textcolor{red}{\ding{55}} & $AdS_5\times S^5~(N=32)$ \\
  $\mathbf{AdS_4}$ & $AdS_4\times S^7~(N=32)$ & $AdS_4\times\mathbb{CP}^3~(N=24)$ & \textcolor{red}{\ding{55}}\\
  $\mathbf{AdS_3}$ & \textcolor{red}{\ding{55}} & \textcolor{red}{\ding{55}} & \textcolor{red}{\ding{55}}\\
  $\mathbf{AdS_2}$ & \textcolor{red}{\ding{55}} & \textcolor{red}{\ding{55}} & \textcolor{red}{\ding{55}}\\
  \bottomrule
\end{tabular}
\label{AdSbckgrlandscape}
\end{table}

It is quite remarkable how sparse the landscape of $N>16$ AdS solutions turns out to be(cf.\ table \ref{AdSbckgrlandscape}). In fact, all the maximally supersymmetric solutions arise as near-horizon limits of stacks of coincident D3-, M2- or M5-branes, respectively. Incidentally, these are also the only configurations of coincident branes giving rise to an AdS factor in the near-horizon limit, therefore it seems as though the existence of more than half BPS AdS backgrounds is linked to supersymmetry enhancement in near-horizon geometries. As for $AdS_4\times \mathbb{CP}^3$, this background can be obtained from M2-branes probing a $\mathbb{C}^4/\mathbb{Z}_k$ singularity \cite{abjm} which corresponds to dividing by $\mathbb{Z}_k$ in the near-horizon geometry, i.e.\ $AdS_4\times S^7/\mathbb{Z}_k$. The $\mathbb{Z}_k$ action makes the $S^1$ in the Hopf fibration $S^1\hookrightarrow S^7 \rightarrow \mathbb{CP}^3$ $k$ times smaller, so as $k$ becomes large, we end up with $\mathbb{CP}^3$ in IIA. This actually illustrates another natural extension of our work. We classified AdS backgrounds up to local isometry, i.e.\ we determined the local geometries of the internal spaces $G/H$. It might be possible to find further solutions by dividing out discrete subgroups, as was also pointed out in section \ref{ads4conclusions}, however this requires a classification of all possible subgroups of $SU(4)$ and $Spin(8)$ which is not available at the time of writing. Lastly, in the spirit of the classification of AdS backgrounds with symmetric spaces as internal manifold in \cite{figueroaa, figueroab, Wulffa, Wulffb}, one could try and classify all $N\leq 16$ homogeneous AdS backgrounds, however the reduced amount of supersymmetry and hence reduced control over the geometry will greatly complicate the analysis. Nevertheless, this would be a fruitful avenue to explore.   


\begin{appendices}

\chapter{Notation and conventions}\label{conventions}
\setcounter{section}{1}
Throughout this thesis, we use the mostly positive metric
\begin{align}
	\eta = (-,+,...,+)~.
\end{align}
Our  conventions for forms are as follows. Let $\omega$ be a k-form, then
\begin{align}
\omega&=\frac{1}{k!} \omega_{i_1\dots i_k} dx^{i_1}\wedge\dots \wedge dx^{i_k}~,~~~\omega^2_{ij}= \omega_{i\ell_1\dots \ell_{k-1}} \omega_{j}{}^{\ell_1\dots \ell_{k-1}}~,\notag\\
\omega^2&= \omega_{i_1\dots i_k} \omega^{i_1\dots i_k}~.
\end{align}
We also define
\bea
{\slashed\omega}=\omega_{i_1\dots i_k} \Gamma^{i_1\dots i_k}~, ~~{\slashed\omega}_{i_1}= \omega_{i_1 i_2 \dots i_k} \Gamma^{i_2\dots i_k}~,~~~\slashed{\gom}_{i_1}= \Gamma_{i_1}{}^{i_2\dots i_{k+1}} \omega_{i_2\dots i_{k+1}}~,
\eea
where the $\Gamma_i$ are the Dirac gamma matrices. The gamma matrices are always taken in an orthonormal frame and we use the notation
\begin{align}
	\Gamma^{i_1...i_k} = \Gamma^{[i_1}\,...\,\Gamma^{i_k]}
\end{align}
for anti-symmetrised products with total weight 1. 
To simplify expressions, we use the shorthand notation
\bea
\omega^n=\omega\wedge\dots\wedge\omega~,~~~\ell^{12\dots n}=\ell^1\wedge \ell^2\wedge\dots\wedge \ell^n
\eea
where $\ell^i$ is a left-invariant frame, and similarly for an orthonormal frame $\bbe^i$.

Our Hodge duality convention is
\bea
\omega\wedge *\omega={1\over k!} \omega^2 d\mathrm{vol}~.
\eea
The inner product $\langle\cdot, \cdot\rangle$ we use on the space of spinors is that for which space-like gamma matrices are Hermitian while time-like gamma
matrices are anti-Hermitian, i.e. the Dirac spin-invariant inner product is $\langle\Gamma_0\cdot, \cdot\rangle$. The norm $\parallel\cdot\parallel=\sqrt {\langle\cdot, \cdot\rangle}$ is taken with respect to $\langle\cdot, \cdot\rangle$, which is positive definite. For more details on our conventions
see \cite{mads, iiaads, iibads}.

\chapter{\texorpdfstring{$\mathfrak{su}(k)$}{su(k)}}\label{su(k)}

Here, we shall collect some formulae that are useful in understanding the homogeneous spaces admitting a transitive action of a group with Lie algebra $\mathfrak{su}(k)$.
A basis over the reals of anti-Hermitian $k\times k$ traceless complex matrices is
\begin{align}
(M_{ab})^c{}_d&=\frac12 (\delta_a{}^c \delta_{bd}-\delta_b{}^c \delta_{ad})~,\notag\\
(N_{ab})^c{}_d&=\frac{\nu(ab)}{2} i (\delta_a{}^c \delta_{bd}+\delta_b{}^c \delta_{ad}-\frac{2}{k} \delta_{ab} \delta_c{}^d)~,
\end{align}
where $\nu(ab)$ is a normalization factor and $a,b, c, d=1,\dots, k$. The trace of these matrices yields an invariant  inner product on $\mathfrak{su}(k)$. In particular,
the non-vanishing traces are
\begin{align}
\mathrm{tr}( M_{ab} M_{a'b'})&=-\frac12 (\delta_{aa'} \delta_{bb'}- \delta_{ab'} \delta_{ba'})~,\notag\\
\mathrm{tr}( N_{ab} N_{a'b'})&=-\frac{\nu(ab) \nu(a'b')}{2}(\delta_{aa'} \delta_{bb'}+ \delta_{ab'} \delta_{ba'}-\frac{2}{k} \delta_{ab} \delta_{a'b'})~.
\end{align}
It is customary to choose the normalization factors $\nu$  such that all generators have the same length. In such a case, they will depend on $k$.  However in what follows, it is  more convenient to choose $\nu=1$.
The Lie brackets of $\mathfrak{su}(k)$ are
\begin{align}
[M_{ab}, M_{a'b'}]&=\frac12 (\delta_{ba'} M_{ab'}+ \delta_{ab'} M_{ba'}- \delta_{aa'} M_{bb'}-\delta_{bb'} M_{aa'})~,\notag\\
[M_{ab},  N_{a'b'}]&=\frac12  (\delta_{ba'} N_{ab'}- \delta_{ab'} N_{ba'}- \delta_{aa'} N_{bb'}+\delta_{bb'} N_{aa'})~,\notag\\
[N_{ab},  N_{a'b'}]&=-\frac12 (\delta_{ba'} M_{ab'}+ \delta_{ab'} M_{ba'}+ \delta_{aa'} M_{bb'}+\delta_{bb'} M_{aa'})~.
\label{sukbra}
\end{align}
We shall proceed to describe the homogeneous spaces in (\ref{homo166}) and (\ref{homo167}) that admit a transitive $SU(k)$ action.

\section{\texorpdfstring{$M^k=\mathbb{CP}^{k-1}=SU(k)/S(U(k-1)\times U(1))$}{CP(k-1)=SU(k)/S(U(k-1)xU(1))}}

To describe the  $\mathbb{CP}^{k-1}$ homogeneous space, we set
\bea
\mathfrak{h}=\mathfrak{s}(\mathfrak{u}(k-1)\oplus \mathfrak{u}(1))=\bR\langle M_{rs}, N_{rs}, N_{kk})\rangle ~,~~~\mathfrak{m}=\bR\langle M_{rk}, N_{sk}\rangle~,
\eea
where $r, s=1,\dots, k-1$.
  The brackets of the Lie subalgebra  $\mathfrak{s}(\mathfrak{u}(k-1)\oplus \mathfrak{u}(1))$ can be read off from those   in (\ref{sukbra}) while those involving
  elements of $\mathfrak{m}$ are
\begin{align}
[M_{rk}, M_{sk}]&=- \frac12 M_{rs}~,~~~[M_{rk},  N_{sk}]=\frac12   N_{rs}-\frac12 \delta_{rs} N_{kk}~,\notag\\
[N_{rk},  N_{sk}]&=-\frac12  M_{rs}~,
\label{sukbra2}
\end{align}
and
\begin{align}
[M_{rs}, M_{tk}]&=\frac12 (\delta_{ts} M_{rk}- \delta_{tr} M_{sk})~,~~~
[M_{rs},  N_{tk}]=\frac12  (\delta_{ts} N_{rk}- \delta_{tr} N_{sk})~,\notag\\
[ N_{rs}, M_{tk}]&=\frac12  (\delta_{ts} N_{rk}+ \delta_{tr} N_{sk})~,~~~ [N_{rs},  N_{tk}]=-\frac12 (\delta_{ts} M_{rk}+ \delta_{tr} M_{sk})~,\notag\\
[N_{kk},  M_{sk}]&= - N_{rk} ~,~~~    [N_{kk},  N_{rk}]= M_{rk}~.
\label{sukbra3}
\end{align}
The left-invariant frame is $\bbl^A m_A= \bbl^r M_{rk}+ \bbl^{\tilde r} N_{rk}$. The most general left-invariant metric can be expressed as
\bea
ds^2=a\, (\delta_{rs}  \bbl^r \bbl^s+ \delta_{\tilde r\tilde s}  \bbl^{\tilde r} \bbl^{\tilde s})~,~~~
\eea
where $a>0$ is a constant.  The left-invariant forms of $\mathbb{CP}^{k-1}$ are generated by the (K\"ahler) 2-form
\bea
\omega= a\, \delta_{r\tilde s} \ell^r\wedge \ell^{\tilde s}~.
\eea
 The non-vanishing components of the curvature of the metric in the orthonormal frame are
\begin{align}
R_{rs, pq}&=-\frac{1}{4a} (\delta_{rq} \delta_{sp}- {1\over a}\delta_{rp} \delta_{sq}) ~,~~~R_{rs, \tilde p\tilde q}=-\frac{1}{4a} (\delta_{r\tilde q} \delta_{s\tilde p}- {1\over a}\delta_{r\tilde p} \delta_{s\tilde q})~,\notag\\
R_{r\tilde s, p\tilde q}&= \frac{1}{4a} (\delta_{r\tilde q} \delta_{\tilde sp}+ \delta_{rp} \delta_{\tilde s\tilde q})+\frac{1}{2a} \delta_{r\tilde s} \delta_{p\tilde q}~,~~~R_{\tilde r\tilde s, \tilde p\tilde q}=-\frac{1}{4 a} (\delta_{\tilde r\tilde q} \delta_{\tilde s\tilde p}- \delta_{\tilde r\tilde p} \delta_{\tilde s\tilde q})~.
\end{align}
This expression of the curvature matches that in (\ref{cp3curv}) for $\mathbb{CP}^3$ up to an overall scale.

\section{\texorpdfstring{$M^k=SU(k)/SU(k-1)$}{SU(k)/SU(k-1)}}

Next, let us turn to the $SU(k)/SU(k-1)$ homogeneous space.  The embedding of $\mathfrak{su}(k-1)=\bR\langle M_{rs}^{(k-1)}, N_{rs}^{(k-1)}\rangle $, where $r,s=1,\dots, k-1$,  into $\mathfrak{su}(k)=\bR\langle M_{ab}^{(k)}, N_{ab}^{(k)}\rangle$ is given by
\bea
M^{(k-1)}_{rs}=M^{(k)}_{rs}~,~~~N^{(k-1)}_{rs}=N^{(k)}_{rs}+{1\over k-1} \delta_{rs} N^{(k)}_{kk}~.
\eea
As $\mathfrak{m}=\bR\langle  M_{rk}^{(k)}, N_{sk}^{(k)},  N^{(k)}_{kk}\rangle$, the  (non-vanishing) commutators involving elements of $\mathfrak{m}$ are
\begin{align}
[M^{(k)}_{rk}, M^{(k-1)}_{sk}]&=- \frac12 M^{(k)}_{rs}~,~~~[M^{(k)}_{rk},  N^{(k)}_{sk}]=\frac12   N^{(k-1)}_{rs}-\frac{k}{2(k-1)} \delta_{rs} N^{(k)}_{kk}~,\notag\\
[N^{(k)}_{rk},  N^{(k)}_{sk}]&=-\frac12  M^{(k-1)}_{rs}~,
\label{sukbra2x}
\end{align}
and
\begin{align}
[M^{(k-1)}_{rs}, M^{(k)}_{tk}]&=\frac12 (\delta_{ts} M^{(k)}_{rk}- \delta_{tr} M^{(k)}_{sk})~,~~~
[M^{(k-1)}_{rs},  N^{(k)}_{tk}]=\frac12  (\delta_{ts} N^{(k)}_{rk}- \delta_{tr} N^{(k)}_{sk})~,\notag\\
[ N^{(k-1)}_{rs}, M^{(k)}_{tk}]&=-{1\over k-1} \delta_{rs} N_{tk}^{(k)}+\frac12  (\delta_{ts} N^{(k)}_{rk}+ \delta_{tr} N^{(k)}_{sk})~,\notag\\
 [N^{(k-1)}_{rs},  N^{(k)}_{tk}]&={1\over k-1} \delta_{rs} M^{(k)}_{tk}-\frac12 (\delta_{ts} M^{(k)}_{rk}+ \delta_{tr} M^{(k)}_{sk})~,\notag\\
[N^{(k)}_{kk},  M^{(k)}_{rk}]&= - N^{(k)}_{rk} ~,~~~    [N^{(k)}_{kk},  N^{(k)}_{rk}]= M^{(k)}_{rk}~.
\label{sukbra3x}
\end{align}
Setting $\bbl^A m_A= \hat\bbl^r M^{(k)}_{rk}+ \hat\bbl^{\tilde r} N^{(k)}_{rk}+\hat\bbl^0 N^{(k)}_{kk}$ for the left-invariant frame, a direct computation reveals that the most general invariant metric is
\bea
&&ds^2=a\, (\delta_{rs}  \hat\bbl^r \hat\bbl^s+ \delta_{\tilde r\tilde s}  \hat\bbl^{\tilde r} \hat\bbl^{\tilde s})+ b (\hat\bbl^0)^2~,~~~
\eea
where $a,b>0$ are constants. Moreover the left-invariant 2- and 3-forms  for $k=4$ are generated by
\bea
\hat\omega=\delta_{r\tilde s} \hat\bbl^r\wedge \hat\bbl^{\tilde s}~,~~~\hat\bbl^0\wedge \hat\omega~,~~~\mathrm{Re}\, \hat\chi~,~~~\mathrm{Im}\, \hat\chi~,
\eea
and their duals, where
\bea
 \hat\chi=\frac{1}{3!}\,\epsilon_{rst} (\hat\bbl^r+ i\hat\bbl^{\tilde r}) \wedge  (\hat\bbl^s+ i\hat\bbl^{\tilde s}) \wedge  (\hat\bbl^t+ i\hat\bbl^{\tilde t})~,
\eea
is the holomorphic (3,0)-form.

 However for convenience, we re-label the indices of the left-invariant frame as
$ \ell^{2r-1}= \hat\ell^r, \ell^{2r}=\hat\ell^{\tilde r}, \ell^7=\hat\ell^0$, $r=1,2,3$  in which case the left-invariant metric can be rewritten as
\bea
ds^2= a\, \delta_{mn} \ell^m \ell^n+ b\, (\ell^7)^2= \delta_{mn}\bbe^m \bbe^n+ (\bbe^7)^2~,
\eea
where we have introduced an orthonormal frame $\bbe^m=\sqrt{a}\, \ell^m, \bbe^7=\sqrt{b}\, \ell^7$, and  $m,n=1,\dots, 6$.
Note also that, up to an overall scale, the left-invariant 2- and 3-forms can be re-written in terms of the orthonormal frame.  In particular, we have
\begin{align}
\omega &= \bbe^{12} + \bbe^{34} + \bbe^{56}~,~~ \bbe^7\wedge \omega~,~~~\mathrm{Re}\, \chi~,~~~\mathrm{Im}\, \chi~,
\end{align}
where
\bea
 \chi= (\bbe^1+ i\bbe^{ 2}) \wedge  (\bbe^3+ i\bbe^4) \wedge  (\bbe^5+ i \bbe^6)~.
\eea
We shall use this orthonormal basis to solve the KSEs for this internal space.

\chapter{The Berger space \texorpdfstring{$B^7={Sp(2)}/{Sp(1)_{\mathrm{max}}}$}{B7=Sp(2)/Sp(1)max}}\label{bergerspace}

To describe the geometry of the Berger space $B^7$, one  identifies   the vector representation ${\bf 5}$ of  $\mathfrak{so}(5)=\mathfrak{sp}(2)$ with the symmetric traceless
representation of $\mathfrak{so}(3)=\mathfrak{sp}(1)$ and then decomposes the adjoint representation of $\mathfrak{so}(5)$ in $\mathfrak{so}(3)$ representations as
${\bf 10}={\bf 3}\oplus {\bf 7}$, where ${\bf 7}$ is the symmetric traceless representation of $\mathfrak{so}(3)$ constructed with three copies of the vector
representation. As a result $\mathfrak{so}(5)=\mathfrak{so}(3)\oplus \mathfrak{m}$, where $\mathfrak{so}(3)$ and $\mathfrak{m}$ are identified with the 3-dimensional
and 7-dimensional representations, respectively.

This decomposition can be implemented as follows.  Consider the basis $W_{ab}$, $a,b,c,d=1,\dots,5$,
\bea
(W_{ab}){}^c{}_d=\delta_a^c \delta_{bd}- \delta_b^c \delta_{ad}~,
\eea
in $\mathfrak{so}(5)$ leading to the commutators
\bea
[W_{ab}, W_{a'b'}]= (\delta_{ba'} W_{ab'}+\delta_{ab'} W_{ba'}- \delta_{aa'} W_{bb'}-  \delta_{bb' } W_{aa'})~.
\eea
Then re-write each basis element using the ${\bf 5}$ representation of $\mathfrak{so}(3)$  as
$W_{rs, tu}$, where $r, s,t,u=1,2,3$.  Decomposing  this into $\mathfrak{so}(3)$ representations, one  finds that
\bea
W_{rs,tu}&=&O_{ru} \delta_{st}+O_{su} \delta_{rt}+O_{rt} \delta_{su}+O_{st} \delta_{ru}
\cr~~~&&+ \epsilon^p{}_{st} S_{pru}+\epsilon^p{}_{rt} S_{psu}+\epsilon^p{}_{su} S_{prt}+\epsilon^p{}_{ru} S_{pst}~,
\eea
where  $O\in \mathfrak{so}(3)$ and $S\in \mathfrak{m}$.  Using this one can proceed to describe the homogeneous space $B^7$.  However, this decomposition
does not automatically reveal the $G_2$ structure which  is necessary in the analysis of the supersymmetric solutions.  Instead, we shall follow an adaptation \cite{difftype} of the
description in \cite{Castellani:1983yg} and  \cite[Appendix A.1]{Haupt:2015wdq}.  For this, use the inner product
\bea
\langle W_{ab}, W_{a'b'}\rangle=-{1\over2} \mathrm{tr} ( W_{ab} W_{a'b'} )~,
\eea
which is $\mathfrak{so}(5)$ invariant and the basis $W_{ab}$, $a<b$, is orthonormal.  In this basis, the structure constants of $\mathfrak{so}(5)$ are skew-symmetric.
Then, we identify the $\mathfrak{so}(3)$ subalgebra of $\mathfrak{so}(5)$ with the span of the orthonormal vectors
\bea
h_1&=& {1\over\sqrt5} (-W_{12}-W_{34}+\sqrt3 W_{35})~,~~~h_2={1\over\sqrt5} (-W_{13}+W_{24}+\sqrt{3} W_{25})~,~~~
\cr
h_3&=&{1\over\sqrt5} (-2W_{14}+W_{23})~.
\eea
We choose the subspace $\mathfrak{m}$ to be orthogonal to $\mathfrak{so}(3)$ and introduce an ortho-normal basis in $\mathfrak{m}$ as
\begin{align}
m_1&={1\over2\sqrt5} (4W_{12}- W_{34}+\sqrt3 W_{35})~,~~~m_2={1\over2\sqrt5} ( 4W_{13}+W_{24}+\sqrt3 W_{25})~,\notag\\
m_3&={1\over \sqrt5} (-W_{14}-2 W_{23})  ~,~~~m_4={1\over2} (\sqrt 3W_{34}+W_{35})~,\notag\\
m_5&={1\over2} (\sqrt 3 W_{24}-W_{25})~,~~~m_6=W_{15}~,~~~m_7=W_{45}~.
\end{align}
Then it is straightforward to  show that
\begin{align}
[h_\alpha, h_\beta]&={1\over\sqrt 5} \epsilon_{\alpha\beta}{}^\gamma h_\gamma~,~~~[h_\alpha, m_i]= k_{\alpha i}{}^j m_j~,\notag\\
[m_i, m_j]&={1\over \sqrt 5} \varphi_{ij}{}^k m_k+ k_{ij}{}^\alpha h_\alpha~,~~~
\end{align}
where $\varphi$ is given in (\ref{g2phi}), the indices are raised and lowered with the flat metric and
\bea
k^1&=&-{3\over 2\sqrt 5} m_2\wedge m_3-{\sqrt 3\over 2} m_2\wedge m_6-{\sqrt 3\over 2} m_3\wedge m_5+{2\over  \sqrt 5} m_4\wedge m_7+{1\over 2 \sqrt 5} m_5\wedge m_6~,
\cr
k^2&=& {3\over 2\sqrt 5} m_1\wedge m_3-{\sqrt 3\over 2}m_1\wedge m_6-{\sqrt 3\over 2} m_3\wedge m_4-{1\over 2 \sqrt 5} m_4\wedge m_6+{2\over  \sqrt 5} m_5\wedge m_7~,
\cr
k^3&=&-{3\over 2\sqrt 5} m_1\wedge m_2-{\sqrt 3\over 2} m_1\wedge m_5-{\sqrt 3\over 2} m_2\wedge m_4+{1\over 2 \sqrt 5} m_4\wedge m_5+{2\over  \sqrt 5} m_6\wedge m_7~.
\nonumber
\eea
So $f_{ij}{}^k= {1\over \sqrt 5} \varphi_{ij}{}^k$, and the Jacobi identities imply that $\varphi$ is invariant under the representation of $\mathfrak{so}(3)$ on $\mathfrak{m}$.  Therefore the  embedding
of $\mathfrak{so}(3)$ in $\mathfrak{so}(7)$ defined by  $(k^1, k^2, k^3)$  factors through $\mathfrak{g}_2$.  This is useful in the analysis of the gravitino KSE.

\chapter{\texorpdfstring{$\mathfrak{so}(5)=\mathfrak{sp}(2)$}{so(5)=sp(2)}}\label{so5=sp2}

In order to describe the various homogeneous spaces we are using which admit a transitive action of a group with Lie algebra $\mathfrak{so}(5)=\mathfrak{sp}(2)$,
 choose a basis  in $\mathfrak{sp}(2) = \mathfrak{so}(5)$ as
\begin{align}\label{gen-sona}
(M_{\tilde a\tilde b})_{\tilde c\tilde d} = \frac12 ( \delta_{\tilde a\tilde c} \, \delta_{\tilde b\tilde d} - \delta_{\tilde a\tilde d}\, \delta_{\tilde b\tilde c})~,
\end{align}
where $M_{\tilde a\tilde b},{} \tilde a,\tilde b=1,...,5$.  The commutators are
\begin{align}\label{son-commuta}
[M_{\tilde a\tilde b}, M_{\tilde a'\tilde b'}] = \frac{1}{2} (\delta_{\tilde a\tilde b'} M_{\tilde b\tilde a'} + \delta_{\tilde b\tilde a'} M_{\tilde a\tilde b'} - \delta_{\tilde b\tilde b'} M_{\tilde a\tilde a'} - \delta_{\tilde a\tilde a'} M_{\tilde b\tilde b'})~.
\end{align}
In what follows, we shall give various decompositions $\mathfrak{so}(5)=\mathfrak{h}\oplus \mathfrak{m}$ for different  choices of a subalgebra $\mathfrak{h}$ and
summarise some of their algebraic and geometric properties that we require in this thesis.

\section{\texorpdfstring{$M^6=Sp(2)/U(2)$}{Sp(2)/U(2)}}

The subalgebra $\mathfrak{h}$ and $\mathfrak{m}$ are spanned as
 \bea
 \mathfrak{u}(2) = \mathfrak{u}(2) \equiv \bR\, \langle T_r,T_{7}\rangle =\bR\, \langle \frac12 \epsilon_r{}^{st} M_{st},  M_{45}\rangle~,
 \label{decsp2u2aaa}
 \eea
 and
\begin{align}
\mathfrak{m}=\bR\,\langle M_{ra} \rangle = \bR\,\langle M_{r4},M_{r5},\rangle~,
\end{align}
respectively,  where $r,s,t=1,2,3$ and  $a, b, c, \ldots = 4, 5$.
 In this basis, the non-vanishing commutators are
\bea
&&[T_r, T_s]= -\frac12 \epsilon_{rs}{}^t T_t~,~~[T_r, M_{sa}]=-\frac12 \epsilon_{rs}{}^t M_{ta}~,~~~[T_7, M_{ra}]=-\frac12\epsilon_{ab} M_{rb}~,~~~
\cr
&&[M_{ra}, M_{sb}]=-\frac12\delta_{ab} \epsilon_{rs}{}^t T_t-\frac12\delta_{rs} \epsilon_{ab} T_7~.
\label{comsp2so3a}
\eea
In fact, this is a symmetric coset space admitting an invariant metric
\bea
ds^2=a\,\delta_{rs} \delta_{ab} \bbl^{ra} \bbl^{sb}=\delta_{rs} \delta_{ab} {\bf e}^{ra} {\bf e}^{sb}~,
\label{sp2u2metra}
\eea
where $a>0$ is a constant, and $\bbl^{ra}$ and ${\bf e}^{ra}=\sqrt{a}\,\bbl^{ra} $ are the left-invariant and orthonormal frames, respectively.
 The curvature of the symmetric space in the ortho-normal frame is
 \bea
 R_{ra\,sb, tc\, ud}=\frac{1}{4a} (\delta_{rt} \delta_{su} -\delta_{ru} \delta_{st}) \delta_{ab} \delta_{cd}+ \frac{1}{4 a}\delta_{rs} \delta_{tu} \epsilon_{ab} \epsilon_{cd}~,
 \label{curvsp2u2}
 \eea
which is instrumental in the investigation of the gravitino KSE in section 3.5.1.

\section{\texorpdfstring{$M^6={Sp(2)}/({Sp(1)\times U(1)})$}{Sp(2)/Sp(1)xU(1)}}

Viewing the elements of $Sp(2)$  as quaternionic $2\times2$ matrices,   ${Sp(1)\times U(1)}\subset Sp(1)\times Sp(1)$ is embedded in $Sp(2)$ along the diagonal. To describe this embedding choose
a basis in $\mathfrak{sp}(2)=\mathfrak{so}(5)$ as in (\ref{gen-sona}) and set
\bea
T^{(\pm)}_r=\frac{1}{2} \varepsilon^{rst} M^{st} \pm M^{r4}~,~~~, W_a=\sqrt{2} M_{a5}~,
\eea
where  $r=1,2,3$ and now $a=1,\dots 4$.
In terms of this basis,  the non-vanishing commutators of $\mathfrak{sp}(2)$ are
\bea
[T^{(\pm)}_r, T^{(\pm)}_s]&=&-\epsilon_{rs}{}^ t T^{(\pm)}_t~,~~~[T^{(\pm)}_r, W_a]=\frac12 (I^{(\pm)}_r)^b{}_a W_b~,~~
\cr
[W_a, W_b]&=&-\frac12\big( (I^{(+)}_r)_{ab} T^{(+)}_r+(I^{(-)}_r)_{ab} T^{(-)}_r\big)~,
\label{so5so3aa}
\eea
where
\bea
(I^{(\pm)}_r)^4{}_s=\mp \delta_{rs}~,~~~(I^{(\pm)}_r)^s{}_4=\pm \delta^s{}_r~,~~(I^{(\pm)}_r)^s{}_t=\epsilon_{rst}~.
\label{Imatrpm}
\eea
Observe that $(I^{(\pm)}_r)$ are bases in the spaces of (anti-)self-dual forms in $\bR^4$ and that
\bea
I^{(\pm)}_r I^{(\pm)}_s=-\delta_{rs} {\bf 1}-\epsilon_{rst} I^{(\pm)}_t~.
\eea
The subalgebra $\mathfrak{h}$ and $\mathfrak{m}$ are spanned by
\bea
\mathfrak{h}=\mathfrak{sp}(1)\oplus \mathfrak{u}(1)=\bR\langle T^{(-)}_r, T^{(+)}_3\rangle~,
\eea
 and
 \bea
 \mathfrak{m}=\bR\langle W_a, T^{(+)}_1, T^{(+)}_2\rangle~,
 \eea
 respectively.
Introducing the left-invariant frame  $\bbl^A m_A= \bbl^a W_a+ \bbl^{\underline r} T^{(+)}_{\underline r}$, where ${\underline r}=1,2$, the left-invariant metric can be written
as
\bea
ds^2= a\, \delta_{ab} \bbl^a \bbl^b+ b \,\delta_{{\underline r}{\underline s}} \bbl^{\underline r}\bbl^{\underline s} = \delta_{ab} \bbe^a \bbe^{b} + \delta_{{\underline r}{\underline s}} \bbe^{{\underline r}} \bbe^{{\underline s}}~,
\eea
with $a,b>0$ and we have introduced the orthonormal frame $\bbe^a=\sqrt{a}\, \bbl^a $, $\bbe^{{\underline r}}=\sqrt{b}\, \bbl^{\underline r}$.

The curvature of this metric in the orthonormal frame is
\begin{align}
R_{ab,cd} &= \left( \frac{1}{2a} - \frac{3b}{16a^2} \right) ( \delta_{ac} \, \delta_{bd} - \delta_{ad} \, \delta_{bc}) \notag\\
&~~~+ \frac{3b}{16a^2} \left( (I^{(+)}_3)_{ab}  (I^{(+)}_3)_{cd} - (I^{(+)}_3)_{a[b} (I^{(+)}_3)_{cd]} \right)~, \notag\\
R_{a\underline{r}, b \underline{s}} &= \frac{b}{16a^2} \delta_{ab}\delta_{{\underline r}{\underline s}} + \left( \frac{1}{4a} - \frac{b}{16a^2} \right) \epsilon_{{\underline r}{\underline s}} (I^{(+)}_3)_{ab}~, \notag\\
R_{ab,\underline{r}\underline{s}} &= \left( \frac{1}{2a} - \frac{b}{8a^2} \right)\epsilon_{{\underline r}{\underline s}} (I^{(+)}_3)_{ab}, \quad R_{\underline{rs},\underline{tu}} = \frac{1}{b} \epsilon_{{\underline r}{\underline s}} \epsilon_{{\underline t}{\underline u}}~.
\label{curvsp2sp1u1}
\end{align}
We shall use these expressions in the investigation of the gravitino KSE in section 3.5.2.

\section{\texorpdfstring{$M^7={Sp(2)}/{\Delta(Sp(1))}$}{Sp(2)/Sp(1) diagonal}}

The decomposition  of the Lie algebra $\mathfrak{sp}(2) = \mathfrak{so}(5)$ suitable for the  description of  this homogeneous space is as in (\ref{decsp2u2aaa}), but now $\mathfrak{h}$ and $\mathfrak{m}$ are spanned by
\bea
\mathfrak{h}= \bR\langle T_r\rangle~,~~~\mathfrak{m}=\bR\langle M_{ra}, T_7\rangle~,
\eea
respectively, where $r=1,2,3$ and $a=4,5$.
Introducing the left-invariant frame as $\ell^A t_A= \ell^{ra} M_{ra}+ \ell^7 T_7$, the left-invariant metric is
\bea
ds^2=\delta_{rs} g_{ab} \ell^{ra} \ell^{sb}+ a_4 (\ell^7)^2~,~~~
\eea
where $(g_{ab})$ is a symmetric constant positive definite $2\times 2$ matrix and $a_4>0$ is a constant.  The curvature of this metric in the left-invariant frame  is
\begin{align}
R_{pc\, qd,}{}^{ra}{}_{sb}=&-{1\over16} a_4^{-1} \delta_p^r \delta_{qs} g^{ae} ((\Delta g)_{ec}- a_4 \epsilon_{ec}) ((\Delta g)_{db}+ a_4 \epsilon_{db})\notag\\
&+{1\over16} a_4^{-1} \delta_q^r \delta_{ps} g^{ae} ((\Delta g)_{ed}- a_4 \epsilon_{ed}) ((\Delta g)_{cb}+ a_4 \epsilon_{cb})\notag\\
&+{1\over8} \epsilon_{cd} \delta_{pq} \delta^r_s g^{ae} \epsilon_{eb} (\delta^{t_1t_2} g_{t_1t_2}-a_4)-{1\over4} \delta_{cd} \delta^a_b ( \delta_{ps} \delta^r_q- \delta_{qs} \delta_p^r)~,
\label{curvso5so3ab1}
\end{align}
and
\begin{align}
R_{7\, ar,}{}^7{}_{bs}=&{1\over16} a_4^{-1} ((\Delta g)_{ad}+ a_4 \epsilon_{ad}) g^{de} \epsilon_{eb} (\delta^{t_1t_2} g_{t_1t_2}-a_4) \delta_{rs}\notag\\
&-{1\over8} a_4^{-1} \epsilon_a{}^d ((\Delta g)_{db}+a_4 \epsilon_{db}) \delta_{rs}~,
\label{curvso5so3ab2}
\end{align}
where
\begin{align}
(\Delta g)_{ab}= \epsilon_a{}^d g_{db}+\epsilon_b{}^d g_{da}~,
\end{align}
 $(g^{ab})$ is the inverse matrix of $(g_{ab})$ and the indices of $\epsilon$ are raised and lowered with $\delta_{ab}$.
 
The Ricci tensor, again in the left-invariant frame, is
\begin{align}
R_{ra\, sb}=&\big[{a_4^{-1} \over 16} g^{dc} (\Delta g)_{da} (\Delta g)_{cb}-{1\over 16} g^{dc}  (\Delta g)_{cb} \epsilon_{da}\notag\\
&+{1\over 16} g^{cd} \epsilon_{ca} \epsilon_{db}
(\delta^{t_1t_2} g_{t_1t_2}-2a_4)+{a_4^{-1} \over 16} (\Delta g)_{ad} g^{dc} \epsilon_{cb}\, \delta^{t_1t_2} g_{t_1t_2}\notag\\
&- {a_4^{-1}\over 8} \epsilon_a{}^d (\Delta g)_{db}+{5\over8} \delta_{ab}\big]\delta_{rs}~,
\end{align}
and
\bea
R_{77}=-{3\over8} {a_4\over \det g} (\delta^{t_1t_2} g_{t_1t_2}-a_4)+{3\over8} a_4 \delta_{ab} g^{ab}-{3\over8} \epsilon_a{}^d (\Delta g)_{db} g^{ab}~.
\eea
%
It is straightforward to compute the Ricci tensor for $(g_{ab})$ diagonal.

\section{\texorpdfstring{$M^7=Sp(2)/Sp(1)$}{Sp(2)/Sp(1)}}

The decomposition  of the Lie algebra $\mathfrak{sp}(2) = \mathfrak{so}(5)$ suitable for the  description of this homogeneous space  is as in (\ref{so5so3aa}), where in this case
\bea
\mathfrak{so}(3)=\bR\langle T^{(-)}_r\rangle~,~~~\mathfrak{m}=\bR\langle W_a, T^{(+)}_r\rangle~,
\eea
with $r=1,2,3$ and $a=1,\dots,4$.
Introducing the left-invariant frame as $\ell^A m_A= \ell^a W_a+ \ell^r T_r^{(+)}$, the most general  left-invariant metric is
\bea
ds^2= a \delta_{ab} \bbl^a \bbl^b+ g_{rs} \bbl^r\bbl^s~,
\label{mettrra}
\eea
where  $a>0$ is a constant and  $(g_{rs})$ is any constant $3\times 3$  positive definite symmetric matrix.
The non-vanishing components of the curvature tensor of this metric in the left-invariant frame is
\begin{align}
R_{cd,}{}^a{}_b&= {a^{-1}\over 16} \Big[ \delta^{ae} \delta^{ pr} \delta^{qs} (I^{(+)}_p)_{ec}  g_{rs} (I^{(+)}_q)_{db}-(d,c)\Big]\notag\\
&~~-{a^{-1}\over8} \delta^{ae} \delta^{ pr} \delta^{qs}  (I^{(+)}_p)_{eb} g_{rs} (I^{(+)}_q)_{cd}+{1\over2} (\delta^a_c \delta_{db}- \delta_d^a \delta_{cb})~,
\label{curvsp2sp1aa}
\end{align}
\begin{align}
R_{rs}{}^a{}_b&={a^{-1}\over4} \delta^{pq} g_{pq} \epsilon_{rs}{}^t (I^{(+)}_t)^a{}_b
- {a^{-2}\over 8} \epsilon^{pqt} (I^{(+)}_t)^a{}_b g_{pr} g_{qs}\notag\\
 &~~-{a^{-1}\over2} \epsilon_{rs}{}^t \delta^{pq} (I^{(+)}_p)^a{}_b  g_{qt}~,
\label{curvsp2sp1ab}
\end{align}
\begin{align}
R_{ra}{}^s{}_b&={1\over8} \big[ g^{sm} \epsilon_{mr}{}^n g_{np}\delta^{pt}+g^{sm} \epsilon_{m}{}^{tn} g_{nr}\big] (I^{(+) }_t)_{ab}
+{1\over8} \epsilon_r{}^{sp} (I^{(+)}_p)_{ab}+ \notag\\
&~~ + {a^{-1}\over 16} \delta_{ab} \delta^{sm} g_{mr}+{a^{-1}\over 16} \epsilon^{smn} g_{mr} (I^{(+)}_n)_{ab}~,
\label{curvsp2sp1ac}
\end{align}
and
\begin{align}
R_{rs, pq}= g_{pl}R_{rs}{}^l{}_q= \epsilon_{rs}{}^m \epsilon_{pq}{}^n  X_{mn}~,
\label{curvsp2sp1ad}
\end{align}
where
\bea
X_{mn}&=&{1\over2} \delta_{mk} \delta_{nl} g^{kl} (\delta^{q_1q_2} \delta^{p_1p_2} g_{q_1p_1} g_{q_2p_2})
\cr
&&
-2 g_{mn} + \delta_{mn} \delta^{pq} g_{pq}-{1\over4} \delta_{mk} \delta_{nl} g^{kl} (\delta^{q_1q_2} g_{q_1q_2})^2~,
\eea
and the matrix $(g^{rs})$ is the inverse of $(g_{rs})$.
The Ricci tensor in the left-invariant frame is
\begin{align}
R_{ab}&= -{a^{-1}\over8}  \delta^{pq} g_{pq} \delta_{ab}+{3\over2} \delta_{ab}~,\notag\\
R_{rs}&={1\over4} a^{-2} \delta^{mn} g_{mr} g_{ns}+(\delta_{rs} \delta_{pq} g^{pq} - \delta_{rp} \delta_{sq} g^{pq} ) \delta^{mn} X_{mn}+ \delta_{rp} g^{pm} X_{ms}\notag\\
&~~+\delta_{sp} g^{pm} X_{mr}- \delta_{pq} g^{pq} X_{rs}-\delta_{rs} g^{pq} X_{pq}~.
\end{align}
It is straightforward to find the Ricci tensor for $(g_{rs})$ diagonal.
This homogeneous space admits two Einstein metrics one of which is the round sphere metric on $S^7$. This will be explored further in the investigation of the gravitino KSE in section \ref{sp2modsp1}.

\chapter{Coset spaces  with  non-semisimple transitive groups}\label{modif}

We have already argued that if $\mathfrak{t}_0$ is simple then the internal spaces $G/H$, with $\mathfrak{Lie}\,G=\mathfrak{t}_0$ can be identified from the classification results of
\cite{Castellani:1983yg,klausthesis,niko6dim,niko7dim,bohmkerr}.  This is also the case for $\mathfrak{t}_0$ semisimple, provided that in addition one considers modifications to the homogeneous spaces as described in section \ref{modifx}.

Here, we shall describe the structure of homogeneous $G/H$ spaces for  $G$ a compact  but  not semisimple Lie group.  As  we consider homogeneous spaces   up to discrete identifications, we shall perform
the calculation in terms of Lie algebras.
 The Lie algebra of $G$ can be written as $\mathfrak{Lie}\,G=\mathfrak{p}\oplus \mathfrak{a}$, where $\mathfrak{p}$
is semisimple and $\mathfrak{a}$ is Abelian.  Suppose now that we have a $G/H$ coset space, where $H$ is a compact subgroup of $G$. Then $\mathfrak{Lie}\,(H)=\mathfrak{q}\oplus \mathfrak{b}$, where $\mathfrak{q}$ is a semisimple subalgebra
and $\mathfrak{b}$ is Abelian.  Let us now focus on the inclusion $i:\mathfrak{Lie}\,(H)\rightarrow \mathfrak{Lie}\,G$.  Consider the projections $p_1:\mathfrak{Lie}\,G\rightarrow \mathfrak{p}$ and $p_2:\mathfrak{Lie}\,G\rightarrow \mathfrak{a}$.  Then, we have that $p_2\circ i\vert_{\mathfrak{q}}=0$, as there are no non-trivial Lie algebra homomorphisms from
a semisimple Lie algebra into an Abelian one.  Thus $p_1\circ i\vert_{\mathfrak{q}}$ is an inclusion.  Furthermore, $p_1\circ i\vert_{\text{Ker} (p_2\circ i\vert_{\mathfrak{b}})}$
is also an inclusion.  Therefore, $\mathfrak{q}\oplus \text{Ker} (p_2\circ i\vert_{\mathfrak{b}})$ is a subalgebra of $\mathfrak{p}$.  As $\mathfrak{p}$ is semisimple
and for applications here $\text{dim}\, G/H\leq 8$, there is a classification of all coset spaces $P/T$ with $\mathfrak{Lie}\, P=\mathfrak{p}$  and
$\mathfrak{Lie}\,(T)=\mathfrak{q}\oplus \text{Ker}\, p_2\circ i\vert_{\mathfrak{b}}$.

Next, consider $\mathfrak{b}_1=\mathfrak{b}/\text{Ker}\, (p_2\circ i\vert_{\mathfrak{b}})$.  Suppose first that $i(\mathfrak{b}_1)$ is contained in both $\mathfrak{p}$ and $\mathfrak{a}$, then, up to a discrete identification, a coset space $W/X$ with $\mathfrak{Lie}\,W=\mathfrak{p}\oplus i(\mathfrak{b})$ and $\mathfrak{Lie}\,X=\mathfrak {q}\oplus \mathfrak{b}$ is a modification of $P/T$ with an Abelian group which has Lie algebra $p_2\circ i(\mathfrak{b})$.  Furthermore the generators
of  $\mathfrak{a}/i(\mathfrak{b})$ commute with $\mathfrak{Lie}\,X$ and are of course not in the image of $i$.  As a result, $G/H$, up to a discrete identification,
can be written as $W/X\times T^k$, where $k=\text{dim}\, \mathfrak{a}/i(\mathfrak{b})$, i.e.\ up to a discrete identification $G/H$ is the product of an Abelian  modification of a coset space with semisimple
transitive group and of an Abelian group. On the other hand, if $i(\mathfrak{b})$ is all contained in $\mathfrak{a}$, then $G/H$ up to a discrete identification is a product
$P/T\times T^m$, where $m=\text{dim}\,\mathfrak {a}$.
In the classification of AdS$_3$ backgrounds in chapter \ref{ads3}, we  use the above results to describe the geometry of the internal spaces,
whenever $\mathfrak{t}_0$ is not a semisimple Lie algebra.

\chapter{ \texorpdfstring{$N^{k,l}=SU(2)\times SU(3)/ \Delta_{k,l}(U(1))\cdot (1\times SU(2))$}{N(k,l) geometry}}\label{appencx2}

The inclusion of $U(1)\times SU(2)$ in $SU(2)\times SU(3)$ is given by
$$
(z, A)\rightarrow \Big( \begin{pmatrix} z^k & 0\\ 0 & z^{-k} \end{pmatrix}, \begin{pmatrix} A z^l &0\\ 0 & z^{-2l} \end{pmatrix}\Big)~.
$$
Consequently in the notation of \cite{ads4Ngr16}, the Lie subalgebra $\mathfrak{h}$ of the isotropy group is identified as
\bea
\mathfrak{h}=\mathbb{R}\langle  M_{12},  N_{12}, N_{11}+{1\over2} N_{33}, 2k \tilde N_{11}-3 l N_{33}\rangle~.
\eea
The generators of the tangent space at the origin $\mathfrak{m}$  of the homogeneous space must be linearly independent from those of $\mathfrak{h}$
and so one can choose
\bea
\mathfrak{m}=\mathbb{R}\langle \tilde M_{12}, \tilde N_{12}, M_{13}, M_{23}, N_{13}, N_{23}, 2k \tilde N_{11}+ 3l N_{33}\rangle~,
\eea
where $\mathfrak{su}(2)=\mathbb{R}\langle \tilde M_{rs}, \tilde N_{rs}\rangle$ , $r,s=1,2$ and $\mathfrak{su}(3)=\mathbb{R}\langle  M_{ab},  N_{ab}\rangle$ , $a,b=1,2,3$.

A left-invariant frame on $N^{k,l}$ is
\bea
\ell=\ell^7 Z+ \ell^5 \tilde M_{12}+\ell^6 \tilde N_{12} + \ell^r M_{r3}+ \hat \ell^r N_{r3}~,
\eea
where $Z=2k \tilde N_{11}+ 3l N_{33}$.

The exterior differential algebra of the left-invariant frame, modulo the terms  that contain the canonical connection and so lie in $\mathfrak{h}\wedge \mathfrak{m}$,  is
\begin{align}
d\ell^7&=-{1\over 4k}\ell^5\wedge \ell^6+ {1\over 8l} \delta_{rs} \ell^r\wedge \hat \ell^s~,\notag\\
d\ell^5&=2k\ell^7\wedge \ell^6~,~~~~d\ell^6=-2k\ell^7\wedge\ell^5~,\notag\\
d\ell^r&=-3l\,\ell^7\wedge \hat\ell^r~,~~~d\hat \ell^r=3l\,\ell^7\wedge \ell^r~.
\end{align}
Note that upon taking the exterior derivative of invariant forms the terms in the exterior derivative of a left-invariant frame that lie in $\mathfrak{h}\wedge \mathfrak{m}$ do not contribute.

The invariant forms  on $N^{k,l}$ are generated by a 1-form $\ell^7$ and the 2-forms
\bea
\omega_1= \delta_{rs} \ell^r\wedge \hat\ell^s~,~~~\omega_2=\ell^5\wedge \ell^6~.
\eea
 Observe that $d\omega_1=d\omega_2=0$.  On the other hand $d\ell^7=-(4k)^{-1} \omega_2+ (8l)^{-1} \omega_1$.  So $H^2(M^7, \bR)$ has one generator as expected.
The invariant 3- and  4-forms are $\ell^7\wedge \omega_1$, $\ell^7\wedge \omega_2$,  and $\omega_1\wedge \omega_1$,  $\omega_1\wedge \omega_2$, respectively.  Both 4-forms are exact as they are the exterior derivatives of  invariant 3-forms.  As a
result $H^4(M^7, \bR)=0$ as expected.  Note though that $H^4(M^7, \bZ)\not=0$.

\end{appendices}

\newpage

\bibliographystyle{utphys}
\bibliography{references}

\end{document}